\documentclass[a4paper,11pt]{article}

\RequirePackage[colorlinks=true
,urlcolor=blue
,anchorcolor=blue
,citecolor=blue
,filecolor=blue
,linkcolor=blue
,menucolor=blue
,pagecolor=blue
,linktocpage=true
,pdfproducer=medialab
,pdfa=true
]{hyperref}

\usepackage{jinstpub} % for details on the use of the package, please
\allowdisplaybreaks

\usepackage{amsmath,amsthm,amsfonts}
\usepackage{graphicx}
\usepackage{subcaption}
\usepackage{dcolumn}	% Align table columns on decimal point
\usepackage{hyperref}      	% hypertext links %%ARXIV
\usepackage{bm}		% bold math
\usepackage{epsfig}
\usepackage{epstopdf}
\usepackage{setspace}
\usepackage[usenames, dvipsnames]{color}
\usepackage{slashed}
\usepackage{comment}
\usepackage{enumitem}
\usepackage{siunitx}
\usepackage{multirow}
\usepackage{datetime}
\usepackage{lineno}
\usepackage{enumitem}
\usepackage{fancyhdr}
\setlist[description]{leftmargin=0.4cm}
\usepackage{tikz}
\usetikzlibrary{positioning,arrows.meta}

\newcommand{\PRE}[1]{{#1}} % Use if preprint style

\newcommand{\be}{\begin{equation}\begin{aligned}}
\newcommand{\ee}{\end{aligned}\end{equation}}
\newcommand{\beq}{\begin{equation}}
\newcommand{\eeq}{\end{equation}}
\newcommand{\beqa}{\begin{eqnarray}}
\newcommand{\eeqa}{\end{eqnarray}}

\newcommand{\ifb}{\text{fb}^{-1}}

\newcommand{\mev}{\text{MeV}}
\newcommand{\gev}{\text{GeV}}

\newcommand{\mm}{\text{mm}}

\newcommand{\ns}{\text{ns}}

\renewcommand{\eqref}[1]{Eq.~(\ref{eq:#1})}

\RequirePackage[normalem]{ulem}

\RequirePackage[normalem]{ulem} 

\begin{document}

\title{
{\Large The FASER Detector
\PRE{\vspace*{0.5in} \\
FASER Collaboration}
}}

%\begin{figure*}[h]
%\centering
%\vspace*{.4in}
%\includegraphics[width=0.6\textwidth]{figs/FaserLogo.pdf}
%\vspace*{.2in}
%\end{figure*}

\author[1]{Henso Abreu}

\author[2]{, Elham Amin Mansour} % exceptoinal author for TDAQ work

\author[2]{, Claire Antel}

\author[3,4]{, Akitaka Ariga}

\author[5]{, Tomoko Ariga}

\author[6]{, Florian Bernlochner}

\author[6]{, Tobias Boeckh}

\author[7]{, Jamie Boyd}

\author[8]{, Lydia Brenner}

\author[2]{, Franck Cadoux}

\author[9]{, David~W.~Casper}

\author[10]{, Charlotte Cavanagh}

\author[11]{, Xin Chen}

\author[12]{, Andrea Coccaro}

\author[7]{, Olivier Crespo-Lopez}% exceptoinal author for cooling unit work

\author[2]{, Stéphane Débieux} % exceptoinal author for TDAQ work

%\author[11]{, Sergey Dmitrievsky}

\author[10]{, Monica D’Onofrio}

\author[7]{, Liam Dougherty}% exceptoinal author for installation work

\author[13]{, Candan Dozen}

\author[14]{, Abdallah Ezzat}% exceptoinal author for tracker work

\author[2]{, Yannick Favre}

\author[15]{, Deion Fellers}

\author[9]{, Jonathan~L.~Feng}

\author[2]{, Didier Ferrere}

\author[2]{, Edward Karl Galantay} % exceptoinal author for TDAQ work

\author[16]{, Jonathan Gall} % exceptoinal author for trench work

%\author[6]{, Francisco Sanchez Galan} % exceptoinal author for TI12 preparation work 

\author[7]{, Enrico Gamberini} % exceptoinal author for TDAQ work

\author[17]{, Stephen Gibson}

\author[2]{, Sergio Gonzalez-Sevilla}

%\author[11]{, Yuri Gornushkin}

\author[10]{, Carl Gwilliam}

\author[4]{, Daiki Hayakawa}

\author[18]{, Shih-Chieh Hsu}

\author[11]{, Zhen Hu}

\author[2]{, Giuseppe Iacobucci}

\author[11]{, Tomohiro Inada}

\author[7]{, Sune Jakobsen}

\author[2]{, Eliott Johnson} % exceptoinal author for TDAQ work

\author[1]{, Enrique Kajomovitz}

\author[5]{, Hiroaki Kawahara }

\author[19]{, Felix Kling}

\author[7]{, Umut Kose}

\author[7]{, Rafaella Kotitsa}

\author[20,21]{, Jesse Krusse} % exceptoinal author for calo/scint work

\author[7]{, Susanne Kuehn}

\author[17]{, Helena Lefebvre}

\author[22]{, Lorne Levinson}

\author[18]{, Ke Li}

\author[11]{, Jinfeng Liu}

\author[2]{, Chiara Magliocca}

\author[2]{, Fulvio Martinelli}

\author[23]{, Josh McFayden}

\author[7,24]{, Sam Meehan}

\author[2]{, Matteo Milanesio}

\author[4]{, Manato Miura}% exceptoinal author for tungsten measurements for FASERnu

\author[7]{, Dimitar Mladenov}

\author[2]{, Théo Moretti}

\author[2]{, Magdalena Munker}

\author[25]{, Mitsuhiro Nakamura}

\author[25]{, Toshiyuki Nakano}

\author[7,2]{, Marzio Nessi}

\author[26]{, Friedemann Neuhaus}

\author[17]{, Laurie Nevay}

\author[7]{, John Osborne} % exceptoinal author for trench work

\author[5]{, Hidetoshi Otono}

\author[2]{, Carlo Pandini}

\author[11]{, Hao Pang}

\author[7,2]{, Lorenzo Paolozzi}

\author[7]{, Brian Petersen}

\author[7]{, Francesco Pietropaolo}

\author[6]{, Markus Prim}

\author[27]{, Michaela Queitsch-Maitland}

\author[7]{, Filippo Resnati}

\author[2]{, Chiara Rizzi}

\author[25]{, Hiroki Rokujo}

\author[26]{, Elisa Ruiz-Choliz}

%\author[2]{, Jihad Saidi}

\author[7]{, Jakob Salfeld-Nebgen}

\author[7]{, Francisco Sanchez Galan}% exceptoinal author for FASER site preparation work

\author[25]{, Osamu Sato}

\author[3,28]{, Paola Scampoli}

\author[26]{, Kristof Schmieden}

\author[26]{, Matthias Schott}

\author[2]{, Anna Sfyrla}

\author[9]{, Savannah Shively}

\author[7]{, Roland Sipos} % exceptoinal author for TDAQ work

\author[18]{, John Spencer}

\author[29]{, Yosuke Takubo}

\author[2]{, Noshin Tarannum}

\author[2]{, Ondrej Theiner}

\author[7]{, Pierre Thonet} % exceptoinal author for magnet work

\author[15]{, Eric Torrence}

\author[7]{, Serhan Tufanli}

%\author[11]{, Svetlana Vasina}

\author[7]{, Camille Vendeuvre} % exceptoinal author for survey work

\author[7]{, Benedikt Vormwald}

\author[11]{, Di Wang}

\author[2]{, Stefano Zambito}

\author[30]{, Gang Zhang}

%%%%%%%%%%%%%%%%%%%%%%%%
\affiliation[1]{Department of Physics and Astronomy, Technion---Israel Institute of Technology, Haifa 32000, Israel}

\affiliation[2]{D\'epartement de Physique Nucl\'eaire et Corpusculaire, 
University of Geneva, CH-1211 Geneva 4, Switzerland}

\affiliation[3]{Albert Einstein Center for Fundamental Physics, Laboratory for High Energy Physics, University of Bern, Sidlerstrasse 5, CH-3012 Bern, Switzerland}

\affiliation[4]{Department of Physics, Chiba University, 1-33 Yayoi-cho Inage-ku, Chiba, 263-8522, Japan}

\affiliation[5]{Kyushu University, Nishi-ku, 819-0395 Fukuoka, Japan}

\affiliation[6]{Universit\"at Bonn, Regina-Pacis-Weg 3, D-53113 Bonn, Germany}

\affiliation[7]{CERN, CH-1211 Geneva 23, Switzerland}

\affiliation[8]{Nikhef National institute for subatomic physics, Science Park 105, 1098 XG Amsterdam, Netherlands}

\affiliation[9]{Department of Physics and Astronomy, 
University of California, Irvine, CA 92697-4575, United States of America}

\affiliation[10]{University of Liverpool, Liverpool L69 3BX, United Kingdom}

\affiliation[11]{Department of Physics, Tsinghua University, Beijing, China}
 
\affiliation[12]{INFN Sezione di Genova, Via Dodecaneso, 33--16146, Genova, Italy}

%\affiliation[11]{Joint Institute for Nuclear Research, Dubna, Russia}

\affiliation[13]{Institut de Physique des 2 Infinis de Lyon (IP2I), 4 Rue Enrico Fermi, 69100 Villeurbanne, France}

\affiliation[14]{Facult\'e de Physique et Ing\'enierie, University of Strasbourg, 3-5 rue de l’Universit\'e, Strasbourg, France, 67000}

\affiliation[15]{University of Oregon, Eugene, OR 97403, United States of America}

\affiliation[16]{Turner \& Townsend, Calverley Ln, Low Hall Rd, Horsforth, Leeds LS18 4GH, United Kingdom}

\affiliation[17]{Royal Holloway, University of London, Egham, TW20 0EX, United Kingdom}

\affiliation[18]{Department of Physics, University of Washington, PO Box 351560, Seattle, WA 98195-1460, United States of America}

\affiliation[19]{Deutsches Elektronen-Synchrotron DESY, Notkestr. 85, 22607 Hamburg, Germany}

\affiliation[20]{National Institute of Standards and Technology, Boulder, Colorado 80305, United States of America}

\affiliation[21]{University of Colorado, Boulder, Colorado 80309-0440, United States of America}

\affiliation[22]{Department of Particle Physics and Astrophysics, Weizmann Institute of Science, Rehovot 76100, Israel}

\affiliation[23]{Department of Physics \& Astronomy, University of Sussex, Sussex House, Falmer, Brighton, BN1 9RH, United Kingdom}

\affiliation[24]{2021-2022 AAAS Science \& Technology Policy Fellow}
% should this say: American Association for the Advancement of Science, 1200 New York Ave, NWWashington,DC20005 202-326-6400

\affiliation[25]{Nagoya University, Furo-cho, Chikusa-ku, Nagoya 464-8602, Japan}

\affiliation[26]{Institut f\"ur Physik, Universität Mainz, Mainz, Germany}

\affiliation[27]{University of Manchester, School of Physics and Astronomy, Schuster Building, Oxford Rd, Manchester M13 9PL, United Kingdom}

\affiliation[28]{Dipartimento di Fisica ``Ettore Pancini'', Universit\`a di Napoli Federico II, Complesso Universitario di Monte S. Angelo, I-80126 Napoli, Italy}

\affiliation[29]{Institute of Particle and Nuclear Studies, KEK, Oho 1-1, Tsukuba, Ibaraki 305-0801, Japan}

\affiliation[30]{College of Electronic Engineering, National University of Defense Technology, Hefei, 230037, China}

\abstract{
FASER, the ForwArd Search ExpeRiment, is an experiment dedicated to searching for light, extremely weakly-interacting particles at CERN's Large Hadron Collider (LHC). Such particles may be produced in the very forward direction of the LHC's high-energy collisions and then decay to visible particles inside the FASER detector, which is placed 480 m downstream of the ATLAS interaction point, aligned with the beam collisions axis. FASER also includes a sub-detector, FASER$\nu$, designed to detect neutrinos produced in the LHC collisions and to study their properties. In this paper, each component of the FASER detector is described in detail, as well as the installation of the experiment system and its commissioning using cosmic-rays collected in September 2021 and during the LHC pilot beam test carried out in October 2021. FASER will start taking LHC collision data in 2022, and will run throughout LHC Run 3.
}

%\pacs{}

\maketitle

%\tableofcontents

\clearpage

%%%%%%%%%%%%%%%%%%%%%%%%%%%%%%%%%%%
\section{Experimental Overview}
\label{sec:introduction}

The ForwArd Search ExpeRiment (FASER~\cite{Feng:2017uoz,Ariga:2018zuc,FASER:2018bac}) is a new experiment at the CERN Large Hadron Collider (LHC) designed to search for light and very weakly-interacting new particles produced in the LHC collisions. With the addition of the FASER$\nu$ sub-detector~\cite{FASER:2019dxq,Abreu:2020ddv}, FASER can also detect and study neutrinos of all flavours produced at the LHC. 

\begin{figure}[hbt!]
    \centering
    \includegraphics[width=0.7\textwidth]{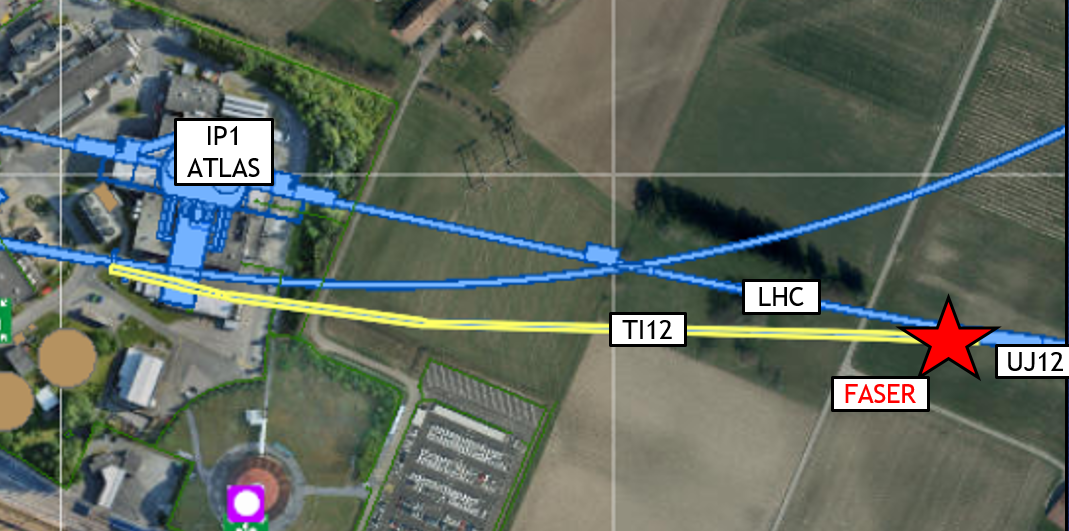}
    \caption{The FASER location: TI12 tunnel, 480~m downstream of the ATLAS interaction point. The detector is located along the beam collision axis line-of-sight.}
    \label{fig:TI12}
\end{figure}

The detector is positioned on the beam collision axis line-of-sight (LOS) 480~m from the ATLAS collision point (interaction point 1, IP1) in an unused service tunnel, TI12. Figure~\ref{fig:TI12} shows the location of the detector in TI12. To allow the detector to be placed on the LOS, a small trench was excavated in the TI12 tunnel. 
The size of this trench sets the overall detector dimensions, including the 10~cm radius transverse-size of the active part of the detector and the total length of about 7~m  (FASER$\nu$ has a transverse size of $25 \times 30$~cm$^2$).
The angular acceptance of FASER
\footnote{The FASER detector uses a cartesian coordinate system with the $z$-axis pointing along the LOS away from IP1, the $y$-axis pointing vertically upwards, and the $x$-axis pointing horizontally towards the LHC machine. The origin of the coordinate system is aligned with the centre of the magnets in the transverse, $x - y$, plane and conventionally at the front surface of the second tracker station in the $z$ coordinate.}
covers the $|\theta| < 0.21$~mrad region (pseudorapidity $\eta>9.2$) around the LOS, with respect to IP1, whereas  FASER$\nu$ extends to angles as large as $|\theta| \simeq 0.41$~mrad ($\eta \simeq 8.5$).  

\begin{figure}[hbt!]
    \centering
        \includegraphics[trim=0cm 0.8cm 0.5cm 1.5cm, clip=true, width=\textwidth]{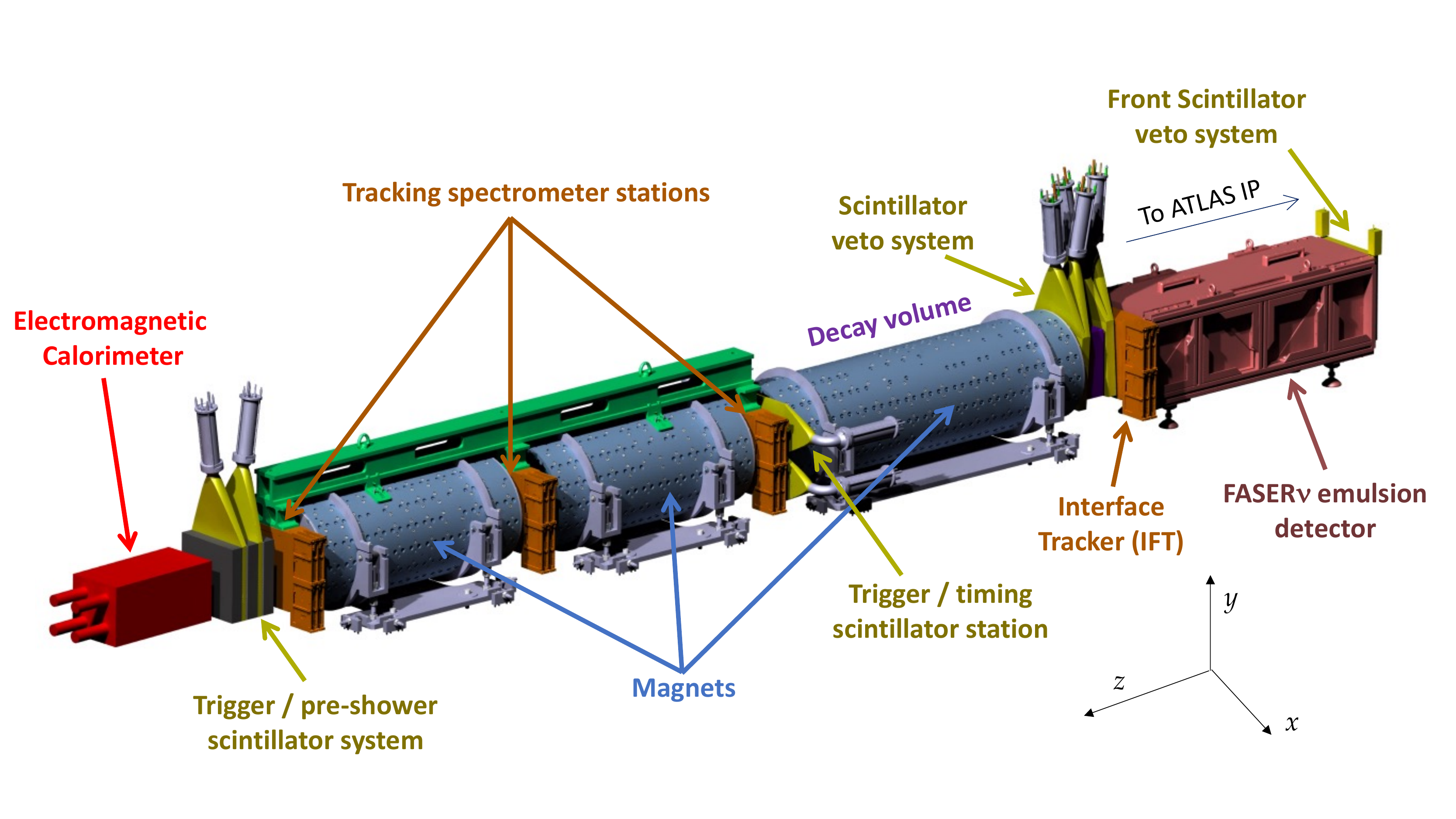}
        \caption{A sketch of the FASER  detector, showing the different sub-detector systems. The FASER coordinate system is also shown. }
    \label{fig:FASER_labels}
\end{figure}

Figure~\ref{fig:FASER_labels} shows a sketch of the FASER detector, with particles originating from collisions in IP1 entering from the right of the picture. From right to left, the detector consists of (i) a front scintillator veto system, (ii) the FASER$\nu$ emulsion detector, (iii) the interface tracker (IFT), (iv) the FASER scintillator veto station, (v) the decay volume, (vi) the timing scintillator station, (vii) the FASER tracking spectrometer, (viii) the pre-shower scintillator system, and (ix) the electromagnetic (EM) calorimeter system. The detector includes three 0.57~T dipole magnets, one surrounding the decay volume and the other two embedded in the tracking spectrometer. 

This paper describes the FASER detector as it was installed into the LHC complex in 2021.  In the remainder of this Section, an executive summary of the various detector components is reported, following a brief description of the expected Run 3 LHC beam conditions and dataset and the FASER physics programme. In the following Sections, details of the various sub-detectors are provided, together with an overview of the installation procedure and the commissioning of the fully integrated detector. 

\subsection{LHC Run 3 beam conditions and data}
\label{sec:run3}
LHC Run 3 follows the first two LHC running periods, and Long-Shutdown 2 (LS2). At the time of the FASER approval, it was scheduled to run from 2021 to 2023, and was expected to deliver 150~fb$^{-1}$ of 14~TeV proton-proton collision data. Following delays accrued due to the COVID pandemic, the Run~3 schedule was updated, to start in 2022 and to run for four years, with an expected total luminosity of more than 150~fb$^{-1}$. Problems observed during the training of the LHC dipole magnets to high energy led to the decision to run the LHC at a centre-of-mass energy $\sqrt{s}=13.6$~TeV during Run~3.~\footnote{Most studies presented in this paper assume $\sqrt{s}=14$~TeV. However, the  difference between 13.6~TeV and 14~TeV is expected to be negligible for FASER physics.}

During LS2 the LHC injector complex was significantly upgraded allowing it to deliver much brighter beams to the LHC, in anticipation of the high-luminosity (HL) LHC phase of operation (starting in Run~4). Since the LHC components and the experiments have not yet been upgraded to be able to run at luminosities higher than $2 \times 10^{34}$~cm$^{-2}$~s$^{-1}$, the luminosity will be levelled at this value. This luminosity corresponds to about 55 interactions per bunch crossing (pileup), however given the large amount of shielding in front of the detector, FASER is not expected to observe signals from multiple  simultaneous interactions.

The main LHC configuration parameters which can effect physics at FASER are the direction and magnitude of the beam crossing angle at IP1, since this moves the LOS compared to its nominal position assuming no-crossing angle ($\theta_\mathrm{cross}=0$) at IP1. In Run 3 the crossing plane in IP1 will be vertical, but the direction (if the beams will be pointing up or down) will be changed during the run in order to distribute the radiation more evenly over the LHC magnets. 
The LOS at FASER moves by 480~m~$\times \sin (\theta_\mathrm{cross}/2)$. In Run~3 the half crossing angle during the physics fills will be in the range 160 - 135 $\mu$rad, and therefore this will move the position of the LOS at FASER by 7.7 - 6.5~cm. 
As discussed in Section~\ref{sec:integration}, the mechanics of FASER are designed to be able to move the detector to be closer to the LOS. Detector movements are planned to be carried out during year end technical stops, when the crossing angle direction is changed.

The expected particle flux at FASER has been calculated assuming a luminosity of $2 \times 10^{34}$~cm$^{-2}$~s$^{-1}$.
Results from FLUKA~\cite{BATTISTONI201510} simulations have been compared to in situ measurements made in 2018 LHC running~\cite{FASER:2018bac}. The FLUKA studies suggest that the only high-energy particles expected to traverse the FASER location are muons and neutrinos, originating from the interactions in IP1. The muons that reach FASER primarily arise from pion decays, where the pions originate from hadronic showers from collision debris interacting with the LHC infrastructure (beam pipe, absorbers, collimators) between the IP and FASER. 
The muon flux on the LOS as estimated by FLUKA is 0.4~Hz cm$^{-2}$. This has been confirmed using a small emulsion detector installed in the FASER location. Measurements of the correlation between the particle flux in this location and the instantaneous luminosity in IP1 showed that these particles are originating from IP1 collisions.

\subsection{Physics Reach for Long-Lived Particles}
\label{sec:darkPhotonPhys}

FASER will search for light, very weakly-interacting particles which may be produced in the collisions at IP1, travel long distances through concrete and rock without interacting and then decay to visible particles in the detector decay volume. The sensitivity reach for FASER for a large number of beyond the Standard Model (BSM) theories characterised by the presence of long-lived particles has been studied in detail and reported in Ref.~\cite{FASER:2018eoc}, with particular emphasis on renormalizable portal
interaction models leading to dark photons. Other scenarios considered include models with dark Higgs bosons, heavy neutral leptons,  light gauge bosons, axion-like particles, and pseudoscalars with Yukawa-like couplings. 

\begin{figure}[hbt!]
    \centering
    \includegraphics[trim=0.cm 0.cm 0.cm 0.cm, clip=true,angle=0,width=0.9\textwidth]{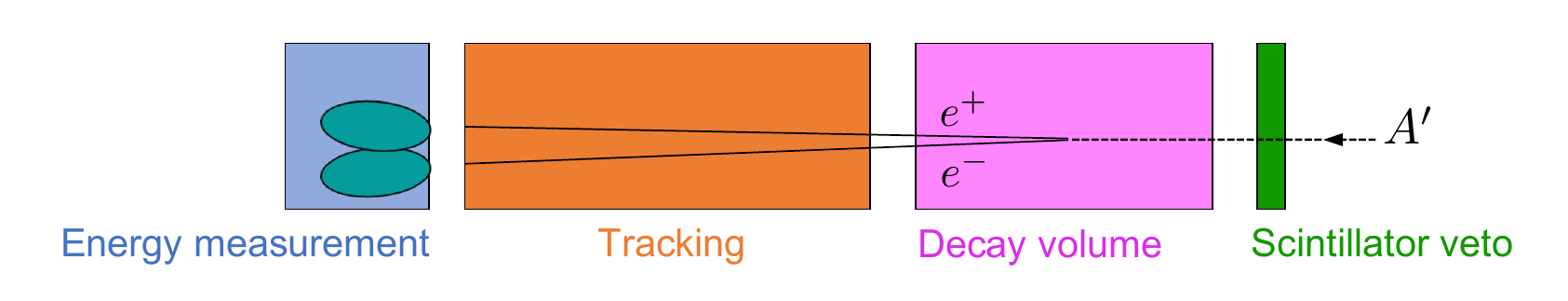}
    \caption{Sketch showing the detector signature of a dark photon ($A'$) decaying to an electron-positron pair inside the decay volume of the FASER experiment. The $A'$ enters the detector from the right.}
    \label{fig:DarkPhotonSketch}
\end{figure}

The dark photon ($A'$) appears in dark (hidden) sector models in which the dark sector contains a new spin-1 vector boson with suppressed couplings (referred to as $\epsilon$) to SM particles via its kinetic mixing with the SM photon. In such models the $A'$ acts as a mediator between the SM and a hypothetical dark matter particle ($\chi$). The $A'$ phenomenology is defined by $\epsilon$ and the $A'$ mass ($m_{A'}$), and there is a large region of viable parameter space yet to be constrained by experiments. If $m_{A'} < 2m_\chi$, the $A'$ decays to SM particles, and FASER is mostly sensitive to the mass range $2m_\mu > m_{A'} > 2m_e$ such that the $A'$ will decay to an electron and positron pair. For sufficiently weak couplings, the $A'$ is long-lived and can travel significant distances before decaying. The relevant parameter space for which FASER has sensitivity is $\mathcal{O}(10^{-6}) < \epsilon <\mathcal{O}(10^{-4})$ with $m_{A'} \lesssim 100~\mev$, such that the $A'$ may travel from IP1 to FASER without interacting and decay in FASER to an electron-positron pair. A sketch of such a detector signature is shown in Figure~\ref{fig:DarkPhotonSketch}. This signature is very clean with no expected non-BSM process able to mimic it. 

The main production process for dark photons relevant for FASER is light meson decays, in particular the decay of $\pi^0$ mesons. Neutral pions may decay to a dark photon and a SM photon, with a branching fraction proportional to $\epsilon^2$. Pion production at the LHC is strongly peaked in the very forward direction, such that $\mathcal{O}{(1\%)}$ of the pions produced with energy $E_{\pi^0}> 10~\gev$ are within the FASER angular acceptance $|\theta| < 0.21$~mrad, despite the fact that this covers only $\mathcal{O}{(10^{-8})}$ of the total solid angle.  
Such pions have a large boost, as much as $\mathcal{O}(1~\mathrm{TeV})$, along the beam direction in the lab frame, allowing dark photons produced in their decay to reach FASER even with relatively short lifetimes. 

\begin{figure}[hbt!]
    \centering
    \includegraphics[trim=0cm 0.cm 0.cm 0.cm, clip=true, width=0.6\textwidth]{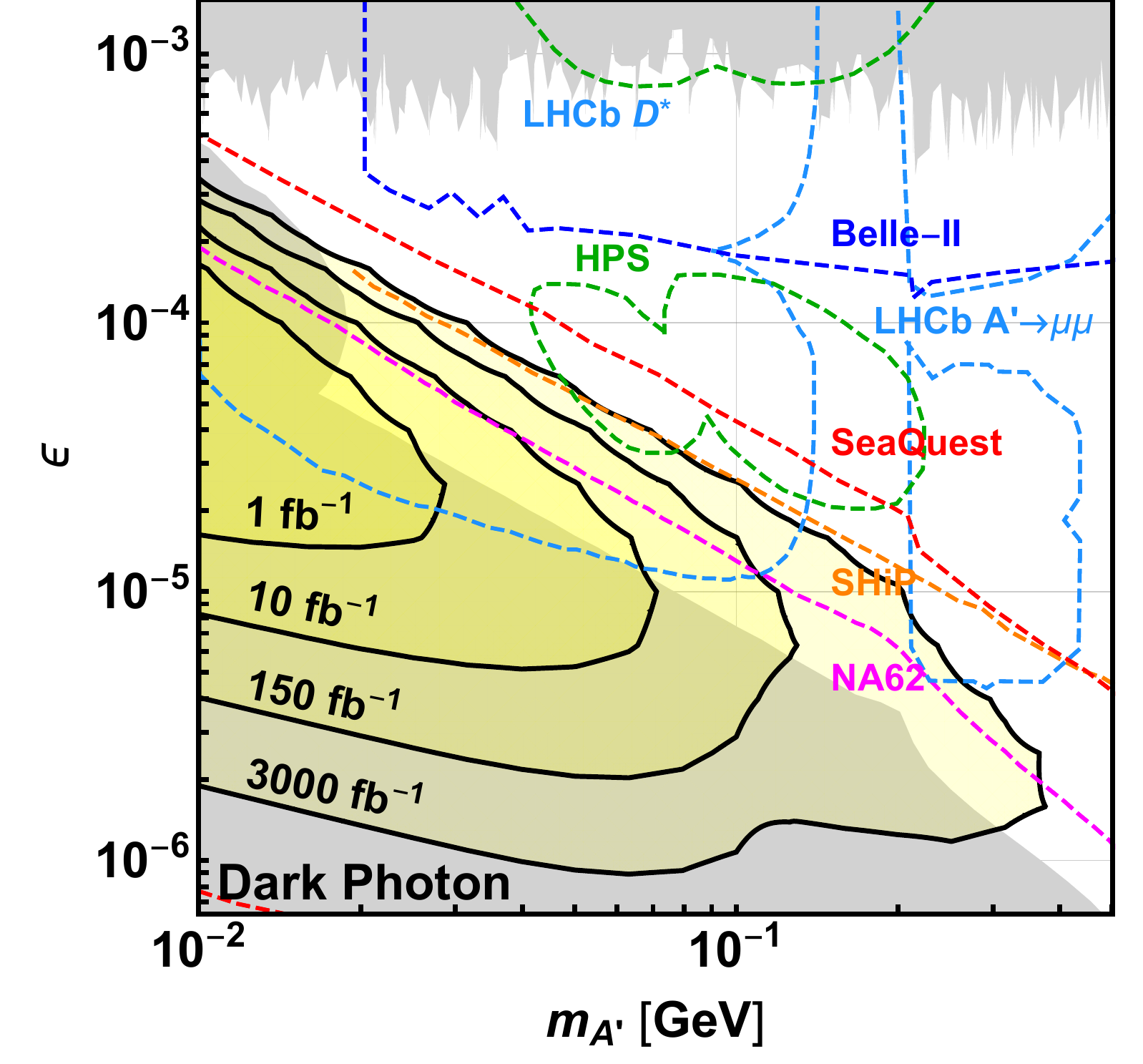}
    \caption{The expected sensitivity of FASER for dark photons as a function of the mass ($m_{A'}$) and coupling ($\epsilon$), for different values of integrated luminosity. The regions of parameter space already excluded by experiment are shown in grey, and the projected reaches of some future experiments are shown in coloured lines. As discussed in the text, 
    the FASER reach assumes full efficiency for selecting dark photons decaying inside the FASER decay volume, and no background. Details of the sensitivity projections shown for other experiments are given in Ref.~\cite{FASER:2018eoc}.
    }
    \label{fig:darkphotonSensitiviy}
\end{figure}

During the Run~3 of the LHC around $\mathcal{O}(10^{14})$  $\pi^0$s are expected to be produced within the FASER angular acceptance. Hence a significant number of signal events can be detected in FASER, even taking into account the large suppression due to the $\pi^0 \to A'$ branching fraction ($\mathcal{O}(10^{-12})$ to $\mathcal{O}(10^{-8})$) and the requirement that the $A'$ decays inside the FASER detector volume. Figure~\ref{fig:darkphotonSensitiviy} shows the expected FASER sensitivity to dark photons for different integrated luminosity scenarios. 
The contours are defined such that at least three signal events pass the kinematic and geometrical requirements and the dark photon decays inside the FASER decay volume. 
The contours assume 100\% efficiency and zero background. It has been shown~\cite{Ariga:2018zuc}
 that a reduction of the assumed efficiency does not affect the sensitivity curves substantially, due to the fact that the number of signal events falls off very rapidly at the edge of the sensitivity boundaries. 
The boundaries of the signal sensitivity contours are set primarily by the production rate falling off at too-small couplings and the $A'$s being too short-lived to decay in FASER at too-large couplings. 
More details on the theoretical aspects of the dark photon model, as well as the FASER sensitivity in other dark-sector models are discussed in Ref.~\cite{FASER:2018eoc}.

\subsection{Neutrino physics programme}
\label{sec:nuetrinoPhys}

A huge number of neutrinos are produced in LHC collisions via hadron decays, and their flux is collimated along the beam collision axis. Table~\ref{tab:neutrino-numbers} summarizes the number of neutrinos expected to traverse, and under-go charged-current (CC) interactions in, the FASER$\nu$ emulsion detector (with a target mass of 1.1~tonnes), assuming a 150~fb$^{-1}$ dataset for the LHC Run~3, as well as the expected average neutrino energy. The table includes all three neutrino flavours (summing $\nu$ and $\overline{\nu}$) and also shows the dominant production process. The numbers reported in the table are obtained using the SIBYLL~2.3d~\cite{Riehn:2019jet} generator to simulate hadron production and the fast neutrino flux simulation introduced in Ref.~\cite{Kling:2021gos} to propagate the SM hadrons through the LHC beam pipe and magnets and to simulate their decays into neutrinos. There are currently large theoretical uncertainties related to very forward hadron production, which translate into large uncertainties on the neutrino flux. As discussed in Refs.~\cite{Kling:2021gos, Bai:2020ukz, Bai:2021ira}, varying the generator or theoretical modelling can lead to changes of the order of $100\%$ ($\nu_e$), 30$\%$ ($\nu_\mu$) and 100$\%$ ($\nu_\tau$). FASER neutrino measurements as a function of energy and rapidity can therefore be used to constrain forward hadron production models. 

\begin{table}[thb]
  \centering
  \begin{tabular}{|l|c|c|c|}
  \hline
    - & \bf{$\nu_e$} & \bf{$\nu_\mu$} & \bf{$\nu_\tau$} \\
  \hline
     Dominant production process & 
     $K \to \nu_e e X$  & $\pi \to \nu_\mu \mu$ & $D_s \to \nu_\tau \tau $ \\
        Number of $\nu$ traversing FASER$\nu$ &
      $3 \times 10^{11}$ & $2 \times 10^{12}$ & $8 \times 10^{9}$ \\
      Number of $\nu$ interacting in FASER$\nu$ (1.1 tonnes) & 
      830 & 4400 & 14 \\
      Average energy of interacting neutrinos (GeV) & 
      820 & 820 & 810 \\
\hline
    \end{tabular}
    \caption{Summary of the number of neutrinos traversing and interacting in FASER$\nu$ assuming a 150~fb$^{-1}$ dataset for the LHC Run~3. The table also shows the dominant production process and the average energy of the interacting neutrinos. Only CC interactions are considered. 
    }
    \label{tab:neutrino-numbers}
\end{table}

\begin{figure}[t]
\centering
\includegraphics[width=0.99\textwidth]{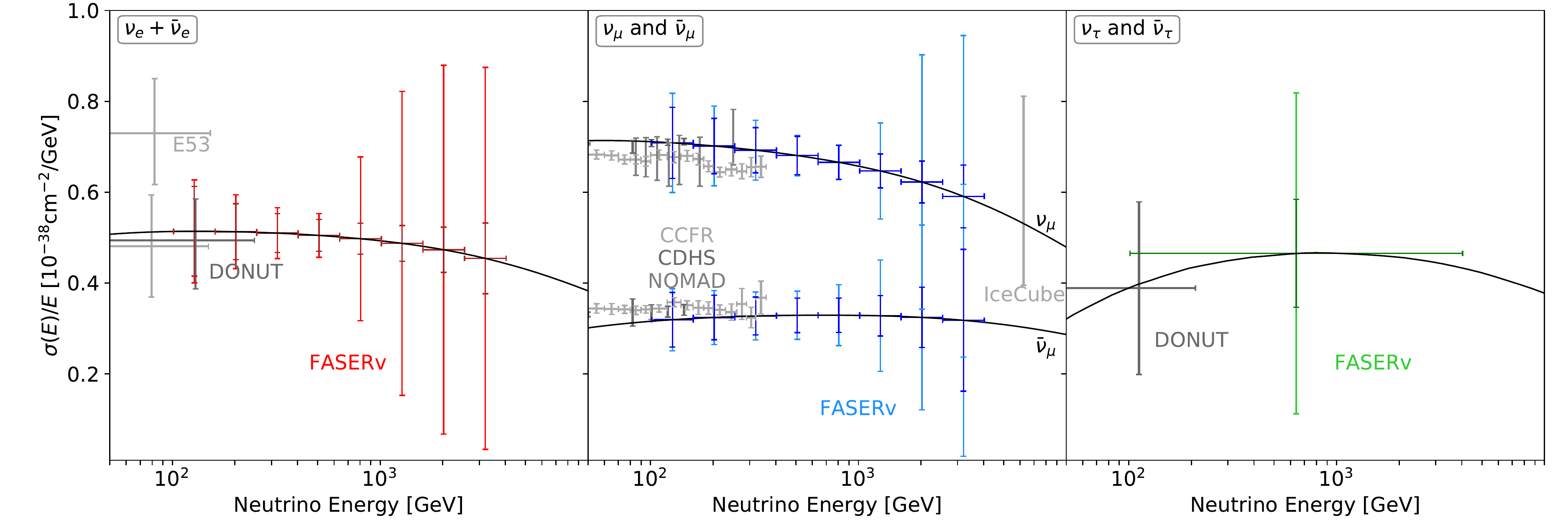}
\caption{FASER$\nu$'s estimated $\nu$-nucleon CC cross section sensitivity for $\nu_e$ (left), $\nu_{\mu}$ (centre), and $\nu_\tau$ (right) for LHC Run 3. Existing constraints are shown in gray~\cite{ParticleDataGroup:2018ovx}. The black curve is the theoretical prediction for the DIS cross section per tungsten-weighted nucleon. The coloured error bars show FASER$\nu$'s cross section sensitivity, where the inner error bars correspond to statistical uncertainties, while the outer error bars show the combination of statistical and flux uncertainties. 
}
\label{fig:xs_measurement}
\end{figure}

Taking advantage of the huge neutrino flux, FASER$\nu$ will measure the neutrino CC interaction cross sections for all three neutrino flavours in an uncovered energy regime. Projections of the expected measurement precision are shown in Figure~\ref{fig:xs_measurement}. Here, the expected statistical uncertainty and the uncertainty related to the neutrino flux, estimated by comparing the SIBYLL~2.3d and DPMJET~3.2017~\cite{Roesler:2000he, Fedynitch:2015kcn} generators,  are shown separately. Note, a number of different generators have also been studied including: \texttt{EPOSLHC}~\cite{Pierog:2013ria}, \texttt{QGSJET~II-04}~\cite{Ostapchenko:2010vb}, and \texttt{Pythia~8}~\cite{Sjostrand:2014zea}, but the envelope of the range of predictions is covered by the difference between SIBYLL and DPMJET predictions. 
Experimental systematic uncertainties are not included, but are expected to be sub-dominant. For $\nu_\mu$, the charge of the muon can be reconstructed in the FASER tracking spectrometer, making it possible to measure both the neutrino and anti-neutrino cross sections separately,~\footnote{Note the FASER spectrometer overlaps with only 42$\%$ of the FASER$\nu$ emulsion detector in the transverse plane, and therefore the number of muon neutrino interactions that can be used to measure the neutrino/anti-neutrino cross sections only corresponds to a target mass of 460~kg.} as shown in the figure. The measured cross sections are interesting in their own right, but can also be used to constrain proton and nuclear parton distribution functions, formation times, colour transparency effects, non-standard neutrino interactions and neutrino oscillations in models with additional sterile neutrinos, as discussed in more detail in Refs.~\cite{FASER:2019dxq, Mosel:2022tqc,Feng:2022inv}. 
In addition, the measurement of the neutrino flux can be used as a probe of forward particle production, which will help validate and improve the underlying hadronic interaction models and provides valuable input for astro-particle physics measurements.
 
 As a proof of principle, the FASER Collaboration placed a small emulsion detector along the LOS for 4 weeks during the Run~2 of the LHC in 2018.  With a fiducial mass of just 11 kg, this pilot detector was able to record the first neutrino interaction candidates at a particle collider~\cite{FASER:2021mtu}.

\subsection{Detector requirements}
The most relevant detector requirements needed to fulfill the FASER physics programme on dark sector searches and neutrinos cross section measurements are listed below. 
\begin{itemize}
    \item Given its transverse size, fixed by the tunnel and trench constraints, the detector must be centred on the LOS to within about 10 cm to maximise the number of signal events, for both light BSM searches and neutrino analyses. This must be done for all possible beam crossing angles that will be used during data taking in IP1.
    Situating the detector on the LOS also maximises the neutrino energy.
    \item In order to be able to efficiently reject physics background events initiated by high-energy muons, the detector must include a system that vetoes charged particles entering the detector. 
    The veto system should have an inefficiency of smaller than $10^{-9}$, since the expected number of muons that will enter FASER during LHC Run~3 is $\mathcal{O}$($10^9$).

    \item The detector must be able to track high energy charged-particles with good precision:
    \begin{itemize} 
        \item For light BSM searches, it must be able to separate very closely spaced charged particles. For a dark photon of mass 100~MeV and 2~TeV momentum, the opening angle of the decay products is only 50~$\mu$rad, hence the magnetic field in the decay-volume and spectrometer must separate the decay products to measurable distances within the detector;
        \item To be able to measure $\nu_\mu$ and $\overline{\nu}_\mu$, the charge of the produced muon must  be measured up to several hundred GeV.
        \end{itemize}
    \item To measure the energy of electron-positron pairs produced in dark-photon decays, or pairs of photons produced in Axion-like-particle decays, the electromagnetic calorimeter must be capable of measuring $\mathcal{O}$(TeV) EM deposits with a few percent resolution.
    \item In order to be able to maximise the number of recorded neutrino interactions taking into account the space limitations, the neutrino detector must have a high-density and large-mass target.
    \item The neutrino detector should have the ability to identify different lepton flavours from neutrino CC interactions. This leads to the following requirements:
    \begin{itemize} 
        \item The detector must have sufficient target material to identify muons; 
        \item The detector needs to  have finely-sampled detection layers to identify electrons and distinguish them from gamma rays; 
        \item The position and angular resolutions must be sufficiently good to be able to detect tau and charm decays. 
 \end{itemize}
    \item The neutrino detector should be able to measure muon and hadron momenta, the energy of electromagnetic showers, and estimate the neutrino energy. 
    \item To enable the combination of information from the passive emulsion-based neutrino detector, and the rest of the FASER (active) detectors, a tracking detector placed immediately after the neutrino detector with sufficient precision is needed to allow matching of charged-particle tracks between the two systems. In the matching, all events triggered in the active detector during the time the emulsion detector was in place need to be considered.  
    Therefore, in order to reduce the matching combinatorics, only triggered events in which the scintillator counters in front of the emulsion detector do not fire are taken into account. This reduces the number of triggered events considered in the matching process by $\mathcal{O}(10^6)$. 
    \item To ensure a good efficiency for collecting rare signal events, the trigger must be highly efficient and the data acquisition (DAQ) system needs to operate robustly with little dead-time. The expected trigger rate from muons produced in the IP1 collisions is $\mathcal{O}$(650~Hz) for a luminosity of $2 \times 10^{34}$~cm$^{-2}$~s$^{-1}$. 
\end{itemize}

As a reminder, the radiation levels at the FASER location are expected to be substantially lower than around IP1.
They have been estimated using FLUKA simulations, and validated using measurements taken during 2018 LHC running~\cite{FASER:2018bac}. The FLUKA estimated dose is less than $5 \times 10^{-3}$ Gy per year and a 1 MeV neutron equivalent fluence of less than $5 \times 10^7$ per year. The radiation level was  measured using a BatMon radiation monitor~\cite{Spiezia:2011jp}, giving a measured high-energy hadron fluence below the device sensitivity (corresponding to $10^6$ /cm$^{2}$), consistent with the expectation from the FLUKA simulation studies. For thermal neutrons the measured flux is $4 \times 10^6$ /cm$^{2}$, to be compared with the simulation estimate of $3 \times 10^6$ /cm$^{2}$. These radiation levels are low for the LHC complex, and enable non radiation-hard electronics to be used in the detector. 

The detector was designed to fulfill the above requirements, but additional constraints have been also taken into account. In particular, the short time available\footnote{FASER was formally approved by CERN in March 2019, with a schedule to install the detector before the end of LHC Long Shutdown 2 (LS2), at that time scheduled for the end of 2020. Since then, LS2 has been extended by 1 year due to the COVID-19 pandemic.} for the design, construction, commissioning and installation implied the use of existing detector components. This allowed the cost of the detector, services and installation works to be minimized. Furthermore, due to the FASER location in the LHC tunnel, the installation of additional services is difficult and it was therefore important to minimize as much as possible the needed services.  Since there is no direct access to the FASER location, and to get there one must walk about 500~m along the LHC tunnel, it was important to make the detector as robust and reliable as possible in order to minimize interventions for maintenance, especially during LHC operations.  

In the remaining part of this Section, a brief overview of each detector sub-system is given, while a specific description of each component is reported in the following dedicated Sections. 

\subsection{The magnet system}
The FASER detector is built around three dipole magnets with a 0.57~T field. Permanent magnets are used to minimise the required services. Each magnet has a 20~cm diameter aperture, and an outer diameter of 43~cm. The first dipole (1.5~m-long) surrounds the decay volume, and the other two (each 1~m-long) are part of the tracking spectrometer. The primary purpose of the magnets is to separate closely spaced charged particles produced in the decay of boosted, light BSM particle decays, and to measure the charge of muons arising from neutrino interactions.

\subsection{The tracking system}
The FASER detector tracking system is composed of two distinct parts, the tracking spectrometer and the IFT,  and a detailed description as well as the results of its commissioning can be found in Ref.~\cite{FASER:Tracker}.
The tracking spectrometer allows the trajectories of charged particles traversing the detector to be reconstructed, and  their position and momentum to be measured. The IFT is placed right after the FASER$\nu$ emulsion detector, and enables  tracks reconstructed in the emulsion to be matched with those in the active detectors. This allows for a time-stamp to be assigned to the reconstructed neutrino vertices, and hence enables the measurement of the charge of muons arising from the neutrino interaction. Once associated to a neutrino interaction vertex candidate, the information from the active detector can also help in background rejection and the neutrino energy reconstruction.

Both detectors in the tracking spectrometer and the IFT are made from the same hardware components. The tracking spectrometer consists of three tracking stations, and the IFT is an identical tracking station. There are therefore four tracking stations in the full FASER detector. Each tracking station consists of three double-layers of single-sided silicon microstrip detectors. A tracking layer is made up of eight silicon strip modules which are spares from the ATLAS experiment’s SCT barrel detector~\cite{Abdesselam:2006wt}, for a total of 96 modules in the full FASER detector. Each module is approximately $6 \times 12$~cm$^2$, and they are arranged in two columns of 4 modules to give a $24 \times 24$~cm$^2$ active detector area which covers the full aperture of the FASER magnets.  The  modules are operated with a bias voltage of 150~V, providing a hit efficiency well above 99\%. Despite using detector modules from the ATLAS experiment, the rest of the tracker system including the mechanics, readout system, cooling system and powering is newly designed and constructed for FASER. 

The SCT modules have a strip pitch of 80~$\mu$m, and a stereo angle between the two sides of 40~mrad, leading to a track resolution of order of 20~$\mu$m in the precision coordinate, and 800~$\mu$m in the other coordinate. In FASER, the modules are aligned such that the precision coordinate corresponds to the magnet bending plane ($y$). The detector is therefore able to resolve closely spaced charged particle tracks which are separated by 100-200~$\mu$m, and to measure the angle of charged particles needed to match tracks in the emulsion detector with the IFT. According to simulation studies, the track angular resolution for the IFT is 250~$\mu$rad and 13~mrad for the $y$ and $x$ coordinates, respectively. 

\subsection{The calorimeter and scintillator systems}
The FASER detector includes four scintillator stations, used for vetoing and trigger purposes, and an electromagnetic calorimeter,  designed to measure the energy of high-energy electrons and photons and provide redundant triggering for signals with large energy deposits. In addition, the calorimeter in conjunction with scintillator counters placed at the back of the tracking spectrometer can provide simple particle identification.

Each of the four scintillator stations is composed of more than one scintillator counter, read out by photomultiplier tubes (PMTs), as detailed below:
\begin{itemize}
\item The first two stations are needed to veto charged particles entering the detector. One is placed in front of the FASER$\nu$ emulsion detector and consists of two scintillator counters. The other is located in front of the FASER decay volume and is made up of four scintillator counters. Each scintillator counter is read out by a single PMT.  The veto scintillator stations must be able to veto charged particles with very high efficiency, to ensure the experiment remains background free with an expected 10$^9$ muons traversing FASER during Run 3 operations. The veto scintillators  are larger than the active transverse size of FASER (given by the aperture of the magnets) in order to be able to veto muons that could enter FASER at an angle with respect to the detector axis.
\item The timing scintillator station is placed after the decay volume. It consists of two scintillator counters, each read out by two PMTs. The station is designed to provide a trigger for charged particles exiting the decay volume, and to give a precise time for triggered events, with a resolution of better than 1~ns. 
\item A pre-shower scintillator station is placed at the back of the tracking spectrometer. It is composed of two scintillator counters, each read out by a single PMT, interleaved with two tungsten absorbers, and graphite blocks. The purpose of the pre-shower is to distinguish between calorimeter signals from neutrino interactions in the calorimeter, and photons. Photon showers will start to develop in the tungsten absorbers (each of about one radiation length), therefore leaving signals in the scintillator counters. The graphite blocks are installed to minimize back-splash from the calorimeter leaving signals in the scintillator counters and the rear tracking station.
\end{itemize}

The calorimeter provides energy measurements with an expected precision at the one percent level for the energy range of interest. It consists of four spare modules from the LHCb experiment's outer electromagnetic calorimeter (ECAL)~\cite{LHCB:2000ab} and is 25 radiation lengths deep. 

Each module is $12 \times 12$~cm$^2$ in the transverse plane and made up of 66 layers of lead and plastic scintillator with wavelength shifting fibers bringing the collected light to a PMT at the back of the module.  Since there is a single PMT per module, the calorimeter provides no longitudinal segmentation, and has a very coarse transverse segmentation provided by the $2\times 2$ module configuration. The calorimeter can therefore not separate the two closely spaced electron showers expected from a dark photon decay, but instead will measure the total electromagnetic energy in the event. 

\subsection{Emulsion detector}  

The FASER$\nu$ emulsion detector is made up of 770 1-mm-thick tungsten plates interleaved with emulsion films. The tungsten acts as the target for the neutrino to interact with, and the emulsion films record the trajectories of charged particles produced in the interaction with excellent position and angular resolution. The detector has a transverse size of $25 \times 30$~cm$^2$ with a total target mass of 1.1 tonnes. Given that the emulsion is a passive detector, the detector needs to be extracted, and the emulsions developed and scanned before particle trajectories and corresponding neutrino interaction vertices can be reconstructed and used in data analysis. 

The detector can be used to identify leptons produced in charged current neutrino interactions. Muons are identified as long tracks penetrating through the upto 8 interaction lengths of the detector. Electrons are identified by the shower they produce in the detector, and their energy can be estimated by the shower size. Finally, taus are identified by observing the short $\tau$ track, with either a kink (for 1-prong $\tau$ decays) or by a secondary vertex (for hadronic $\tau$ decays). It is also possible to estimate the momentum of charged particle tracks by the effect on the measured trajectory due to multiple scattering in the detector.

Since the detector records all charged particle trajectories passing through the emulsion films, it must be replaced before the track multiplicity becomes so high that the track and vertex reconstruction performance is significantly degraded. Given the exceptional track resolution, a multiplicity of less than $\mathcal{O}(10^6)$ tracks per cm$^2$ is acceptable, meaning that the detector needs to be replaced after exposure to 30-50~fb$^{-1}$ of collision data.  The operation can be done during scheduled technical stops of the LHC, which are typically three times per year and lasting a few days. 

\subsection{Trigger and DAQ}
The FASER detector is triggered by signals in any of the scintillator stations or the calorimeter. The expected trigger rate is about 650~Hz at a luminosity of $2 \times 10^{34}$~cm$^{-2}$ s$^{-1}$, which is dominated by muons traversing the detector from IP1. The measured muon flux on the LOS corresponds to about 150~Hz of muons within the magnet aperture. However, the scintillator counters are designed to cover a larger transverse size of up to $40\times 40$~cm$^2$ leading to an expected rate of 650~Hz.

The raw calorimeter and scintillator counter PMT signals are sent to a commercial digitizer card, which issues trigger signals if the waveforms are above pre-defined thresholds. These digitizer trigger signals are sent to a custom FPGA-based Trigger Logic Board (TLB) which can combine inputs to form a global trigger (Level 1 Accept, L1A). The TLB can pre-scale and  inhibit triggers and run monitoring algorithms. The L1A signal is sent to the tracker readout boards (TRBs) and the digitizer, to initiate read out of the detector data. On receiving a L1A signal, the different types of readout boards send data over a dedicated fiber network to a software event builder DAQ process running on a server situated on the surface (about 600~m away). The system allows monitoring of the data by software processes running on the DAQ server, while the DAQ electronics are situated in the TI12 tunnel close to the FASER detector.
The trigger and data acquisition (TDAQ) hardware operates on the LHC clock, which is received, processed and distributed via a dedicated electronics board. 

A detailed description of the FASER TDAQ system, and its commissioning can be found in Ref.~\cite{FASERTDAQ:2021}.

\subsection{Computing and software}
Computing is crucial for the success of the FASER experiment and development has to adhere to international standards and employ commercial solutions wherever possible. For this reason, the software framework, called Calypso~\cite{calypso}, follows as closely as possible that of the ATLAS experiment, and it is based on the Gaudi~\cite{gaudi} and Athena frameworks~\cite{athena}.

The raw data from the experiment will be written to tape in the CERN Computing Centre using the CERN Tape Archiving system~\cite{CTA}. The data will be promptly reconstructed at CERN to produce the {\it xAOD} analysis format~\cite{Buckley:2015tjh} which can be accessed directly in the ROOT~\cite{root} analysis software. The prompt reconstruction software will account for masking dead and noisy channels through calibration.  The track reconstruction is based on the common ACTS~\cite{ACTS} framework. The typical raw event size is 22~kBytes/event and the reconstructed data at xAOD level is currently about 10~kBytes/event although this will be significantly reduced once information needed for detector commissioning can be removed. 

The detector simulation is based on the GEANT4~\cite{G4} package. It incorporates a detailed description of the detector geometry including material in the non-sensitive parts of the detector, implemented in GeoModel~\cite{GeoModel} as well as the magnetic field map. Simulated particles are tracked through the detector, recording hits in the sensitive elements. The output of the simulation is passed through a digitization process which simulates the detector electronics response (for example adding in detector noise), and producing an output that is equivalent to the raw detector data. The software also includes an event display program based on VP1~\cite{VP1} which allows the reconstructed events to be visualized. 

\subsection{Outline of this paper}

This paper is structured as follows. 
Sections~\ref{sec:tracker},~\ref{sec:caloScint} and ~\ref{sec:tdaq} outline the details of the design, construction and standalone commissioning of the tracking system, the calorimeter and scintillator system  and the trigger and data acquisition system, respectively. 
This is followed by the description of the magnet system in Section~\ref{sec:magnets}. The detector control and safety systems are described in Section~\ref{sec:dcs}. Section~\ref{sec:fasernu} outlines the emulsion-based neutrino detector, before the integration and in-situ commissioning of the full experiment is detailed in Sections~\ref{sec:integration} and Section~\ref{sec:commissioning}. Finally, Section~\ref{sec:conclusions} provides the summary.

%%%%%%%%%%%%%%%%%%%%%%%%%%%%%%%%%%%
\clearpage
\section{The tracking system}
\label{sec:tracker}

The tracking spectrometer and the interface tracker, IFT, are the two components of the FASER detector tracking system. 
The tracking spectrometer is composed of three tracker stations separated by two dipole magnets, and is designed to detect two oppositely charged particles arising from the decay of a light, long-lived, hypothetical new particle inside the FASER decay volume. 
The pair of charged particles would have high momentum and would be extremely collimated, hence the tracking system must be able to resolve two charged particles with a spatial separation down to $\mathcal{O}$(100~$\mu$m).
For example, for a signal particle with mass $m = 100$~MeV and energy $E = 2$~TeV decaying inside the decay volume, the separation between the decay products at the first tracker station is  $\mathcal{O}$(200~$\mu$m). The tracking spectrometer is composed of three tracker stations, with a total of nine planes of silicon strip modules that were originally the spares for the ATLAS SCT barrel detector~\cite{Abdesselam:2006wt}.  All other parts of the tracker, including dedicated support frames, detector control system (DCS) and cooling system, readout electronics and services were newly developed for FASER. It aims to allow the reconstruction of all tracks that penetrate at least one full station.

The IFT has an identical design to the single tracking station of the tracking spectrometer and can be considered as a fourth tracker station. It enables tracks from a neutrino interaction in the emulsion detector to be matched to events in the spectrometer tracker stations. Both the IFT and the spectrometer fully cover the aperture of the magnets in the transverse ($x$-$y$) plane.

This Section briefly describes the characteristics and specifications of the SCT modules, then it details the mechanical design of the tracker stations, alignment and metrology, and performance obtained during the commissioning. A more detailed description of the FASER tracking detector can be found in Ref.~\cite{FASER:Tracker}.

%%%%%%%%%%%%%%%%%%%%%%%%
\subsection{SCT module}
%%%%%%%%%%%%%%%%%%%%%%%%
An SCT barrel module consists of four identical single-sided silicon microstrip detectors glued as pairs on the two sides of a central base board. The copper/polymide flex hybrid with the front-end ASICs (ABCD3TA chips~\cite{Campabadal:2005rj}) is attached to the sensor assembly. The sensors use $p$-in-$n$ technology where the sensor substrate is $n$-type with 285~$\mu$m thickness and the $p^{+}$ implants are AC-coupled to aluminium readout strips via a silicon dioxide layer. Each sensor has the dimension of $64.0\times 63.6$~mm$^2$ with 768 readout strips at a constant pitch of 80~$\mu$m. 

The ABCD3TA chip contains 128 readout channels that consist of a preamplifier, shaper and discriminator. The signal delivered by the preamplifier-shaper circuit has a peaking time of 25~ns, which is enough to ensure a discriminator time-walk of less than 16~ns and a double-pulse resolution better than 50~ns as required for operation at the ATLAS experiment. 
A common threshold is applied across the entire chip. However there is a spread in threshold offsets between different channels and thus a per-channel 4-bit chip parameter (the TrimDAC) is tuned as a threshold trim to retain a uniform response.
The hit information is provided as 3-bit binary data on reception of a trigger.

\begin{figure}[th]
\centering
\includegraphics[width=0.8\textwidth]{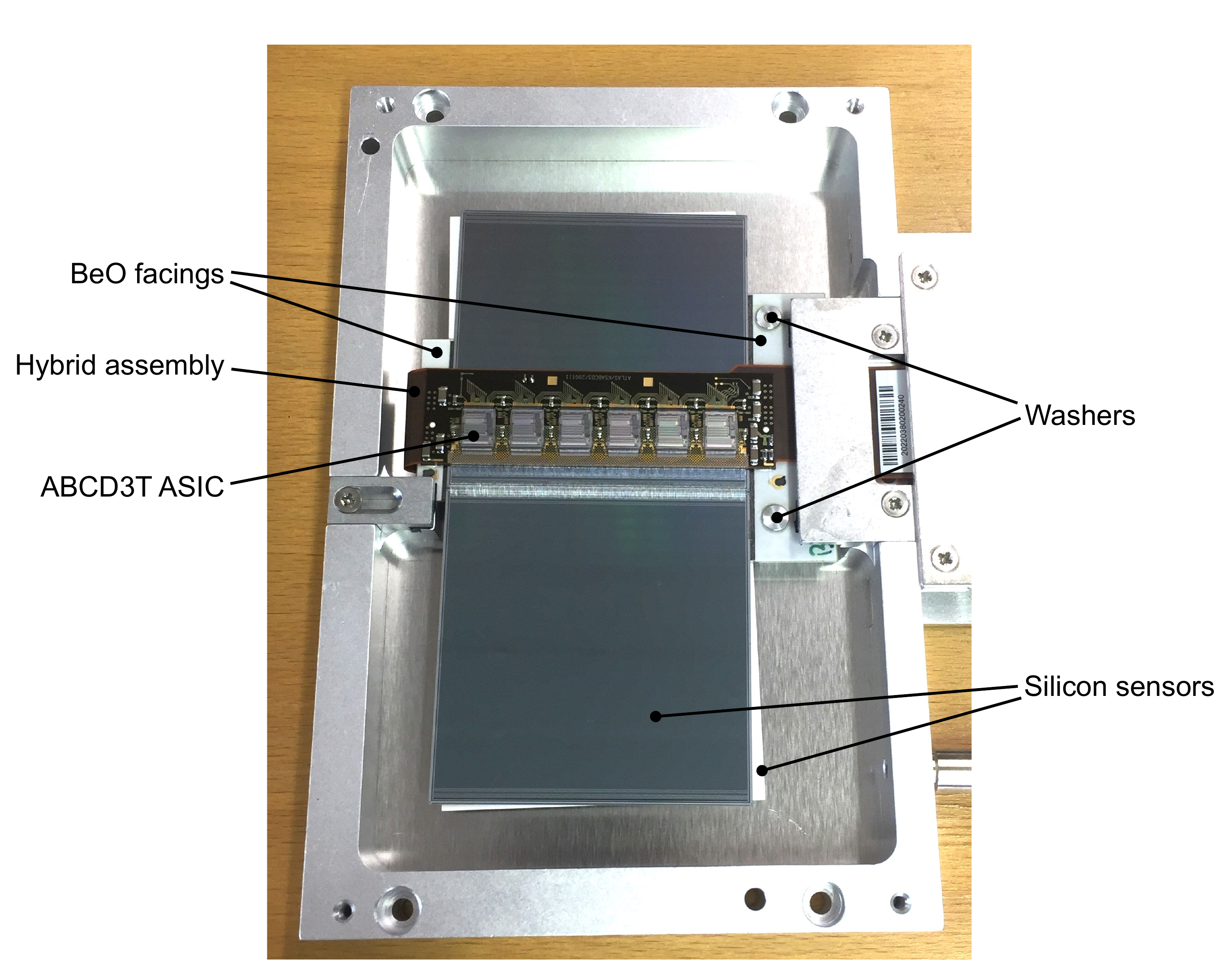}
\caption{Photograph of a SCT barrel strip module.}
\label{f:SCT-photo}
\end{figure}

Figure~\ref{f:SCT-photo} shows a SCT barrel module in an aluminium test-box. On the front and back side of the module, the two sensors are bonded edge-to-edge to create about 12.4~cm-long readout strips. The sensors on each side are placed with a 40~mrad stereo angle so that the hit position can be identified with about 16~$\mu$m resolution in the precision coordinate, and 816~$\mu$m in the other coordinate. The flex hybrid with six ABCD3TA chips per side is bridged over the sensors via a carbon-carbon substrate. The hybrid is attached to beryllia (BeO) facing plates located on the two ends of the baseboard made of Thermal Pyrolytic Graphite (TPG) with an excellent in-plane thermal conductivity and low radiation length, which provides the mechanical support to the sensors and allows for the heat generated by the ABCD3TA chips to be dissipated. 

The strip modules used for the FASER tracker have been selected among the existing spares of the SCT barrel modules. Since completion of the production in 2004, the modules were stored in individual sealed bags. Electrical tests were performed to select the modules to be used in  the FASER tracker. The modules were selected based on the behaviour of the  leakage current as a function of bias voltage (High Voltage, HV) applied to the sensor, and to minimise the number of strips with large noise, low efficiency and cross-talk. In total, 96 modules are used for the four tracker stations.

%%%%%%%%%%%%%%%%%%%%%%%%
\subsection{Tracker plane}
%%%%%%%%%%%%%%%%%%%%%%%%

\begin{figure}[tb]
\centering
\includegraphics[width=12cm]{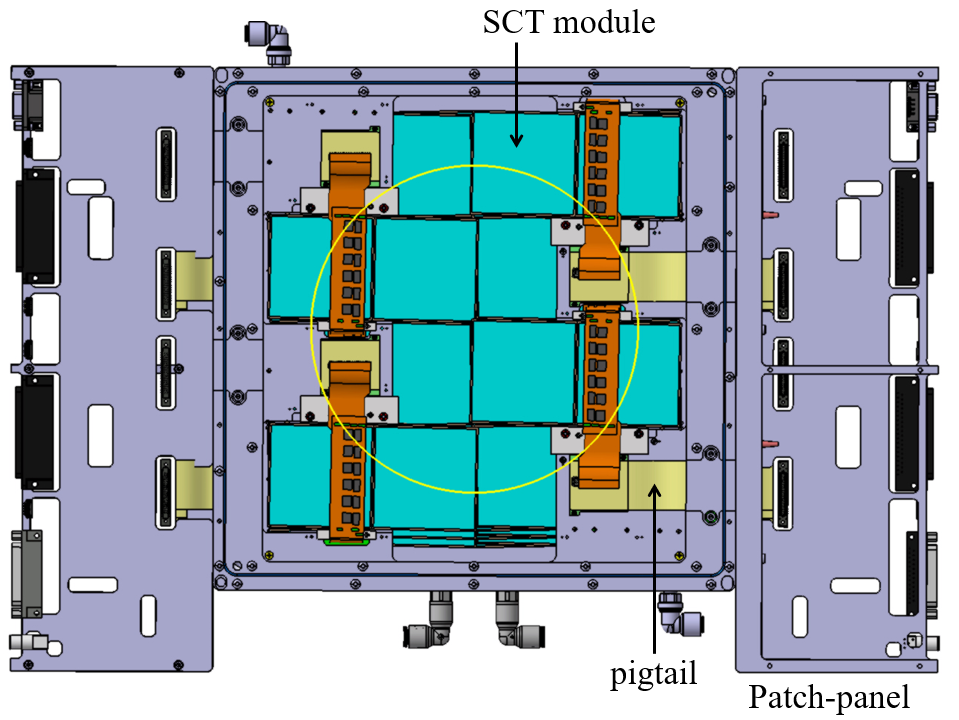}
\caption{Schematic view of the tracker plane. The pigtail is connected to each hybrid on the SCT module. Note that adjacent SCT modules are mounted on different side of FASER module frame. The four pigtails in one side are connected to one patch panel. The circle represents the 200~mm-diameter magnet aperture.}
\label{fig:FPCB-PP}
\end{figure}

\begin{figure}[th]
\centering
\includegraphics[width=0.75\textwidth]{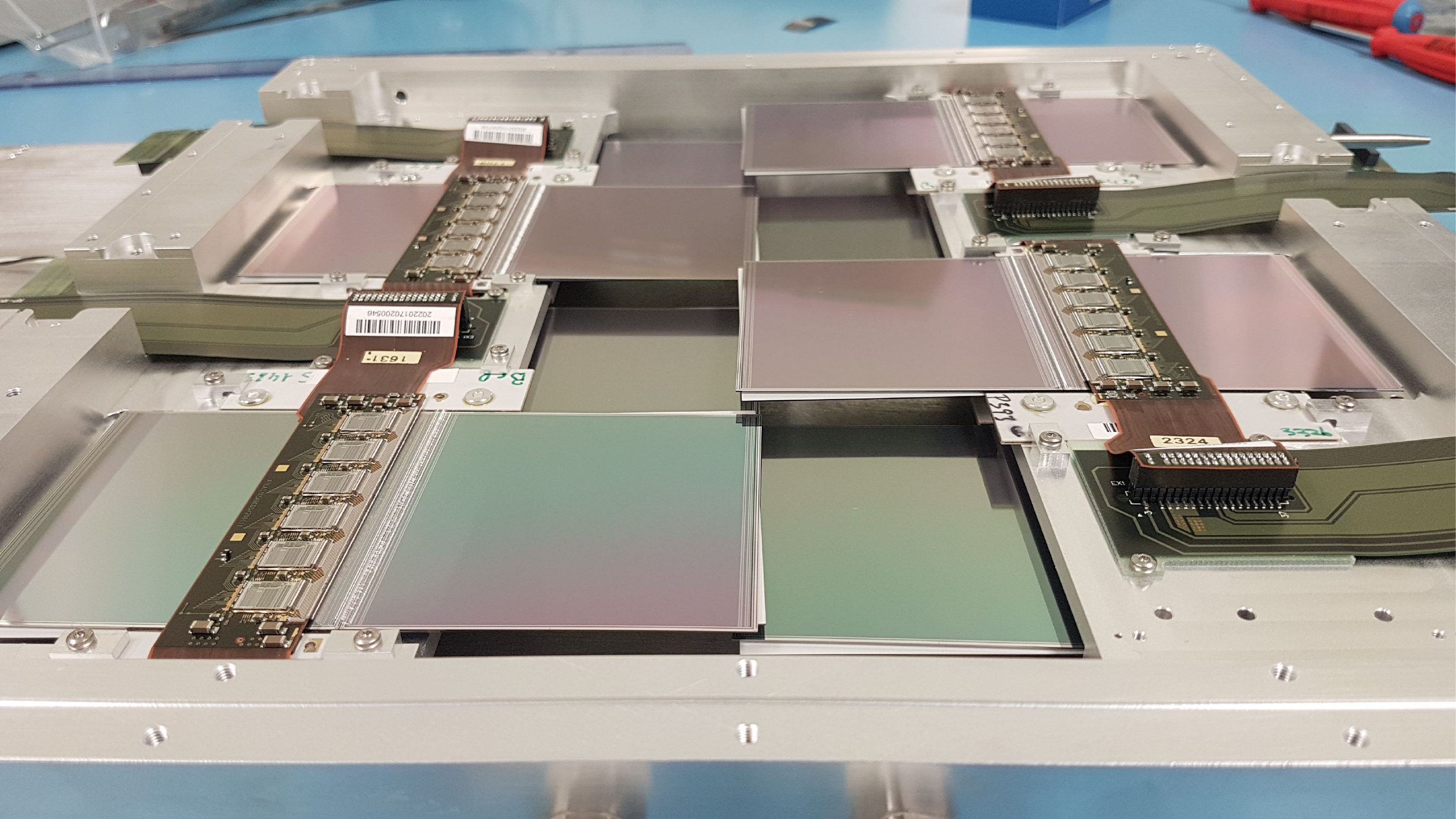}
\caption{Photograph of a tracker plane with all eight SCT modules installed. The beam axis is perpendicular to the plane.}
\label{f:plane-photo}
\end{figure}

Figure~\ref{fig:FPCB-PP} shows a schematic view of the tracker plane. Each plane consists of eight SCT barrel modules within an AW-5083 aluminium frame. Four modules are located on each side (front and back) of the frame as shown in Figure~\ref{f:plane-photo}. The distance between closest sensors in the modules along the out-of-plane direction is 2.4~mm, and the active area overlap along the strip-length is 2~mm. A flexible printed circuit board (called the "pigtail") is attached to each module and routes the electric lines to the outside of the frame. Four pigtails (one per module) on each side of the tracker plane are connected to a single patch panel. The patch-panel is used as the interface between the DAQ and powering systems.

The aluminium frames were produced with CNC (Computer Numerical Control) machining. The size of the frame is 320~mm $\times$ 320~mm $\times$ 31.5~mm. The frame is cut out for most of the active area within the acceptance of the magnet aperture to minimize the material (Figure~\ref{fig:FPCB-PP}). An inner cooling channel with 5~mm diameter was integrated into the frame for the water cooling which extracts heat generated by the ABCD3TA chips on the modules. Details of the water cooling system are described in Section \ref{sec:trk_cooling}. A heat conducting thermal paste~\footnote{Electrolube HTCP-20S} is used at the contact surface between the BeO facing plates of the SCT modules and the aluminium frame for a good thermal contact. An inlet for dry-air is also included in the frame to keep a low relative humidity inside.

%%%%%%%%%%%%%%%%%%%%%%%%
\subsection{Tracker stations}
\label{sec:trackerstation}
\begin{figure}[tb]
\centering
\includegraphics[width=0.48\textwidth]{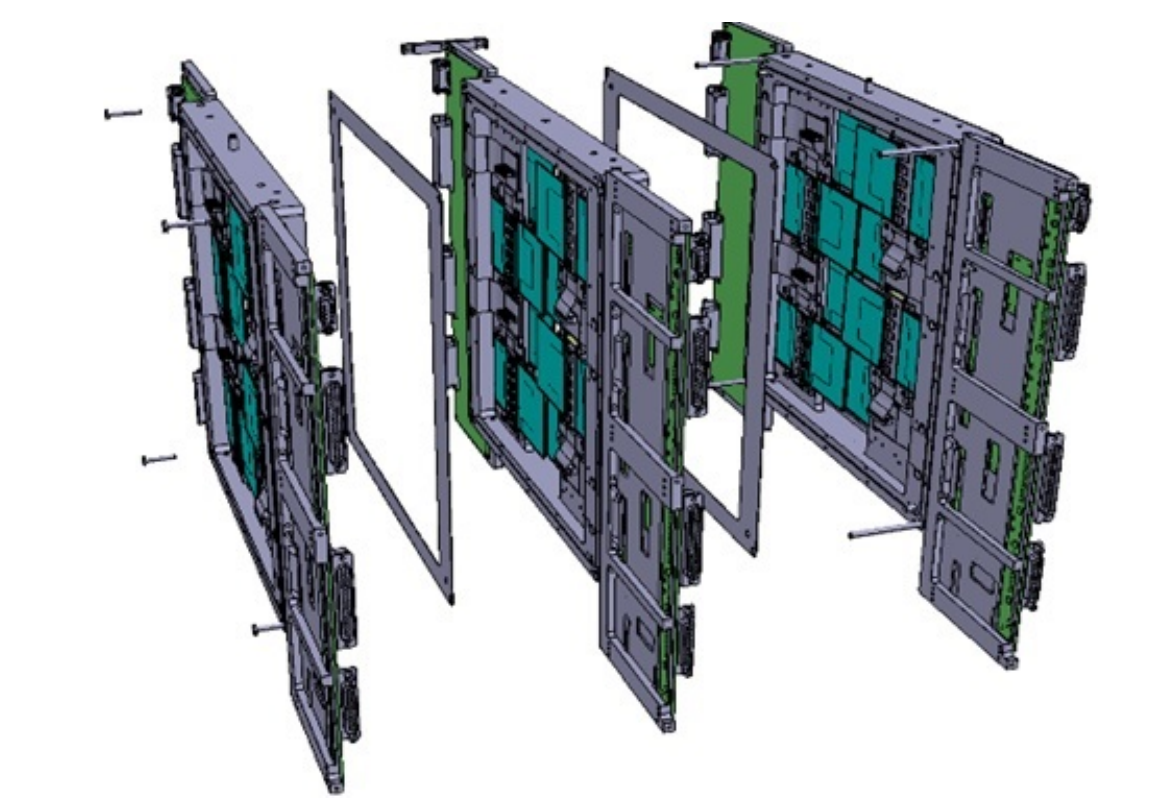}
\includegraphics[width=0.48\textwidth]{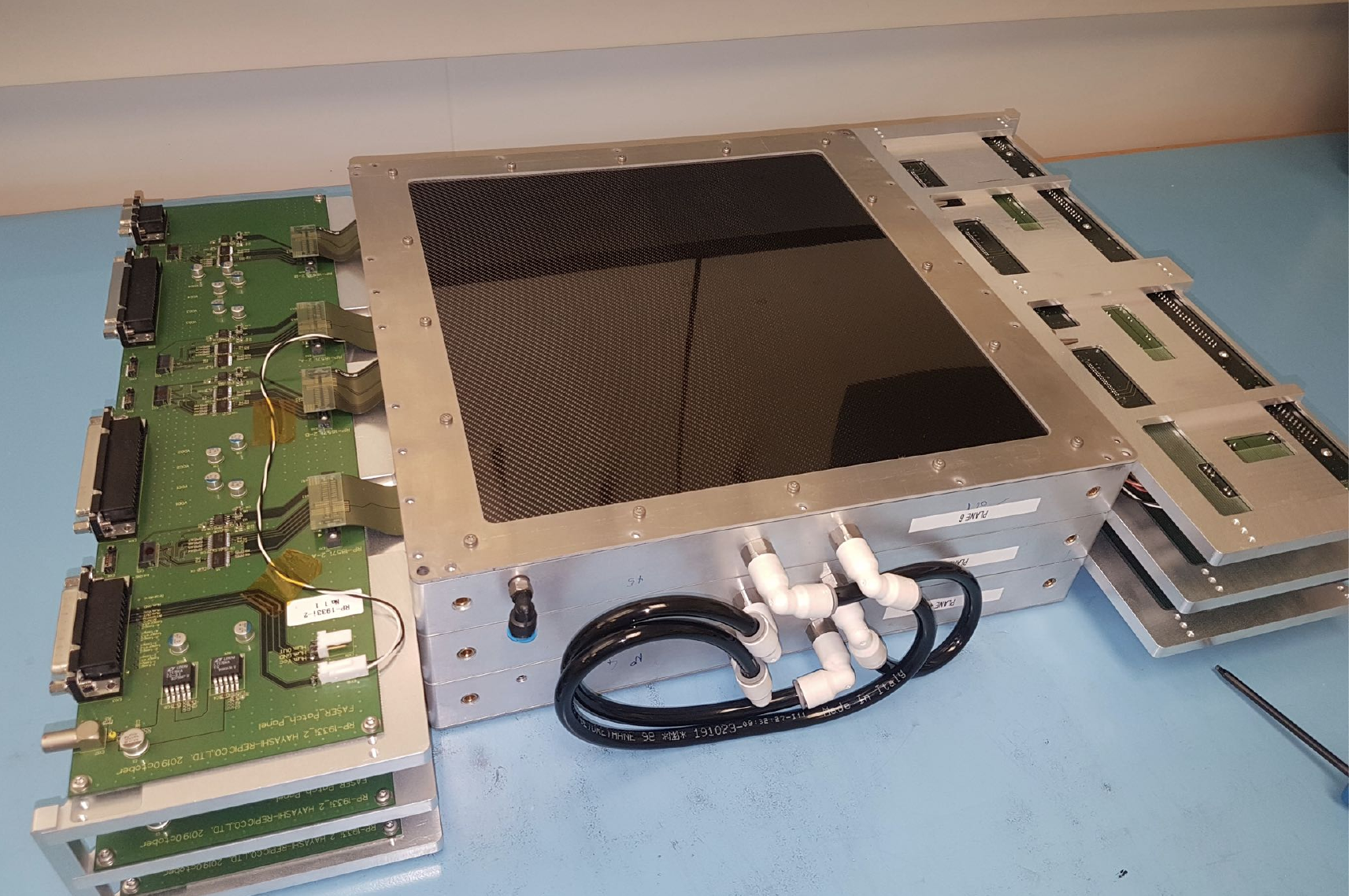}
\caption{(left) Exploded CAD view of a tracker station and (right) photograph of a fully assembled station. The black cover on top of the station is a carbon-fibre plate. The cooling loops of each plane are connected together, so that per station there is only one inlet and one outlet for the cooling fluid.}
\label{f:station}   
\end{figure}

%%%%%%%%%%%%%%%%%%%%%%%%
The tracker station is an assembly of three planes as shown in the computer-aided drawing (CAD) view  and picture of Figure~\ref{f:station}. The distance between the sensor cut edge and the sensitive region is 1~mm. For an individual plane, this results in a dead region of about 2~mm in between modules along the vertical direction. To overcome this, the three planes are staggered along the vertical direction with a relative shift of the middle (last) plane of $+5$~mm ($-5$~mm) with respect to the first plane. This  ensures that there are at least two 3D reconstructed hit points for a track crossing the station. 
An additional dead region corresponds to the vertical slice in the centre of each module. This represents 1.6\% of the active area, and is accounted for in the detector description in the simulation and reconstruction.

Each station volume is closed by two end-covers made of carbon-fibre plates with 400~$\mu$m thickness (standard T300 fibres).
To prevent corrosion by the corona discharge processes that might occur  after putting the frames in contact during the station assembly, a post-treatment with SURTEC-650 was performed for all aluminium parts. 
In addition, an O-ring sealing joint was attached between the frames for a good tightness and to keep the humidity inside the station as low as possible (typically $\sim 1$\%). The total weight of one station is about 15~kg without cables.

The thermal performance was investigated with various Finite Element Analysis (FEA) simulations. The temperature measured with a thermistor on the flex hybrid of the SCT module is required to be less than 35~$^{\circ}$C, which corresponds to the glass transition of the epoxy glue used for the module assembly. Keeping the coolant temperature at 15~$^{\circ}$C, a water flow of 3~$\ell$/min (considering a heat transfer coefficient for water of 500~W/m$^{2}$), and the outside air convection at 23~$^{\circ}$C, the FEA gives a maximum temperature on the ABCD3TA chips of $\sim 28$~$^{\circ}$C, neglecting the temperature rise within the water channel due to the heat load. The latter is estimated to be $+0.6$~$^{\circ}$C for 3~$\ell$/min. These  results are in good agreement, within 2-3~$^{\circ}$C, with measurements taken during commissioning, hence validating the simulation. The FEA simulation was then used to estimate the temperatures on the silicon sensors and predicted 21-23~$^{\circ}$C, which is well within the specifications for the epoxy glue.

Table~\ref{tab:TrackerMaterial} summarizes the material budget in a tracker station. The central region with the least amount of material in the station, {\it i.e.}, six silicon sensors and two carbon-fibre covers, accounts for a total of 2.1\% of a radiation length (X$_{0}$). 
The worst case is when the particle penetrates the edge region that consists of six SCT modules including sensors, TPG baseboard, flex hybrid with carbon-carbon bridge and ABCD3TA chips as well as the aluminium frames and station covers. In such a case, the material budget becomes 21.5\%~X$_{0}$.

Simulation results based on a dark photon benchmark model with  m$_{A'} = 100$~MeV and $\epsilon$ = 10$^{-5}$ show that 70\% of the dark photons are contained within the low-material central region of the tracker. 
Finally, given the high-momentum spectrum expected for the signals of interest the effect of multiple scattering from the traversed material will be negligible. 

\begin{table}[thb]
    \centering
    \begin{tabular}{|l|c|c|c|c|}
  \hline
Component & Material & Number & \multicolumn{2}{c|}{$X_0$ (\%)}  \\
 & & / station & Central region & Edge region \\
\hline
Silicon sensor & Si  & 6 & 1.8\% & 1.8\%  \\

 Station Covers & CFRP  & 2 & 0.3\%  & 0.3\%   \\

SCT module support & TPG  & 3 & -  & 0.6\%  \\

C-C Hybrid & C (based) & 3 & - & 2.2\%  \\

ABCD chips & Si & 3 & - & 6.5\%   \\

Layer frame & Al & 3 & - & 10.1\%   \\
\hline
\bf{Total / station} & \bf{-} & \bf{-} & \bf{2.1\%} & \bf{21.5\%} \\
\hline
    \end{tabular}
    \caption{Amount of material in $X_0$ in the active area of a tracking station for two regions: i) the central region (|$x$| < 4 cm) with only the silicon sensor material and ii) the edge region. Details of the material in the SCT module are given in Table 8 of Ref.~\cite{Abdesselam:2006wt}. The numbers are calculated directly from the CAD description of the tracking station.}
    \label{tab:TrackerMaterial}
\end{table}

The three spectrometer tracker stations are mounted into the FASER detector with an AW-5083 aluminium structure (called the "backbone") whilst  the IFT is fixed on an independent support structure. The details are described in Section~\ref{sec:integration}.

%%%%%%%%%%%%%%%%%%%%%%%%
\subsection{Alignment and metrology}
\label{sec:trkalignment}
%%%%%%%%%%%%%%%%%%%%%%%%
\begin{figure}[th]
\centering
\includegraphics[width=0.65\textwidth]{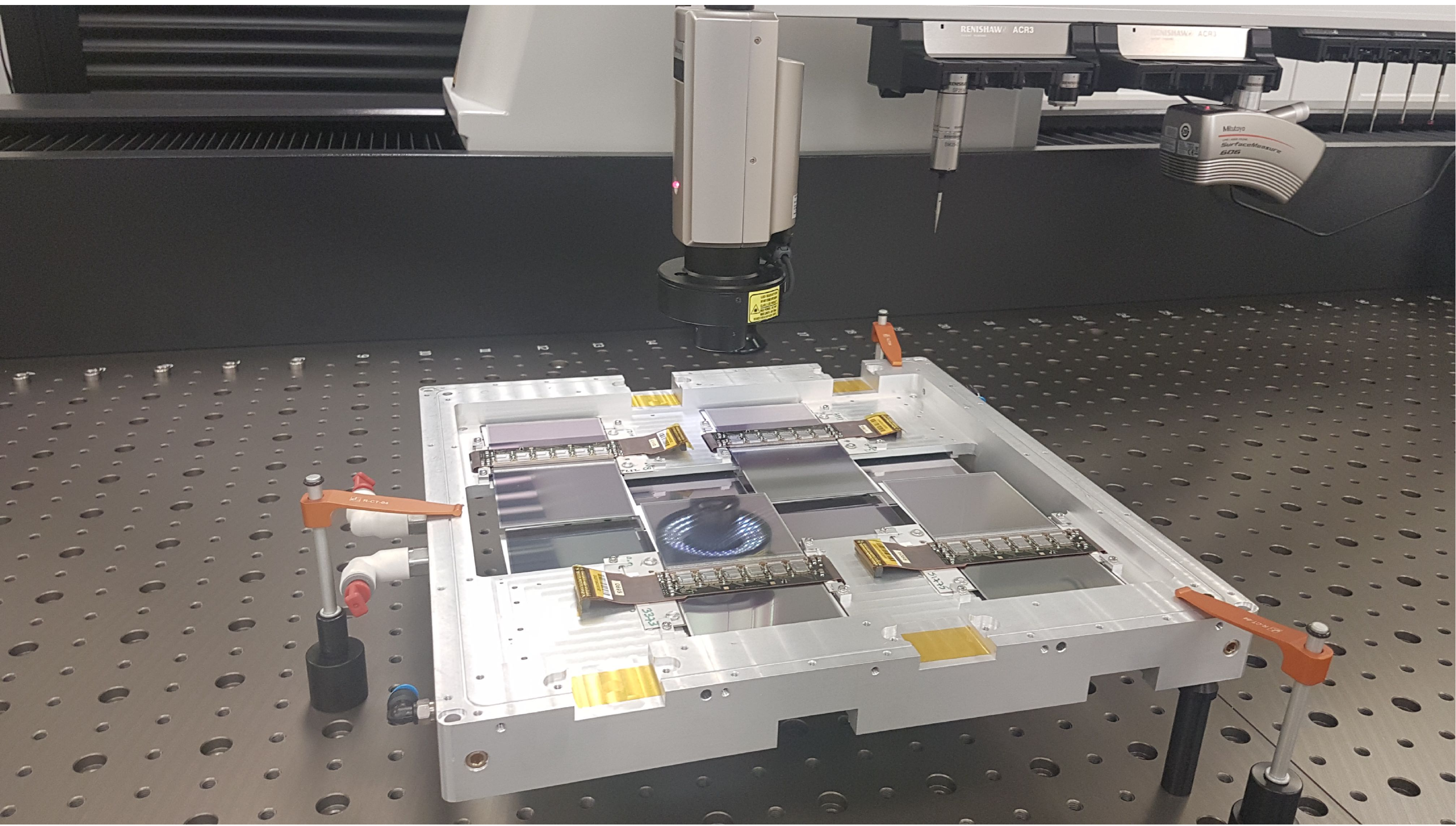}
\caption{Metrology of a fully assembled tracker plane.}
\label{f:metrology}
\end{figure}

Metrology was performed for all assembled planes and stations at the University of Geneva using a Mitutoyo CRYSTA-Apex S CNC coordinate-measuring machine with an automatic probe changer (Figure~\ref{f:metrology}). Measurements were performed with a mechanical touch-probe and an optical camera. 

There are four stainless steel targets on each frame (one in each corner) to define the plane reference coordinate system. The targets are visible from both sides, allowing measurements done on each side of the plane separately to be correlated. The silicon sensors of the SCT modules have a set of fiducial marks that were used for the mechanical alignment during module assembly. Some of the fiducial marks were measured in 3D with respect to the plane reference system. The precision for in-plane and out-of-plane measurements was 5~$\mu$m and 10-15~$\mu$m, respectively. All frames satisfied the required tolerances ($\pm 20$~$\mu$m) with respect to the CAD manufacturing drawings. The maximum deviation was 100~$\mu$m in positioning the SCT modules, which will be corrected for using the metrology measurements in the data events reconstruction.

%%%%%%%%%%%%%%%%%%%%%%%%
\subsection{Standalone Tracking Commissioning}
\label{sec:trk-commissioning}
%%%%%%%%%%%%%%%%%%%%%%%%
A detailed overview of the commissioning of the tracking detector is given in Ref.~\cite{FASER:Tracker} and a brief summary is provided below. 
Various aspects of the electrical performance of the silicon sensors such as noise levels,  number of bad strips, HV, and thermal behaviour were investigated at each stage of construction of the tracker.  Individual SCT modules were qualified by using a test system developed at Cambridge University which is used for muon spectroscopy~\cite{Keizer:2018nju}. The planes and stations were tested with standalone operation of the tracker readout board (TRB) and with a similar DCS and cooling system as used in the completed FASER tracker. 

\begin{figure}[tbh]
\begin{center}
\includegraphics[width=0.7\textwidth]{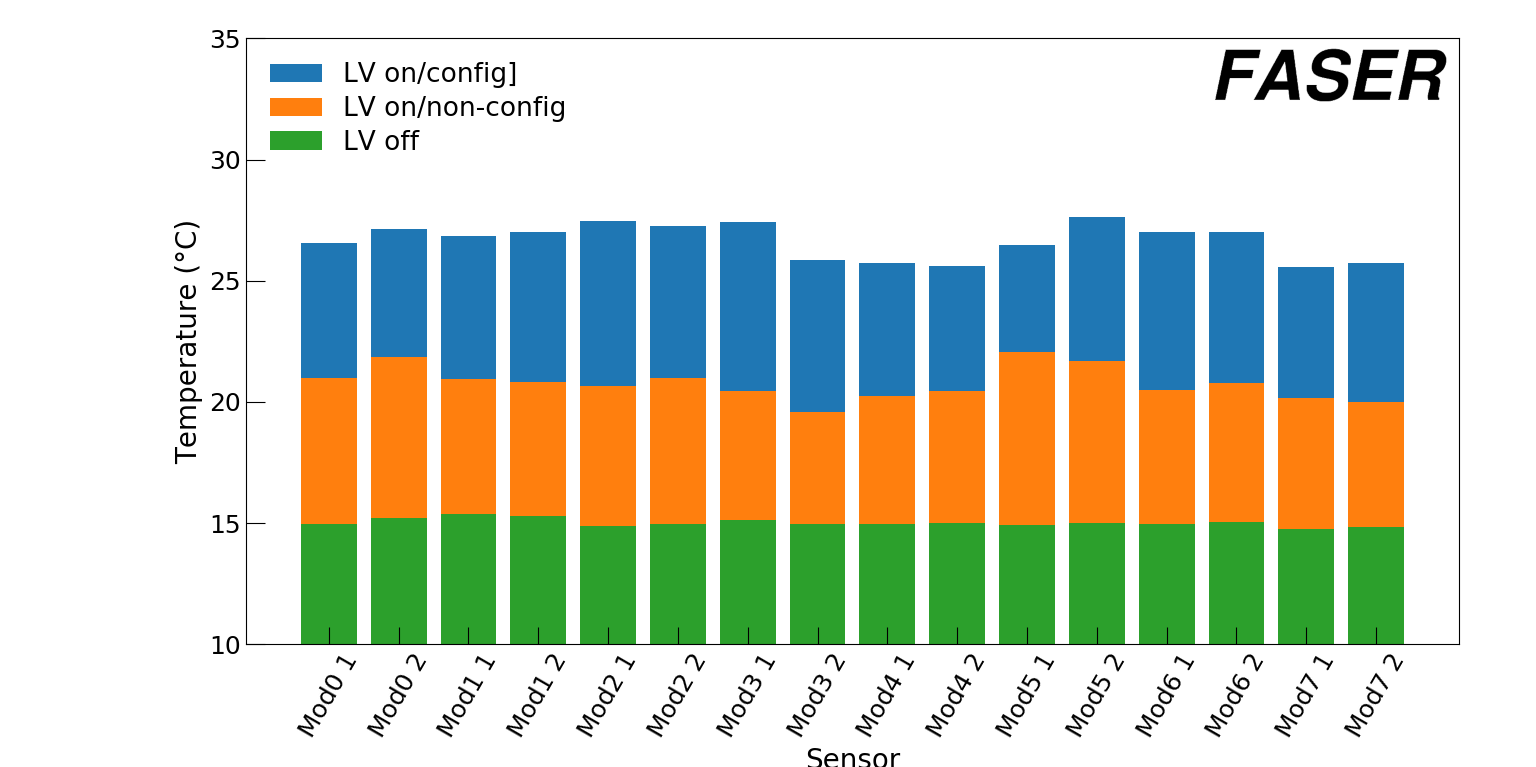}
\caption{Thermal test results of the first station during surface commissioning. Temperature values of the SCT module NTCs for power-off (green), after powering the modules (orange) and after powering and configuring (blue).}
\label{fig:tracker_temperature}
\end{center}
\end{figure}

Following the installation of the tracker  in TI12, the total number of dead channels was measured to be  about 0.5\% of the total, including the region outside the main magnet acceptance (see also Figure~\ref{fig:FPCB-PP}). The tracker is operated with a bias voltage of 150~V for the silicon sensors and at 15$^{\circ}$C as set by the cooling system. Figure~\ref{fig:tracker_temperature} shows the temperature measured by NTC sensors on each module (two sensors per module) in one tracker plane, during the surface commissioning. The temperature without powering corresponds to the coolant temperature of 15~${^{\circ}}$C. When powered and configured, the module temperatures are kept well below 31~${^{\circ}}$C, the threshold temperature for the software interlock.

The calibration procedure of the ABCD3TA chips is performed by injecting a test pulse generated in the chips. The calibration steps consist in identifying noisy and dead strips, making adjustments of the timing between the test pulse and the level-1 trigger, and trimming of the threshold offsets. 
In addition, during combined detector data-taking, the tracker detector needs to be synchronized with the FASER trigger system triggering on a physics signal. Global time-tuning parameters for the tracker are configured on the TRB. More details are provided in Section~\ref{sec:TRB}. 
The results shown in this Section are obtained considering all four tracking stations (tracking spectrometer and IFT) after installation in TI12 and setting the temperature to 15$^{\circ}$C.

The hit occupancy, defined as the fraction of injected charge above threshold, is computed for each readout channel at various threshold points. Since the signal amplitude is convoluted with Gaussian electronics noise, the hit occupancy does not follow an ideal step-function but is smeared to give rise to the so-called $s$-curve. The threshold at which the hit occupancy is 50\% is called the $vt_{50}$ point and corresponds to the amplitude of the test pulse. 

\begin{figure}[tbh]
\begin{center}
\includegraphics[height=0.45\textwidth]{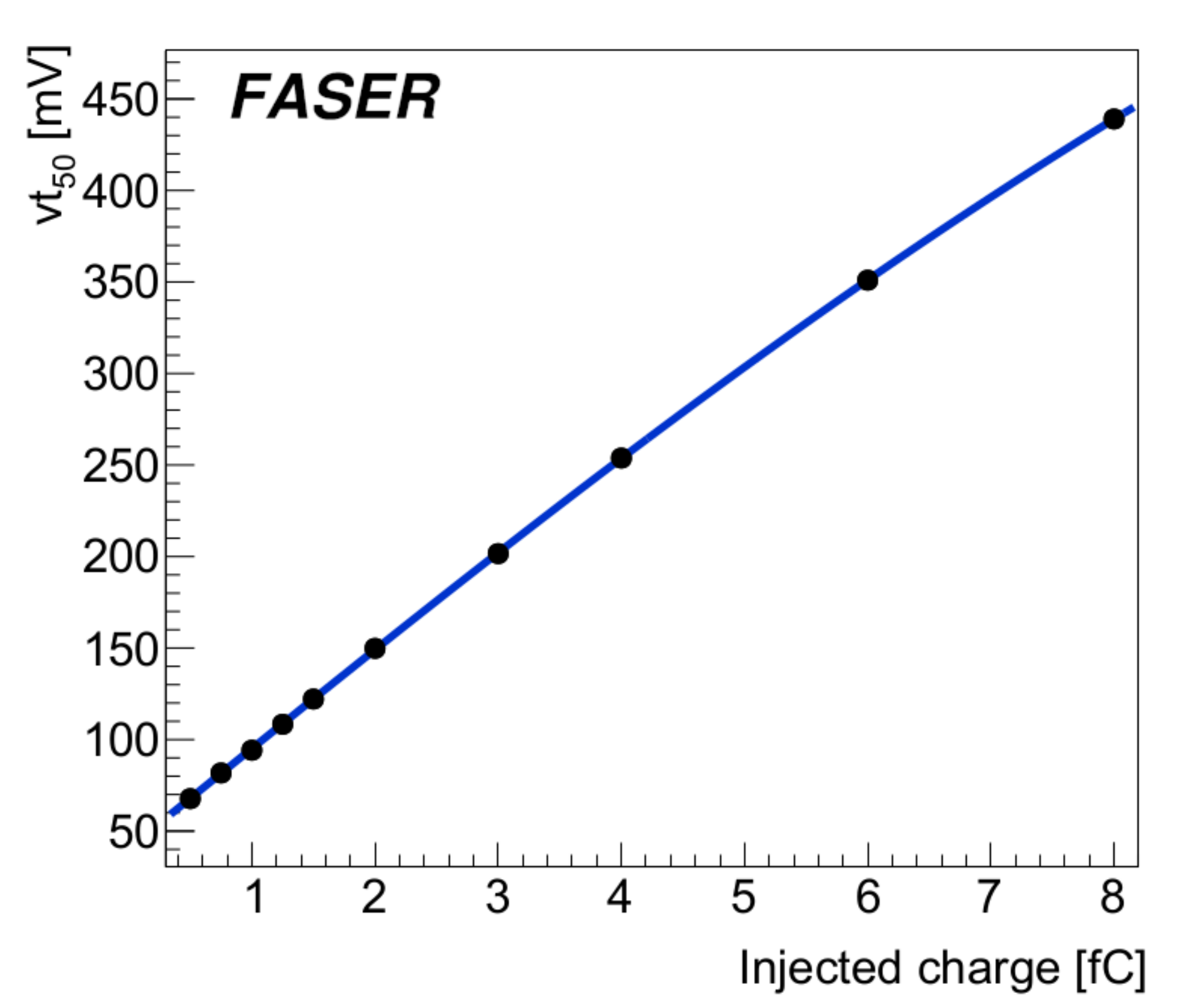}
\hspace{0.5cm}
\includegraphics[height=0.45\textwidth]{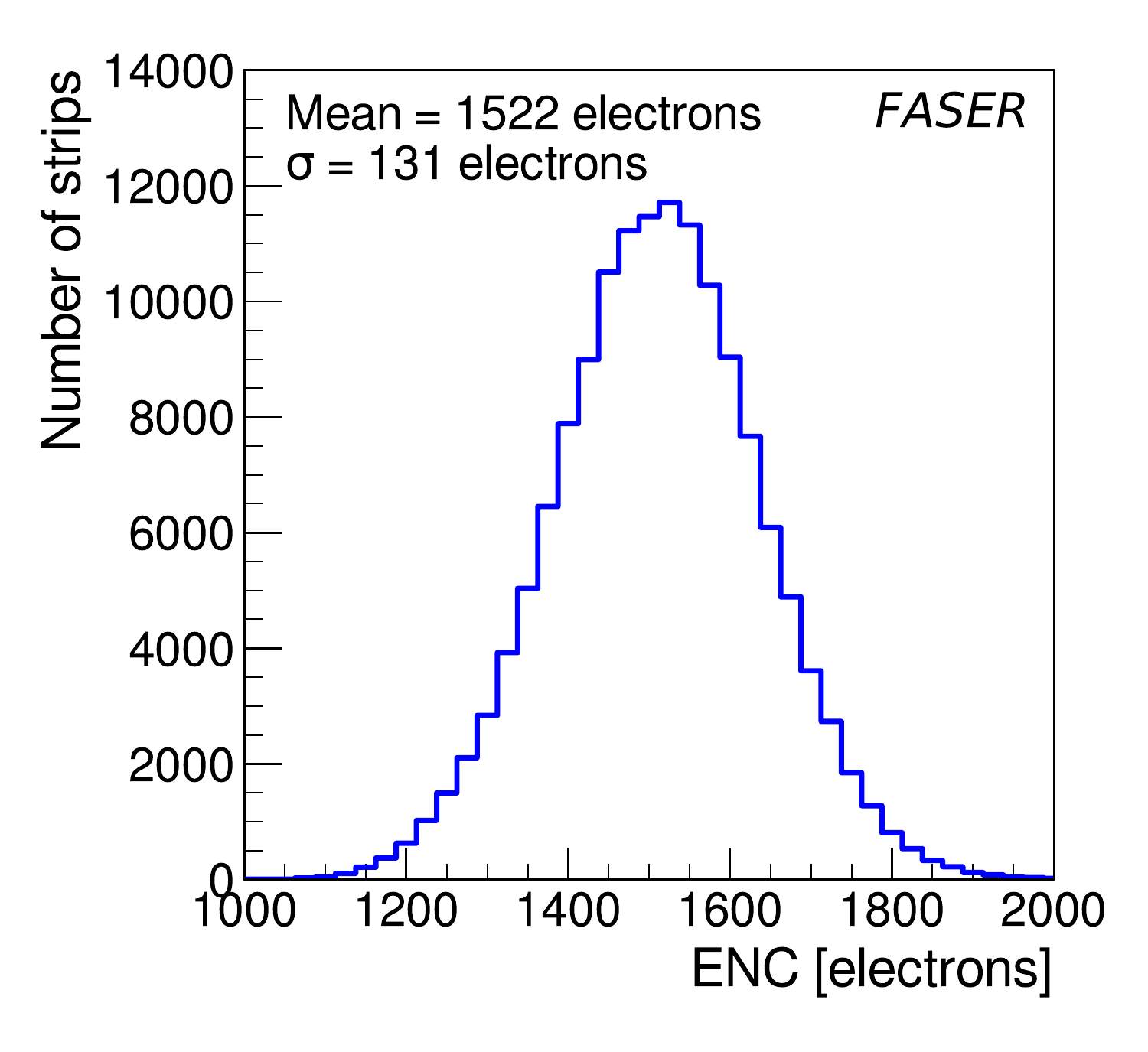}
\includegraphics[height=0.45\textwidth]{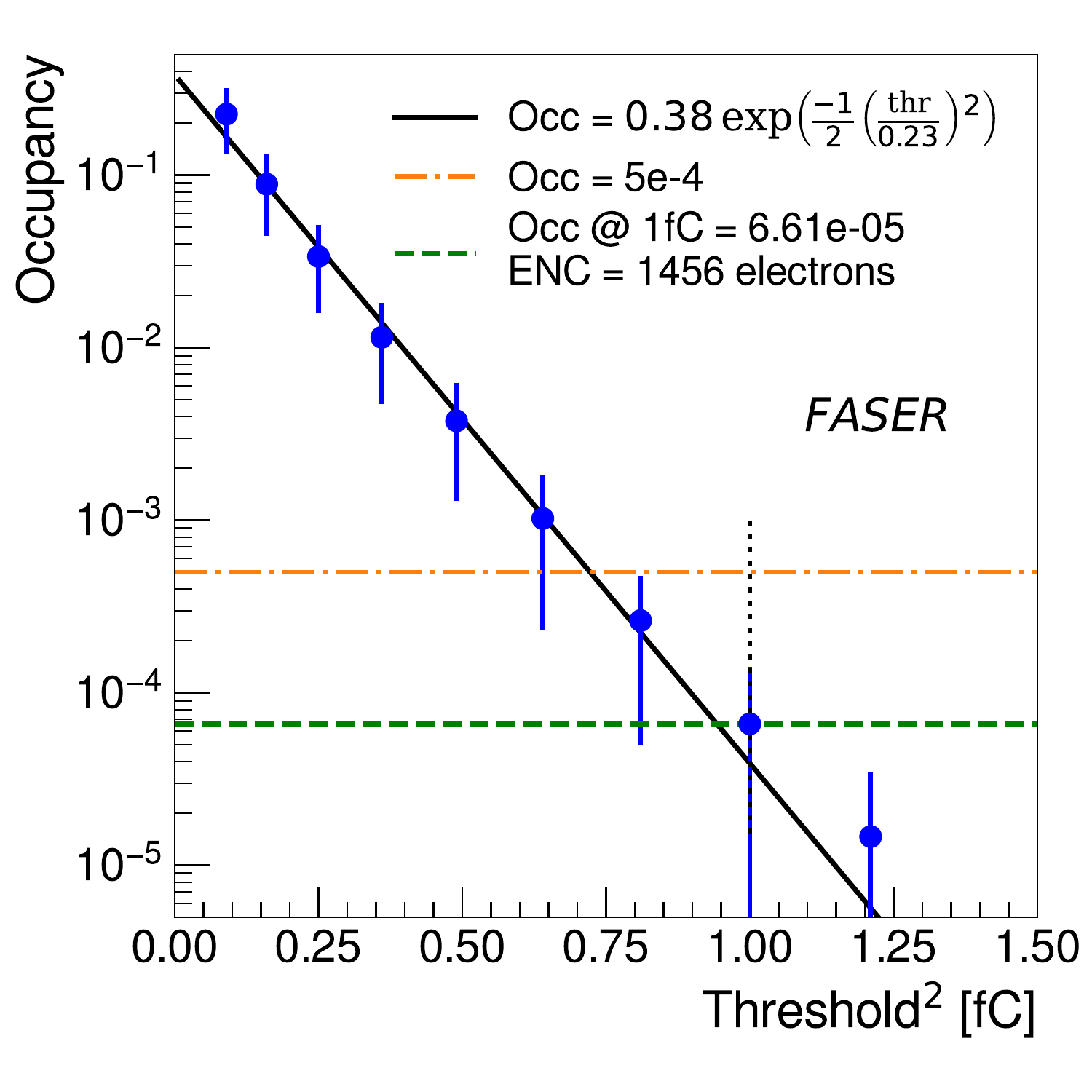}
\caption{(top) $vt_{50}$ as a function of the injected charge, (bottom left) measured ENC of all strips in the four tracking stations, (bottom right) and the noise occupancy scan results for an example module.}
\label{fig:tracker_response}
\end{center}
\end{figure}

Figure~\ref{fig:tracker_response} (top) shows the typical curve of the $vt_{50}$ as a function of the injected charge. The gain of the amplifier (typically 50 mV/fC) is derived by fitting the $vt_{50}$ values at three different charges (1.5, 2.0 and 2.5~fC) with a linear function. Then, the Equivalent Noise Charge (ENC) is calculated by dividing the standard deviation of the $s$-curve, extracted  in a threshold scan with 2~fC injected charge, by the gain as shown in Figure~\ref{fig:tracker_response} (bottom left). The average ENC was evaluated to be 1522 electrons across all strips in the tracker. The noise hit occupancy at 1~fC threshold was measured to be $6.61 \times 10^{-5}$ (Figure~\ref{fig:tracker_response} (bottom right)) which is well below the specification requirement of $<5 \times 10^{-4}$. In addition, the ENC was extracted to be 1456 electrons from the fitting, that is similar to the measured value with the injected charge (1522 electrons). 

Finally, a hit efficiency above 99.8\% was confirmed in the 2021 testbeam which took place at the H2 beamline at the CERN-SPS (Super Proton Synchrotron)~\cite{testbeampaper}.

%%%%%%%%%%%%%%%%%%%%%%%%%%%%%%%%%%%
\clearpage
\section{The calorimeter, pre-shower and scintillator systems}
\label{sec:caloScint}
The FASER experiment has four scintillator stations with multiple scintillator counter layers in each station as well as a lead-scintillator electromagnetic calorimeter, all of which use photo-multiplier tubes  to detect the scintillation signals. This Section presents a detailed description of each component, as well as results from pre-installation and standalone commissioning tests. 

\subsection{Scintillator system}

\label{sec:scintsystem}

To achieve high detection efficiency, especially for the veto stations, all scintillator layers are constructed from 10 or \SI{20}{mm} thick EJ-200 plastic scintillator material \cite{eljen}. The light output of this material is about $22,000$~photons/cm for muons~\cite{TKACZYK201896} and the thickness corresponds to 2.5--5\% of a radiation length. The light in each scintillator counter layer is transmitted to one or two Hamamatsu PMTs through a wavelength shifting (WLS) rod or plastic light guides depending on the station type. The combined arrangement of scintillator, light guide/WLS rod and PMT(s) is referred to as a \emph{scintillator module}. 
The scintillator counter layers, light guides and WLS rods are wrapped in \SI{0.5}{mm}-thick aluminum foil to ensure light tightness and improve fire safety. In front of each PMT is placed an open-ended optical fiber for injecting light pulses from the calibration system, see Section~\ref{sec:LEDcalib}.

\begin{figure}[hbt!]
    \centering
     \begin{subfigure}[b]{0.45\textwidth}
         \centering
         \includegraphics[width=\textwidth]{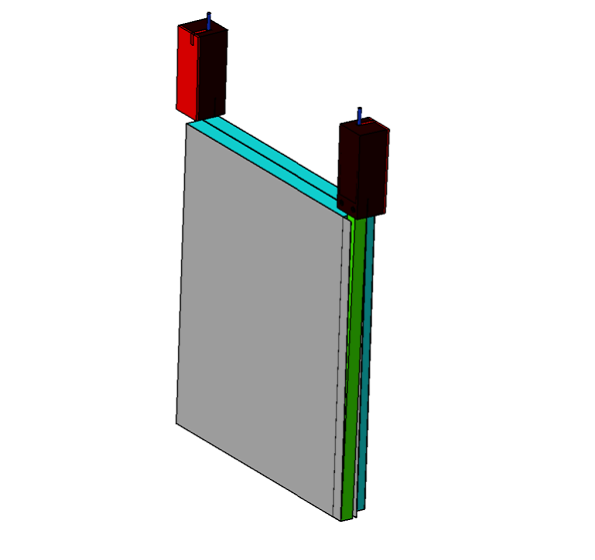}
         \caption{First veto station.}
         \label{fig:scint1}
     \end{subfigure}
     \hfill
      \begin{subfigure}[b]{0.45\textwidth}
         \centering
         \includegraphics[width=\textwidth]{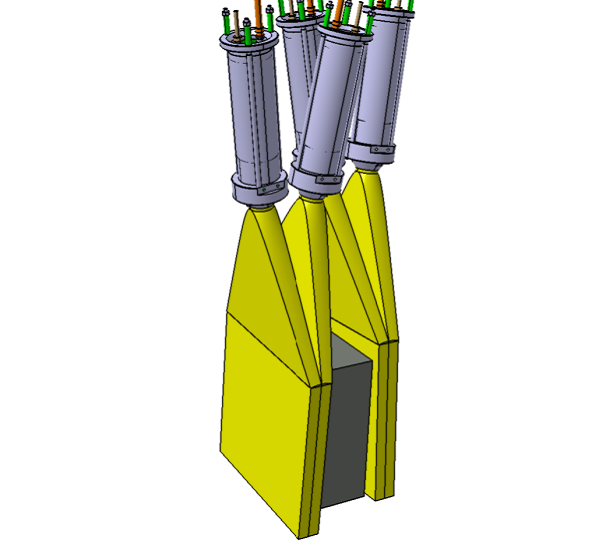}
                 \caption{Second veto station.}
         \label{fig:scint2}
     \end{subfigure}
     \vskip\baselineskip
     \begin{subfigure}[b]{0.45\textwidth}
         \centering
         \includegraphics[width=\textwidth]{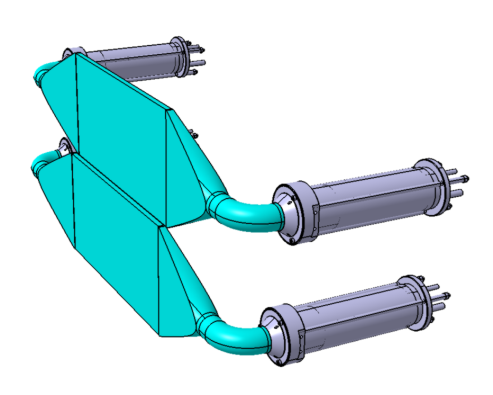}
         \caption{Timing station.}
         \label{fig:scint3}
     \end{subfigure}
     \hfill
      \begin{subfigure}[b]{0.45\textwidth}
         \centering
         \includegraphics[width=\textwidth]{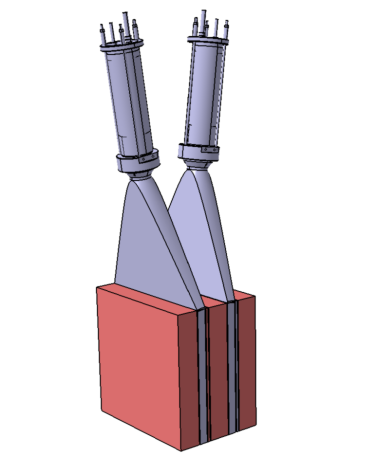}
                 \caption{Preshower station.}
         \label{fig:scint4}
     \end{subfigure}
 \caption{Drawings of the different scintillator stations. The PMTs are located at top and shown in grey (red for the first veto station) except for the timing station. The first station uses WLS rods shown in green to transport the light to the PMTs, while the others use triangular light-guides. For the second veto station and the preshower station, the absorbers are included in the drawings as the grey and red blocks, respectively.}
        \label{fig:scintstations}
\end{figure}

The first veto station is positioned upstream of the FASER$\nu$ emulsion detector to  veto incoming muons and thus to provide discrimination between the muon-induced background and the neutrino interaction events. The design of the first scintillator veto station is heavily constrained by the limited space available around FASER$\nu$. Therefore, its design is different to that of the other stations. It is constructed from two modules placed back-to-back. Each module is made up of a 30~cm~$\times$~35~cm, 2~cm thick EJ-200 plastic scintillator connected by a 1.5~cm~$\times$~1.5~cm~$\times$~37.5~cm EJ-280 plastic wavelength shifting rod \cite{eljen} to a Hamamatsu H11934-300 PMT~\cite{Hamamatsu11934} as shown in Figure~\ref{fig:scint1}. The H11934-300 PMT is a very compact 12 dynode-stage head-on PMT with a \SI{23}{mm}~$\times$~\SI{23}{mm} sensitive photocathode, a typical gain of $1.2\times10^6$, a fast rise time of \SI{1.3}{ns} and low transit-time-spread of \SI{0.27}{ns}, though the signal time precision will be limited by the long emission time of the WLS rod.

The three downstream scintillator stations have a  different design. EJ-200 plastic scintillator plates are again used but are connected via light guides to Hamamatsu H6410 PMT(s)~\cite{Hamamatsu6410}. The H6410 PMT is a large 12 dynode-stage head-on PMT, with a \SI{47}{mm} sensitive aperture, a typical gain of  $3\times10^6$, a rise time of \SI{2.7}{ns} and transit-time-spread of \SI{1.1}{ns} for a single photo-electron. The head of the PMT assembly is mounted directly on to the end of the light-guide under spring-load to keep good optical contact between them. 
Two layers of permalloy tube protection surround the PMT to reduce the impact of magnetic fields.

The second veto station, shown in Figure~\ref{fig:scint2} is located in front of the decay volume magnet.  It features two pairs of modules, with the modules in a pair placed back-to-back for redundancy and improved veto efficiency. The station's primary purpose is to suppress events with incoming SM particles, mostly high-energy muons passing through the FASER decay volume. To avoid energetic photons arising from muon bremsstrahlung in front of the detector entering undetected, an absorber block of \SI{10}{cm}-thick lead is placed between the two pairs of modules. This will either absorb the photons completely or generate a shower that is detectable by the second pair of modules in this station. To suppress background related to muons from the interaction point, each pair of modules is required to detect more than 99.99\% of the incoming muons. This is achieved by having a large scintillation signal and efficient light collection as well as a large coverage with respect to the detector aperture. Each module therefore consists of a 2~cm-thick, 30~cm~$\times$~30~cm scintillator layer. The light guide and PMT are oriented vertically above the scintillator to minimize the transverse size with the light guides tilted from vertical by $\pm$\SI{140}{mrad} in order for the PMTs of the two back-to-back layers not to overlap. The two pairs of modules are positioned either side of the lead blocks and each pair has a \SI{65}{mrad} angular rotation around the LOS, again to avoid interference between neighbouring PMTs.

The third scintillator station, shown in Figure~\ref{fig:scint3}, is primarily responsible for providing trigger and timing information for FASER. It is located after the decay volume magnet and before the first tracking station. It detects the appearance of a charged particle pair from the decay of a LLP in the decay volume and thus provides the primary trigger signal for physics analysis. It is also used to precisely measure the arrival time of the signal with respect to the $pp$ interaction at the ATLAS interaction point. This timing information is used to suppress non-collision backgrounds, for which a resolution of better than \SI{1}{ns} is required. Another design constraint on this station is that the material should be minimised, while maintaining a signal trigger efficiency above 98\%. Finally, the active area of this station must be large enough to cover most of the magnet front surface in order to detect muons coming in at an angle, without passing through the veto stations or the first two tracker stations, but causing an electromagnetic shower in the magnet material and thus a detectable shower in the downstream calorimeter, as this could mimic a photon-only signature. 
The station is constructed from two 1-cm~thick, 40-cm~wide and 20-cm~high scintillator layers, stacked vertically with a \SI{5}{mm} overlap in the middle as shown in Figure~\ref{fig:scint3}. Light guides on both sides of each scintillator layer connect to the same H6410 PMT type of assembly as used in the second veto station, but the light guide has 90$^\circ$ bend to minimize the transverse size of the station which is limited by the tunnel wall and trench width. The timing precision is optimized by reading out the scintillation signal on both ends of each scintillator as the horizontal time-walk cancels out when averaging the arrival time for the two PMTs.

The fourth and final scintillator station, shown in Figure~\ref{fig:scint4} is used both as an additional trigger station, which, if needed, can be used in a coincidence with the first trigger station to reduce the rate of non-physics triggers, and as the active sensor in the preshower detector. This station is located after the last magnet and tracker station and in front of the calorimeter system. The two scintillator modules are identical to the ones used in the second veto station. Both modules are preceded by a \SI{3}{mm}-thick layer of radiator (tungsten) to create a simple preshower detector. This helps distinguish a physics signal of two close-by energetic photons,  which would otherwise leave only large energy deposition in the calorimeter, from deep inelastic scattering of high-energy neutrinos. 
This is needed because the calorimeter does not have any longitudinal segmentation. To reduce backsplash from the calorimeter and preshower radiator into the last tracking station, a 5~cm-thick low-Z absorber material (graphite) is placed in front of each layer of tungsten and between the final scintillator module and the calorimeter. The scintillator modules are oriented vertically and the graphite absorber and tungsten radiator layers have the same 30~cm~$\times$~30~cm transverse size as the  scintillator modules. In total the preshower station has about 2.5 radiation lengths of material in front of the calorimeter.

\subsection{Calorimeters}

\label{sec:calodescription}

\begin{figure}
    \centering
    \includegraphics[width=\textwidth]{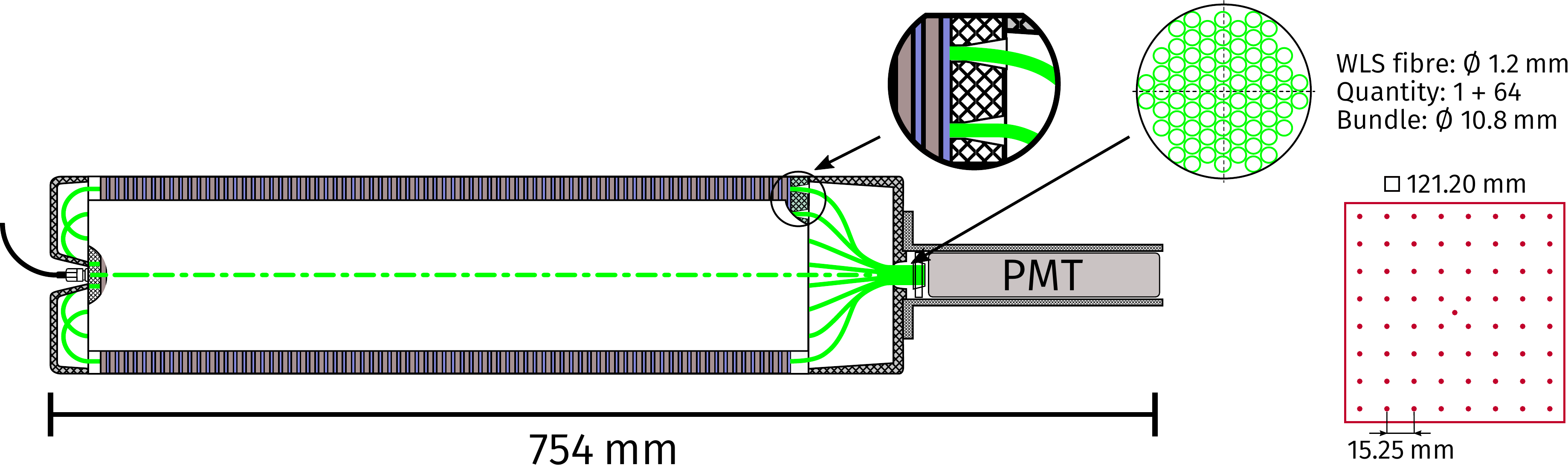}
    \caption{Design of a FASER (LHCb outer) calorimeter module.}
    \label{fig:CalorimeterDrawing}
\end{figure}

The calorimeter is constructed from spare LHCb outer ECAL modules~\cite{LHCB:2000ab}, shown in Figure~\ref{fig:CalorimeterDrawing}. The LHCb Collaboration has kindly agreed to allow FASER to use eight of these modules on indefinite loan.  
The modules are of so-called Shashlik-type  with interleaved scintillator and lead plates and 64 wavelength shifting fibers penetrating through the whole module and delivering the scintillation light  to a single PMT situated in a steel tube at the centre of the calorimeter's modules.
A single clear fiber penetrates  through the middle of the calorimeter in order to provide a light path for the LED calibration system, see Section~\ref{sec:LEDcalib}. The modules are \SI{754}{\mm} long, including the PMT, and have transverse dimensions of $\SI{121.2}{\mm}\times \SI{121.2}{\mm}$.  Each module has a total depth of 25 radiation lengths and consists of 66 layers of \SI{2}{\mm} lead and \SI{4}{\mm} plastic scintillator,  and 120 $\mu$m-thick Tyvek reflective paper.

\begin{figure}
    \centering
    \includegraphics[width=0.8\textwidth]{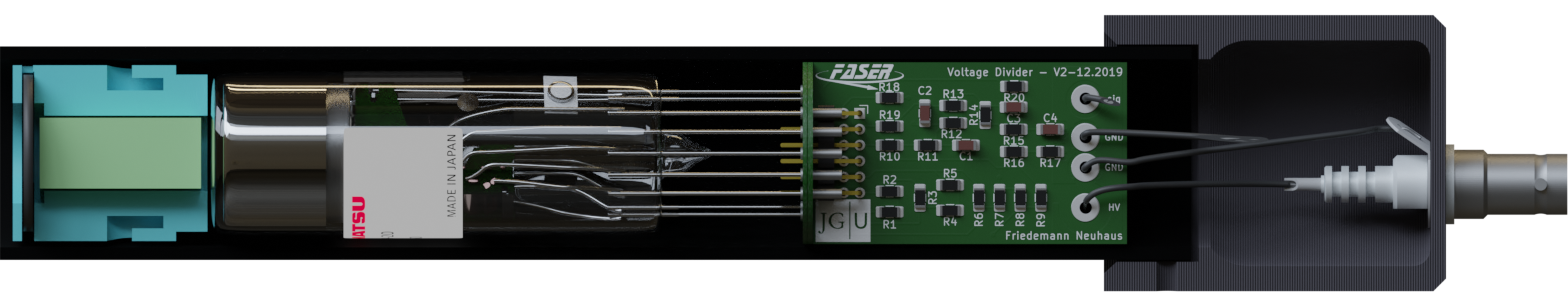}
    \caption{Design of a FASER PMT assembly with voltage divider
    and optical filter holder.}
    \label{fig:PMTAssemblyDrawing}
\end{figure}

The light from a calorimeter module is measured using the same Hamamatsu R7899-20 PMT~\cite{HamamatsuR7899} type as used by LHCb. This is a ten dynode-stage head-on PMT with a cathode diameter of \SI{22}{\mm} and a typical gain of up to $2\times10^6$. The voltage divider for the PMT was custom-built for FASER following the Hamamatsu recommendations for a tapered voltage-divider circuit in order to maintain good linearity for large pulses. The PMT and the voltage divider are situated inside the steel tube as illustrated in Figure ~\ref{fig:PMTAssemblyDrawing}. The PMT is in addition surrounded by a permalloy protection tube to reduce the impact of magnetic fields. In front of that is a \SI{32}{\mm} long, \SI{8}{\mm} wide rectangular polystyrene light mixer to reduce the non-uniformity of the PMT response. An absorptive neutral density filter with \SI{10}{\percent} transmission efficiency can be installed in the PMT assembly in front of the light mixer. This allows the PMT to be operated at higher gain where the non-linearity is small, see Section~\ref{sec:pmttests}, without saturating the readout electronics for energy deposits up to \SI{4}{\TeV}.

\begin{figure}
    \centering
    \includegraphics[width=0.75\textwidth]{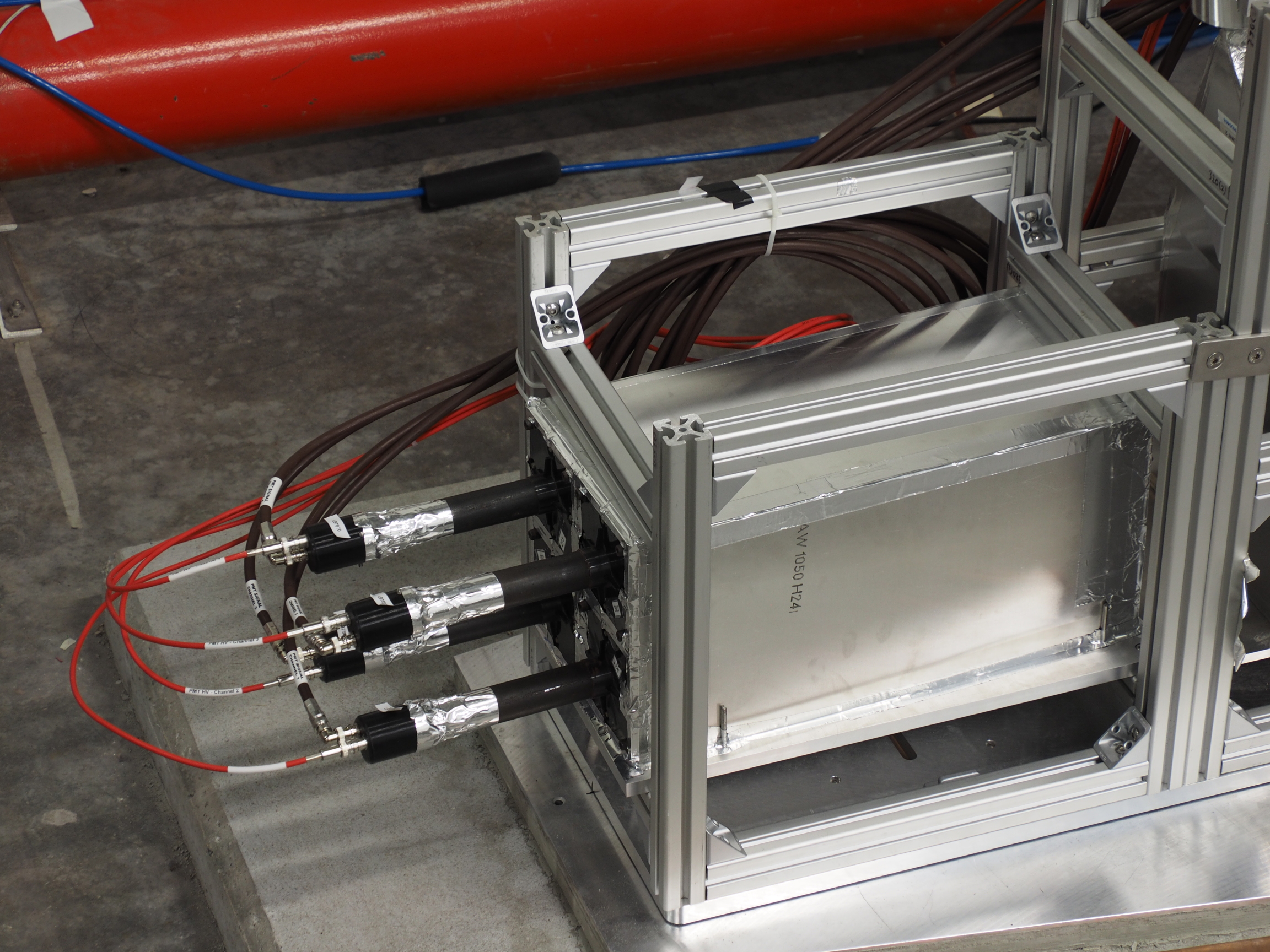}
    \caption{Assembled FASER calorimeter.}
    \label{fig:CalorimeterPhoto}
\end{figure}

The full FASER acceptance is covered by four calorimeter modules in a $2\times 2$ configuration. To avoid insensitive regions along the scintillating fibers and the gaps between modules, the modules are tilted horizontally and vertically by \SI{50}{mrad} with respect to the LOS. To avoid light ingress into the calorimeter modules, all sides of the calorimeter are covered in 0.5\,mm thick aluminium plates and gaps between the modules at the end are covered in aluminium tape as shown in Figure~\ref{fig:CalorimeterPhoto}. After installation in TI12, it was found that a measurable amount of electronics noise was picked up in the PMT assemblies from a nearby 4G GSM antenna which is part of the tunnel safety system. A Faraday-cage made from aluminium plates and connected to the detector ground plane was therefore constructed and installed around the
calorimeter PMTs.

\begin{figure}
    \centering
    \includegraphics[width=0.7\textwidth]{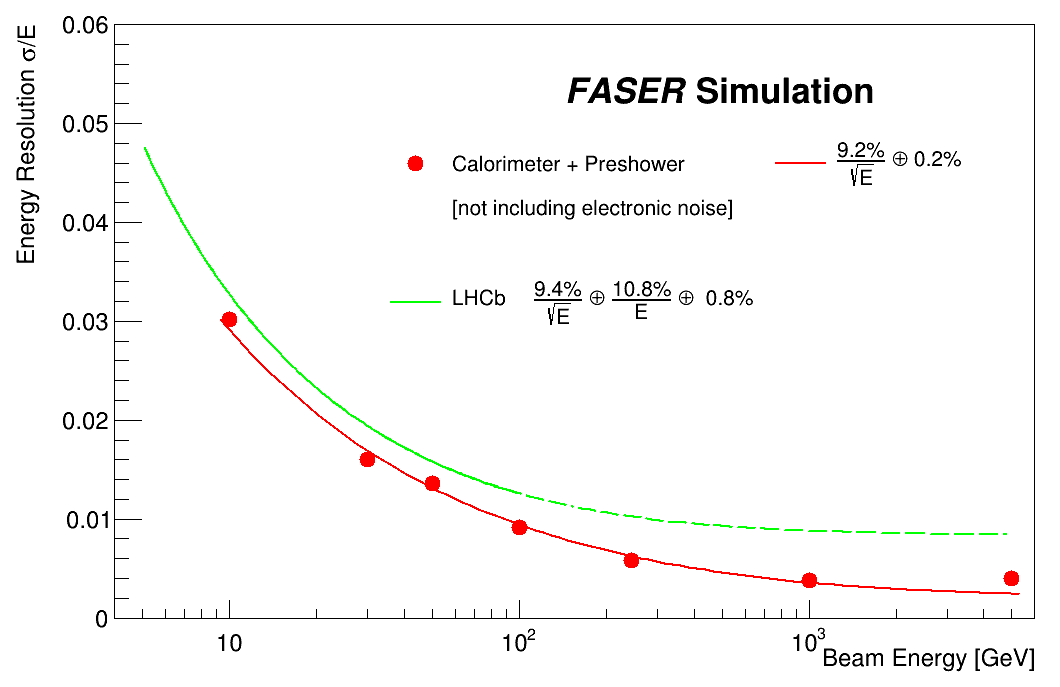}
    \caption{FASER calorimeter electron energy resolution from simulation as a function of electron energy compared to LHCb measurements (solid green line) and extrapolations (dashed green line). The FASER simulation does not include contributions from electronic noise. }
    \label{fig:CalorimeterResolution}
\end{figure}

From previous LHCb measurements~\cite{Arefev:2007zz} and simulation studies about 2500 photo-electrons (250 when the neutral-density filter is installed) are expected per GeV of electron or photon energy. Figure~\ref{fig:CalorimeterResolution} shows the predicted energy resolution estimated from simulation. Corrections for energy deposits in the extra material from the preshower station in front of the FASER calorimeter are also taken into account. The simulation results are compared to those from LHCb, where the solid line indicates the energy range probed by real measurements and the dashed line extrapolates the behaviour to higher energy. The predicted FASER  energy resolution is given by
\begin{equation}
    \frac{\sigma_E}{E} = \frac{9.2\%}{\sqrt{E}} \oplus 0.2\%
\end{equation}
 The simulation does not fully capture the expected 1\% constant term and does not include contributions from  electronics noise, which will partially depend on the energy range the system is operated over. 
The resolution degrades at energies above \SI{1}{\TeV} due to leakage out of the back of the calorimeter. At \SI{1}{\TeV}, about 1.6\% of electrons are expected to leak more than $3\%$ of their energy, while for \SI{5}{\TeV} electrons 6.5\% of them will lose more than 3\% of the energy out of the back.
In addition there will be a significant systematic uncertainty on the overall energy scale as there are no high-energy electromagnetic signals in-situ which can be used for an absolute scale calibration.

\subsection{Calibration system}

\label{sec:LEDcalib}

A common practice for calibrating the energy response of calorimeters is using energy deposits from
known processes e.g. $Z\to e^+ e^-$ in ATLAS~\cite{atlascollaborationElectronPhotonEnergy2019} and CMS~\cite{Sirunyan_2021}.
The calorimeter of the FASER experiment does not have this possibility with the only
charged particles crossing the detector at high rates being muons. 
However, as they are minimum ionizing, the signal from muons is quite low and can not be detected at
the low gains the PMT will be operated at. The only way to use muons to calibrate the energy is
to run at higher gains and then extrapolate down to the nominal operating range.
The accuracy of this extrapolation is ensured by injecting a known amount of light into the module while lowering the PMT bias voltage in discrete steps and tracking the change in amplitude.
To realize this, a dedicated LED-based calibration system was constructed. It is connected to the calibration port in the front of the calorimeter modules as can be seen in  Figure ~\ref{fig:CalorimeterDrawing}.
The calibration system is also connected to all scintillator counter PMTs where it will be used to
generate test pulses and monitor the long-term stability through regular dedicated calibration runs when there is no beam in the LHC.

In order to produce short light pulses, blue LEDs~\footnote{MULTICOMP PRO OVL-5523} are driven by a
commonly used circuit originally proposed by J.S. Kapustinsky~\cite{KAPUSTINSKY1985612}.
The circuit works by quickly discharging a capacitor through the LED resulting in a short flash of
light.
To reduce the duration of the pulse, an inductor is put in parallel with the LED, producing  a current opposite to the current from the discharging capacitor. 
The capacitor and inductance values are chosen such that the pulses have sufficient amplitudes to
feed all four calorimeter modules using a single LED. The values chosen are \SI{10}{\nano\farad} and
\SI{220}{\nano\henry}.

The calibration system features two independent LEDs with one channel used to drive the calorimeter
and the second channel used for the scintillator counters. To ensure good light yield, a custom fibre
bundle with four \SI{1}{\milli\meter} diameter fibres is used to guide the light to the calorimeter
modules. As the scintillator stations are operated at much higher gains, commercial off-the-shelf
optical fibres are used. To split the light from the LED into multiple outputs, the LED is
positioned in front of an MPO-to-LC fan-out. The individual scintillator counters are then connected using
LC-LC fibres. 

\begin{figure}
    \centering
    \includegraphics[width=0.7\textwidth]{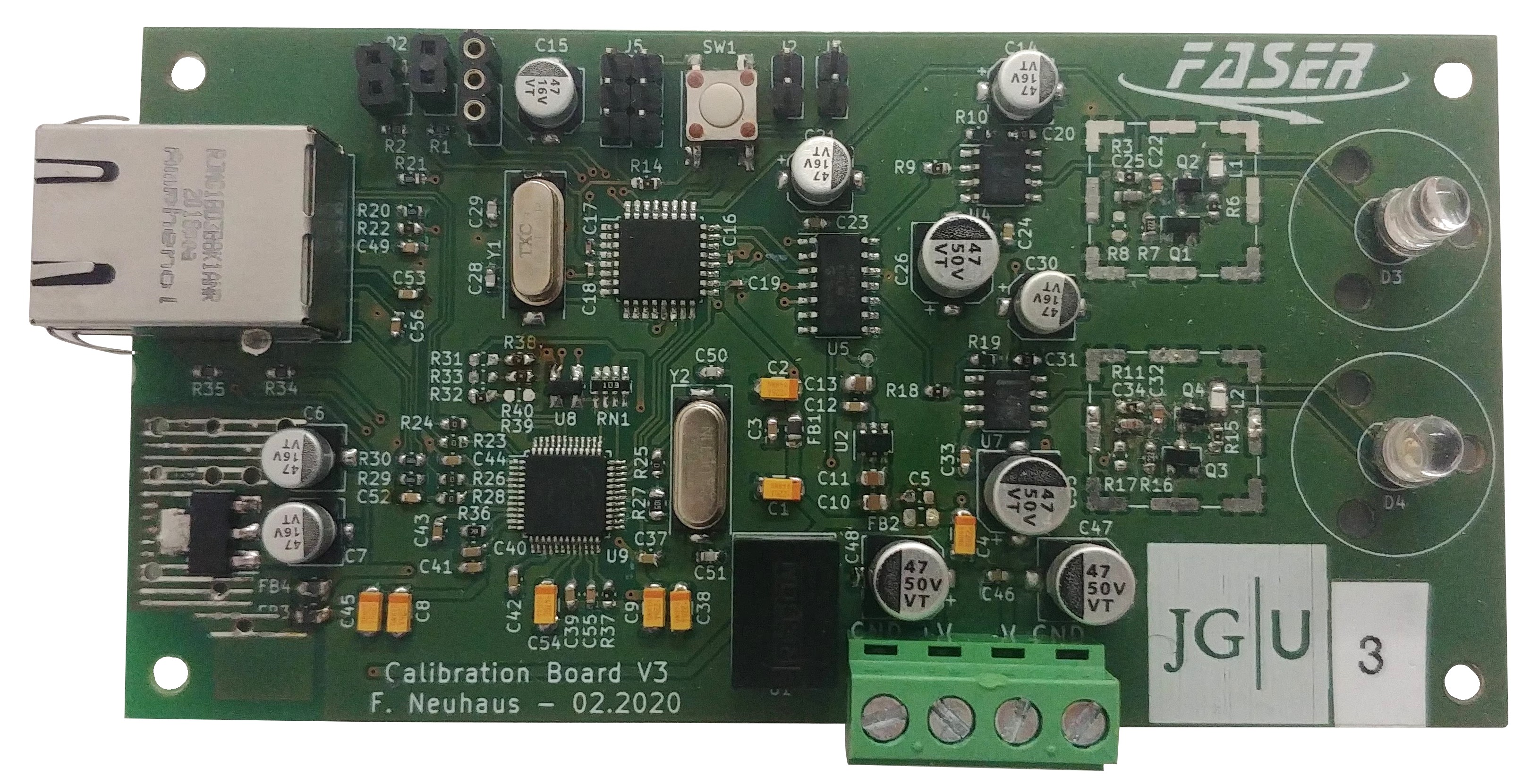}
    \caption{Photograph of the calorimeter calibration system board.}
    \label{fig:calibration-system-board}
\end{figure}

The calibration system is controlled by an 8-bit microcontroller.~\footnote{Microchip ATmega328p}
The system provides a simple HTTP-API to set and query all relevant parameters such as pulse frequency and amplitude as well as to enable and disable the two outputs. 
The frequency can be set in a range from \SI{0.25}{\hertz} to a few hundred \si{\hertz}. For every
pulse an additional TTL-signal is sent to the TDAQ system for triggering on the calibration pulses.
The bias voltages are set by a 12-bit two channel DAC in combination with inverting op-amps, yielding a range between \SIrange{0}{-22.5}{\volt}.
A picture of the calibration board is shown in Figure~\ref{fig:calibration-system-board}.

In addition, the calibration board can be used to produce a random trigger for testing the DAQ behaviour. In this mode the LED drivers are disabled and only TTL-signals are generated.

\subsection{Pre-installation Commissioning}

Before installation, all PMTs, scintillator counters, calorimeter modules and the calibration system, including spares, underwent a series of tests to ensure they were fully functional and to select the best performing PMTs/modules. The tests used a combination of light and gamma ray radiation sources as well as cosmic rays.~\footnote{The calorimeter performance was also characterized at the CERN SPS test beam using a combination of electron, muon and pion beams, the results of which will be presented in an upcoming paper~\cite{testbeampaper}.}

\subsubsection{Standalone PMT tests}
\label{sec:pmttests}

\begin{figure}
    \centering
    \includegraphics[width=0.65\textwidth]{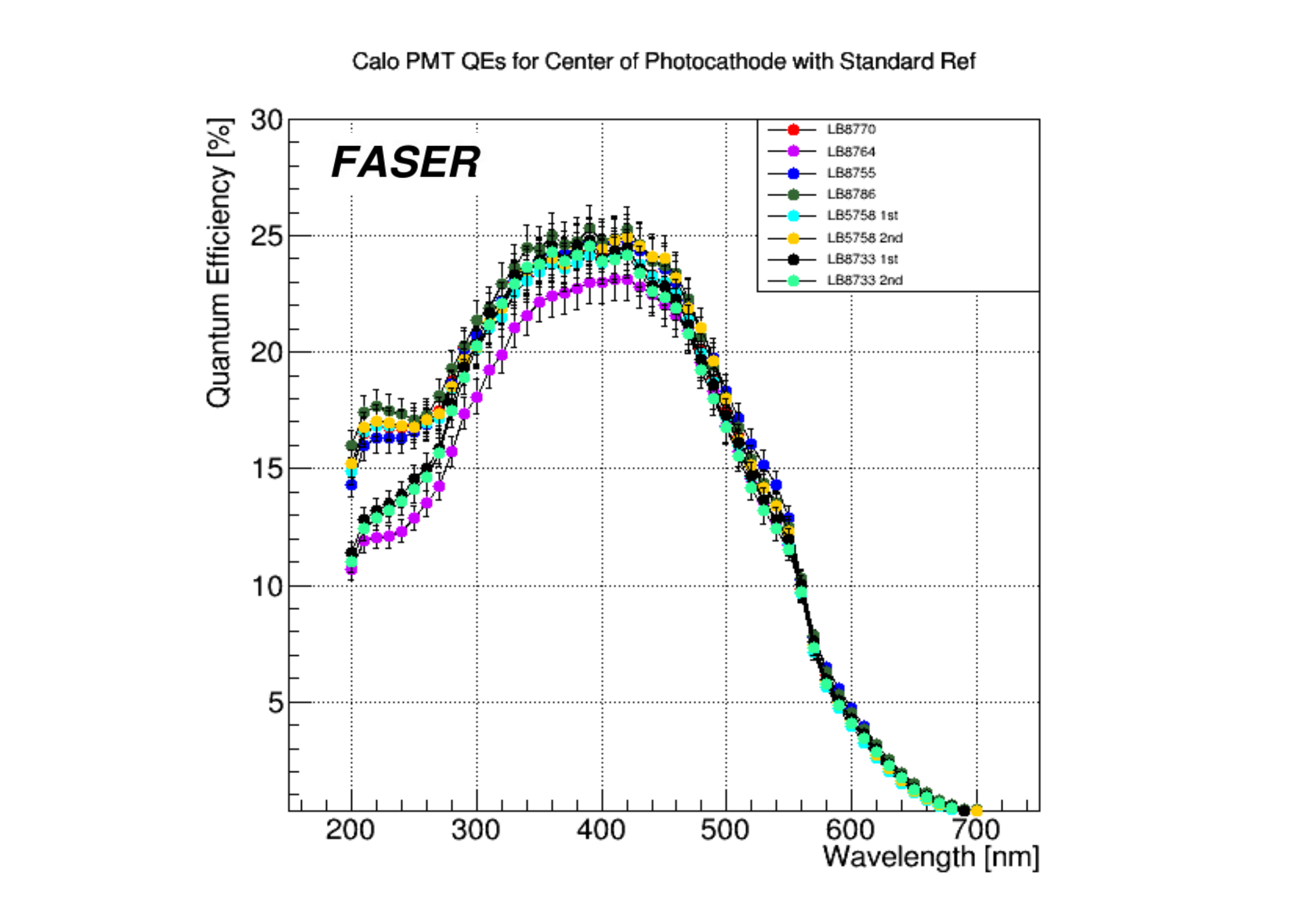}
    \caption{Measurement of the calorimeter PMT quantum efficiency as a function of photon wavelength. Two PMTs were measured twice to check the reproducibility of the measurement.}
    \label{fig:caloquantumefficiency}
\end{figure}

Seven PMTs were characterized for use in the calorimeter. The quantum efficiency of each PMT as a function of the photon wavelength was measured using a dedicated setup at CERN. For most wavelengths, the quantum efficiency was consistent between all but one PMT (LB8764) as shown in Figure~\ref{fig:caloquantumefficiency}. The emission peak for the Kuraray wavelength shifting fiber \cite{kuraray} is found at about \SI{495}{\nm} where the efficiency for all PMTs is around \SI{17}{\percent}. The quantum efficiency for the scintillator counter PMTs was not measured as this would have required unsoldering the voltage divider from the pre-assembled PMTs.

\begin{figure}
    \centering
    \includegraphics[width=0.5\textwidth]{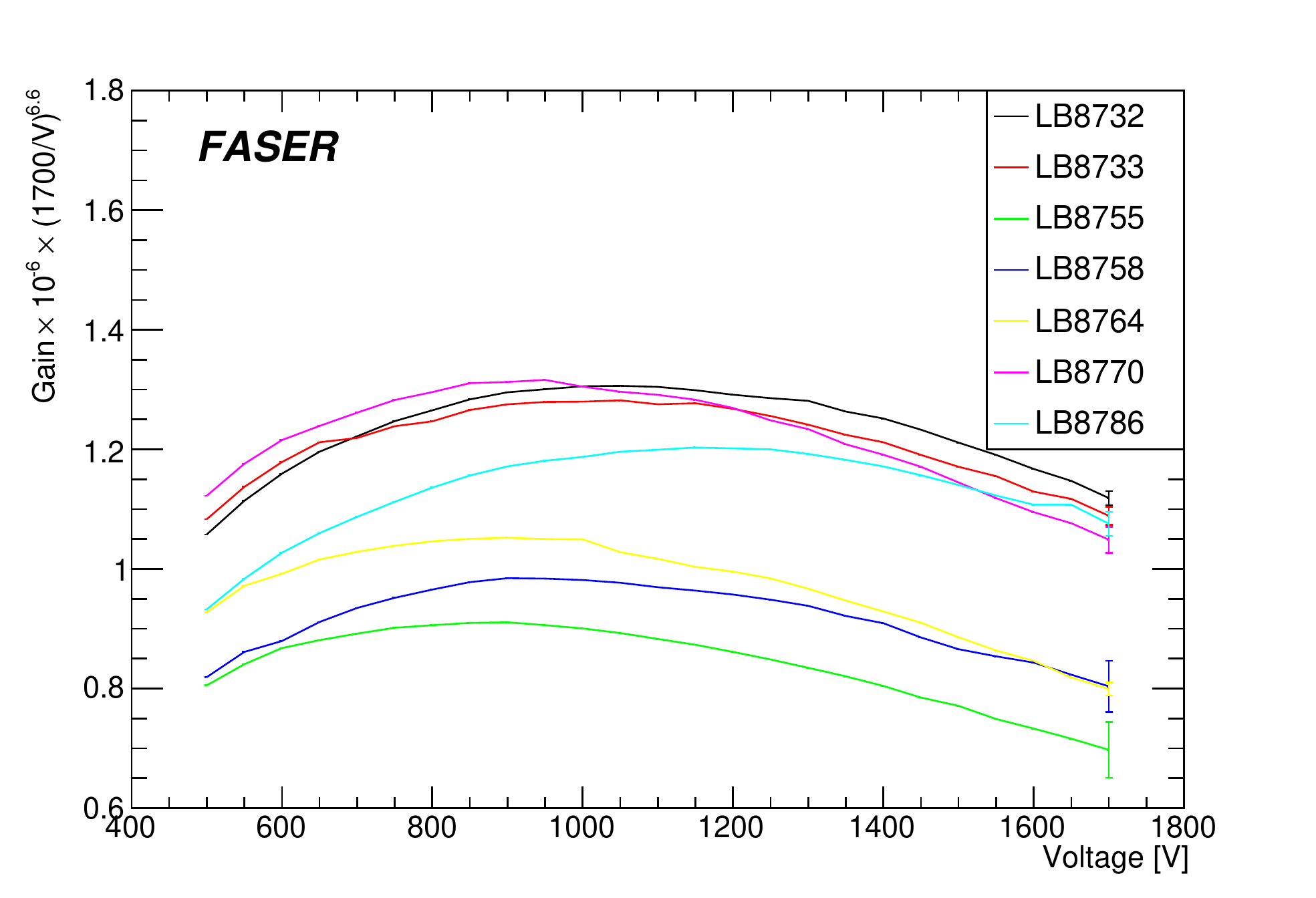}
    \caption{Measured calorimeter PMT gain as a function of applied voltages for the different PMTs measured using LED pulses. The absolute gain is measured at  \SI{1700}{\volt} using single photon-electron signals and extrapolated to lower voltages using a larger fixed pulse size. The gains are scaled by $10^{-6}\times\left(\frac{\SI{1700}{\volt}}{V}\right)^{6.6}$ in order to easily compare the behaviour over the full measurement range. The measurement uncertainty of the absolute gain is shown with the error bars at \SI{1700}{V} and applies to the full curve.}
    
    \label{fig:pmtgaincurves}
\end{figure}

The absolute gain of the PMTs was measured using single photons from a flashing, very low intensity LED. 
Results are shown in Table~\ref{tab:scintresults} for the scintillator counter PMTs considering minimum ionizing particle (MIP) signals. All measured gains are within a factor of two of the specified typical gain. 
For the calorimeter PMT, Figure ~\ref{fig:pmtgaincurves} shows the measured gain as a function of applied voltages for the different PMTs measured using LED pulses. The absolute gain is measured at  \SI{1700}{\volt} using single photon-electron signals and extrapolated to lower voltages using a larger fixed pulse size.  
The PMTs show the same voltage dependence within 15\%. The voltage dependence will be monitored in-situ using the LED calibration system discussed in Section~\ref{sec:LEDcalib}. 
The four calorimeter PMTs with gain above $10^6$ were selected for final installation, while for the scintillators, PMTs with similar gain were paired for neighbouring modules to provide better uniformity before voltage adjustments for a given station.

\begin{table}[]
    \centering
    \begin{tabular}{|c|cc||c|c|}
    \hline
        Station & Module & Gain  & \multicolumn{2}{c|}{MIP signal} \\
        & & ($\times 10^6$) & Efficiency & Most probable signal \\ \hline
         Veto station 1 & 1 & $4.85\pm0.07$& $>99.99\%$ & 205 p.e.\\
                        & 2 & $4.10\pm0.17$& $>99.95\%$ & 200 p.e.\\ \hline
         Veto station 2 & 1 & $2.69\pm0.02$& $>99.985\%$ & 285 p.e.\\
                        & 2 & $3.30\pm0.04$& $>99.995\%$ & 380 p.e.\\ 
                        & 3 & $4.19\pm0.07$& $>99.996\%$ & 360 p.e.\\ 
                        & 4 & $4.27\pm0.08$& $>99.991\%$ & 305 p.e.\\ \hline
         Timing station & 1, PMT 1 & $1.44\pm0.03$& $>99.7\%$ & 85 p.e.\\
                        & 1, PMT 2 & $1.74\pm0.03$& $>99.8\%$ & 135 p.e.\\ 
                        & 2, PMT 1 & $2.20\pm0.04$& $>99.8\%$ & 135 p.e.\\ 
                        & 2, PMT 2 & $2.48\pm0.05$& $>99.8\%$ & 115 p.e.\\ \hline
         Preshower      & 1 & $3.96\pm0.04$& $>99.96\%$ & 330 p.e.\\
                        & 2 & $4.73\pm0.05$& $>99.97\%$ & 370 p.e.\\ \hline
    \hline
    \end{tabular}
    \caption{Scintillator counter PMT gain measured with single photons at \SI{950}{V} for Veto station 1 and \SI{1700}{V} for the rest. The table also lists the scintillator module efficiency (at 95\% CL) for MIP signals and the most probable MIP signal expressed in photo-electrons as measured with cosmic ray muons. The efficiencies are estimated with a threshold of half the most probable MIP signal. The precision of the efficiency measurements varies between PMTs as some measurements were done with
    significantly larger samples of cosmic rays.}
    \label{tab:scintresults}
\end{table}

\begin{figure}
    \centering
    \includegraphics[width=0.55\textwidth]{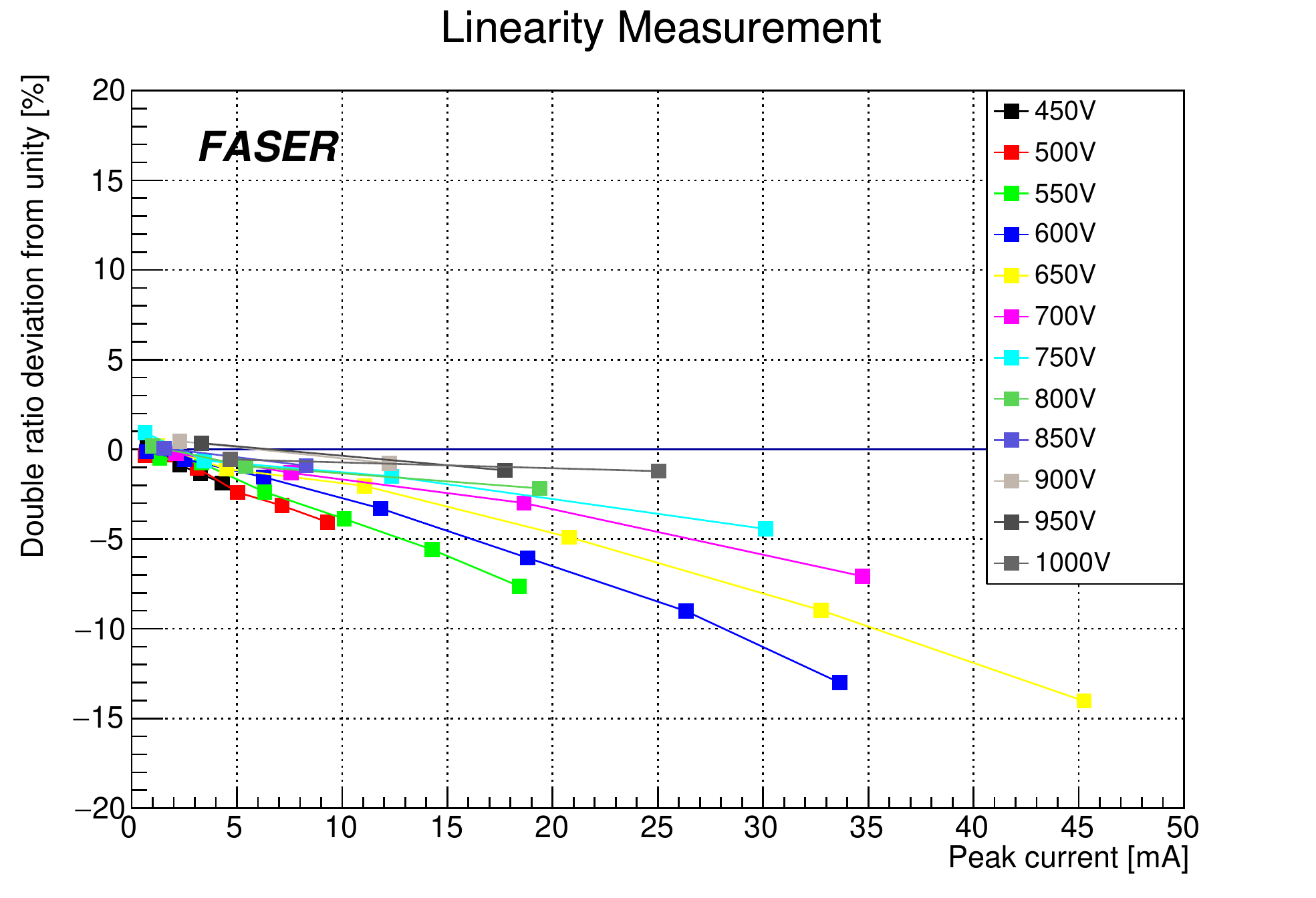}
    \caption{Measurement of the calorimeter PMT linearity for one PMT (LB8733) using a double pulsing system as explained in the text.}
    \label{fig:calolinearity}
\end{figure}

In calorimetry, a linear signal response is critical for an accurate measurement of deposited energy. 
The linearity of the PMT response as a function of the applied voltage was measured for each calorimeter PMT using two LED pulses separated by \SI{550}{\ns}, with the second pulse a fixed factor larger than the first pulse. The linearity is measured from the relative response to the two pulses as a function of the peak current of the larger pulse, normalized to the ratio measured in a linear regime. An example measurement is shown in Figure~\ref{fig:calolinearity} for one PMT and one pulse ratio. Only minor variations were found between the different PMTs, but $>5\%$ non-linearity was observed at the maximum readout range (peak current $<\SI{40}{\milli\ampere}$) for voltage settings below \SI{800}{\volt}. This motivates the use of an optical filter to enable physics operation at higher voltage. This measurement was carried out using the same digitizer as used in the final experiment (see Section~\ref{sec:digitizer}), thus demonstrating that non-linearity effects of the PMT are dominant as good linearity is seen for pulses measured at large high voltage settings.

Finally, the dark rate was measured as a function of signal threshold for different voltage settings and found to be very low. At \SI{1500}{\volt}, all but one of the calorimeter PMTs had a rate of less than \SI{10}{\hertz} for a threshold set to that expected for a three photo-electron signal. For the H6410 and H11934 scintillator counter PMTs, the rate was at or below \SI{10}{\hertz} for a one photo-electron signal threshold at \SI{2000}{\volt} and \SI{850}{\volt}, respectively.

\subsubsection{Scintillator counter measurements}
\label{sec:scintillatormeasurements}

\begin{figure}[hbt!]
    \centering
    \includegraphics[height=0.3\textwidth]{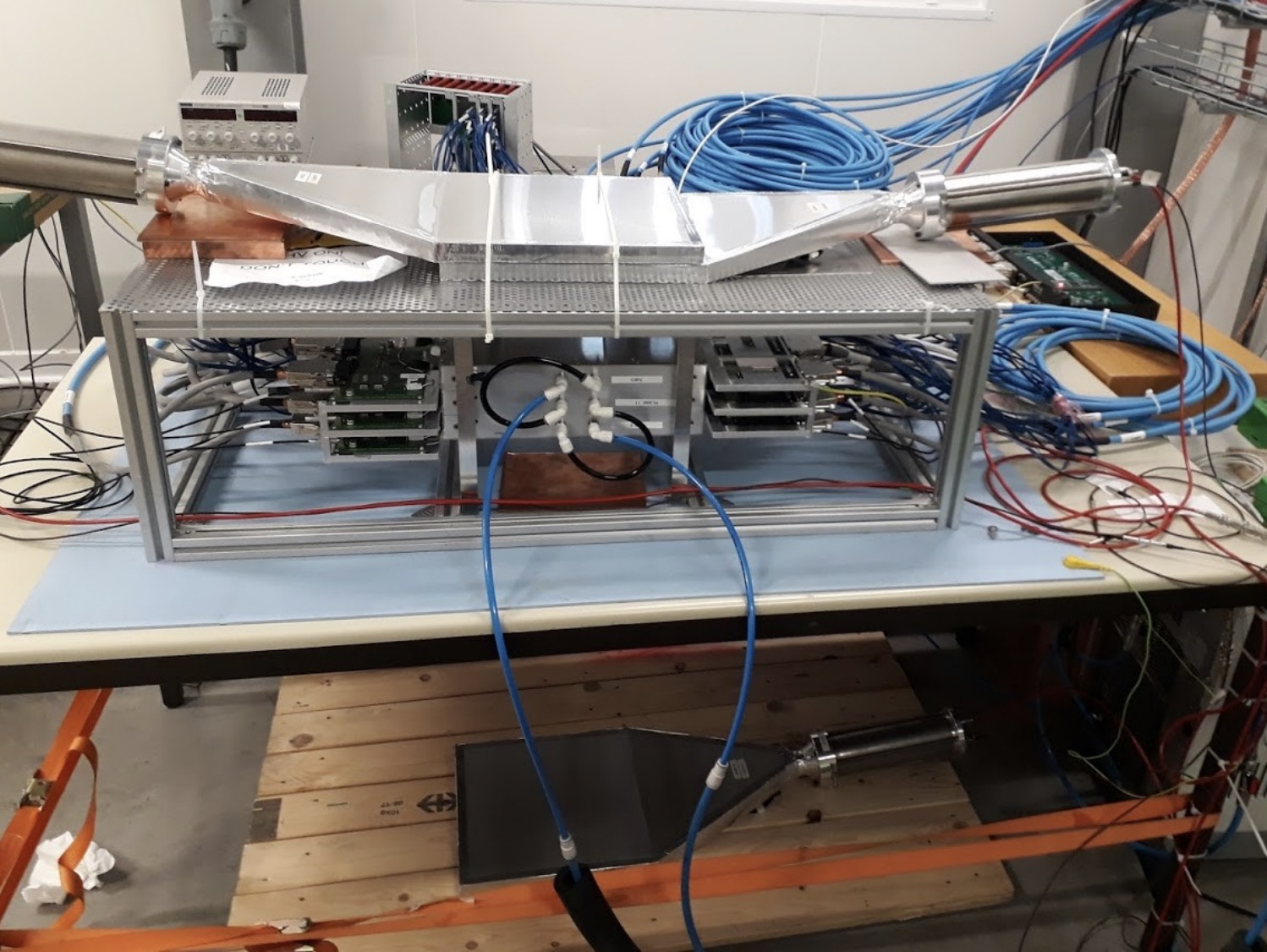}
    \includegraphics[height=0.3\textwidth]{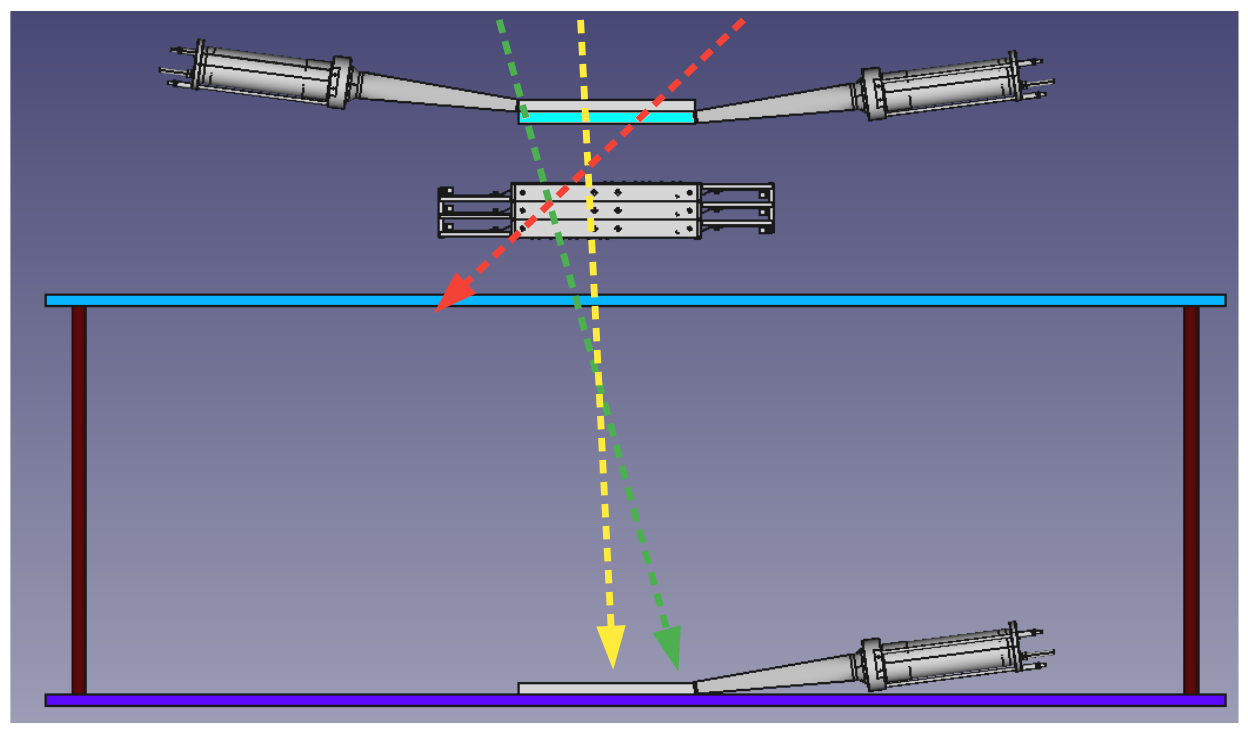}
\caption{Left: Detector setup for efficiency measurements. Right: Schematic diagram of the setup with incident cosmic ray muons. Two scintillator modules are placed above a tracker station, all elevated by 90~cm above a third scintillator counter at the bottom used to select mostly approximately vertical cosmic ray muons. The scintillator module under test is highlighted in cyan colour.}
    \label{fig:scint_effs}
\end{figure}

\begin{figure}
    \centering
    \includegraphics[width=0.65\textwidth]{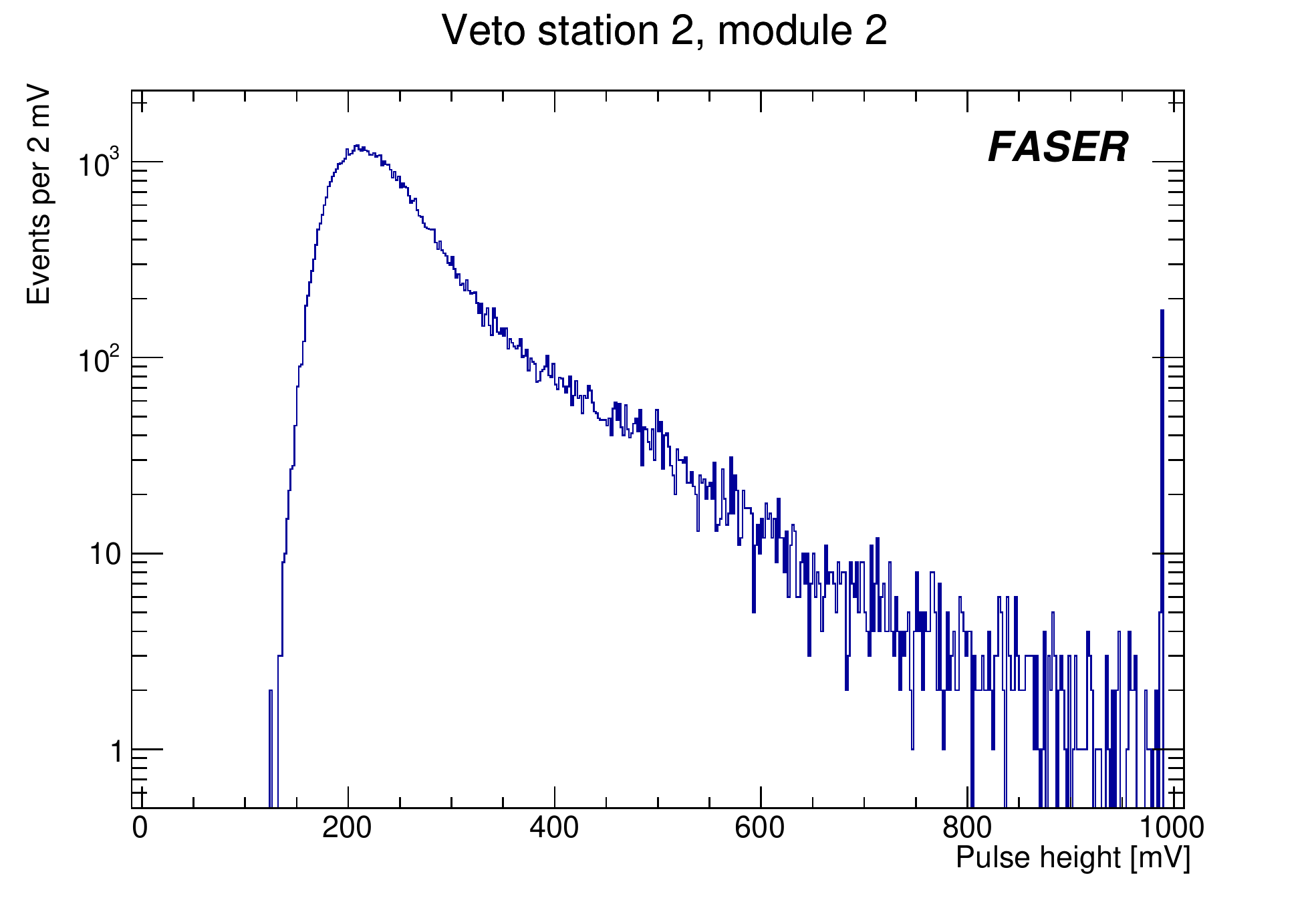}
    \caption{Pulse height distribution in veto station 2, module 2 PMT measured for cosmic ray muons selected independently using two other scintillator counters and a tracker station. The readout for the pulse height saturated at \SI{1000}{mV}.}
    \label{fig:scintEff}
\end{figure}

The performance of each scintillator module was measured using cosmic ray muons. Multiple scintillator counters were stacked vertically above and below the scintillator module under test with up to 5~cm of copper shielding placed above to select a clean sample of events.
For the veto station modules, additional and more precise measurements were performed by adding one tracker station below the scintillator module, as illustrated in Figure~\ref{fig:scint_effs}, to select a very pure sample of single particle cosmic rays. For each module, more than 50,000 events containing cosmic ray muons passing through the scintillator counters were collected. These samples were used to measure the muon signal in the scintillator counter and the detection efficiency for muons for a detection threshold of half of the most probable minimum ionizing particle signal. An example distribution of the measured signal is shown in Figure~\ref{fig:scintEff} where no events below the selected threshold are observed out of more than 65,000 events. The resulting limits on the scintillator counter efficiency are shown in Table~\ref{tab:scintresults} along with an estimate of the most probable light signal for a minimum ionizing particle. Only lower limits are shown since for all measurements the number of non-detected events are zero or just a few events. In the latter case, these events are consistent
with being due to non-muon background events with the exception of module 1 in veto station 2, where the measured efficiency is
$99.992\pm0.004\%$. This module has also a lower light yield than the rest of that type of scintillator modules. In all cases, 
the resulting efficiency when combining two modules in a station are well above those given by the requirements in Section~\ref{sec:scintsystem}.

\begin{figure}[hbt!]
    \centering
    \includegraphics[width=0.45\textwidth]{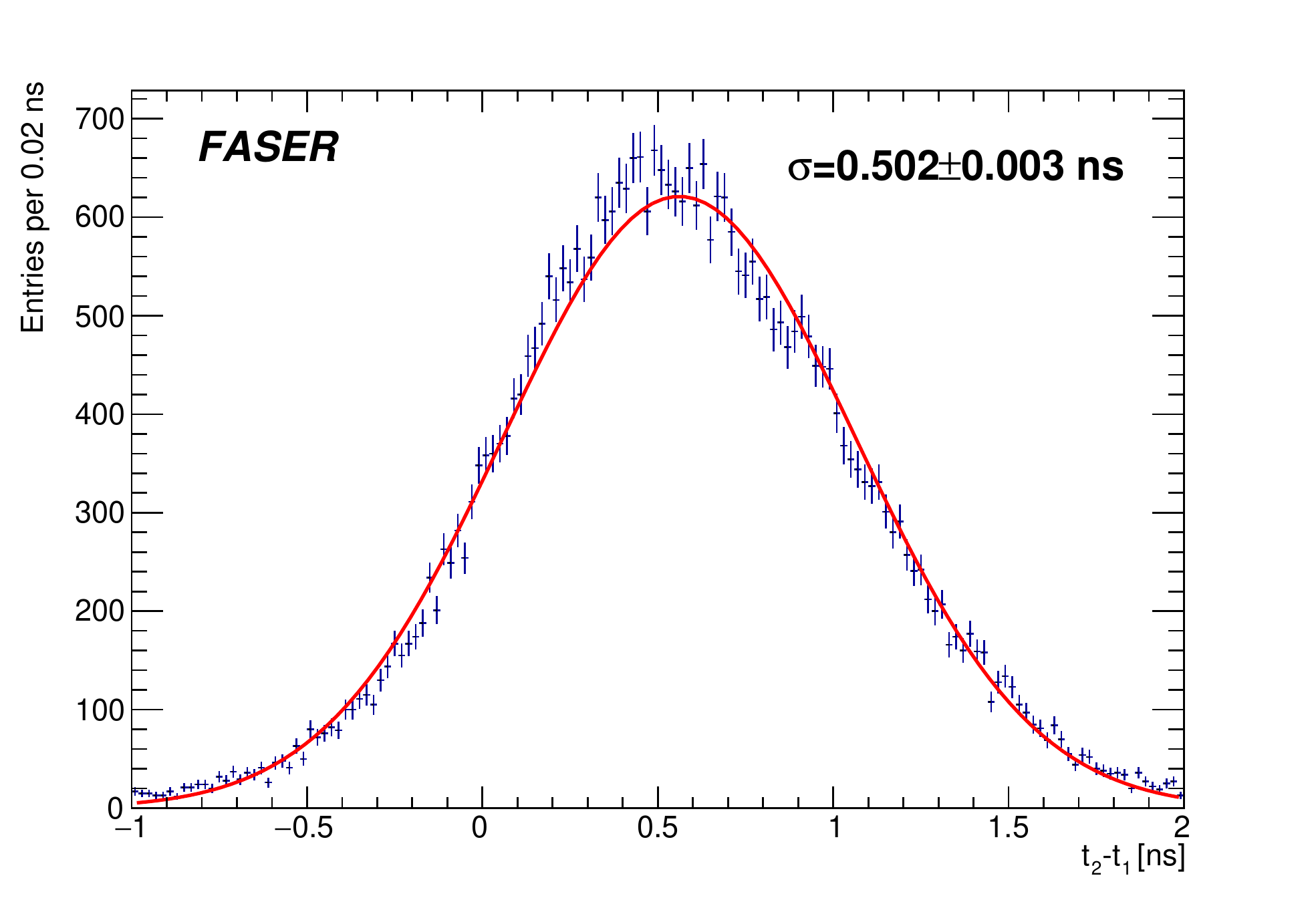}
    \includegraphics[width=0.45\textwidth]{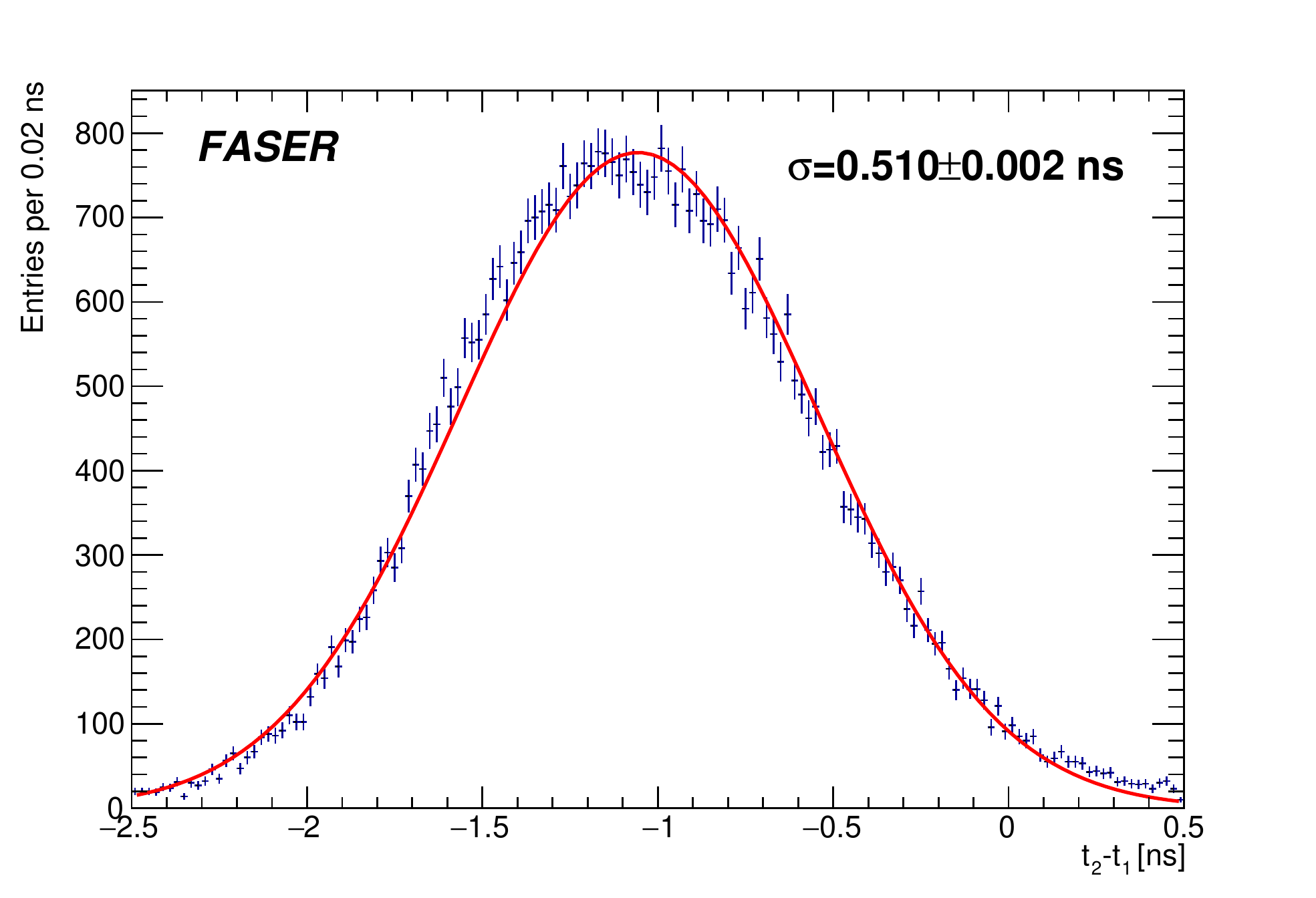}
\caption{The time difference in the reconstructed arrival time for the left and right PMT in a timing module
for cosmic rays \SI{15}{cm} (left) and \SI{28}{cm} (right) from the left edge (PMT 1) of the 40-cm wide scintillator  plate. The distribution have been fitted with a Gaussian function shown in red to measure the timing resolution.}
    \label{fig:timingspread}
\end{figure}

The timing resolution of the timing station modules was tested using cosmic ray muons in a similar setup. The other scintillator counters were arranged with a minimal overlap between them in order to select muons in a narrow region of the module. The time for a signal to arrive at the PMT at each end of the module depends on the location at which the muon traversed the scintillator counter. By using the time difference between the PMTs on each  side it is possible to extract the time resolution. 
The spread in time difference between PMTs was measured at two known points along the length of the timing scintillator. As shown in Figure~\ref{fig:timingspread}, it is about \SI{500}{ps}. This implies a precision on a single PMT timing measurement of about \SI{350}{ps} and therefore, assuming measurements are uncorrelated, an expected precision of the average signal time measurement from both PMTs of \SI{250}{ps}. This is well below the \SI{1}{ns} requirement and close to the intrinsic \SI{180}{ps} spread in the collision time.

\subsubsection{Standalone calorimeter tests}

\begin{figure}
  \centering
  \definecolor{JGU_red}{RGB}{193,0,42}
  \begin{tikzpicture}[
    outer sep=0, JGU_red,
    labelBox/.style={ text opacity=1, JGU_red, fill=white, fill opacity=0.7, inner sep=5pt, align=center },
    m/.style={ -Circle, JGU_red, thick }
  ]
    \def\factor{0.47}
    \node [inner sep=0] (image) {\includegraphics[width=\factor\textwidth]{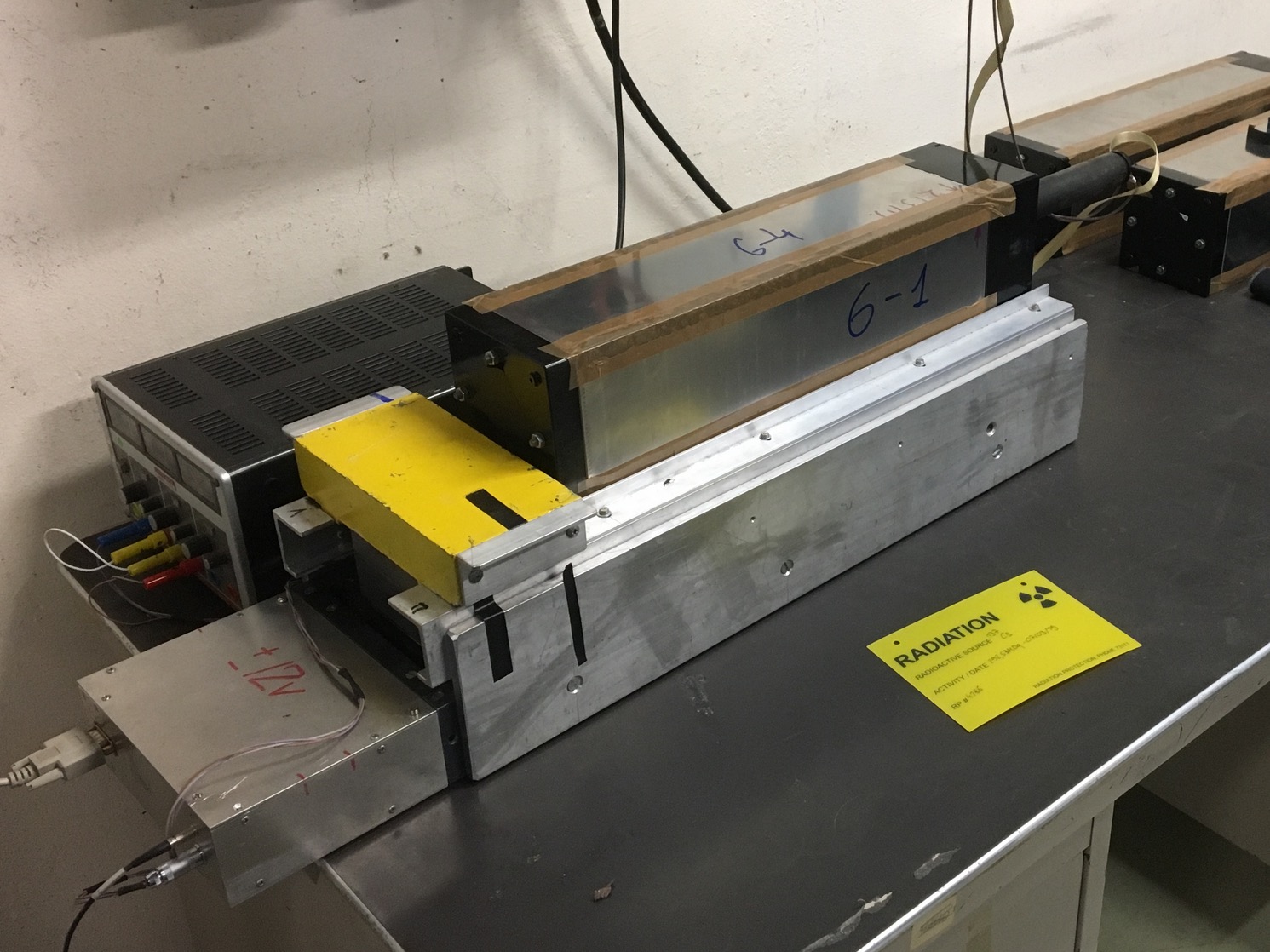}};
    \node [labelBox, anchor=north, below left=\factor*0.5cm and \factor*0.17cm of image.north]%
      (module_under_test) {Module under test};
    \draw [m] (module_under_test) -- (-\factor*0.17cm, \factor*2);
    \node [labelBox, align=center, anchor=east, above=\factor*0.5cm of image.south, xshift=\factor*0.1cm]%
      (source) {$^{137}\textrm{Cs}$-source\\(\SI{662}{\keV} $\gamma$)};
    \draw [m] (source.north) -- (-\factor*2.2cm, \factor*0.3cm);
    \node [labelBox, align=center, anchor=south east, above=\factor*4cm of image.south east,
    xshift=-\factor*2cm]%
      (pmt) {PMT};
    \draw [m] (pmt) -- (\factor*5.1cm, \factor*3.45cm);
    \node [below=0.12cm of image.south, inner sep=0pt] {};
  \end{tikzpicture}\hfill%
  \begin{tikzpicture}[JGU_red, outer sep=0, scale=1, label/.style={scale=0.8, anchor=west}]
    \node [anchor=center, scale=1] {%
      \includegraphics[width=0.47\textwidth]{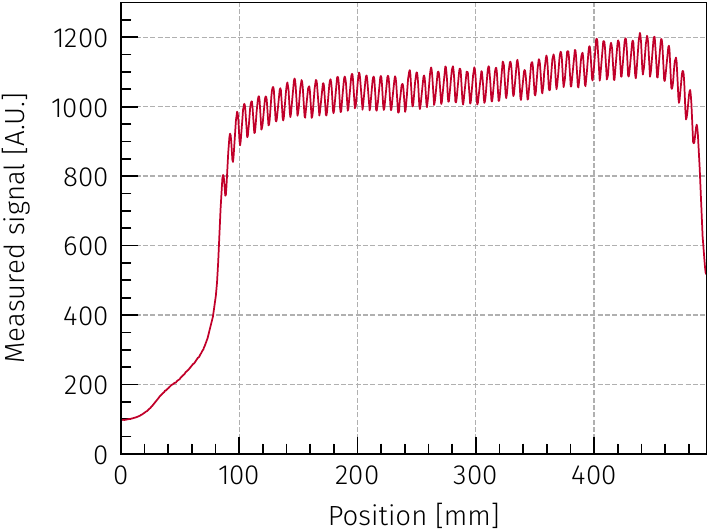}
    };
    \node [label] at (-1.4cm, -1.6cm) (source) {Source entering active area};
    \draw [-Latex, thick] (source) -- (-1.5cm, -1cm);
  
    \node [label] at (-2.3cm, 2.4cm) (layer) {Scanning the layers};
    \draw [-Latex, thick] (layer.south) -- (-0.3cm, 2.1cm);
  
    \node [label, align=center] at (0.3cm, 0.5cm) (rise) {Rise due to lower\\light attenuation};
    \draw [-Latex, thick] (rise) -- (2.5cm, 1.8cm);
  \end{tikzpicture}
  \caption{%
    LHCb test setup used to scan the calorimeter modules (left) and the measured response for one side of a calorimeter module (right).
  }\label{fig:lhcb-calorimeter-scan}
\end{figure}
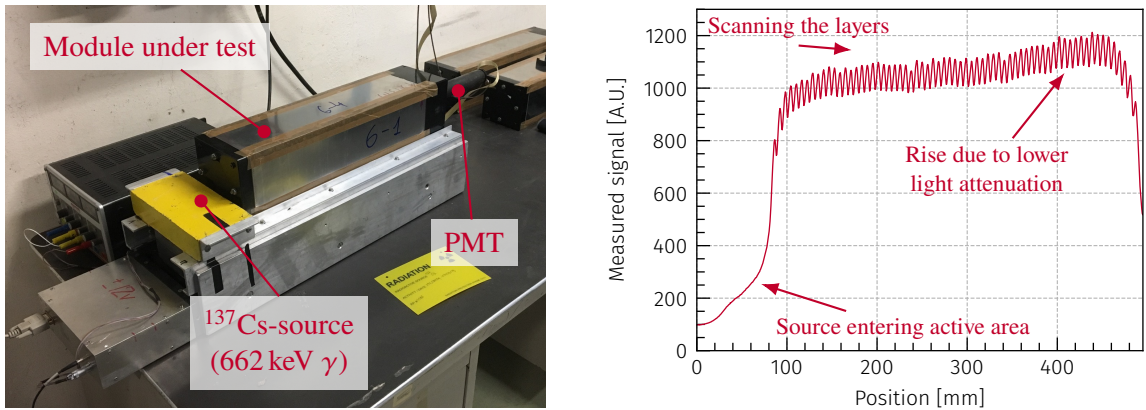

Eight calorimeter modules were tested to select the four final modules. An initial test was done using a $^{137}\text{Cs}$ radioactive source scanned along each side of the calorimeter module to confirm that all modules were fully functional after more than ten years in storage. The setup, along with the results for one side of a module, are shown in Figure~\ref{fig:lhcb-calorimeter-scan}. 

\begin{figure}
    \centering
    \includegraphics[width=0.25\textwidth]{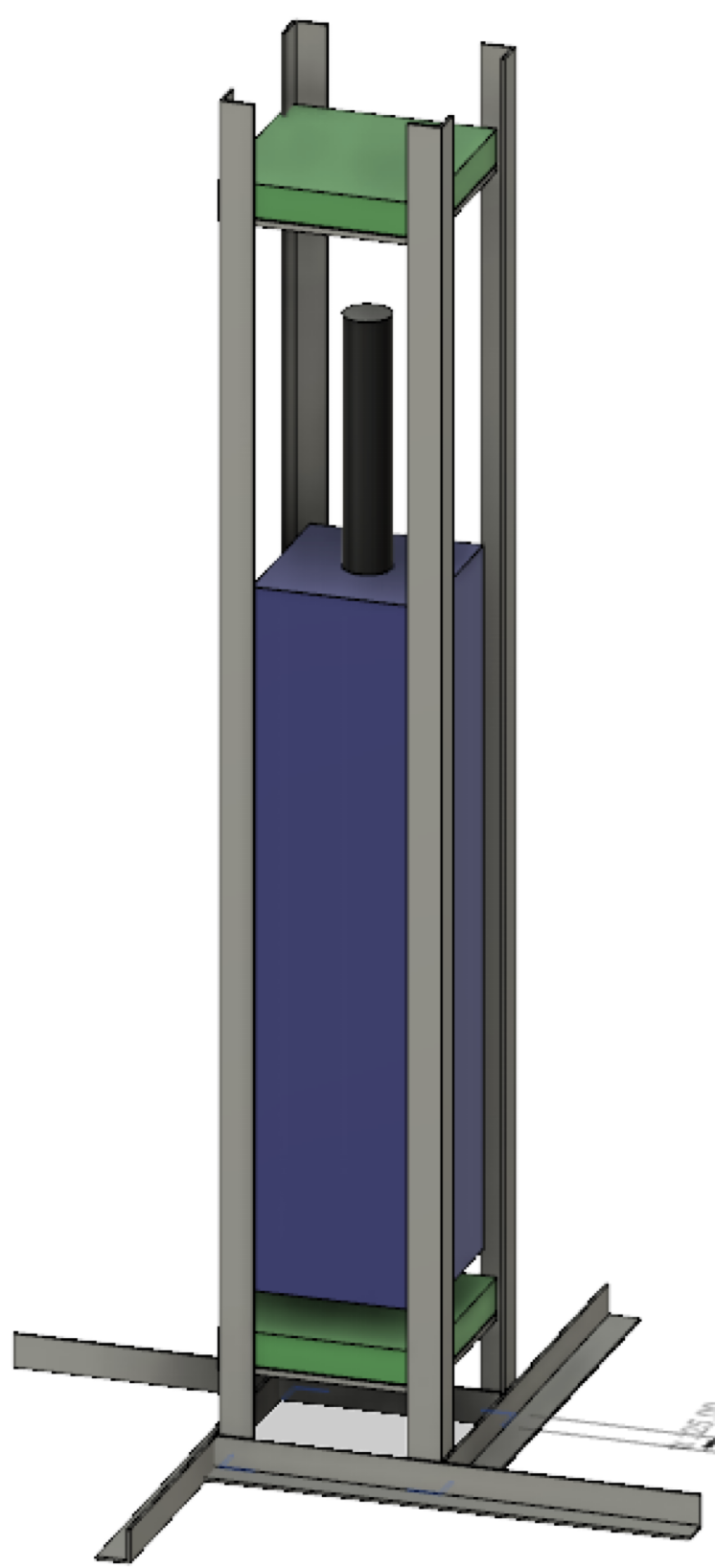}
    \includegraphics[width=0.67\textwidth]{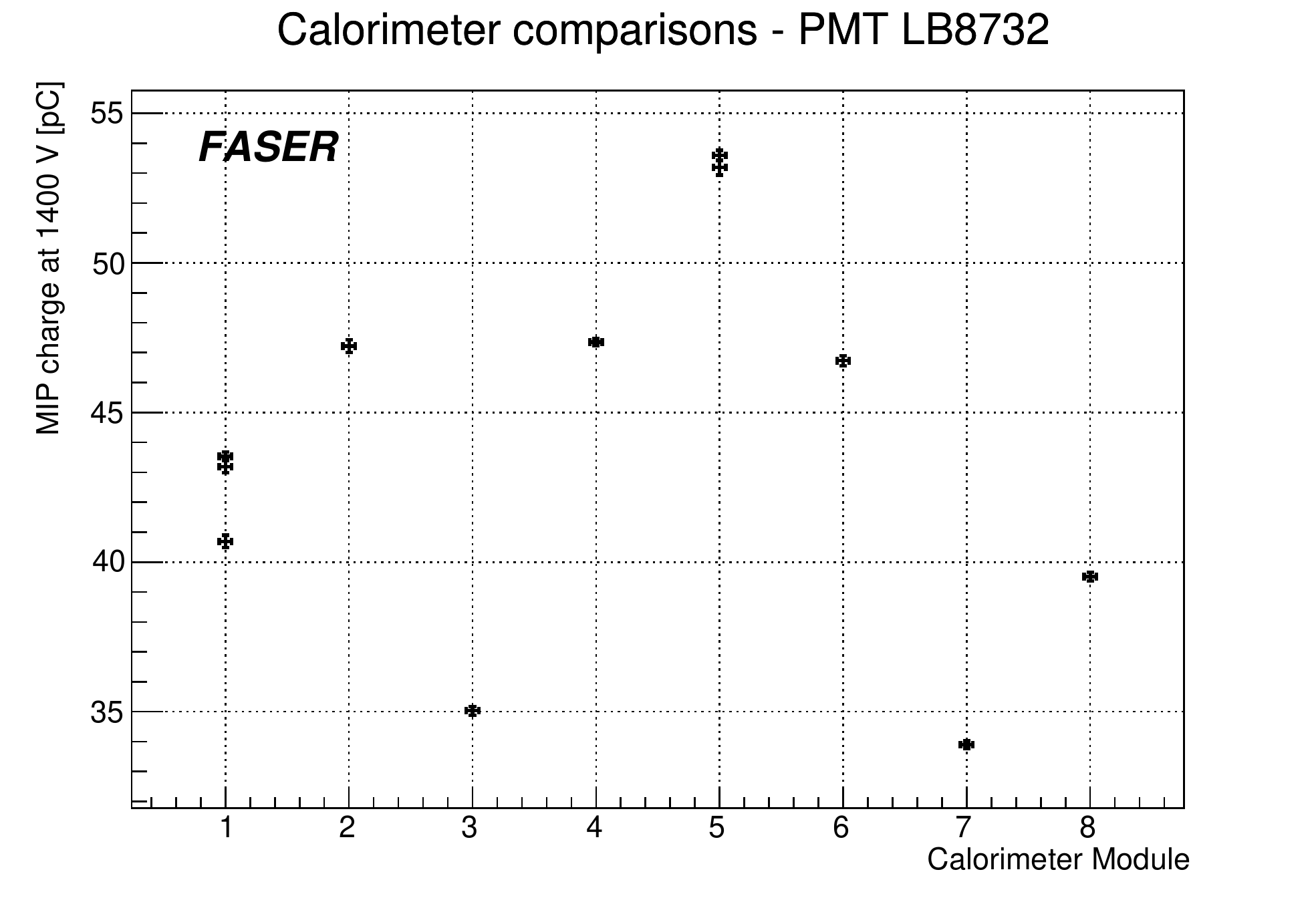}
    \caption{Left: Calorimeter cosmic ray test stand with scintillator counters for triggering shown in green above and below the module under test. Right: Calorimeter signal for cosmic ray muons as measured with the same PMT at 1400 V. For module 1 and 5, several measurements where done with the PMT removed and reinserted at a different rotation between measurements.}
    \label{fig:calocosmics}
\end{figure}

\begin{figure}
    \centering
    \includegraphics[width=0.65\textwidth]{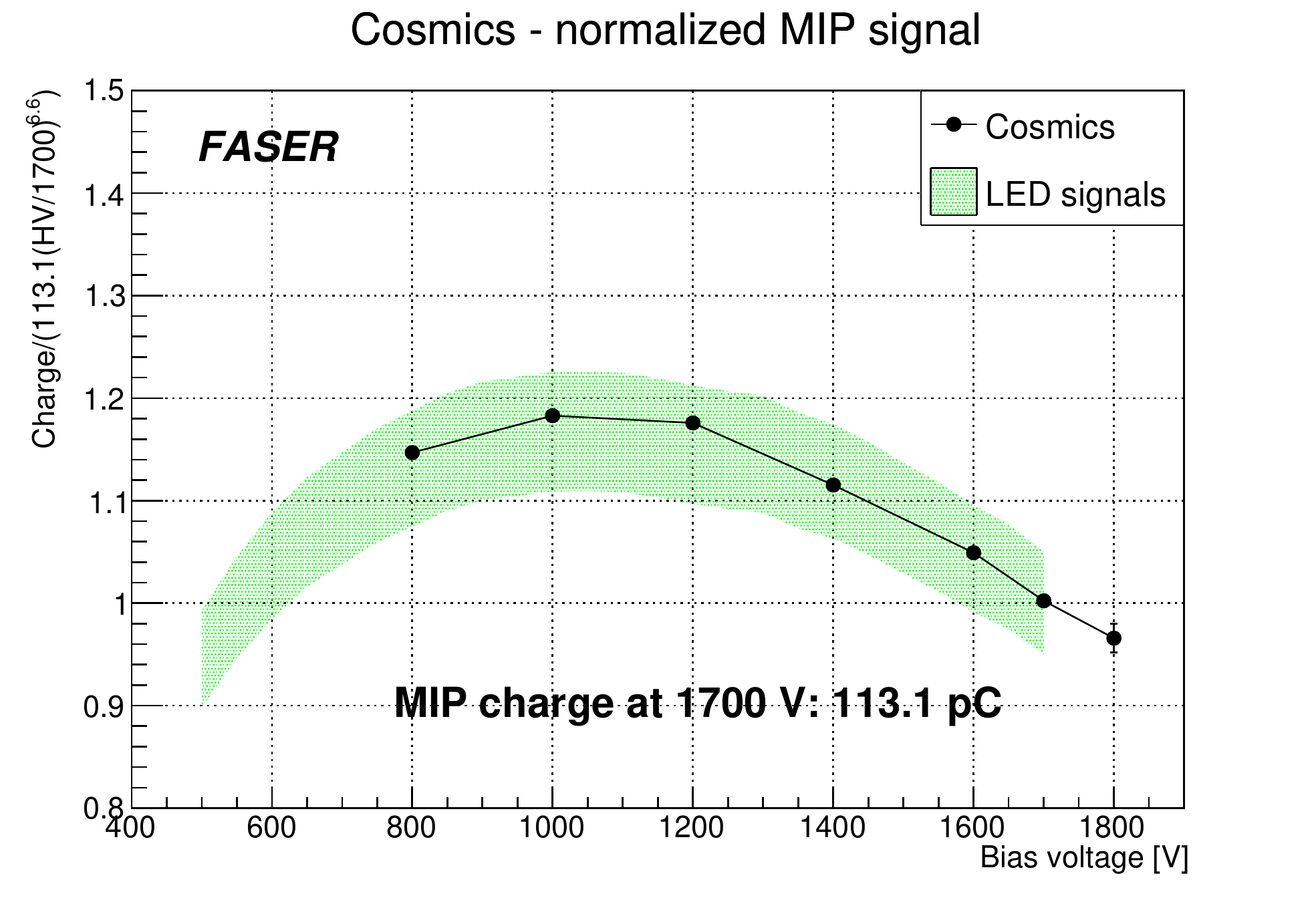}
    \caption{Calorimeter signal for cosmic ray muons as measured with the one PMT at different bias voltages compared to voltage dependence measured using LED signals. The signals have been normalized to 1 at 1700~V and the $V^{6.6}$ dependence divided out to ease comparison. The green band corresponds to the 5\% uncertainty assigned to the LED measurement systematic uncertainty.}
    \label{fig:calocosmicsvsHV}
\end{figure}

The calorimeter performance was measured using cosmic rays with the setup shown in Figure~\ref{fig:calocosmics} (left). 
Scintillator counters above and below the calorimeter module were used to trigger on vertical cosmic rays and the deposited charge was measured at different PMT voltage settings. Figure~\ref{fig:calocosmics} (right) presents a comparison of the response for the eight modules, all measured using the same PMT at \SI{1400}{\volt}. All modules demonstrated good performance and modules 2, 4, 5 and 6 were selected for installation as they had the most similar response. The dependence on the PMT voltage and the comparison to the dependence measured using LED signals is shown in Figure~\ref{fig:calocosmicsvsHV} for one specific PMT, LB8732. Agreement between cosmic ray and LED signals is found to be better than 5\% for all modules.

\subsubsection{LED calibration standalone tests}

\begin{figure}
    \centering
    \includegraphics{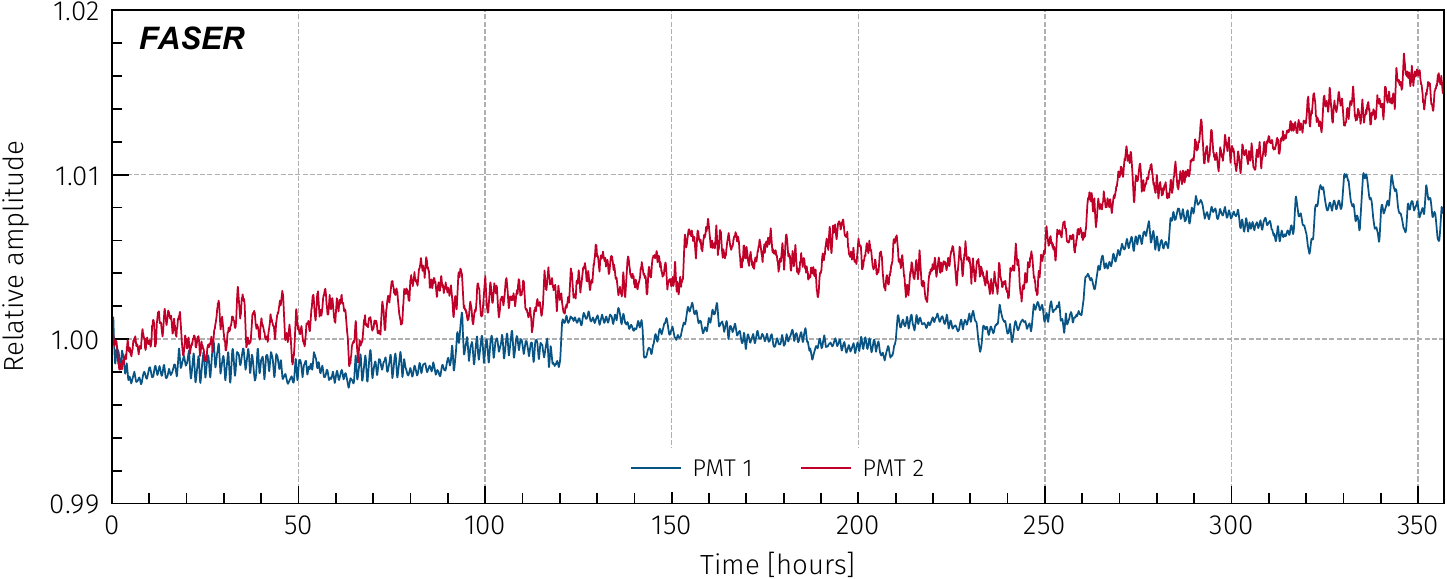}
    \caption{Measurement of the signal amplitude for calibration pulses over a duration of 15 days to evaluate the long-term stability. The amplitude is normalized to the first few pulses.}
    \label{fig:calibration-longterm}
\end{figure}

To evaluate the stability of the calibration system, the response to calibration pulses was measured using two photomultipliers over a duration of 15 days with a frequency of \SI{100}{\hertz} and a fixed pulse amplitude. The relative amplitude is shown in Figure~\ref{fig:calibration-longterm}. The relative deviation was found to be around \SI{1.5}{\percent}. In addition, no degradation of the LED light yield was observed over the two week period. As the in-situ calibration procedure is performed over a period of about 30 minutes, the drift is small enough to not significantly influence the calibration which does not rely on the absolute intensity of the calibration pulses, but the drift will limit the ability to precisely monitor the PMT stability over extended periods. 

%%%%%%%%%%%%%%%%%%%%%%%%%%%%%%%%%%%
\clearpage
\section{The trigger and data acquisition system}
\label{sec:tdaq}
The FASER trigger and data acquisition (TDAQ) system hardware and software are designed to be lightweight but robust as the experiment aims to capture a potentially extremely rare signal over four years of data-taking. The complete trigger logic takes place in hardware installed in a rack for electronics next to the detector itself. The number of hardware components is minimal to reduce the amount of cabling and equipment, which are inaccessible during data-taking.
Data from each sub-component is transferred via a 1 Gbit/s link to the FASER TI12 Ethernet switch and thereafter via a 10 Gbit/s optical fiber line directly to a single DAQ PC above surface,  where software combines the data of each sub-component to build complete events and record them to file. 
The experiment will be triggering on any high-energy particle traversing its detector volume.
The expected trigger rate in the scintillator counters for an instantaneous luminosity of $2\times 10^{34}$~cm$^{-2}$s$^{-1}$  is roughly 650~Hz, predicted by simulation and in-situ measurements~\cite{FASER:2018bac}. This is dominated by muons originating from the ATLAS IP. Around 5~Hz of energetic signatures will be deposited in the calorimeter. The event size is almost 22~kBytes/event, thus the expected data recording rate is around 14~MBytes/s.
The FASER experiment is planned to be operated remotely with no shifters populating a control room. The TDAQ design therefore emphasises abundant monitoring and alerts.

The current Section provides an overview of the TDAQ  core hardware (Section~\ref{sec:tdaq_hw_overview}), description of individual readout components (Section~\ref{sec:readout_hw}), and finally an overview of the software (Section~\ref{sec:tdaq_sw}). A complete and detailed description of the TDAQ system can be found in Ref.~\cite{FASERTDAQ:2021}.

\subsection{TDAQ core hardware overview} \label{sec:tdaq_hw_overview}

At the core of the TDAQ system is the FASER Trigger Logic Board (TLB), the central trigger logic processor that manages trigger signals by combining them and regulating their rate via prescales and vetoing, and the FASER clock board that provides a stable clock to drive the system. The detector components used for triggering are the four calorimeter modules and the four scintillator stations. 
The PMT pulses are first digitized by a CAEN digitizer board (described in Section~\ref{sec:digitizer} below), which transmits trigger signals for pulses exceeding a preset threshold to the TLB.

The core system runs on the LHC clock signal with a frequency of 40.08\,MHz, corresponding to the frequency of proton bunch-crossings. The period of one clock cycle is thus equal to the spacing between bunch-crossings, or  25\,ns. The clock signal as well as the LHC orbit signal at 11.245\,kHz, are part of the beam synchronous timing (BST) system, transmitted over optical fibers to beam instrumentation equipment around the LHC using the TTC system\,\cite{TTC}. For FASER, this signal is received by a legacy VME system, the BST receiver interface for beam observation system (BOBR)\,\cite{BOBR}, produced by the LHC beam instrumentation group. However, the LHC clock provided by the BOBR has a non-negligible jitter, mainly due to noise in its power module, changes during the energy ramp of the LHC, and is not guaranteed to be continuous when there is no beam in the LHC. For this reason, the FASER TDAQ incorporates the FASER clock board that provides a high-quality, uninterrupted reference clock with a constant phase with respect to the LHC clock across power-cycles.
The clock board cleans the jitter of the BOBR  to less than 4~ps and a zero-delay feature of the jitter cleaner guarantees that the output FASER clock is aligned to the LHC clock with its phase unchanged across resets and power-cycles.

The TLB is a custom GPIO board with an adapter card. 
The GPIO board was developed as a general readout board utilising a CYCLONE V A7 FPGA (Field Programmable Gate Array). The FPGA is nominally driven by the external LHC clock input but can be driven by a 40~MHz oscillator on the GPIO board itself.

\begin{figure}[h]
    \centering
    \includegraphics[trim=1cm 0cm 9cm 0cm, clip=true, width=\textwidth]{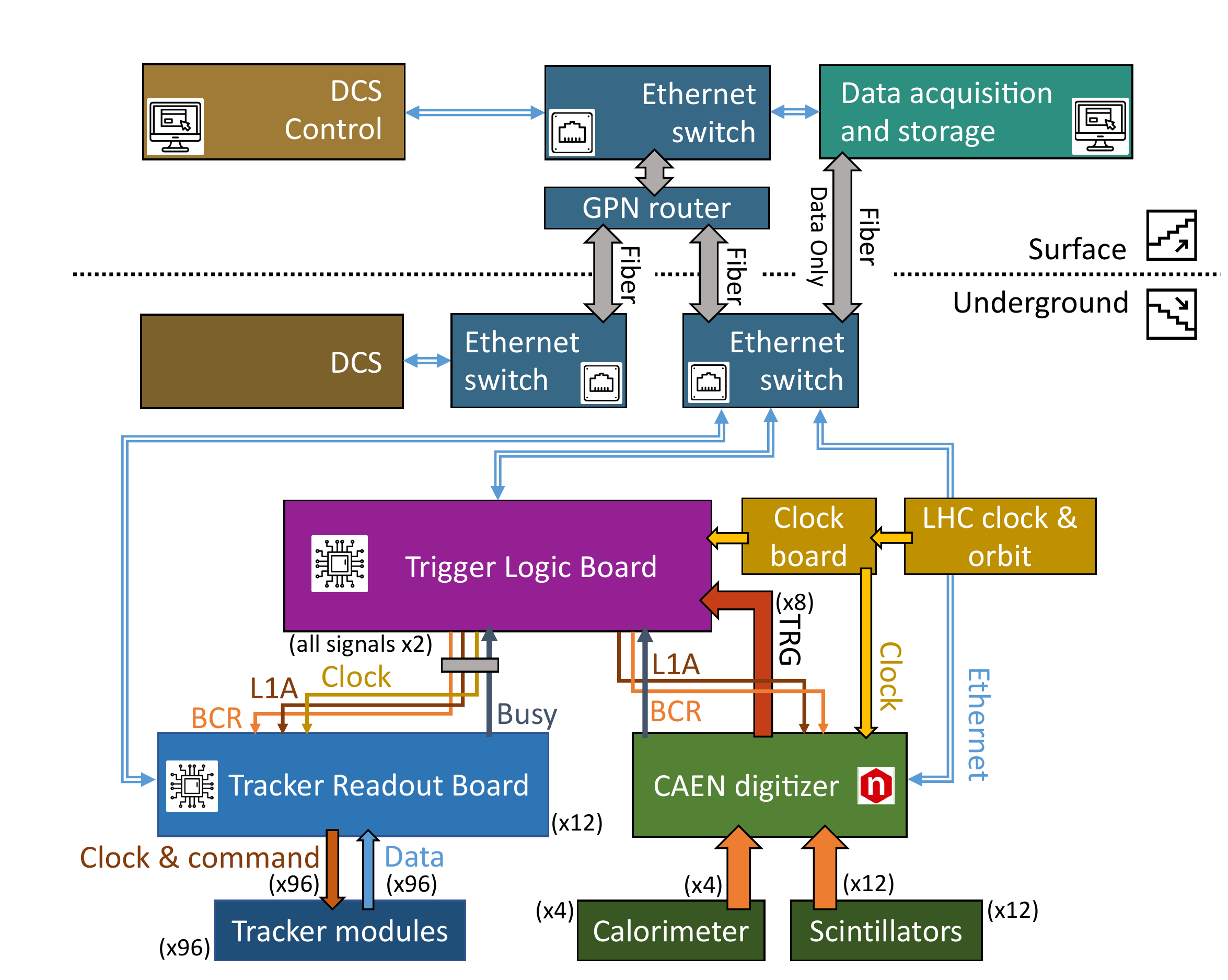}
    \caption{A simple schema of the FASER TDAQ architecture. The numbers in parentheses indicate the number of channels or lines. The blue double-line arrows indicate the connections via Ethernet. The grey thick arrow indicates fibers from the TI12 tunnel to the surface. The digitizer, clock board and TLB are housed in a VME crate. The TRBs are located in two dedicated mini-crates just above the tracker structure, one of the two housing the IFT TRBs.}
    \label{fig:FASER_TDAQ_schema}
\end{figure}

When an event satisfies a trigger logic, the TLB distributes a global Level-1 accept (L1A) signal to all readout components (the Tracker Readout Boards (TRBs) (Section~\ref{sec:TRB}) and the digitizer (Section~\ref{sec:digitizer})) to initiate the readout of an event. It may also send a L1A based on an internal trigger or the LED calibration signal. Both the TLB and TRBs run on the LHC clock, while the digitizer runs on its own internal 500~MHz clock; nevertheless it receives the LHC clock to record as a reference clock so that the relative timing of PMT pulses can be calculated in software. In addition, the TLB receives the LHC orbit signal that it distributes to the TRBs and the digitizer. The TLB and TRB both have an on-board bunch counter that is incremented with every clock cycle and reset with the orbit signal, otherwise known as the bunch counter reset (BCR) signal. A bunch counter for the digitizer is calculated in software based on a timestamp reset by the orbit signal, and the expected LHC clock frequency.
Figure~\ref{fig:FASER_TDAQ_schema} provides an overview of the TDAQ hardware system.

The TLB includes tunable delays for the L1A and BCR signals to ensure that data from the correct event is read out. Triggers may be vetoed due to several sources. A settable simple deadtime window will veto any triggers for \textit{N} clock cycles following a L1A. Both the TRBs and the digitizer can send a busy signal to the TLB to halt triggers while reading out data. A TRB is only able to read out one SCT event at a time, thus it activates a busy signal for the duration of its event readout. The digitizer activates a busy signal based on the occupancy of its readout buffers. Further sources of deadtime are an on-board rate limiter, which limits the output trigger rate to 2.2~kHz to guard against noise bursts, and a bunch-crossing reset veto which disallows L1As for the duration of a bunch-counter reset on the SCT modules.
Upon a L1A, the TLB will send out a data packet containing trigger information for that event. The TLB regularly publishes monitoring data packets containing monitoring numbers such as trigger item counts and veto counts.

\subsection{Readout hardware}\label{sec:readout_hw}
In the following the readout components that receive the global L1A signal, the tracker readout boards and the digitizer, are described. 

%%%%%%%%%%%%%%%%%%%%%%%%
\subsubsection{Tracker readout board}
\label{sec:TRB}
\begin{figure}[tb]
\begin{center}
\includegraphics[width=9cm]{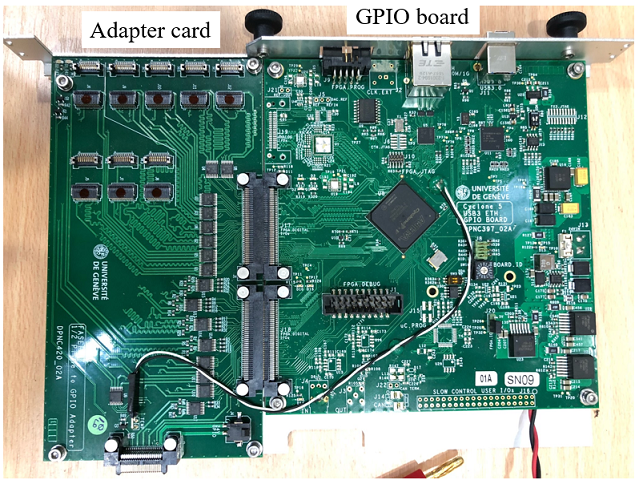}
\caption{FASER tracker readout board (TRB) consisting of a GPIO board and adapter card.}
\label{fig:trb}
\end{center}
\end{figure}

%%%%%%%%%%%%%%%%%%%%%%%%
The Tracker Readout Board operates and reads out the SCT modules. It consists of a GPIO board of the same design as the TLB described above and its adapter card as shown in Figure~\ref{fig:trb}. 
 The adapter card works as an interface between the tracker patch-panel and the TLB and is directly attached to the GPIO board.

The TRBs are housed on the detector in a custom-made mini-crate, with the mini-crate backplane providing 24~V power as well as the TLB signals for the GPIO boards. The TRB is connected to the TLB via a RJ45 connector located on the rear side of the mini-crate backplane.

One TRB reads out eight SCT modules corresponding to one tracker plane. Therefore, a total of nine TRBs are used for the FASER tracking spectrometer, and a further three TRBs are used for the IFT. The TRB adapter card is connected with the tracker plane patch-panel via 8 Samtec twinax Firefly cables.

In order to configure the tracker modules as well as transmit to them timing and trigger signals, the TRB must send configuration commands and data 
encoded in a bit-stream to the tracker modules. The TRB firmware therefore has the functionality to prepare the bit-stream understood by the SCT modules from 
commands and data received from the host PC.
When in data taking mode, the TRB can be operated both in standalone mode and within the global system. The standalone mode is used during calibration scans, generating an internal trigger with a self-regulated rate and running on its own internal 40~MHz clock. Once included in the global system in data taking mode, the TRB receives and is driven by the L1A signals, BCR signals and external clock signal provided by the TLB as described in the previous section.

The time tuning of the L1A signal is especially important in the case of the tracker readout, as the SCT hit readout window extends only  across 3 clock cycles, equivalent to 3 proton bunch-crossings.
During physics data taking the L1A and BCR signals arrive at the SCT modules with a fixed latency. The tracker readout has an internal pipeline of 132 clock cycles, so that the TLB L1A signal must arrive at the SCT modules 132 clock ticks after the charged particle traversed the tracker plane. There are inherent delays in signal propagation from the occurrence of an event to the arrival of the L1A at the TRB. To cover the full 132 clock cycle latency, both the TRB and TLB have settable coarse time delays in increments of clock ticks for the L1A/BCR signals. A signal delay of up to 127 clock ticks on the TLB and a further adjustment of 0 to 7 clock ticks on a TRB are possible, where the latter allows one to adjust for different particle times-of-flight at each tracker station location in the final detector setup. 
Tracker hits are sampled at the rate of the input clock and 3 samples (corresponding to 3 clock cycles or bunch-crossings) are ultimately recorded. The timing of the tracker sampling of hits can be adjusted with a resolution of 390~ps via an adjustment of the input clock phase on the TRB. The fine time tuning ensures the silicon hits are sampled on the pulse peak, maximising the hit detection efficiency.

\subsubsection{CAEN Digitizer}
\label{sec:digitizer}
The readout electronics of the calorimeter and scintillator detectors need to sample the signals from the calorimeter and scintillator counter PMTs, provide trigger input signals to the TLB and buffer the calorimeter and scintillator counter data, which are read out upon a trigger signal.
This functionality is accomplished by using a VME-based system consisting of a single 16-channel, 14-bit CAEN VX1730 digitizer board~\footnote{VX1730 / VX1730S 16/8 Channel 14 bit 500 MS/s Digitizer. \url{https://www.caen.it/products/vx1730}} 
    that is controlled and read out via Ethernet using the Struck Innovative Systems SIS3153 VME interface board.~\footnote{Struck innovative systems, SIS3153 USB3.0 and Ethernet to VME interface, \\ \url{https://www.struck.de/sis3153.html}} The dynamic range of a digitizer channel is 2~V with a programmable offset of up to $\pm1$ V that can be set independently on each channel. The signals on each channel are continuously digitized at 500~MHz and stored in a circular buffer to be read out on receiving a trigger signal. The full waveform for each channel is read out for a user-defined period of up to 1.2 $\mu$s. The long readout window allows for a detailed offline analysis of the waveform, in particular in the case of any anomalous signal, but can be later reduced after experience with beam data is collected. The on-board trigger logic has, for each channel, a user-defined, fixed, absolute threshold that can be set to generate an over/under-threshold signal. 
The scintillator counter trigger threshold will be below that of a single minimum ionizing particle, while the calorimeter threshold will be set to trigger on electromagnetic showers depositing more than about 20~GeV of energy.
The trigger channels are combined in pairs into eight trigger signals using a logical AND or OR, depending on the need. The trigger signals are propagated and further combined in the TLB as described above.

\subsection{TDAQ software}\label{sec:tdaq_sw}

Event building and writing to file is hosted on a single DAQ server on the surface. The DAQ server is a 24-core, 48-thread AMD CPU SuperMicro server, with 64 Gigabyte memory, two 500 Gigabyte SSDs for the operating system and two 8 TB hard-drives for data storage.
FASER DAQ, FASER's event building and writing software framework, relies on DAQling~\cite{daqling,daqling-sw}, a lightweight open-source DAQ framework designed for the data acquisition of small and medium sized experiments. The framework builds DAQ modules written in C++ and enables synchronised Finite State Machine (FSM)-like control of all module processes.
DAQ modules are connected via an asynchronous messaging communication layer for the throughput of data packets. 
The configuration of modules is stored in a JSON format, while control is handled via python applications.

\begin{figure}[h]
    \centering
    \includegraphics[trim=1cm 0.5cm 2cm 1cm, clip=true, width=\textwidth]{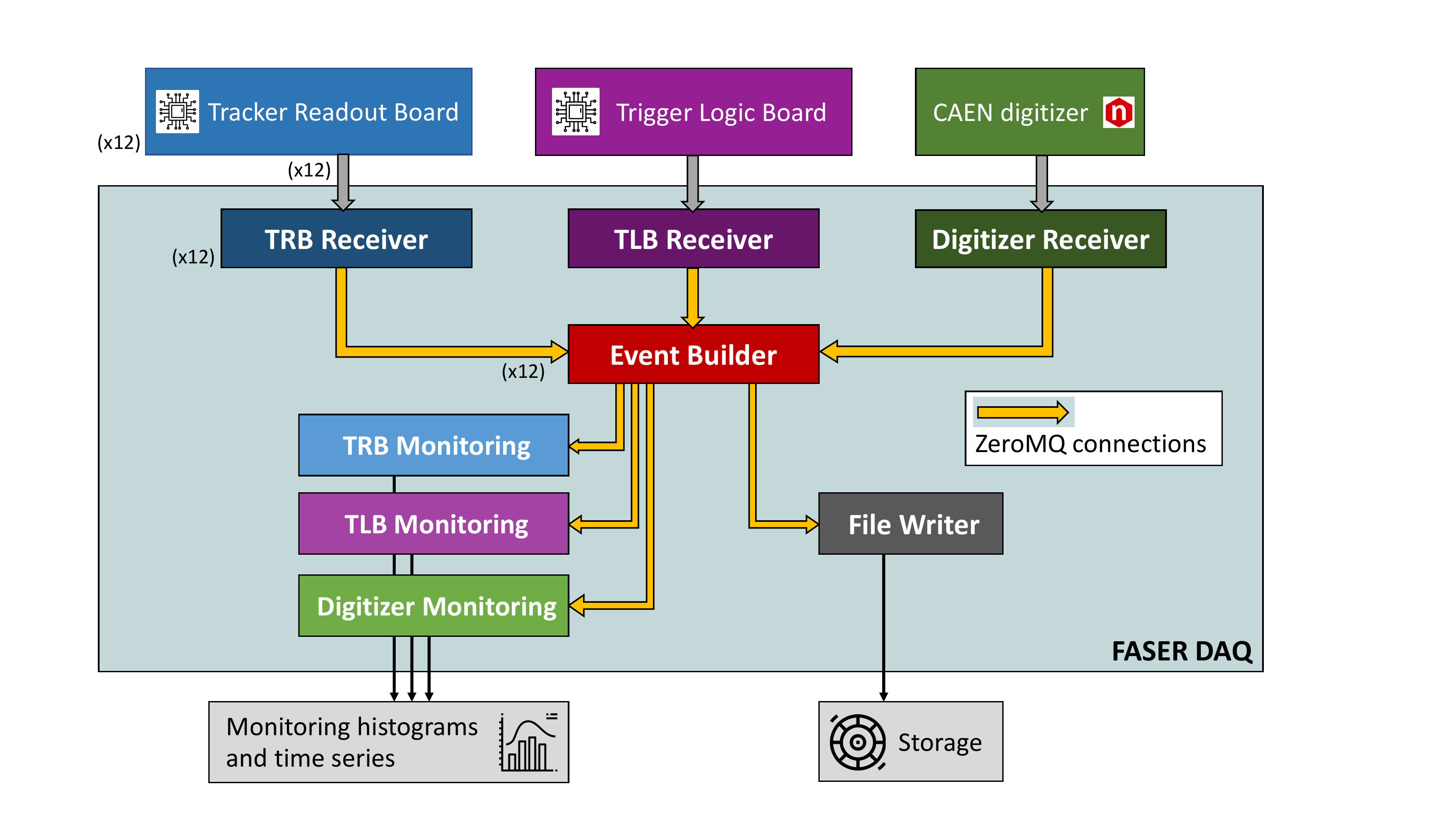}
    \caption{Overview of the FASER DAQ software processes.  }
    \label{fig:DAQSoftwareDiagram}
\end{figure}

Figure~\ref{fig:DAQSoftwareDiagram} presents the structure of the FASER DAQ~\cite{faser-daq-sw}. During physics data taking, the complete framework runs 16 DAQ modules as the main event handlers: 12 receiver modules (one for each TRB/tracker plane), a receiver module each for the TLB and digitizer, an event builder module, and a file writer. Receiver modules communicate with the readout boards to retrieve event data fragments and dress these with a header that includes the event ID and bunch counter ID (BC ID). The event builder stores the fragments received from each receiver module and in parallel matches them based on a common event ID to form a complete event within a configurable time before timeout. Events are given a header that includes a time stamp, event ID, BC ID and trigger bit information, and sent to the file writer, which buffers events and eventually writes them to file. Several recording streams exist to pipe complete events, incomplete (timed-out), corrupted and duplicate events to separate files.
In addition, another $\mathcal{O}(20)$ monitoring modules are run online, dedicated to monitoring specific event features in events from the event builder. Monitoring metrics and histograms are regularly published to a Redis database~\cite{redis}.

A Run Control web interface allows a user to choose the run configurations and control the FASER DAQ FSM. Separate web pages visualise the monitoring metrics. Metric values defined in DAQ modules are regularly sent to an InfluxDB~\cite{influxDB} database and presented in Grafana~\cite{grafana} dashboards, while a custom online histogramming API polls the Redis database to display histograms live.

\subsection{Pre-installation commissioning}

The commissioning of the TDAQ system before installation in the TI12 tunnel began with the qualification of the hardware functionality of each standalone component in 2019. Throughout 2020, communication and readout was tested by combining components, culminating in a cosmic-ray test stand, consisting of a FASER tracker station and FASER scintillator counters for triggering, in a dedicated surface area at the CERN Prevessin site. This allowed testing of the TLB coincidence triggering and L1A delay to the tracker readout (so that tracker hits from the cosmic muon interaction appear in the readout window), as well as testing of the run control, monitoring and alerts. More details of the TDAQ commissioning is described in Ref. ~\cite{FASERTDAQ:2021}. Further commissioning of the TDAQ system as part of combined running with the installed detector is detailed in Section~\ref{sec:commissioning}.

%%%%%%%%%%%%%%%%%%%%%%%%%%%%%%%%%%%
\clearpage
\section{The magnet system}
\label{sec:magnets}
Three dipole magnets are installed in the FASER detector, based on a Halbach array permanent magnet (PM) design~\cite{Halbach:1979mv}. The longest, 1.5-m long, surrounds the decay volume, and is followed by two 1-m long dipoles installed along the tracking spectrometer. They produce a field of 0.57 T inside an aperture diameter of 200 mm. The main parameters are listed in Table~\ref{tab:Mag-param}.

\begin{table}[hbt!]
    \centering
    \begin{tabular}{|l|c|c|c|}
  \hline
Parameter & Short Model & Long Model & Unit \\
 \hline
 Aperture diameter &	200 &	200 & 	mm\\
Length	& 1000 &	1500 &	mm\\
Outer diameter	& 430 &	430 &	mm\\
Mass	& 914 &	1331 &	kg\\
Mass of permanent magnet &	606	& 909 &	kg\\
Nominal field at the centre	& 0.57 &	0.57 &	T\\
Good field region (GFR) radius	& 100 &	100 &	mm \\
Field homogeneity in GFR &	$\le \pm 3$ &	$\le \pm 3$ &	\% \\
Permanent magnet material &	Sm$_2$Co$_{17}$ &	Sm$_2$Co$_{17}$  & -	\\
 \hline
    \end{tabular}
    \caption{Main design parameters of the FASER dipoles.}
    \label{tab:Mag-param}
\end{table}

The main advantage of the Halbach design is to produce a strong and homogeneous field inside a relatively large aperture while keeping compact overall dimensions. The FASER location in the TI12 tunnel gives tight constraints on the magnet dimensions. Due to the limited depth of the trench, only 250~mm are available between the trench floor and the dipole central axis when aligned on the beam collision axis LOS. In addition, at the back of the detector there is only 250~mm in the horizontal plane between the magnet central axis and the tunnel wall.  

\begin{figure}[hbt!]
    \centering
    \includegraphics[width=0.9\textwidth]{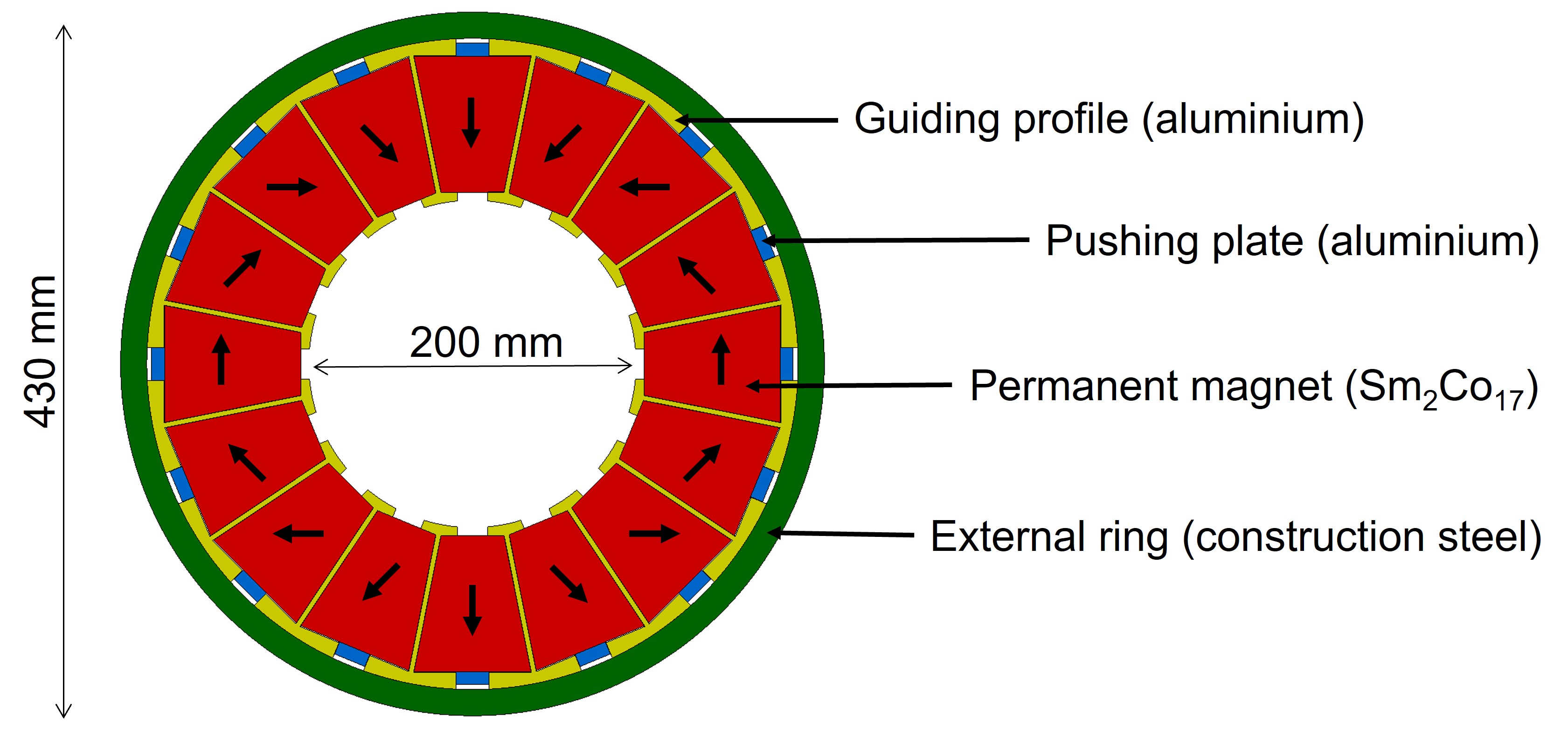}
    \caption{Cross section of FASER dipole. The arrows indicate the direction of the magnetic field in each of the permanent magnetic blocks.}
    \label{fig:mag1}
\end{figure}

The design is based on a Halbach array with 16 magnet sectors. The number of sectors was defined to keep the permanent magnet blocks to reasonable dimensions and to provide a field homogeneity of better than ±3\% inside the whole dipole aperture. The cross section of the dipole is identical for both models and is shown in Figure~\ref{fig:mag1}. 
The permanent magnet blocks, made of rare earth Samarium Cobalt Sm$_{2}$Co$_{17}$, have a trapezoidal shape, with five different easy axis~\footnote{The easy axis defines the direction for which the spontaneous magnetization of the material is easiest.} orientations to shape the dipolar field inside the aperture. The blocks are 83.3 mm long, and are arranged in 12 (18) rings of 16 blocks each for the short (long) magnet(s). They are installed inside a structure made of aluminium guiding profiles attached to an external steel ring. The PM blocks are locked in position with aluminium pushing plates.

\subsection{Magnetic Design}
\begin{figure}[hbt!]
    \centering
    \includegraphics[width=0.6\textwidth]{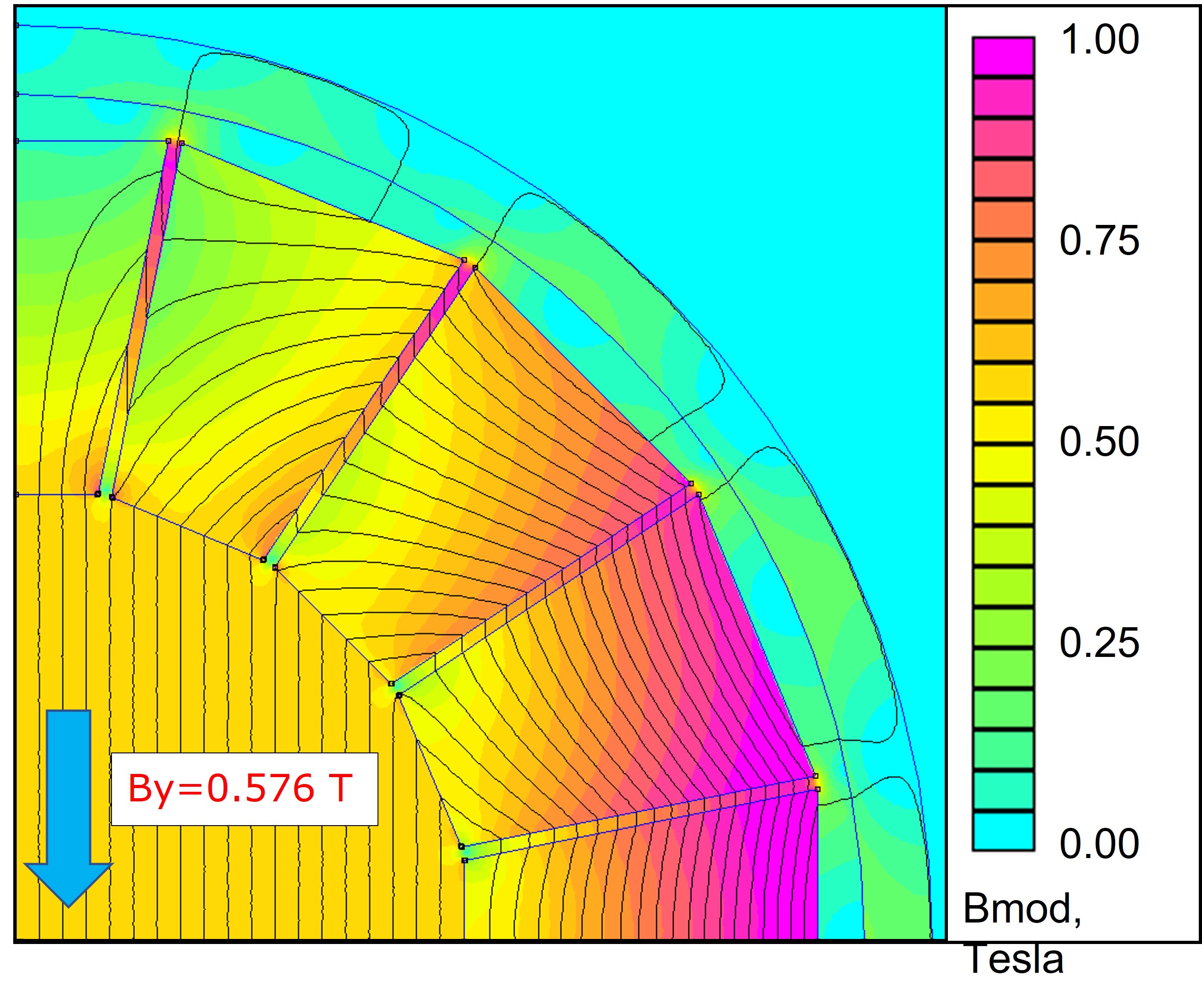}
    \caption{2D magnetic field distribution.}
    \label{fig:mag2}
\end{figure}

The main magnet design parameters such as the number of sectors in the array, the PM grade and geometry, and the dipole dimensions were defined using the 2D magnetic design shown in Figure~\ref{fig:mag2}, using FEMM~\cite{Magnet:FEM}. The maximum clearance between magnet sectors has been set to 3 mm to limit the impact on field homogeneity and, at the same time, ensure a good rigidity of the aluminium guiding profiles.

\begin{figure}[hbt!]
    \centering
    \includegraphics[width=1.\textwidth]{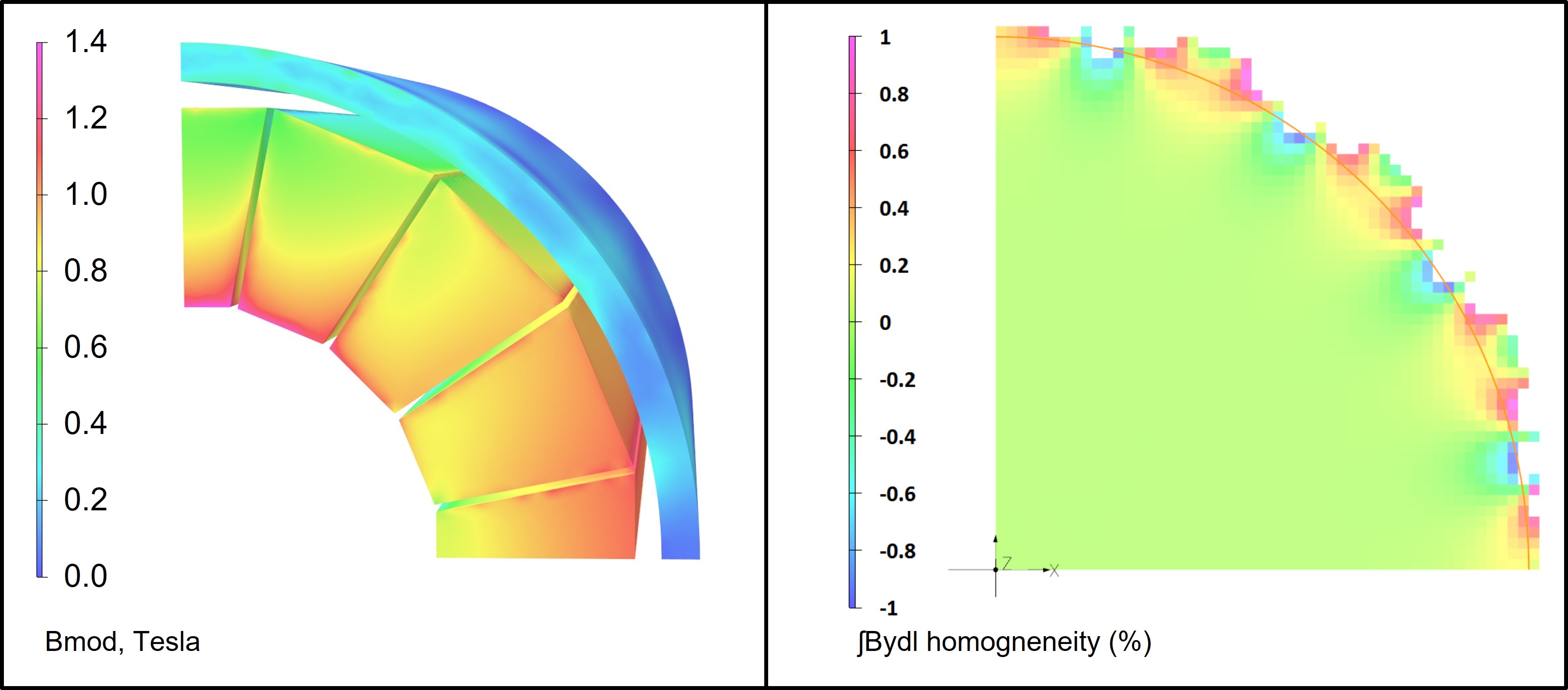}
    \caption{left: The magnitude of the magnetic field (Bmod) distribution (T), right: The integrated field homogeneity (\%).}
    \label{fig:mag3}
\end{figure}

The integrated field homogeneity inside the good field region was evaluated with the 3D magnetic design using Opera 3D/TOSCA~\cite{Magnet:OPERA} as shown in Figure~\ref{fig:mag3}.
The tolerances on the PM blocks were specified to keep the integrated field homogeneity within 3\% in the good field region. This requires the dimensions of the PM blocks to be within ±0.025 mm and a maximum deviation of the field axis direction below ±3°.

\subsection{Manufacturing and assembly}
The PM blocks were produced with Samarium Cobalt grade YXG32H. 
The good temperature stability and high intrinsic coercivity minimizing the risk of local demagnetization during assembly were the main reasons to use Samarium Cobalt. Each PM block was made of two parts glued together with the joint plane parallel or perpendicular to the magnetization direction. In total 690 PM blocks with five different easy axis orientations were produced, with a total mass of 2130 kg.

The external ring was made of construction steel grade S355JR. Non-magnetic material such as stainless steel could also be used, but soft magnetic material has the advantage of creating a radial magnetic shielding and, therefore, more stability for magnetic forces during assembly, as the PM blocks are attracted to this external ring until the insertion of the last magnet sector.

\begin{figure}[hbt!]
    \centering
    \includegraphics[width=0.7\textwidth]{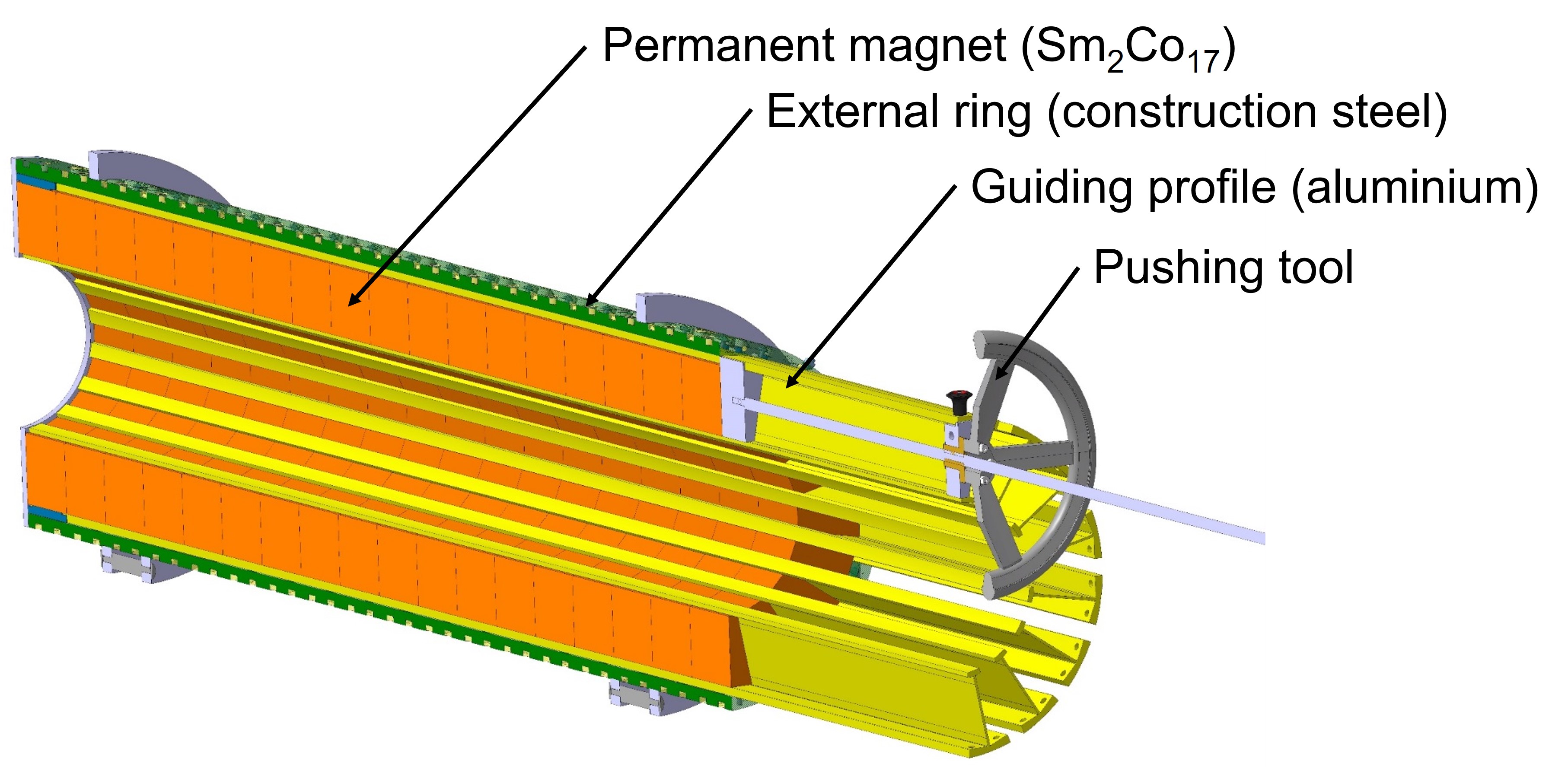}
    \includegraphics[width=0.7\textwidth]{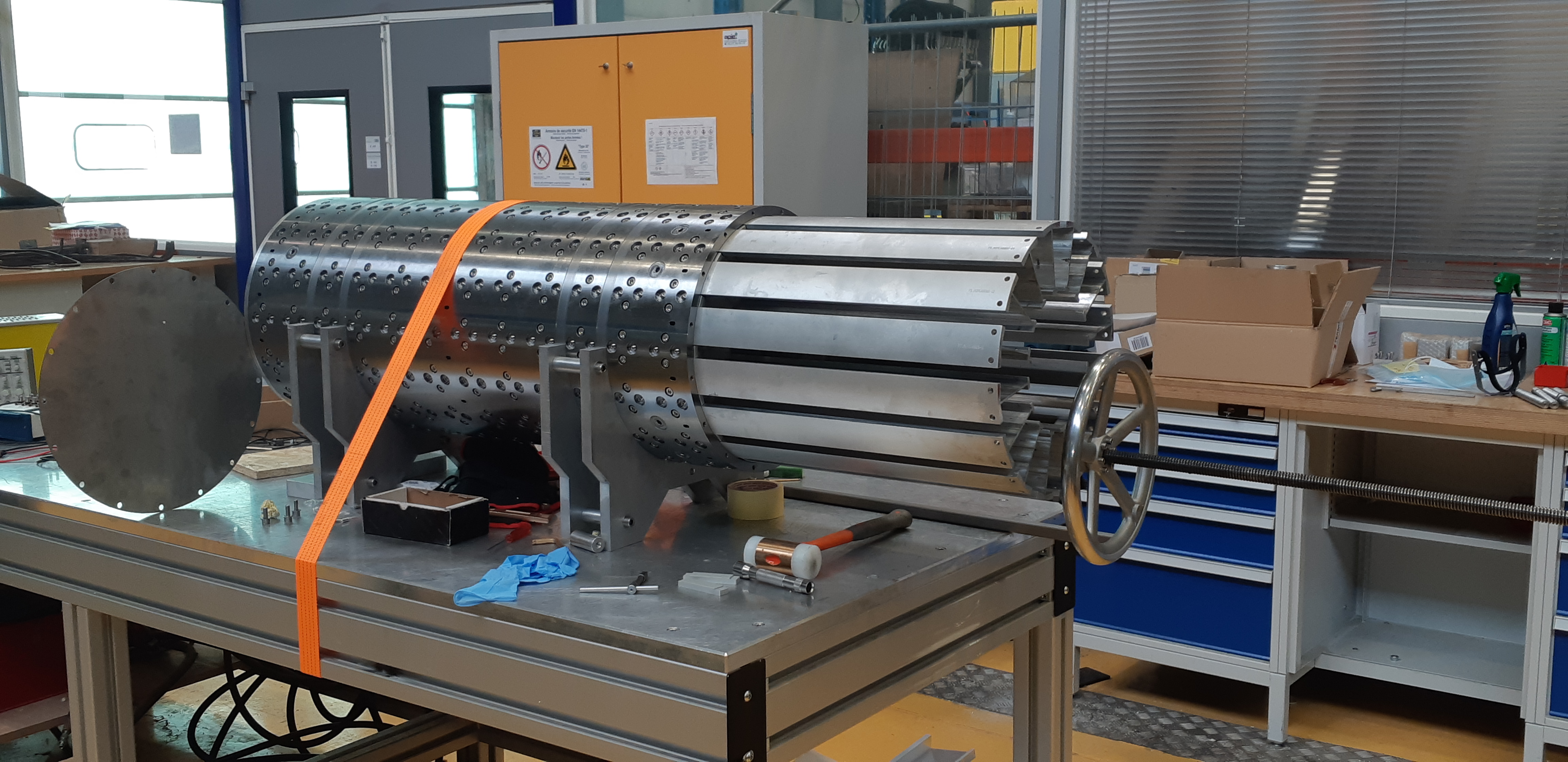} 
    \caption{Top: Tooling for permanent magnet insertion. Bottom: Insertion of a permanent magnet block.}
    \label{fig:mag4}
\end{figure}

The guiding profiles were machined from aluminium grade 6082. 
In the mechanical structure of the dipoles, the guiding profiles holding the PM blocks in the assembly are also used for the PM block insertion as shown in Figure~\ref{fig:mag4}. The profiles are manufactured to be 0.5 m longer than the external ring. Each PM block is inserted between two profiles without external forces and pushed inside the dipole assembly with a dedicated tooling. The magnetic forces during assembly were calculated to define the optimal insertion sequence. At the end of the assembly the extra length of guiding profile was trimmed off and a protective cover was installed.

\subsection{Magnetic measurements}
The 690 permanent magnet blocks were individually characterized at CERN. The magnetic moment and the deviation of the magnetization direction were measured to avoid polarity errors and significant field inhomogeneity, which is almost impossible to correct once the  dipole is assembled. The measurements were carried out with a three-dimensional Helmholtz-Coil~\cite{Magnet:Helmholtz}. 
An average magnetic moment of 331.3 Am$^2$ was measured over the magnet production with a peak-to-peak variation of $\pm$2.0\% ($\sigma$ = 0.6\%). The deviation of the magnetization direction was within $\pm$1.44$^{\circ}$ ($\sigma$ = 0.39$^{\circ}$) in the horizontal plane and $\pm$1.9$^{\circ}$ ($\sigma$ = 0.49$^{\circ}$) in the vertical plane. As all PM blocks were within the specified tolerances, it was not necessary to apply local field corrections in the dipole assembly by assigning a specific position to each PM block.

A number of magnetic measurements were performed during the assembly process, mainly to avoid the risk of positioning or polarity errors when inserting a PM block. 
A dedicated tooling based on a rotating Hall probe and an angular encoder was developed, to allow errors to be caught in the assembly process while they could still be corrected.

\begin{figure}[hbt!]
    \centering
    \includegraphics[width=0.8\textwidth]{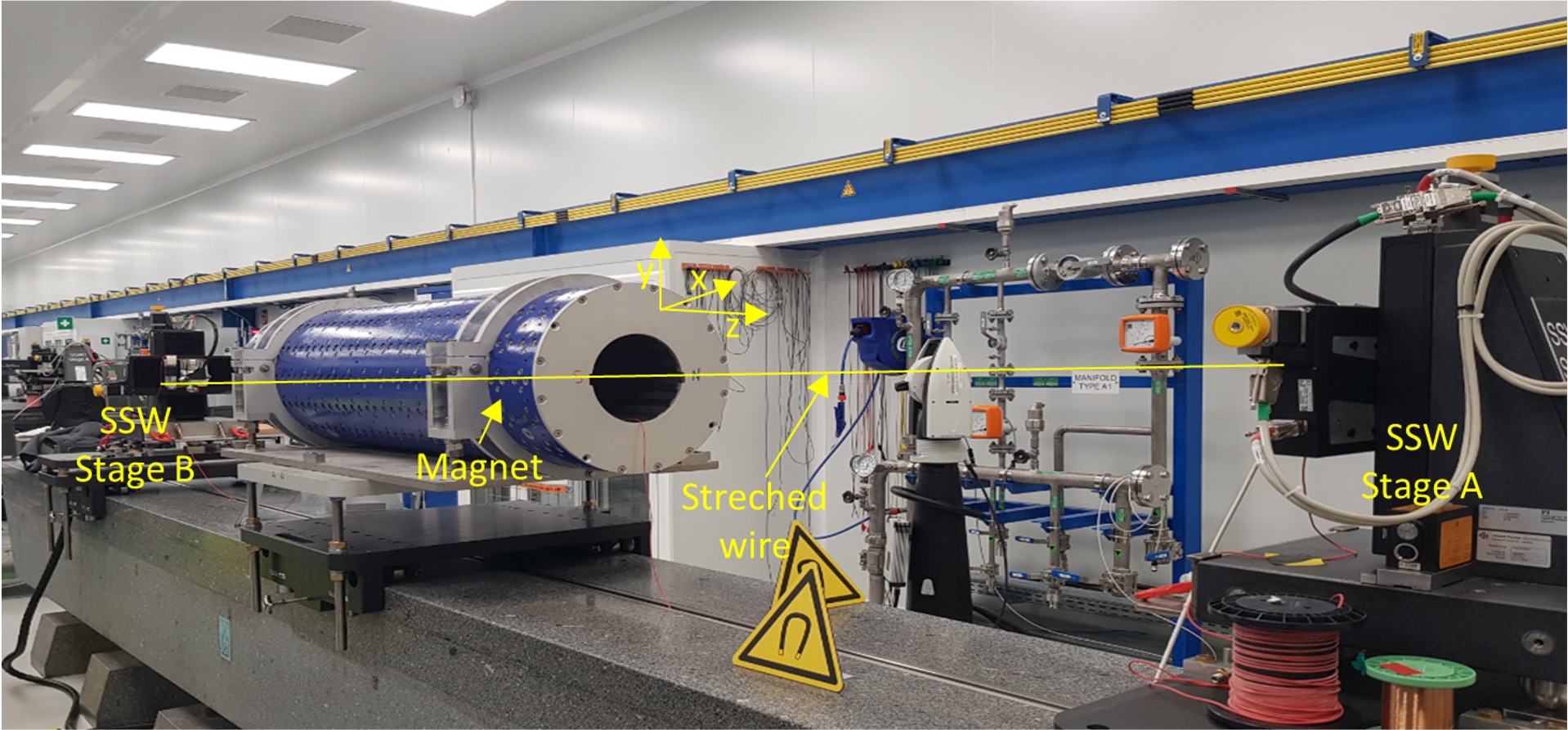}
    \includegraphics[width=0.5\textwidth]{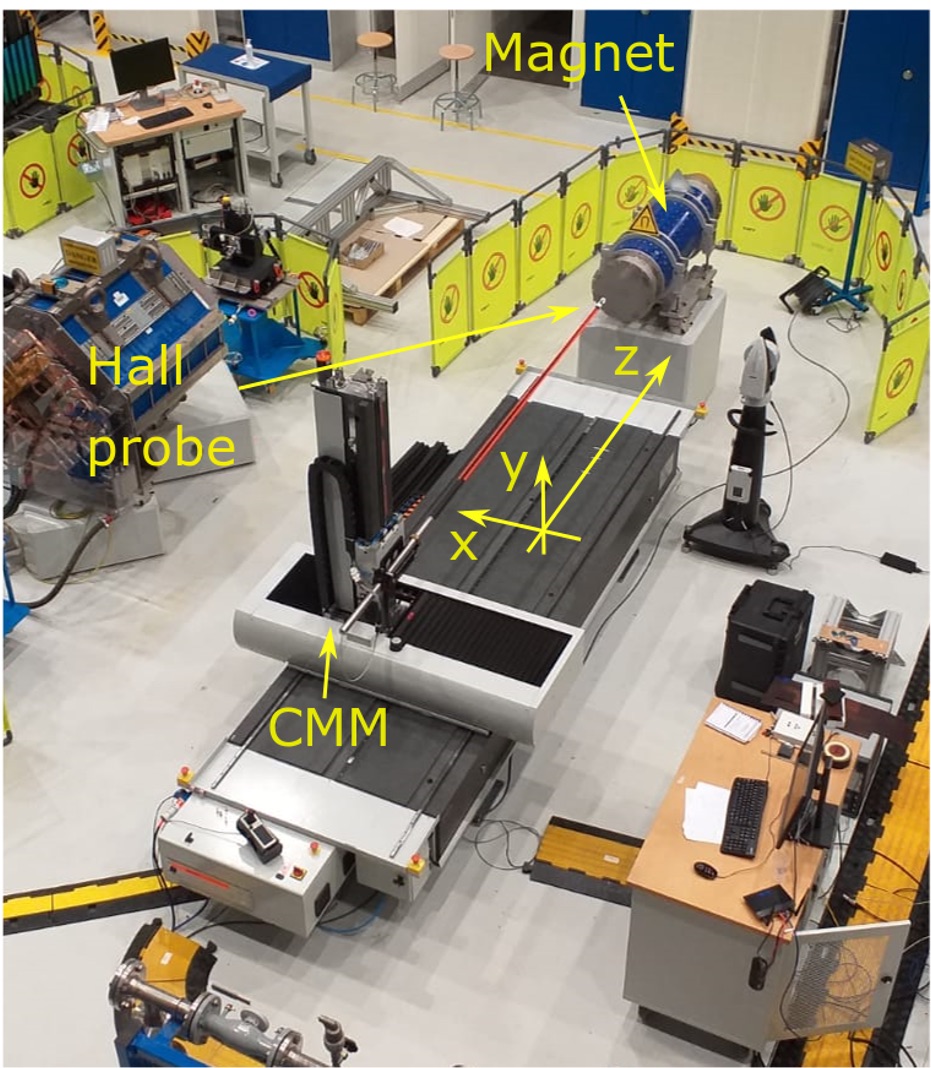}
    \caption{Top: One of the FASER dipoles under measurement using the single-stretched wire at CERN. Bottom: One of the FASER dipoles under measurement using the 3D Hall probe mapper at CERN.(This figure is taken from~\cite{mag-measurements}. )} 
    \label{fig:mag-measurements}
\end{figure}

The assembled dipoles were measured~\cite{mag-measurements} with a single-stretched wire (SSW) and the 3D Hall probe mapper, as shown in Figure~\ref{fig:mag-measurements}. The integrated field, the integrated 2D field homogeneity and field orthogonality, relative to the optical reference targets measured with a laser system, were measured with the SSW. These measurements are summarized in  Table~\ref{tab:Mag-meas}.

\begin{figure}[hbt!]
    \centering
    \includegraphics[width=0.9\textwidth]{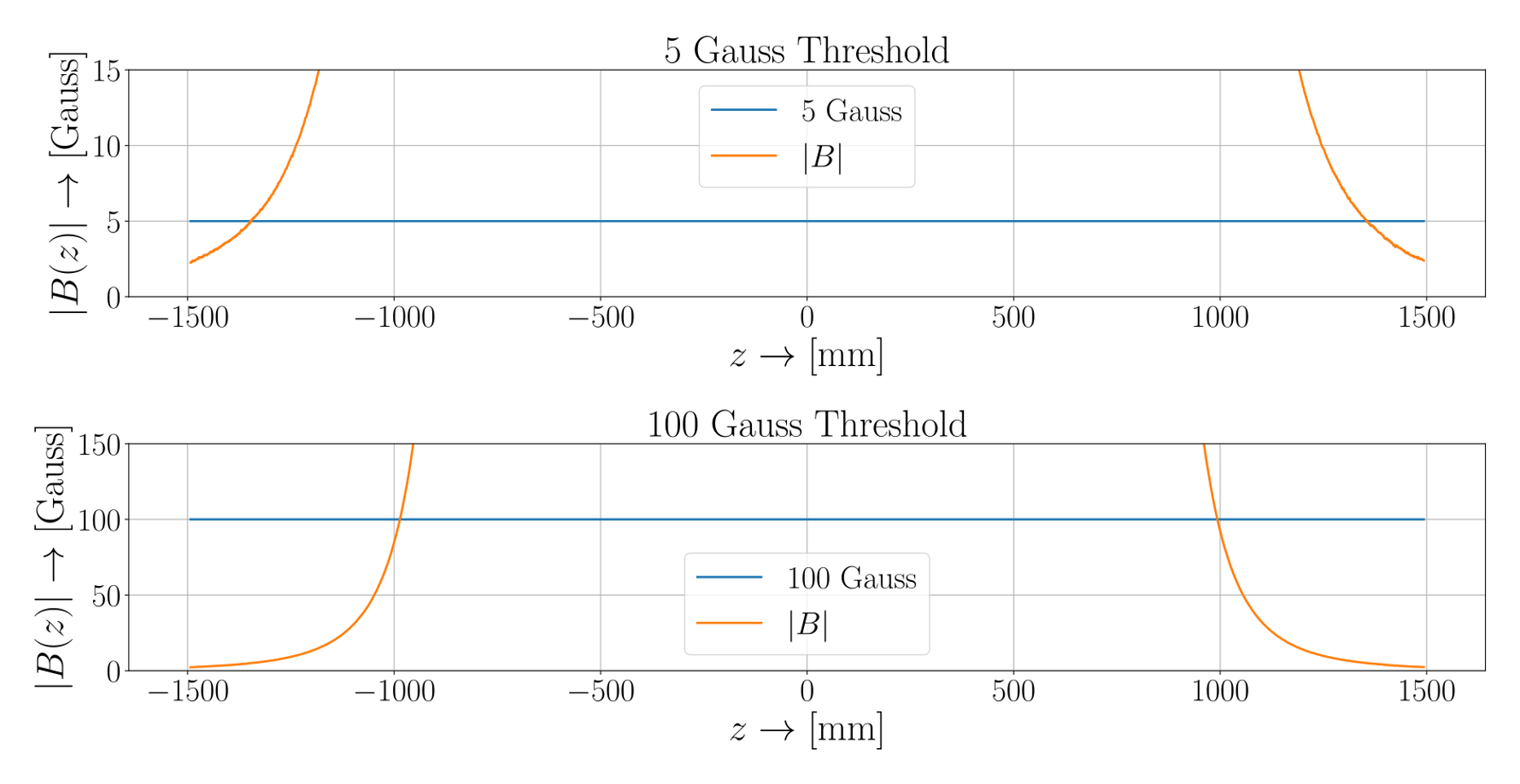}
    \caption{The measured stray field along the magnet central axis for the 1.5-m long dipole. The magnet is placed between z=-750 - +750 mm and the measurements show that the stray field drops below 100 Gauss (10 mT) about 250 mm from the magnet aperture (top), and below 5 Gauss (0.5 mT) about 600 mm from the magnet aperture (bottom). (This figure is taken from Ref.~\cite{mag-measurements}).} 
    \label{fig:stray-field}
\end{figure}

Measurements of integrated higher-order field harmonics were also made with the SSW, at the reference radius r = 67 mm up to order N = 15. A skew sextupole (N = 3) harmonic of 30 to 40 units is present in all three dipoles. According to past studies on Halbach arrays [6], this irregularity is due to a small clearance between some of the magnets created by magnetic forces. Nevertheless, the magnetic measurements of the three FASER dipoles are within specified values. In addition, the local field homogeneity was measured with the 3D Hall probe mapper and was found to be within the specifications.
Finally, the stray magnetic field was measured. The stray field outside the sides of the magnet is zero. Figure~\ref{fig:stray-field} shows the stray field in the central axis of the magnet outside the aperture, this shows that the field drops below 10 mT at a  distance of 250 mm from the end of the magnet.

\begin{table}[thb]
    \centering
    \begin{tabular}{|l|c|c|c|c|}
  \hline
Magnet & Dipole 1  & Dipole 2 & Dipole 3 & Unit \\
  & (short) & (short) & (long) & \\
 \hline
 $\int$ Bx dl &	-0.57692 &	-0.57840 &	-0.86150 &	Tm \\
$\int$ By dl &	0.00021	& 0.00040 &	-0.00250 &	Tm \\ 
Roll Angle &	1.57045 &	1.57008 &	1.57366 &	rad \\
 \hline
    \end{tabular}
    \caption{Measured integrated field and field orthogonality for the three dipoles.}
    \label{tab:Mag-meas}
\end{table}

\subsection{Alignment}
\label{sec:magnetSupport}
\begin{figure}[hbt!]
    \centering
    \includegraphics[width=0.6\textwidth]{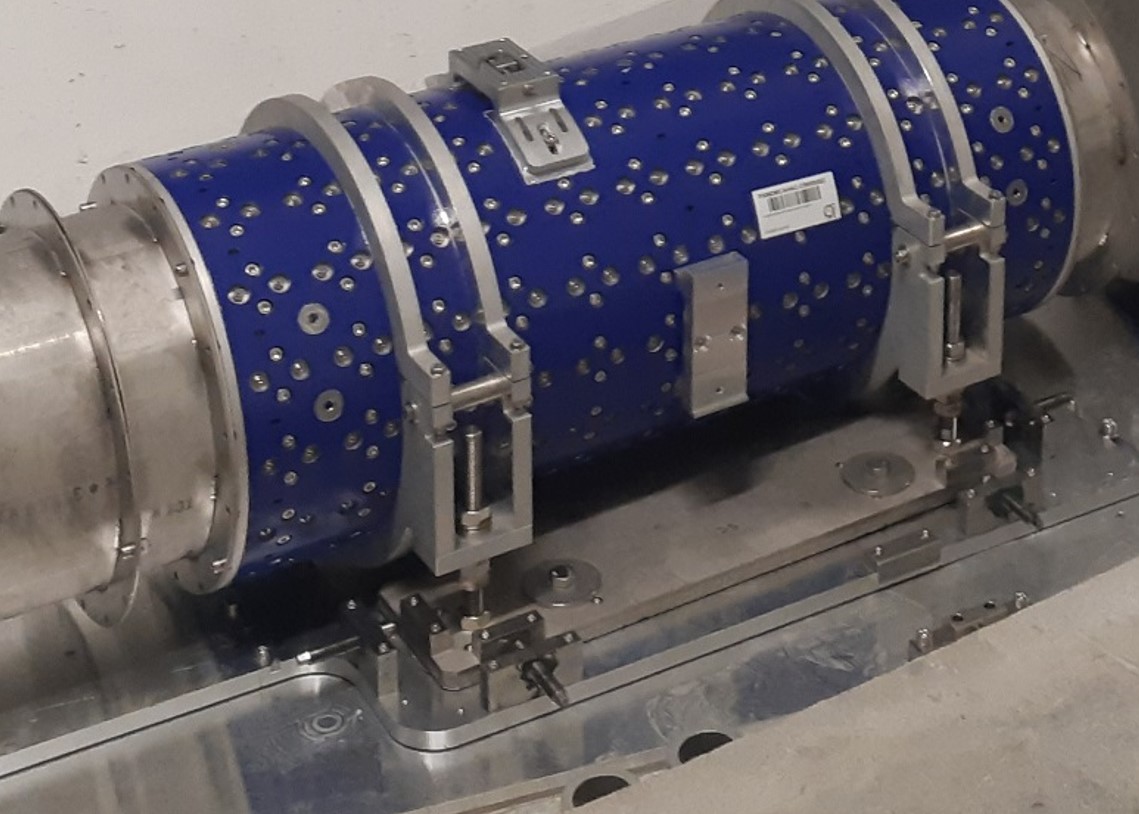}
    \caption{Support and alignment structure.}
    \label{fig:mag8}
\end{figure}

Optical references, installed on the external ring are related to the dipole mechanical centre and are used to align the dipole to the beam collision axis LOS. The supporting structure shown in Figure~\ref{fig:mag8} allows movement of the magnet in the three directions. The roll angle is measured with a reference surface located on the side of the dipole.  

\subsection{Magnet Covers}
\label{sec:magnetCovers}
To ensure no metallic objects get stuck inside the magnet aperture,  magnet covers are installed on both ends of each magnet. For the two spectrometer magnets, plastic covers which can be opened during physics operations are used. This minimizes the material in the active detector volume for physics, while the covers are closed during any work needed on the detector apparatus. For the decay volume magnet, there is not sufficient room for a magnet cover that can be opened, and instead permanent covers of plastic (on one end) and reinforced carbon fiber (on the other end) are used.

%%%%%%%%%%%%%%%%%%%%%%%%%%%%%%%%%%%
\clearpage
\section{Detector Control and Safety Systems}
\label{sec:dcs}
The detector control system, DCS, of the FASER experiment controls and monitors the parameters of the experiment's power system, monitors the environmental conditions in the detector and the cavern, monitors and configures the safety-interlocks of the detector, and implements automatic procedures and alerts to ensure the safe operation of the experiment. 

\begin{figure}[h]
     \centering
     \includegraphics[trim=1cm 0.5cm 1.5cm 3.5cm, clip=true, width=0.7\textwidth]{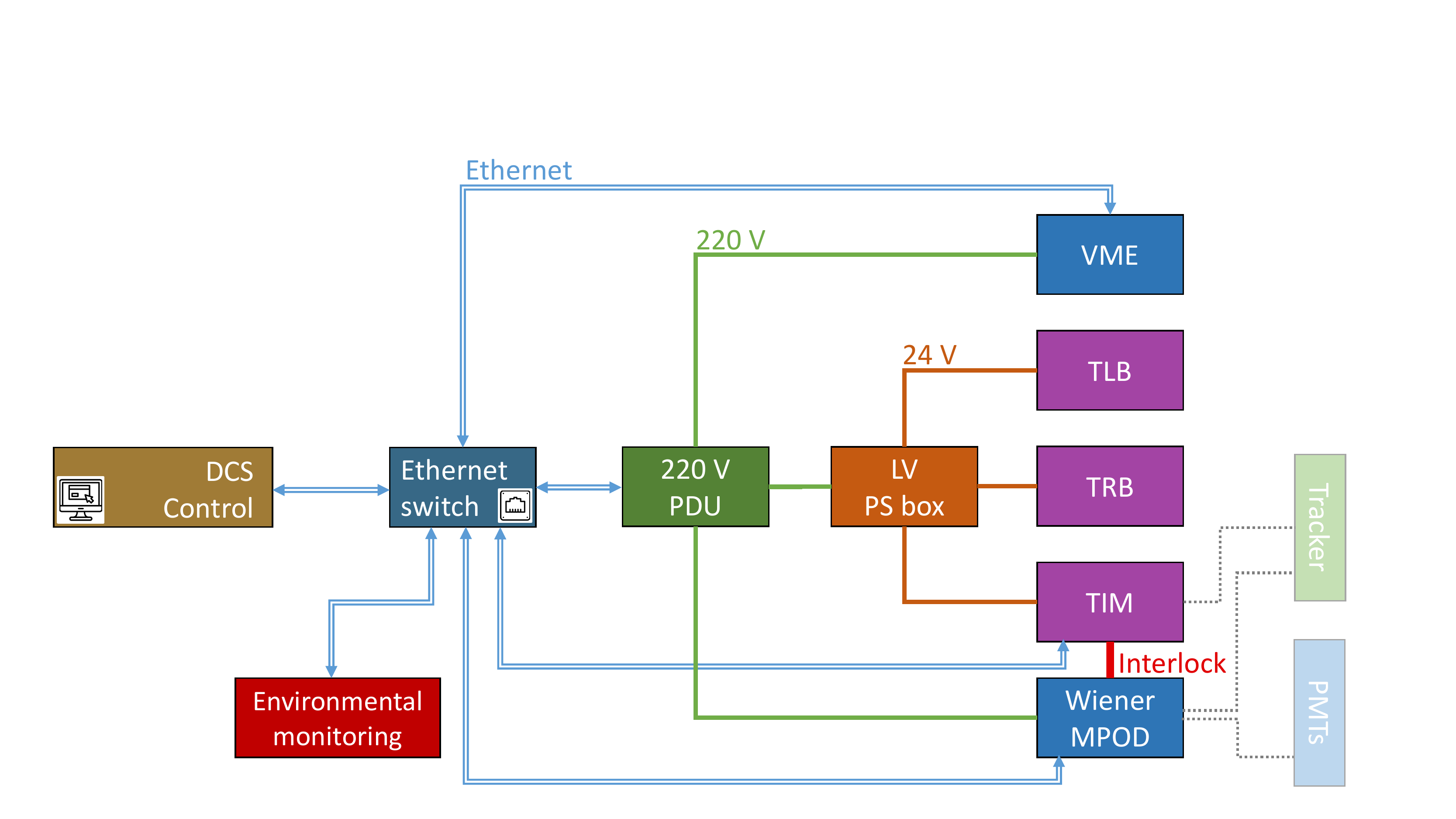}
    \caption{The DCS control hierarchy of the power system components of the FASER experiment and the connections between them. MPOD Weiner indicates the LV/HV Power Supply System. The DCS connections to the detector are also shown.
    }
     \label{fig:DCS}
 \end{figure}

The diagram in Figure~\ref{fig:DCS} shows the components handled by the DCS and their corresponding connectivity. These items can be grouped in the following categories:

\begin{itemize}
    \item HV and LV detector power systems
    \item Safety interlock (Tracker)
    \item Cooling system - via Detector Safety System (DSS) (Tracker)
    \item Power Distribution Units (PDUs)
    \item VME crate 
    \item Environmental conditions
\end{itemize}

%%%%%%%%%%%%%%%%%%%%%%%%
\subsection{Power systems} 
%%%%%%%%%%%%%%%%%%%%%%%%

A total of four HV\footnote{EHS 84 05p manufactured by Iseg Spezialelektronik GmbH} and 24 LV power-supplies\footnote{MPV 8008I manufactured by W-IE-NE-R Power Electronics GmbH} are used to bias the silicon sensors and power the ABCD3TA chips in the SCT modules, respectively, and  are stored in three 19-inch rack mountable crates called the MPOD LV/HV Power Supply System\footnote{MRAAH2500A2H manufactured by W-IE-NE-R Power Electronics GmbH}. An additional HV \footnote{EHS F030n by Iseg Spezialelektronik GmbH} power supply is used to power the PMTs for the calorimeter, pre-shower, veto and timing scintillator stations. 
For other electronics, one 19-inch rack mountable box (called the PSbox) is also installed, which holds fifteen 24~V power supplies\footnote{TXL 035-24S manufactured by Traco Power}. Two custom-made PCBs, the HV splitter board and the LV protection board, are placed between these power supplies and the detector. The HV splitter board is used to divide one HV channel to the four SCT modules in one patch-panel, corresponding to half a tracker plane. The LV protection board equips an integrated circuit\footnote{LTC4365 manufactured by ANALOGUE DEVICE} which can protect against over-voltage possibly caused by single event effects of radiation in the LV power supply. The patch-panel and other electronics are powered via the power distribution units which are remotely controlled via Ethernet.~\footnote{NETIO PowerPDU 4C}

\subsection{Tracker Cooling System} \label{sec:trk_cooling} 
The FASER cooling system was designed and built by the CERN cooling and ventilation group (EN-CV). It consists of two air-cooled water chillers\footnote{HRS030-AF-20MT manufactured by SMC Corporation} in which one is running to cool the detector and the other acts as a hot spare. The system with all instruments is mounted on a single support structure. An additional water reservoir is also installed to allow refilling the water tank in the chillers automatically. Each chiller has the cooling capability of about 1.8~kW at 15~$^{\circ}$C with a temperature difference, $\Delta T = 3$~$^{\circ}$C between inlet and outlet temperature. Given that one SCT module consumes 6~W, leading to a required total cooling power of 576~W for the four tracker stations, the cooling capability is sufficient.

In case of a failure of the running chiller, the other can take over the cooling automatically by controlling the valves. Under usual condition, both chillers are running, where one is connected to the detector and the other is bypassed. If both chillers are not operational, the power supply system is forced to be turned off through a hardware interlock signal.
%without software intervention.

%%%%%%%%%%%%%%%%%%%%%%%%
\subsection{Tracker Safety Interlock}
\label{sec:TIM}
%%%%%%%%%%%%%%%%%%%%%%%%
Two stages of the interlock system based on hardware and software mechanisms protect the detector from electrical and thermal damage. Electrical failure can be caused by over-voltage and over-current on the detector electronics and silicon sensors. Over-heating damages the ABCD3TA chips (for $T>40 ^{\circ}$C \cite{Campabadal:2005rj}) and can cause problems in the mechanical integrity and alignment of the modules due to the glass-transition of the glue used for the module assembly (for $T>35 ^{\circ}$C \cite{Poley:2015zza}).

\begin{figure}[tbh]
\begin{center}
\includegraphics[width=0.8\textwidth]{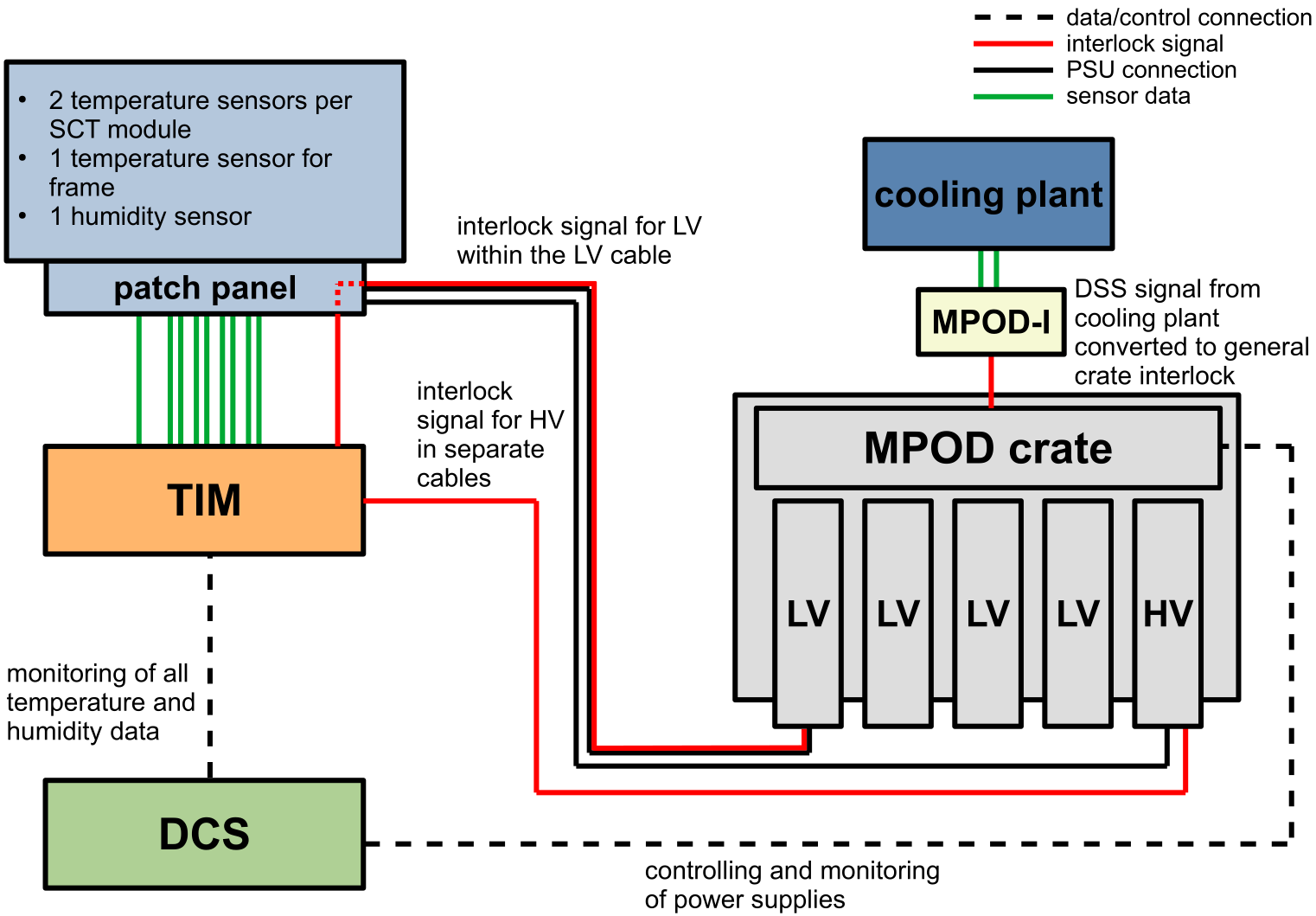}
\caption{Schematic overview of the interlock system of the FASER tracker.}
\label{fig:interlock_overview}
\end{center}
\end{figure}

Figure~\ref{fig:interlock_overview} shows an overview of the interlock system of the FASER tracker. Two NTC-10k thermistors are mounted on each SCT module. In addition, there is one NTC-10k thermistor attached to the mechanical frame as well as one humidity sensor (HIH-4000) inside the plane. They are electrically connected to one TIM (Tracker Interlock and Monitoring board) through the patch-panel, which is equipped by a AM335X micro-controller and three comparators. Each of the four tracker stations has an associated TIM, which digitizes signals from the temperature and humidity sensors and provides the information to the DCS via an Ethernet connection. An analog comparator circuit uses the frame temperature to generate the hardware interlock signal to the LV and HV power supplies. The signal for the HV power supply is provided with a dedicated cable while that for the LV power supply is propagated with the LV cable via the patch-panel. The LV protection board placed in the LV line gives additional protection from over-voltage for each channel.

The DCS controls all of the power supplies and centrally monitors the temperature and humidity measurements provided from the TIM as well as the voltage and current of the HV and LV, storing the information into a database. In addition, it can turn off the power supplies in case a deviation from the detector operation parameters is detected. 

The Wiener MPOD (Multichannel Power Supply System) Interlock (MPOD-I) PCB receives the Detector Safety Signal (DSS) from the cooling plant and provides the interlock signal to turn off the entire set of MPOD crates in case of a failure in the cooling plant.

An automatic software action  shuts down the tracker in cases that the temperature measured by the NTCs on the module reaches 31~${^{\circ}}$C. The hardware interlock takes action at temperature on the mechanical frame of the plane above 25~${^{\circ}}$C.

\subsection{DCS software}

The DCS software is implemented with the SIEMENS Simatic WinCC Open Architecture, which is a SCADA system used at CERN by the LHC experiments.

The communication between the different devices handled by the DCS and the DCS back-end is routed via a commodity switch using standard 1~Gbit/s Ethernet in a network secured from the general CERN network. The communication between the DCS and HV/LV MPOD modules is handled through the OPC-UA protocol, the communication between the DCS and the TIM board and the PDU is handled through the MODBUS protocol, and the DCS monitoring of the cooling system is done through a subscription to the DIP~\cite{DIP} publication by the CERN cooling and ventillation group. 

To ensure the safety of the detector, the DCS triggers alerts and automatic emergency procedures and controls the safety interlocks of the experiment. The DCS also implements autonomous integrity checks and diagnostics and provides a user interface to inform FASER on-call experts on the state and status of the experiment. In addition, the DCS is the interface for configuring and monitoring the parameters of the detector interlock systems, as discussed in Section~\ref{sec:TIM}. 

\begin{figure}[h]
    \centering
    \includegraphics[width=0.7\textwidth]{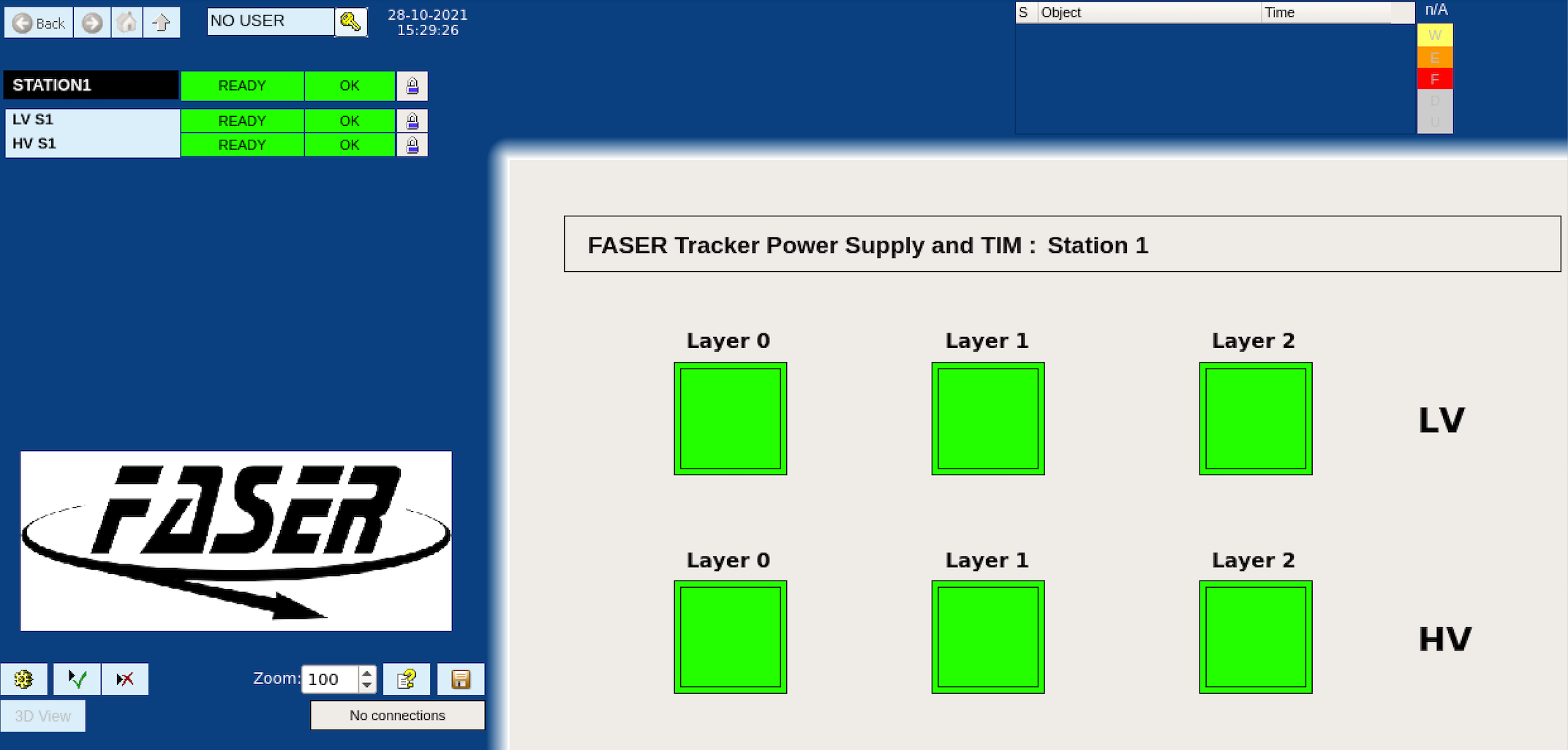}
     \caption{A DCS FSM panel via which the low and high voltage of a tracker station is controlled.}
     \label{fig:dcs_fsm}
 \end{figure}

Transitions between the different operational states of the experiment are implemented as a Finite State Machine (FSM) (see Figure~\ref{fig:dcs_fsm}). Transitions are initiated either by a user input or by an autonomous protocol. The action of the transition is initiated by propagating a transition signal through a sub-system hierarchy which in turn triggers a series of autonomous actions that configure the necessary parameters to achieve the transition between the different operational states. 

\begin{figure}[h]
     \centering
     \includegraphics[width=0.7\textwidth]{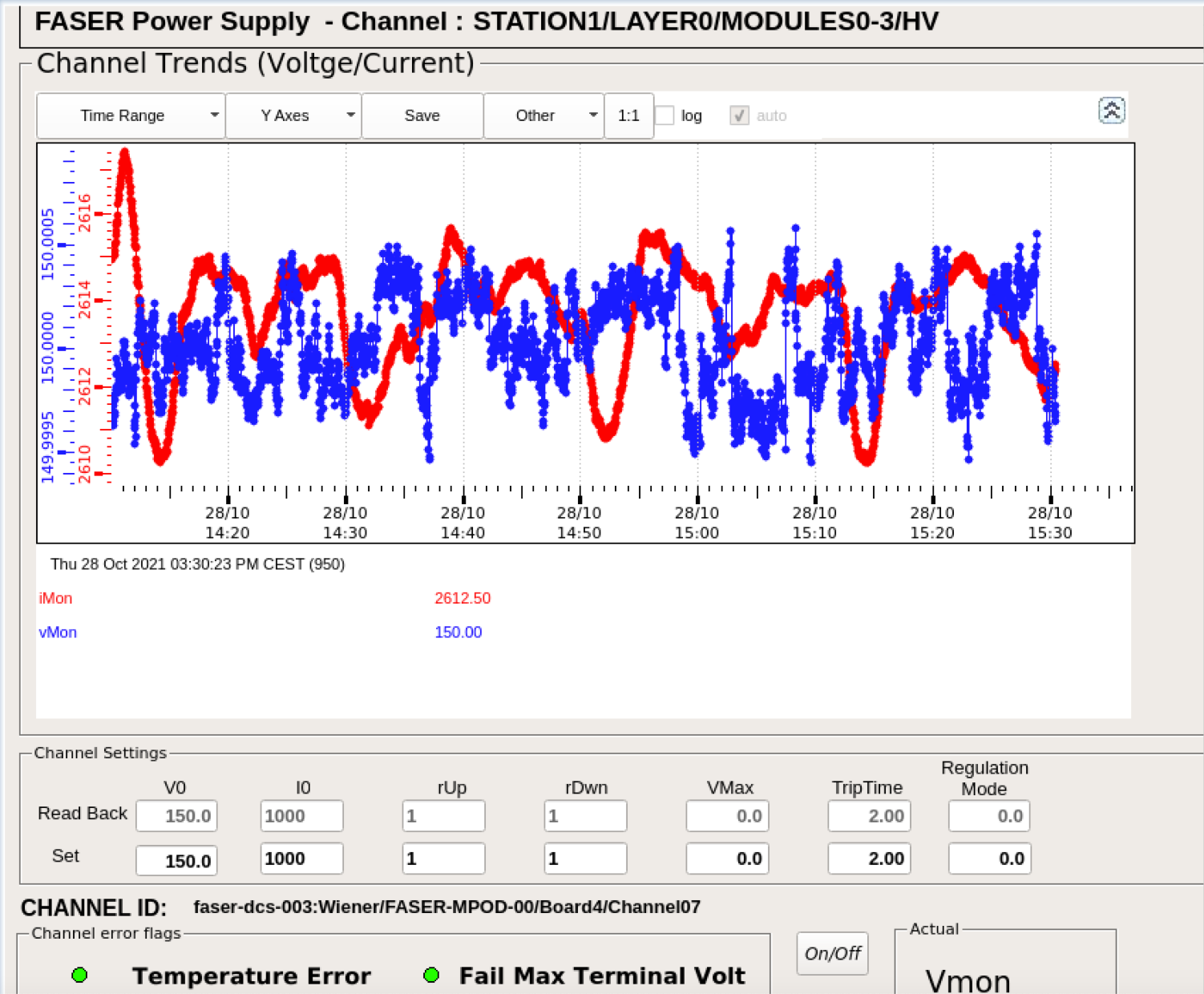}
     \caption{Example of the time series for high-voltage channel controlled and monitored by the FASER DCS.}
    \label{fig:dcs_timeseries}
 \end{figure}

The DCS also records time series of the detector parameters, as well as the operational state and the detector status. The time series are persistently archived for later retrieval in an Oracle database, which allows for offline data quality checks.
%which ensures the quality of the data collected at a particular time. 
Figure~\ref{fig:dcs_timeseries} shows an example of a time series for values monitored by the DCS. 
The DCS monitors approximately 600 parameters at a 1~Hz rate, collecting data at a rate of 1.3~kBytes/s. 

The Oracle DCS data is sampled once per minute and archived in an InfluxDB database along with DAQ monitoring, see Section~\ref{sec:tdaq_sw}. Host system monitoring (CPU/memory usage, disk status, etc.) for the DAQ and DCS servers are regularly archived as well. This data can easily be monitored from remote using Grafana dashboards and various alert messages are generated for on-call experts in case of unusual values before the automatic actions of the DCS system are activated.

%%%%%%%%%%%%%%%%%%%%%%%%%%%%%%%%%%%
\clearpage
\section{The emulsion detector of FASER$\nu$}
\label{sec:fasernu}

\begin{figure}[hbt!]
\centering
\includegraphics[width=0.95\textwidth]{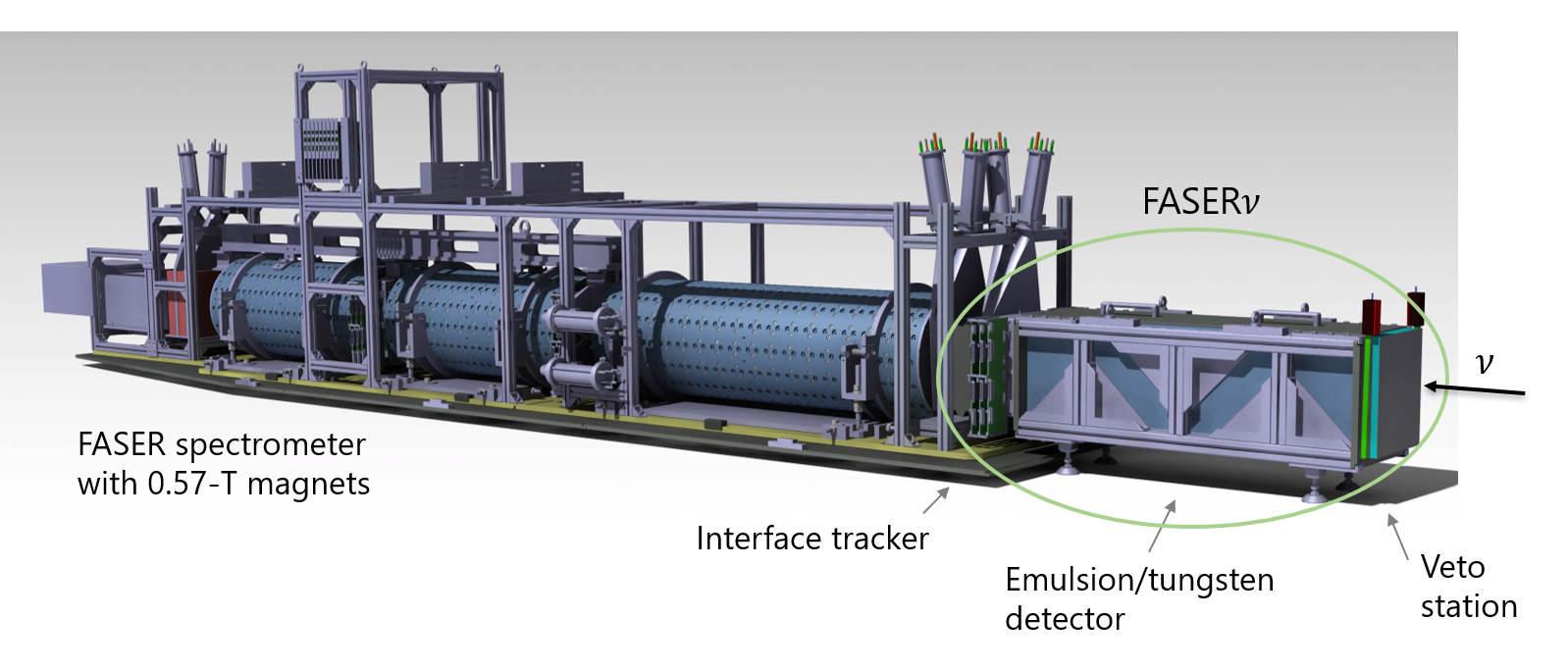}
\caption{A sketch of the FASER detector, highlighting the FASER$\nu$ detector.}
\label{fig:FASERnu_detector}
\end{figure}

The FASER$\nu$ detector~\cite{FASER:2019dxq,Abreu:2020ddv} is located in front of the main FASER detector along the beam collision axis to maximize the neutrino interaction rate of all three flavours. The detector includes an emulsion detector, a veto station and an interface tracker to the FASER spectrometer, as shown in Figure~\ref{fig:FASERnu_detector}. The IFT is placed downstream of the emulsion detector, and along with the veto scintillators, placed upstream of the emulsion detector, enable a global analysis that links information from FASER$\nu$ with the FASER spectrometer and makes muon charge measurements possible. This Section provides an overview of the emulsion detector for FASER$\nu$, including results from a pilot analysis carried out in 2018. The IFT is discussed in Section~\ref{sec:tracker}, and the veto scintillators in Section~\ref{sec:caloScint}.

%%%%%%%%%%%%%%%%%%%%%%%%%%%%
\subsection{Detector design}
%%%%%%%%%%%%%%%%%%%%%%%%%%%%

The emulsion detector is made of a repeated structure of emulsion films interleaved with 1-mm-thick tungsten plates. The emulsion film is composed of two emulsion layers, each 65 $\mu$m thick, that are poured onto both sides of a 210-$\mu$m-thick plastic base. The whole emulsion detector contains a total of 770 emulsion films with the dimensions of 25~cm $\times$ 30~cm, and a total tungsten mass of 1.1 tons. The total tungsten length is 770 mm, corresponding to 220 radiation lengths and 7.8 hadronic interaction lengths. 
The emulsion detector has the ability to identify different lepton flavours: sufficient target material to identify muons; finely sampled detection layers to identify electrons and to distinguish them from gamma rays; good position and angular resolutions to detect tau and charm decays. The detector can also measure the momenta of muons and hadrons, the energy of electromagnetic showers, and estimate the energy of neutrinos.

The emulsion detector readout and reconstruction works for a track density up to $\sim 10^6$~tracks/\si{cm^2}. To keep the detector occupancy sufficiently low, the emulsion films will be replaced during every technical stop of the LHC, which will take place about every three months. This corresponds to $10-50~\ifb$ of data in each data-taking period. 
In 2018  \textit{in situ} measurements were performed~\cite{Abreu:2020ddv} and measured a charged particle flux of 3 $\times$ 10$^4$~/cm$^2$/fb$^{-1}$ at the FASER location. When the emulsion films are removed, the track density will be roughly 0.3--1.5 $\times$ 10$^6$~tracks/cm$^2$. 
The experience with these in situ measurements demonstrates the ability to analyze emulsion films in this environment. 
The detector will be replaced 12 times during LHC Run 3 (three replacements in each of 2022, 2023, 2024, and 2025). For the first installation in March 2022, only 210 emulsion films, about 30\% of the following replacements, were included since less than a few $\ifb$ of data is expected in the data-taking period until July 2022, when the full complement of emulsion films will be installed.

%%%%%%%%%%%%%%%%%%%%%%%%%%%
\subsection{Emulsion films}
%%%%%%%%%%%%%%%%%%%%%%%%%%%

The emulsion sensitive layers consist of silver bromide crystals, which are semiconductors with a band gap of 2.684 eV, dispersed in a gelatine substrate. The diameter of the crystals which will be used for FASER$\nu$ is approximately 200~nm. When a charged particle passes through the crystal, electrons are excited through electromagnetic interaction to the conduction band, trapped in lattice defects, and groups of silver atoms (so-called latent images) are formed with interstitial silver ions. They can then be amplified and fixed by specific chemical development. An emulsion detector with 200~nm crystals has a spatial resolution of 50~nm. The two-dimensional intrinsic angular resolution of a double-sided emulsion film with 200-nm-diameter crystals and a base thickness of 210 $\mu$m is therefore 0.35 mrad. More details on the emulsion technology are summarized in Ref.~\cite{Ariga2020}.

The emulsion gel and film production is performed at a large-scale production facility established in Nagoya University. The left panel of Figure~\ref{fig:gel} shows an electron microscope photo of the produced silver bromide crystals. The sensitivity of the emulsion layers was checked by exposing the produced emulsion to electrons with several tens of MeV at the UVSOR Synchrotron Facility (Okazaki, Japan), measuring $\sim$45 grains per 100 $\mu$m for minimum ionizing particles (the right panel of Figure~\ref{fig:gel}). This sensitivity is sufficient for detecting minimum ionizing particles by setting the emulsion thickness to 65 $\mu$m. The produced emulsion gel is then used to produce films (65 $\mu$m emulsion layers deposited on both sides of 210 $\mu$m plastic base) using the coating system shown in Figure~\ref{fig:coating_system}. The production of emulsion gel and films are scheduled a few months before each installation. The 770 emulsion films produced in each batch correspond to a total area of $\sim$58 m$^2$.

\begin{figure}[hbt!]
\centering
\includegraphics[width=0.7\linewidth]{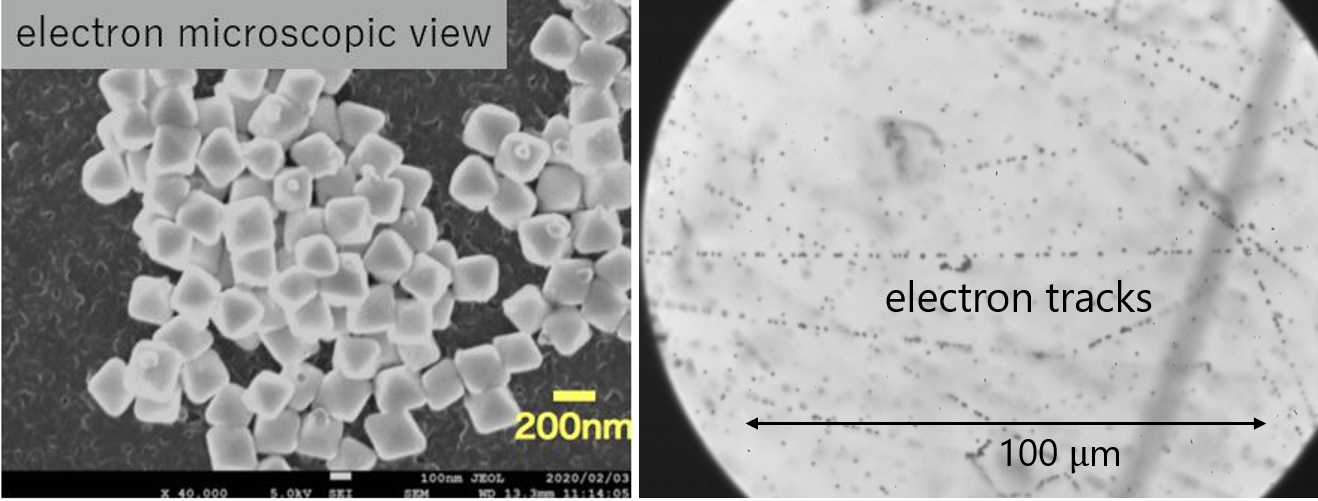}
\caption{Left: Microscopic view of silver bromide crystals. Right: $\beta$-ray tracks in an emulsion layer.}
\label{fig:gel}
\end{figure}

\begin{figure}[hbt!]
\centering
\includegraphics[width=0.6\linewidth]{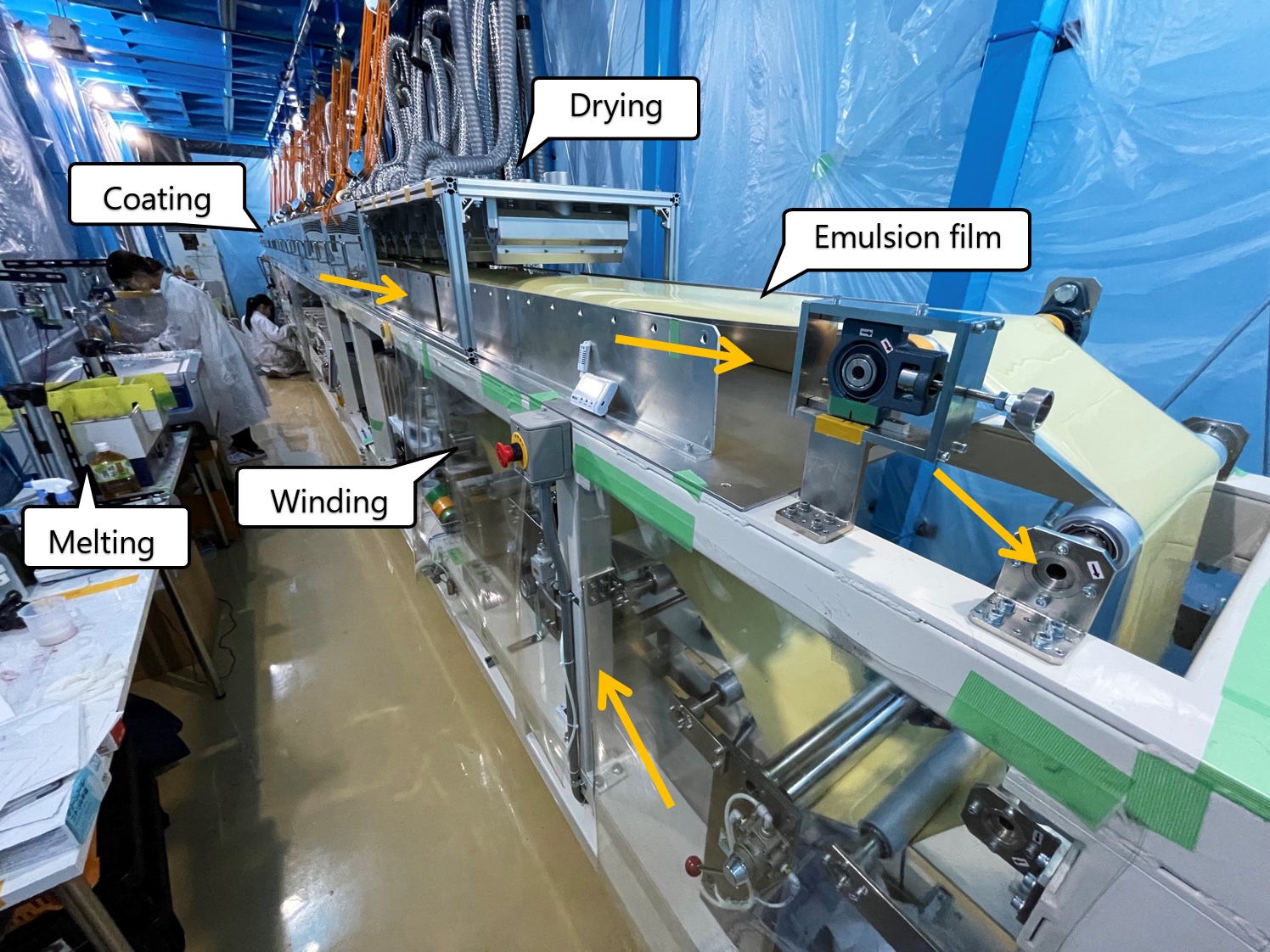}
\caption{A picture of the emulsion film coating system.}
\label{fig:coating_system}
\end{figure}

%%%%%%%%%%%%%%%%%%%%%%%%%%%%
\subsection{Tungsten target}
%%%%%%%%%%%%%%%%%%%%%%%%%%%%

Table~\ref{table:target_material} shows the properties of possible target materials. Tungsten was chosen as the target material for the following reasons. 
First, its high density allows for a higher neutrino interaction rate, keeping the detector small. Space for the detector along the beam collision axis is limited by the size of the FASER trench, and it is important to make the detector size small, which also lowers the cost of the  emulsion. Second, its short radiation length is good for a higher performance both in electromagnetic shower reconstruction, keeping shower tracks to a small radius, and in momentum measurement using multiple Coulomb scattering. Last, radioactivity levels are sufficiently low to guarantee the safe use of the emulsion films.

The thermal expansion coefficient of tungsten is very small, $\alpha=4.5\times 10^{-6}$/K. 
The temperature in the TI12 tunnel was monitored in 2018 and its variation was found to be very small, namely $\sim$0.1~\si{\degree C} RMS. The linear thermal expansion of 25~cm of tungsten is then expected to be 0.1~$\mu$m. 
Since the thermal expansion coefficient is very different between emulsion films ($\alpha\sim 10^{-4}$/K) and tungsten, it is necessary to exert a large mechanical pressure on the emulsion films and tungsten plates in such a way that the soft emulsion films follow the thermal expansion of the tungsten plates.

A total of 1600 1-mm-thick tungsten plates were purchased to be used for the FASER$\nu$ detector. A dedicated device for measuring and mapping the thickness was prepared to check the tungsten plate thickness uniformity. The thickness was measured semi-automatically at 24 points on each plate, and the maximum difference among the 24 points was checked. The plates with a difference smaller than 80 $\mu$m are used to construct the emulsion detector, corresponding to about 97\% of the measured tungsten plates. The average thickness of the qualified tungsten plates is 1093$\mu$m, with an RMS of 25~$\mu$m.

\begin{table}[htbp]
\centering
\begin{tabular}{|c|c|c|c|c|c|}
\hline
Material & Atomic & Density & Hadronic Interaction  & Radiation length & Thermal expansion \\
\ & number & [g/cm$^3$] & [cm] & length [mm] & $\alpha$ [\si{\times 10^{-6}K^{-1}}]\\ \hline 
Iron     & 26 & 7.87 & 16.8 & 17.6 & 11.8\\ 
Tungsten & 74 & 19.30 & 9.9 & 3.5 & 4.5\\ 
Lead     & 82 & 11.35 & 17.6 & 5.6 & 29\\  \hline
\end{tabular}
\caption{Properties of possible target materials.
}
\label{table:target_material}
\end{table}

%%%%%%%%%%%%%%%%%%%%%%
\subsection{Mechanical structure}
%%%%%%%%%%%%%%%%%%%%%%
The mechanical structure was designed to keep the emulsion films aligned. The position alignment has to be kept within sub-micrometer accuracy during data taking so that particle momenta can be measured by the multiple Coulomb scattering (MCS) coordinate method described in Ref.~\cite{Kodama:2002dk}. 

To put sufficient pressure on the emulsion films and tungsten plates, the following steps are taken. First, 10 emulsion films and 10 tungsten plates are vacuum-packed with aluminum-laminated foils to create a modular structure in the detector, as shown in Figure~\ref{fig:FASERnu_module_structure}. The pressure on each module is given by atmospheric pressure. All the 77 modules are then installed in a mechanical structure, which presses all the modules to one side, again, to keep the position relation between the modules. The mechanical support, including a presser with a thickness of 3.5~cm, is called the FASER$\nu$ box (in order to allow the replacement of the emulsion detector in a short access, two FASER$\nu$ boxes have been constructed). 
The design and a picture of the box are shown in Figure~\ref{fig:FASERnu_box}. The inner dimensions of the box are 27~cm $\times$ 30~cm $\times$ 104~cm, and the outer dimensions are 39~cm $\times$ 37~cm $\times$ 116~cm. The upstream and downstream walls are 2~cm thick stainless steel, and the side walls are 1~cm thick aluminum plates. Additional structures with 5~cm $\times$ 5~cm aluminum frames are placed on the sides and bottom of the box for strength. The total weight of the empty box is approximately 190~kg. The height of the box above the trench floor can be adjusted depending on the LOS position, which will vary from year to year due to the change of the beam crossing angle as discussed in Section~\ref{sec:run3}. The 77 vacuum-packed modules are housed in the structure, and a force of 7500~N is imposed by the presser located upstream.

To avoid temperature fluctuations, which may cause a mis-alignment between the emulsion modules, an insulating wall is placed between the IFT and the FASER$\nu$ box. It is made of an aerogel-based insulation blanket~\footnote{SPACETHERM A1 from the Proctor Group.} wrapped with aluminum foil. In order to keep the thermal stability of the FASERnu box, the part of the trench where this is installed is covered by an aluminium cover, lined with the same thermal insulation. Being isolated from the rest of the electronic detectors, the temperature of the FASER$\nu$ box reaches an equilibrium with the trench wall. Temperatures at various positions around the FASER$\nu$ box will be monitored via temperature sensors during data taking.

\begin{figure}[hbt!]
\centering
\includegraphics[width=0.8\textwidth]{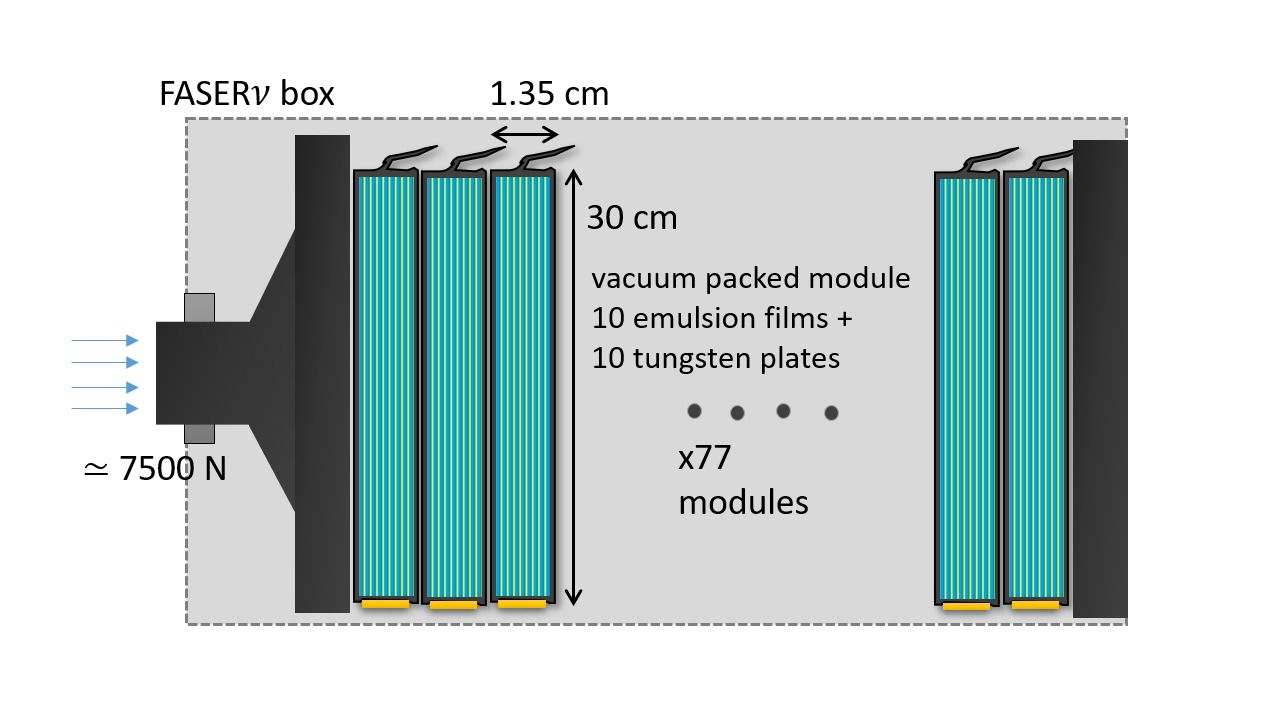}
\caption{Module structure of the emulsion/tungsten detector.
}
\label{fig:FASERnu_module_structure}
\end{figure}

\begin{figure}[hbt!]
\centering
\includegraphics[height=5.5cm]{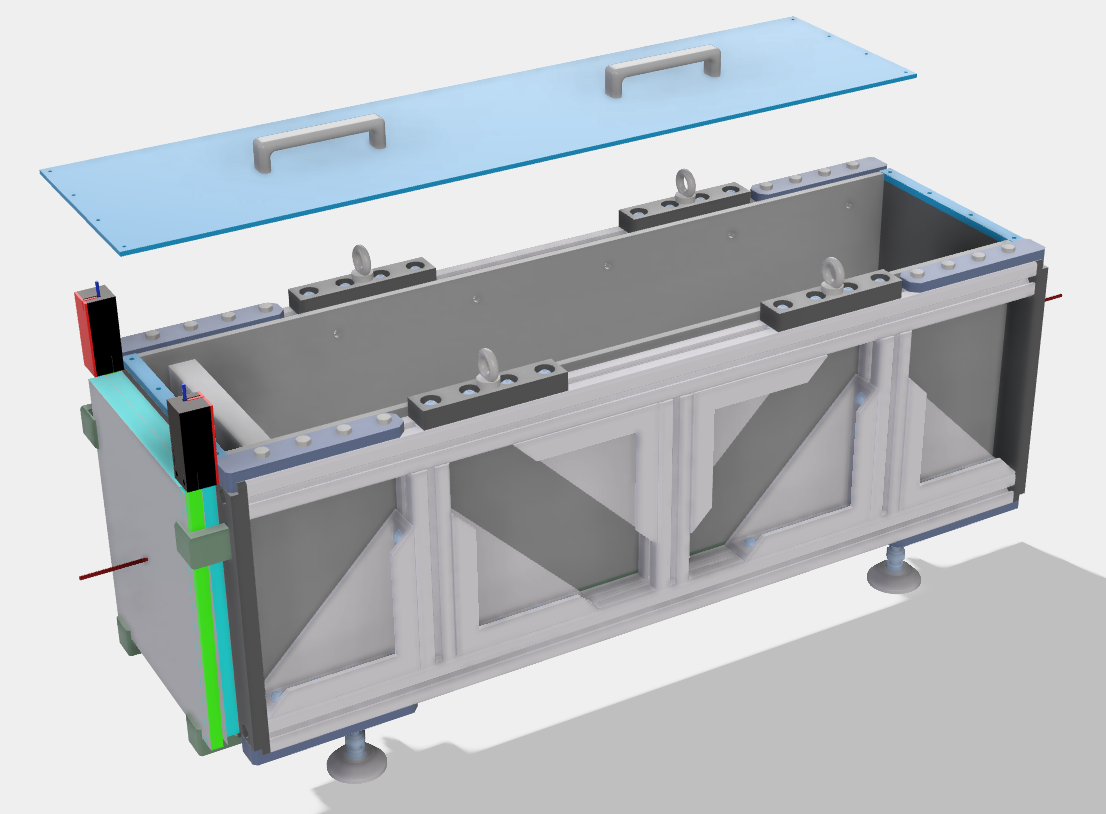}
\hspace{1cm}
\includegraphics[height=5.5cm]{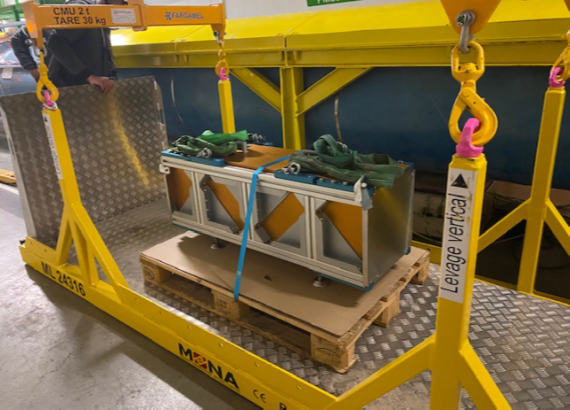}
\caption{Design of the FASER$\nu$ box (top) and a picture of the produced box (bottom).}
\label{fig:FASERnu_box}
\end{figure}

%%%%%%%%%%%%%%%%%%%%%%%%%%%%%%%%%%%%%%%%%%%%%%
\subsection{Operational procedure}
%\subsection{Assembly/installation/replacement/chemical development}
%%%%%%%%%%%%%%%%%%%%%%%%%%%%%%%%%%%%%%%%%%%%%%

The emulsion/tungsten detector will be assembled in the dark room at CERN just before each installation. The detector will then be transported to the experimental site in one piece. This minimizes the amount of underground work under restricted conditions.

For the installation of each FASER$\nu$ box, the detector will be brought down to the LHC tunnel using the elevator at Point 1 where the ATLAS interaction point is located (see Figure~\ref{fig:TI12}). It will be transported along the LHC beamline on an electric cart, and then carried over the LHC in UJ12 using the crane employed for the main FASER detector installation. A protection device was installed under this crane, with dimensions similar to those of the detector and a 1.5 tonne load capability. The detector will be installed into the FASER trench in front of the main FASER detector by using the crane installed in TI12. Figure~\ref{fig:installed} shows a photo of the installed FASER$\nu$ box.

The detector will be replaced during planned technical stops. 
The exchange procedure steps are: 
(1) construction of the new emulsion modules using the second (unused) set of tungsten plates. These modules are assembled into the unused FASER$\nu$ box;
(2) extraction of the exposed FASER$\nu$ box and installation of the newly assembled FASER$\nu$ box ; 
(3) disassembly of the emulsion films and their chemical development.

\begin{figure}
    \centering
    \includegraphics[width=0.6\linewidth]{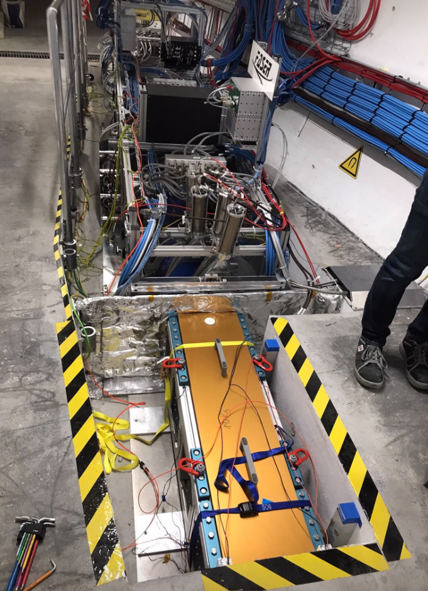}
    \caption{A picture of the FASER$\nu$ box installed in the FASER trench. The cover of the trench where FASER$\nu$ is placed was not installed when this picture was taken.}
    \label{fig:installed}
\end{figure}

Specific chemical development will amplify the recorded signals in silver bromide crystals.
The chemical solutions needed for the development are described in Table~\ref{tab:solutions}. In the developer solution, filaments of metallic silver start to grow from the latent image speck and become visible as dots under optical microscopes. The amplification gain is about $\mathcal{O}(10^8)$, and depends on the temperature and duration of the treatment.
The chemical solutions will be prepared and the development will be carried out in the dark room at CERN, which will be equipped so that 100 films/day can be processed.

\begin{table}[hbt!]
\centering
\small
\begin{tabular}{|l|p{2.2cm}|p{4.4cm}|p{4.4cm}|p{1.4cm}|}
\hline

Solution & \raggedright Time and temperature & Function & Chemical & Amount /58 \si{m^2}\\
\hline

Developer & \raggedright 20 min at \si{20 \pm 0.1 \degree C} & Chemical amplification of signal with a gain of $\mathcal{O}(10^8)$& \raggedright OPERA Dev (Fujifilm), RD-90s starter (Fujifilm) & 370 L\\

Stopper & 10 min & Stop chemical amplification & Acetic acid & 180 L\\

Fixer & 90 min & Resolve unused silver bromide crystals & UR-F1 (Fujifilm) & 1100 L\\

Wash & $>$300 min & Wash out all chemicals & Running water & \\

Thickener & 20 min & \raggedright Control emulsion layer thickness & \raggedright Glycerine, Drywell (Fujifilm) & 50 L\\

Drying & $\sim$1 day & Dry films & Air at R.H.=50--60\% & \\

\hline
\end{tabular}
\caption{Solutions required for the emulsion chemical development.}
\label{tab:solutions}
\end{table}

%%%%%%%%%%%%%%%%%%%%%%%%%%%%%
\subsection{Facility at CERN for the emulsion detector handling}
%%%%%%%%%%%%%%%%%%%%%%%%%%%%%

Emulsion films are sensitive to light, therefore an assembly of the emulsion/tungsten detector and chemical development of the emulsion films will be performed in the dark room facility at CERN. The facility was originally set up for the CHORUS experiment and has been used by several experiments since then. In preparation for LHC Run 3, the facility is being refurbished, to allow several emulsion-based experiments to use it in parallel.

While the refurbishment is ongoing, operation tests were performed using the available space in October 2021 together with the NA65/DsTau experiment~\cite{dstau}, by going through assembly and chemical development. Figure~\ref{fig:dark_room} shows a picture of the current facility. 

\begin{figure}
    \centering
    \includegraphics[width=0.7\linewidth]{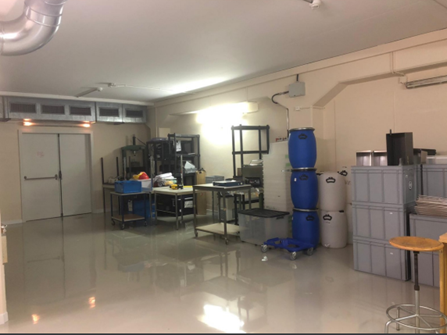}    
    \caption{Dark room facility at CERN available for the FASER$\nu$ detector assembly and chemical processing. The facility was recently refurbished and is ready for the first installation.}
    \label{fig:dark_room}
\end{figure}

%%%%%%%%%%%%%%%%%%%%%%%%%%%%%
\subsection{Emulsion readout}
%%%%%%%%%%%%%%%%%%%%%%%%%%%%%

The emulsion readout system takes a sequence of tomographic images by changing the focal plane through each emulsion layer. The digitized images are then analyzed to recognize sequences of grains as a track segment. The FASER$\nu$ event analysis will be based on readout of the full emulsion detector by the Hyper Track Selector (HTS) system~\cite{Yoshimoto:2017ufm}. A picture of the HTS system is shown in Figure~\ref{fig:hts}. The HTS includes a dedicated lens, camera, XYZ-axis stage, and computer cluster for image processing. It takes 22 tomographic images, and 16 successive images in the emulsion layer are used for track recognition. 
The HTS system makes use of a custom-made objective lens with a very large field of view of 5.1 mm $\times$ 5.1 mm and a magnification of 12.1. The optical path is split 6-fold. Correspondingly, the image is projected onto six mosaic camera modules to cover the large field of view with high resolution. Each mosaic camera module consists of 12 2.2-Mpixel image sensors. In total, 72 image sensors work in parallel to build the large field of view. The raw image data throughput from 72 image sensors amounts to 48 GBytes/s, which is then processed in real time by 36 tracking computers with two GPUs each. The readout speed of the HTS system is 0.45~m$^2$/hour/layer. 
Currently, an upgraded HTS system (HTS2, which will be about 5 times faster) is under commissioning. The baseline plan for FASER$\nu$ is to use the HTS system, since its performance, such as the readout speed and resolution, is already proven. The total emulsion film surface to be analyzed in FASER$\nu$ is 174 m$^2$/year implying a readout time of 770 hours/year. Assuming some hours of machine time each day, it will be possible to finish reading out the data taken in each year within a year. The HTS system was also used for the readout of the 2018 pilot detector, which led to the observation of the first neutrino interaction candidates at the LHC~\cite{FASER:2021mtu}.

\begin{figure}[hbt!]
\begin{center}
\includegraphics[height=8cm]{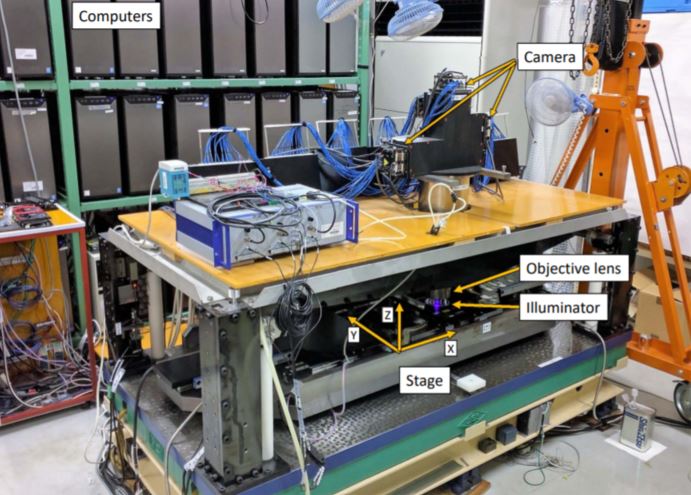}
\caption{The fast emulsion readout system HTS~\cite{Yoshimoto:2017ufm}, with a readout speed of 0.45~m$^2$/hour/layer.
}
\label{fig:hts}
\end{center}
\end{figure}

\subsection{Pilot analysis with the 2018 data}

In 2018, a pilot emulsion detector was installed in the TI18 tunnel. An integrated luminosity of 12.2~fb$^{-1}$ was collected during four weeks of data taking from September to October with $pp$ collisions at 13~TeV centre-of-mass energy. 
The analysis of the pilot detector demonstrated that the emulsion readout and reconstruction can work in the actual experimental environment. The data analysis is based on the readout of the full emulsion films by the HTS system. Data processing was divided into sub-volumes with a maximum size of 2~cm~$\times$~2~cm~$\times$~25 emulsion films. After a precise alignment procedure, tracks are reconstructed in multiple films with a dedicated tracking algorithm for high-density environments~\cite{Aoki:2019jry}.

The majority of the tracks observed in the detector are expected to be background muons and related electromagnetic showers. These background charged particles were analysed using a unit of 10 emulsion films. In this case the angular resolution is expected to be 0.05~mrad. Figure~\ref{fig:pilot_analysis} (left) shows the observed angular distribution peaked in the direction of the ATLAS IP. 
There are two peaks separated by 2.5~mrad. The reason for the two-peak structure is not understood, but simulation studies are ongoing, which could inform future data measurements. It's clear that the angular resolution should be better than the angular spread of the peaks. For example, the horizontal angular spread of one of the peaks (left in the figure) is 0.6~mrad, equivalent to the multiple Coulomb scattering of 700~GeV particles through 100~m of rock. 
The charged particle flux within 10~mrad from the peak angle is measured to be $(1.7 \pm 0.1) \times 10^4$~tracks/cm$^2$/fb$^{-1}$, which is consistent with the values previously reported~\cite{FASER:2018bac,FASER:2019dxq} and also consistent with the FLUKA prediction of $2.5\times10^4$~ tracks/cm$^2$/fb$^{-1}$ for $E_\mu>10$~GeV. The expected uncertainty on the FLUKA estimate is of the order of 50\%.

For the neutrino analysis, vertex reconstruction was performed by searching for converging patterns of at least five tracks with a impact parameter to the vertex within 5~$\mu$m. Additional topological cuts were applied to these vertices to select high-energy interactions and suppress neutral hadron backgrounds. Vertices are categorized as charged or neutral based on the presence or absence of charged parent tracks. Within the fiducial volume, 18 neutral vertices passed the vertex selection criteria. Figure~\ref{fig:pilot_analysis} (right) shows a selected neutral vertex. A multivariate discriminant was then applied to distinguish neutrino signal from neutral hadron background, resulting in a 2.7$\sigma$ excess of the neutrino-like signal. A more detailed description of the pilot detector analysis can be found in Ref.~\cite{FASER:2021mtu}.

The above measurements of the charged particle flux and results detecting the first neutrino interaction candidates proved FASER$\nu$'s ability to study neutrinos at the LHC.

\begin{figure}[htb]
\centering
\includegraphics[height=4.5cm,keepaspectratio,clip]{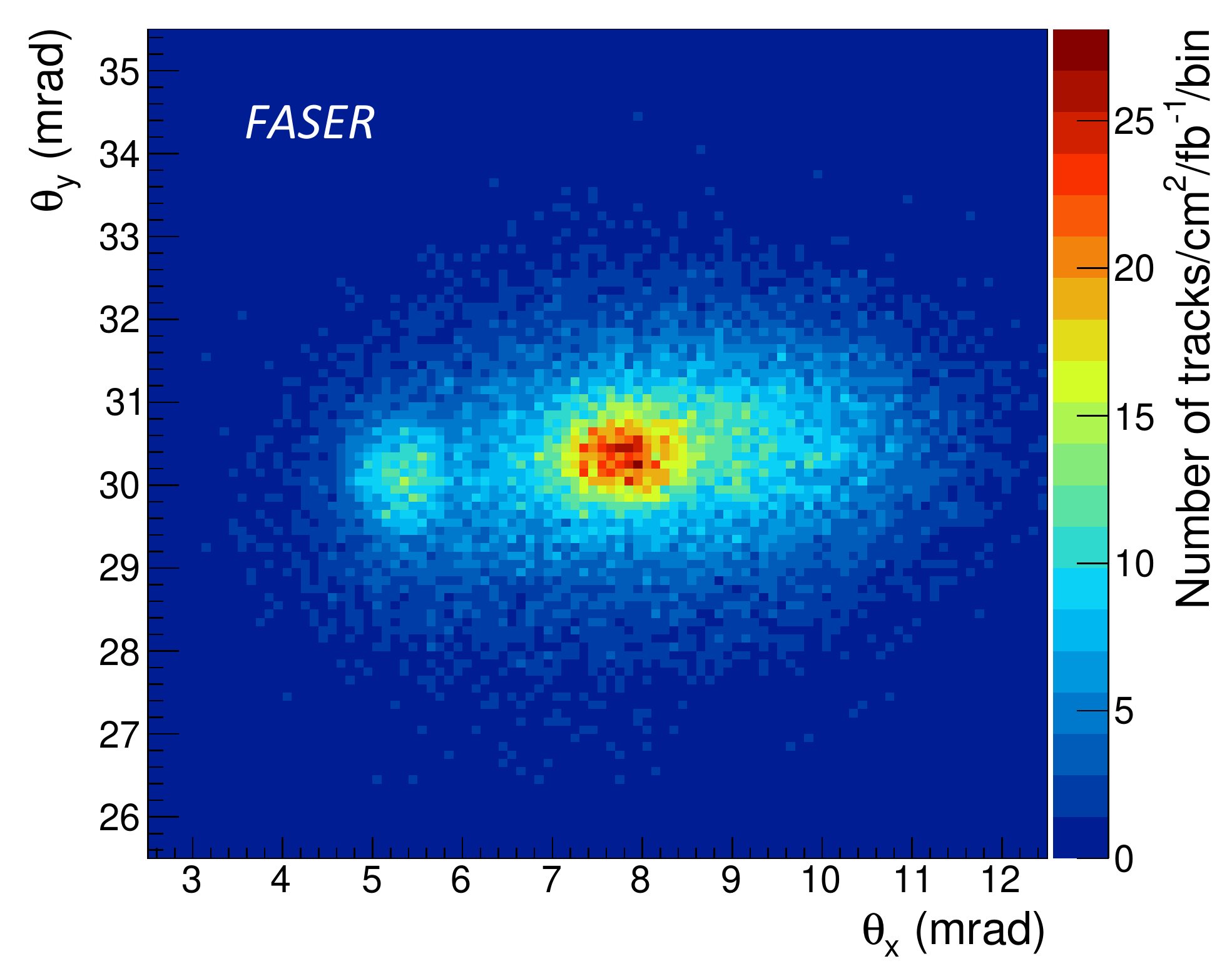}
\includegraphics[height=4.5cm,keepaspectratio,clip]{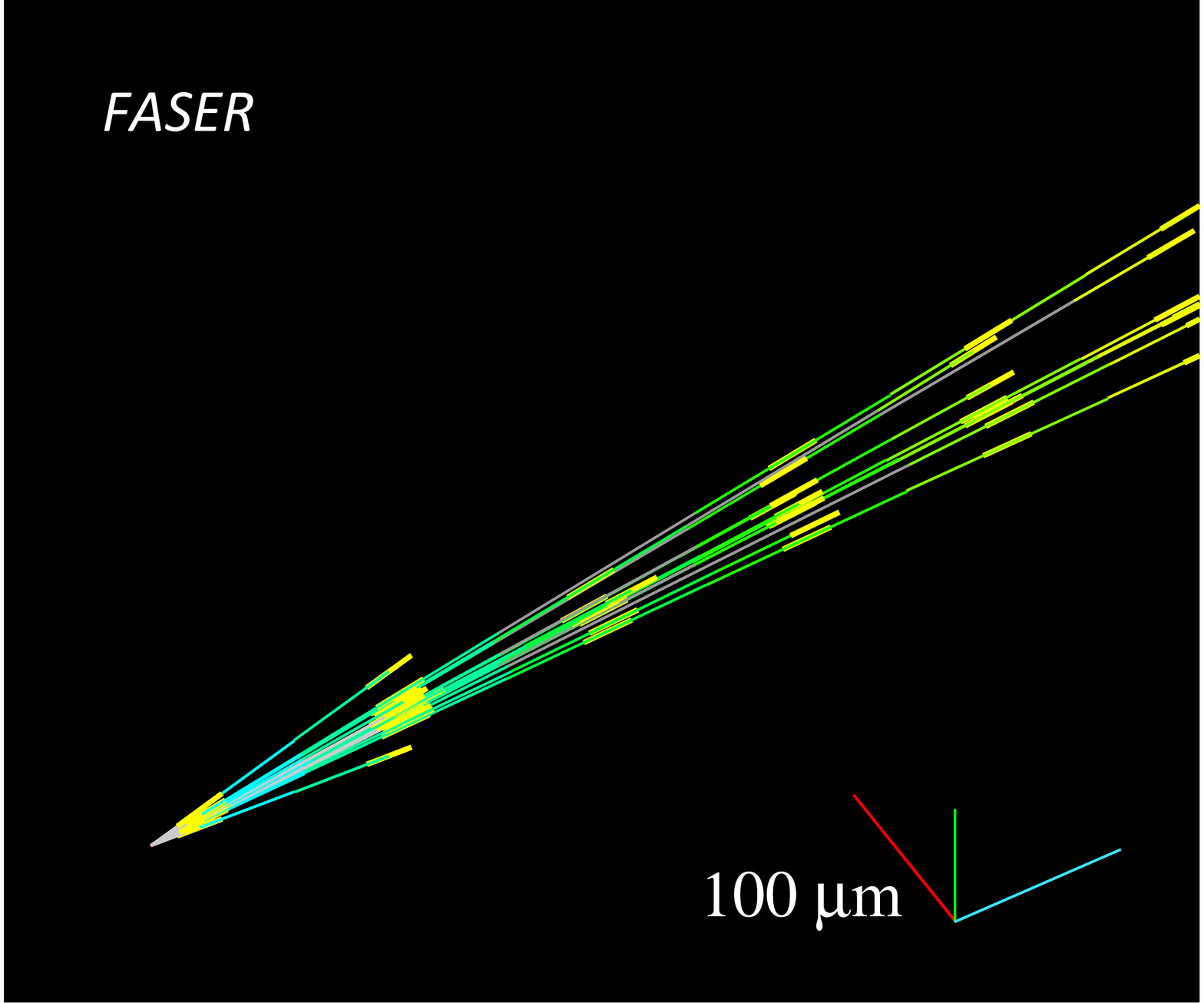}
\caption{(left) A zoom of the angular distribution observed in the pilot run module. The uncertainty in the angle of the installed detector with respect to the LOS means that the measured angular peak is compatible with pointing back to the IP. (right) Event displays of one of the neutral vertices in a tilted view.}
\label{fig:muon}
\label{fig:pilot_analysis}
\end{figure}

%%%%%%%%%%%%%%%%%%%%%%%%%%%%%%%%%%%
\clearpage
\section{Detector Integration}
\label{sec:integration}
The integration and installation of the FASER detector in its final location started in 2020 with preparation of the TI12 tunnel and engineering work on detector supports.  
This is described in this section, together with the installation procedure and the detector and magnet alignment procedure.   

\subsection{Preparation of TI12 for FASER}

In 2018 the TI12 tunnel had no working services installed, and contained substantial obsolete ventilation and electrical equipment left over from the former LEP collider, as can be seen in Figure~\ref{fig:TI12beforeafter} (left). During the first half of 2019 this equipment was removed, to allow the start of the FASER works.

The FASER trench is needed to allow the detector to be positioned on the LOS, which was precisely mapped out at the mm-level by the CERN survey team. It was designed to minimize the amount of civil engineering works that needed to be carried out in the tunnel. Particular attention was paid to keeping the structural integrity of the tunnel, minimizing the amount of dust produced, and minimizing the time needed for the works to be compatible with the overall LS2 schedule. The trench design is shown in  Figure~\ref{fig:trench_3DCAD} and in Figure~\ref{fig:trench_dwg1}. The main trench is about 5.5~m long and 1.4~m wide, and is 60~cm deep at the front of the FASER detector and 20~cm deep at the back. This compensates for the TI12 tunnel sloping up away from the LHC. The bottom of the trench is parallel to the LOS at an angle of about 1\% to the horizontal. At the front of the main trench is an additional, more narrow part where FASER$\nu$ is installed. Figure~\ref{fig:TI12beforeafter} (right) shows a photo of the  trench, which was completed in May 2020.

After the civil engineering works were done, 3D laser scanning was performed by the CERN survey group to check for potential deviations with respect to the design. A rendering of the 3D scan is shown in Figure~\ref{fig:scan3D}, and shows the comparison with the CAD model by superimposing the two shapes in CAD software. The maximum deviation was found to be within 10~mm and located on the edges of the trench floor, further small civil engineering works were done to correct for this deviation and to allow the lower baseplate to be installed in the correct position as described in Section~\ref{sec:BPs}. 

During the summer of 2020, the needed infrastructure for FASER was installed in the TI12 tunnel. Figure~\ref{fig:TI12-infrastructure} shows an integration drawing of the tunnel highlighting most of the installed infrastructure. This included installation of electrical power, lighting, handling equipment for the detector installation, compressed air connections, optical fiber connections to the surface, and electrical racks.

\begin{figure}[hbt!]
    \centering
    \includegraphics
    [trim=3cm 3.5cm 4cm 3cm,clip,width=1.0\textwidth]{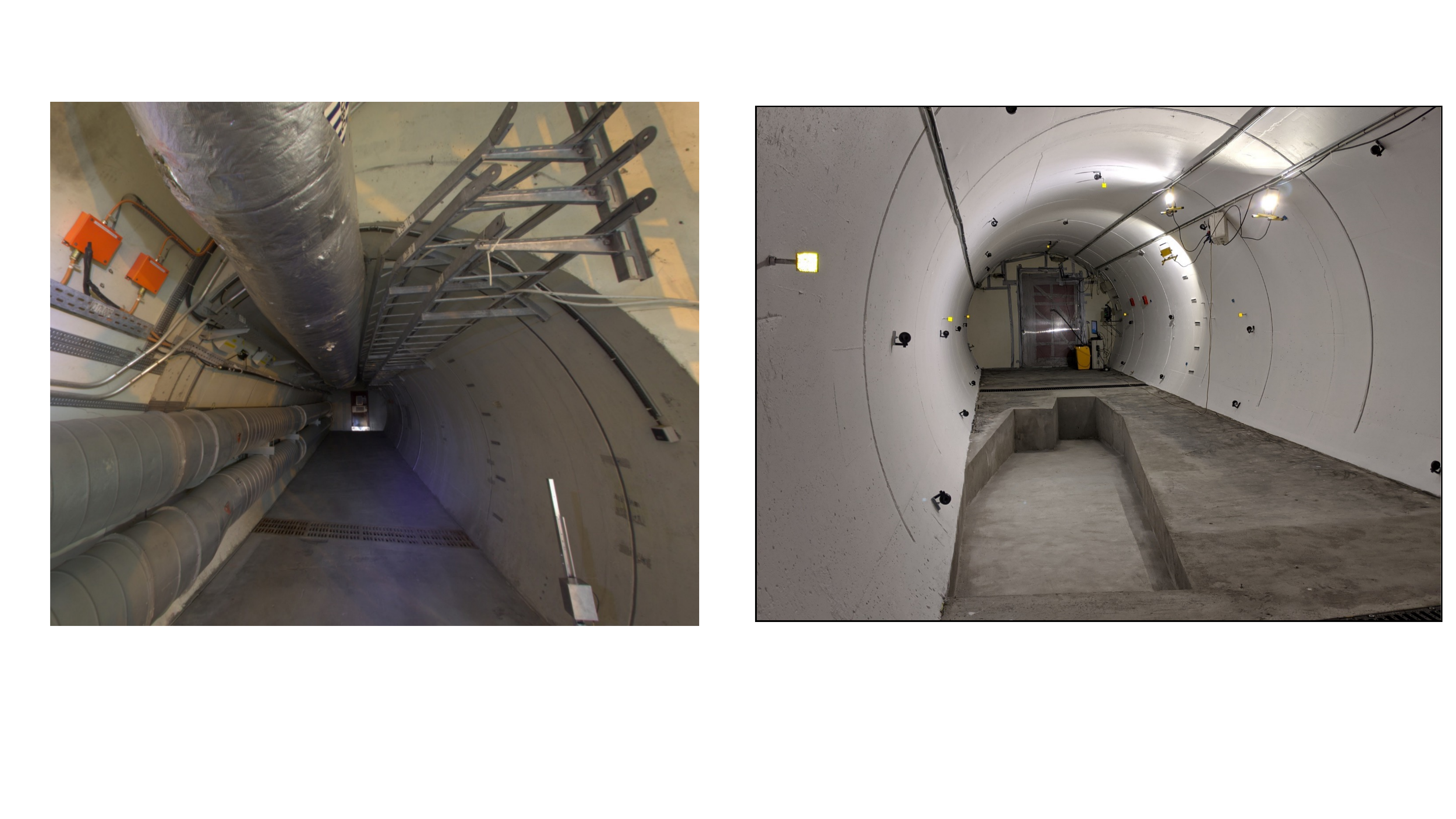}
    \caption{The TI12 tunnel (left) before the FASER preparation work (right) after the tunnel has been cleared out and the FASER trench dug in the tunnel floor.}
    \label{fig:TI12beforeafter}
\end{figure}

\begin{figure}[hbt!]
    \centering
    \includegraphics[width=0.9\textwidth]{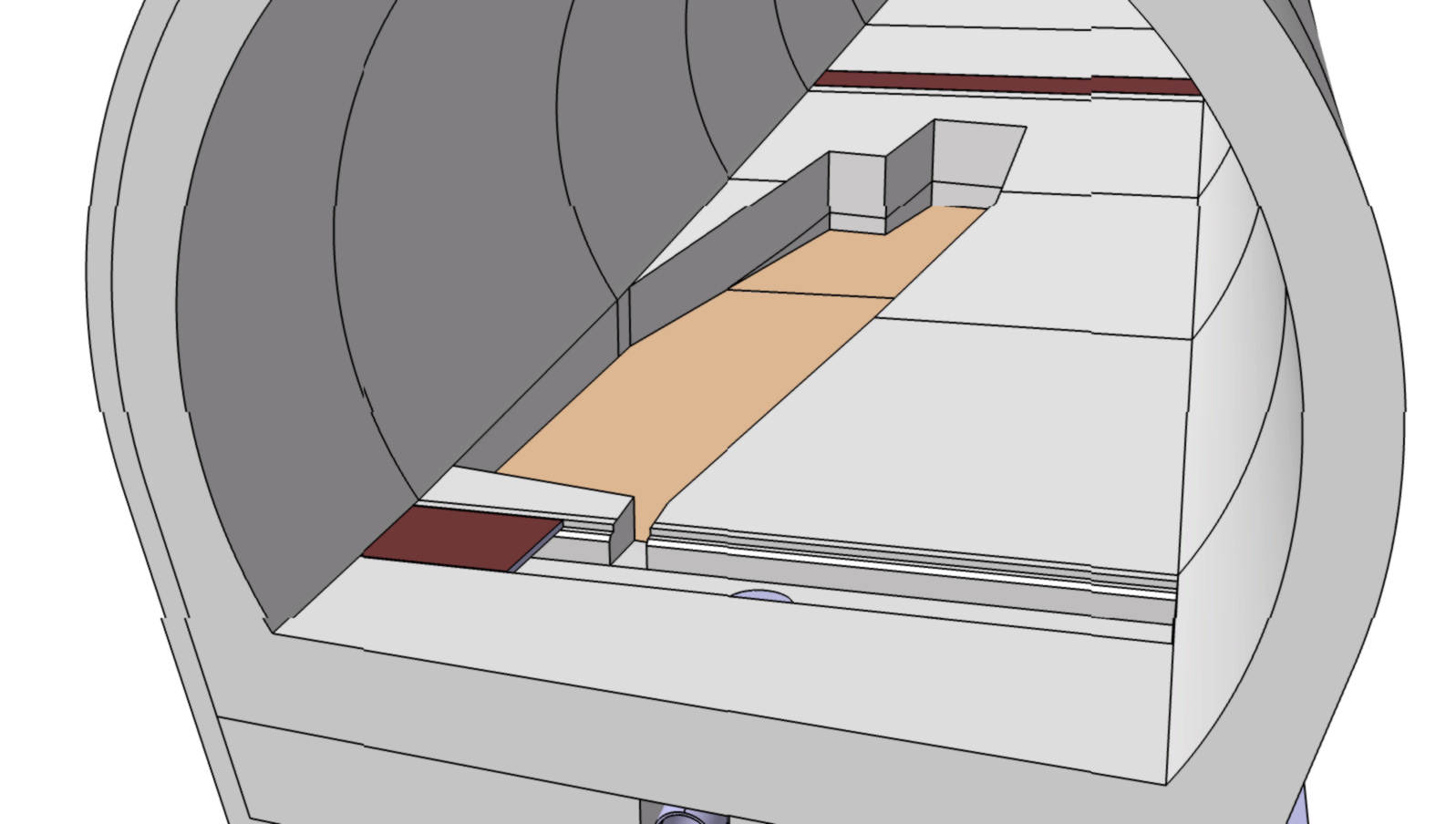}
    \caption{The 3D CAD model of the FASER trench design in TI12.}
    \label{fig:trench_3DCAD}
\end{figure}

\begin{figure}[hbt!]
    \centering
    \includegraphics[width=1.\textwidth]{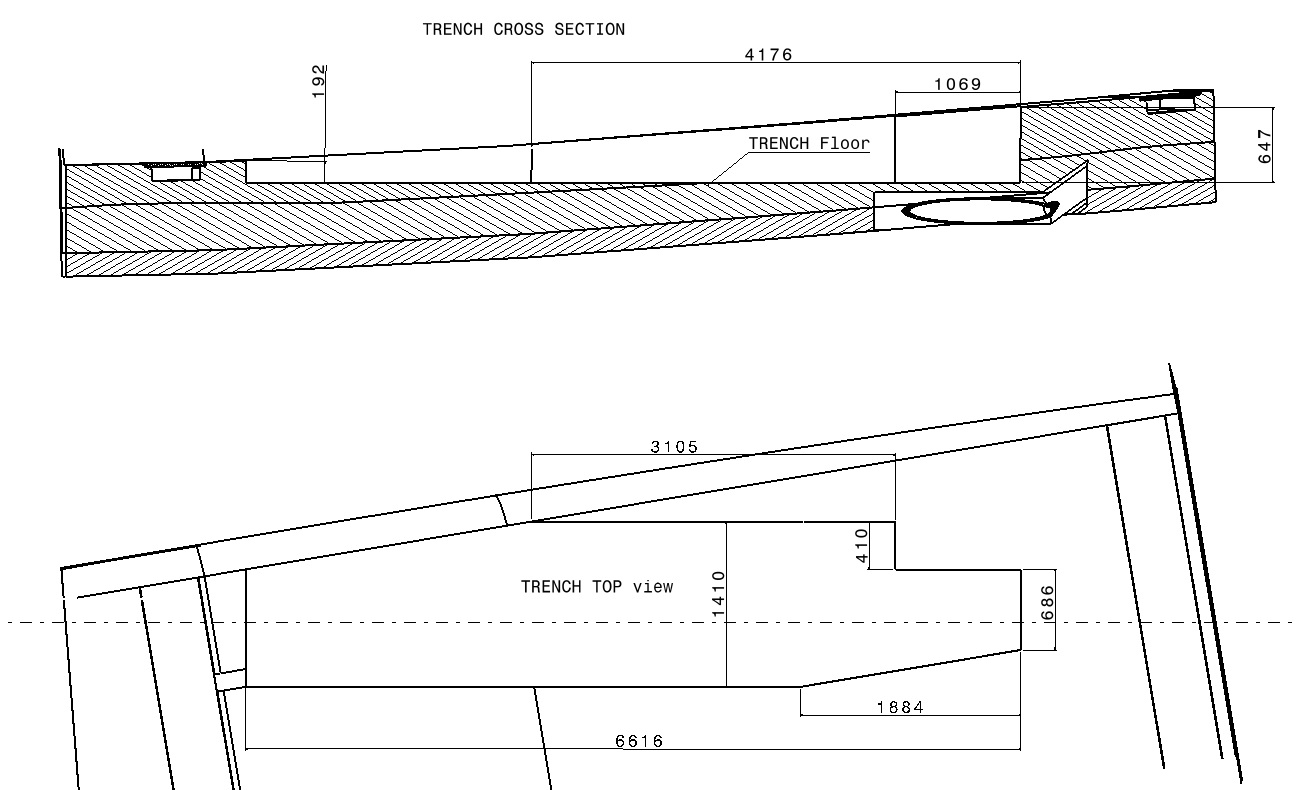}
    \caption{The design of the FASER trench (top: side view; bottom: plan view), with key dimensions shown in mm.}
    \label{fig:trench_dwg1}
\end{figure}

\begin{figure}[hbt!]
    \centering
    \includegraphics[width=1. \textwidth]{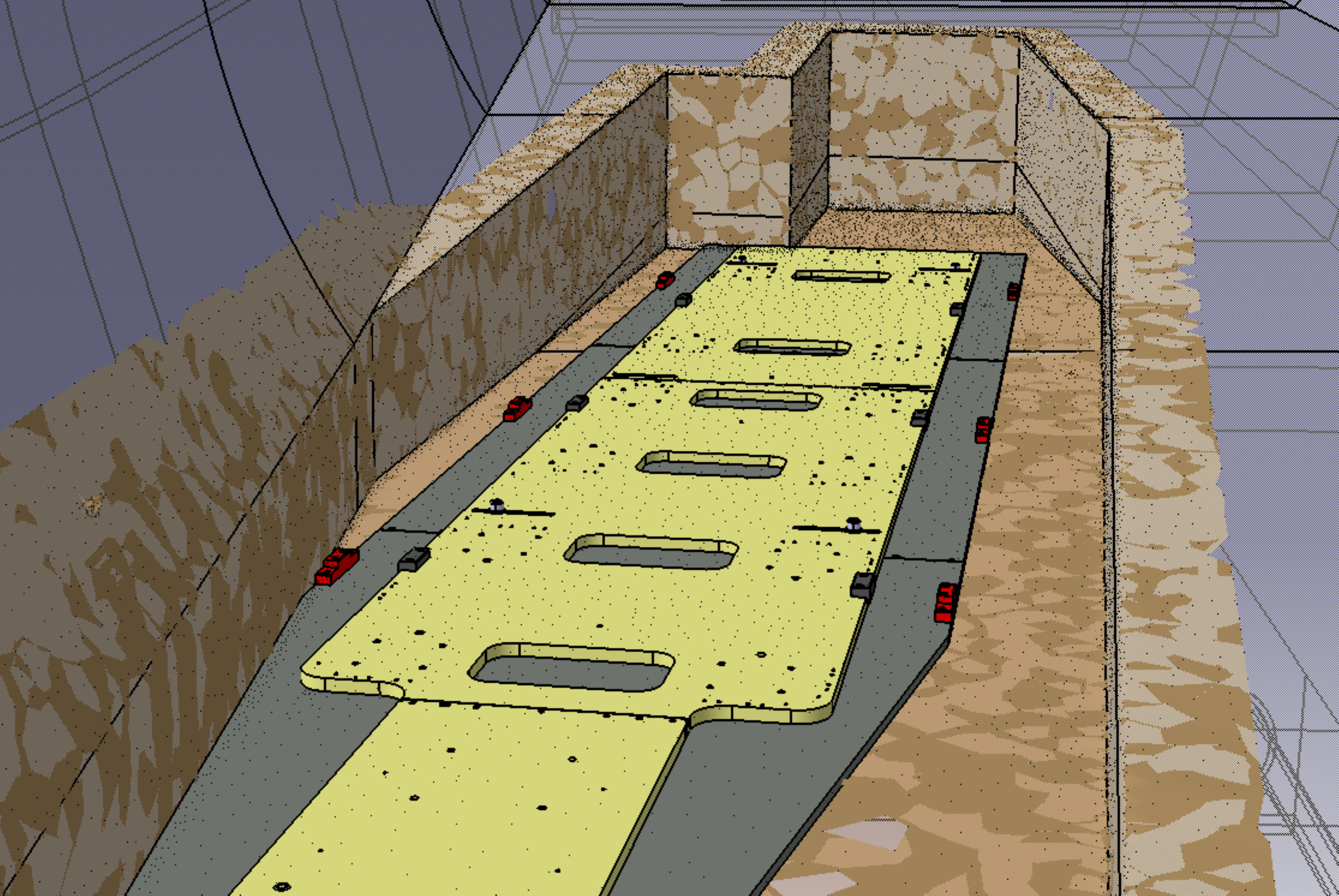}
    \caption{Comparison between the 3D scan of the FASER trench and the CAD model, also showing the lower and upper baseplates. 
    The solid lines are from the CAD model, and dots are the measurements from the laser scan that have been smoothed in the trench region to give the brown coloured surfaces. The yellow and dark-grey colours show the upper and lower baseplate positions from the CAD model.}
    \label{fig:scan3D}
\end{figure}

\begin{figure}[hbt!]
    \centering
    \includegraphics[trim=4cm 1cm 4cm 3cm,clip,width=1.\textwidth]{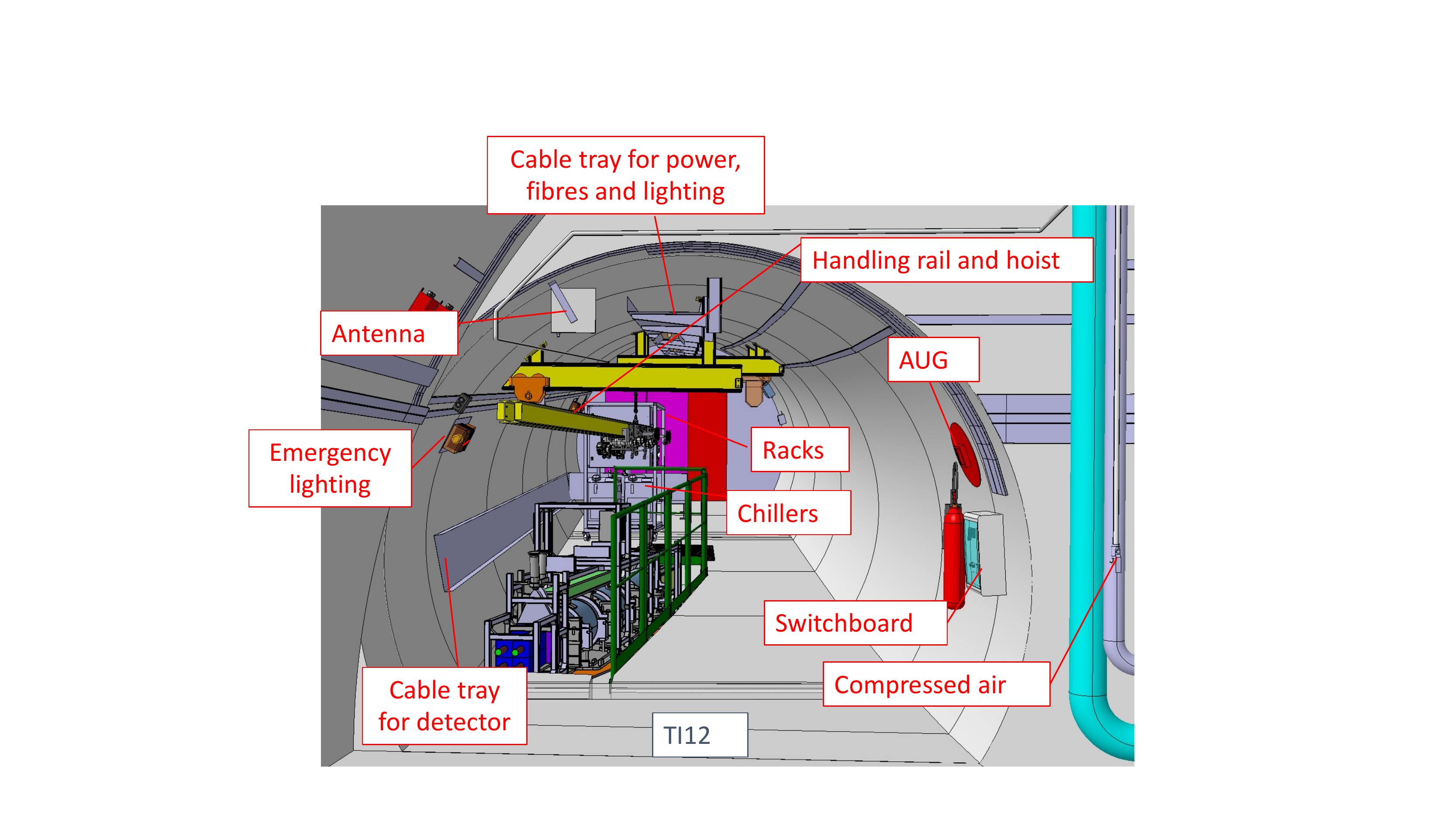}
    \caption{An integration picture of the TI12 tunnel showing the different infrastructure installed for FASER.}
    \label{fig:TI12-infrastructure}
\end{figure}

\newpage

\subsection{The detector support}

\subsubsection{The lower and upper baseplates}
\label{sec:BPs}
As an interface between the base of the FASER detector and the floor of the trench, two aluminium baseplates are used. These baseplates allow the full detector to be moved horizontally (perpendicularly to the LOS) such that the detector can follow changes in the LOS due to a horizontal beam crossing angle at IP1 (as described in Section~\ref{sec:run3}). Due to the position of the LOS with respect to the TI12 tunnel wall, the detector will only be able to move around 5~cm towards the wall. Although this is less than the maximum movement of the crossing angle, a movement of 5~cm will still give a significant increase in signal acceptance for such a scenario.

The lower baseplate is machined from AW5083 stabilized aluminium and is 5460~mm long, 1030~mm wide, and 15~mm thick.
It was constructed in three sections for CNC machining purposes, which are connected together with nine stainless steel screws (M6 size). The lower baseplate  was grouted to the trench floor using around a 10~mm thickness of grout.~\footnote{{\it SikaGrout-314 N} grout.} During the grouting process the lower baseplate was supported and aligned with the LOS using 12 jackscrews (M12 size). The screws were adjusted during a careful alignment procedure to ensure the flatness of the baseplate and that it was parallel to the theoretical LOS. The alignment was carried out by the CERN survey team, using a Leica laser tracker. Once aligned, the grouting was carried out by pouring liquid grout through dedicated holes in the baseplate and placing weights on the baseplate during the three day drying period. The whole grouting proccess was tested on the surface at CERN using a second baseplate identical to that used in TI12. During the tests, several attempts were made to to ensure the correct fluidity of the grout and define the final procedure. After the grouting, the baseplate was surveyed to determine the final position. The key areas below the magnet fixations stayed within 1.3~mm (with a flatness of around 0.5~mm) with respect to the CAD nominal position. Some displacements at the very end of the lower baseplate were observed, of the order of 3-4~mm in the vertical direction. All the in-plane surveyed points stayed within 2~mm of the design. Overall, this is considered to be acceptable given the size of the plates. The flexibility of aluminium ensures that the upper baseplate, resting on top of the lower baseplate, will adjust to the residual non-flatness. 

The technical design of the upper baseplate is shown in Figure~\ref{fig:upper-BP}. It is 5250~mm long, 760~mm wide, 20~mm thick, and machined from AW5083 stabilized aluminium (as for the lower baseplate). It was constructed in three sections which are connected together with seven stainless steel screws of M8 size. A total of 15 sliding pads in Bronze CuSn8 are attached to the bottom side of the upper baseplate, which allow it to smoothly slide over the lower baseplate  minimizing the friction between the two pieces. The sliding is done by applying a lateral force simultaneously at three sliding positions using a hydraulic jack pushing system.~\footnote{Enerpac manual pump, single speed, with three compact jacks, 5 tons maximum total force} The sliding system was successfully tested on the surface, with parts of the detector installed onto the upper baseplate, and the rest of the detector weight faked with a concrete block. When in position, the upper baseplate is secured to the lower one by seven M16 stainless steel screws attached  through slotted holes in the upper baseplate. The upper baseplate has also a number of threaded holes to allow the supports of the three magnets (described in Section~\ref{sec:magnetSupport}) to be secured, as well as holes to install the detector upper frame discussed in the next Section. 
Figure~\ref{fig:BPs} shows a photo of the two baseplates installed in the FASER trench in TI12, 

\begin{figure}[hbt!]
    \centering
    \includegraphics[width=1.\textwidth]{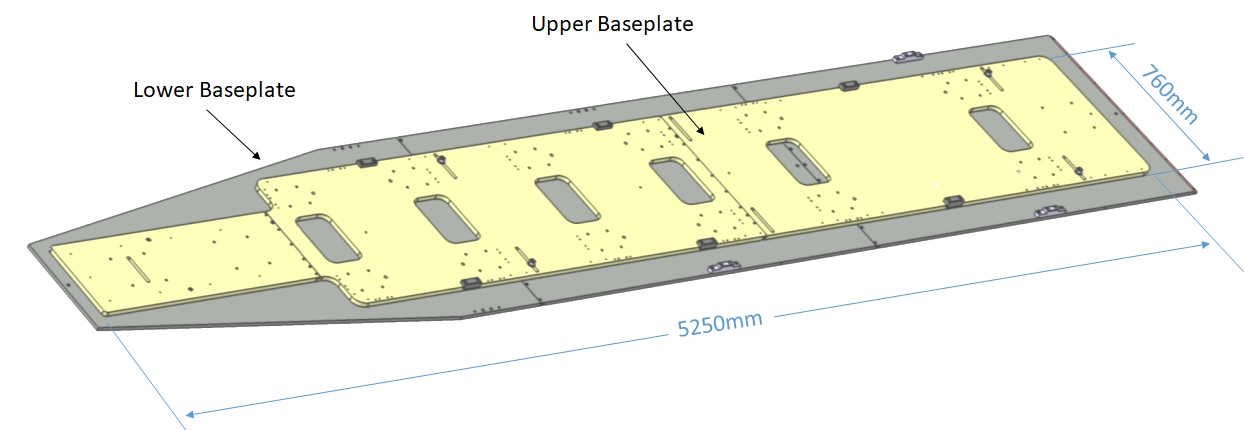}
    \caption{The design of the upper baseplate, shown resting onto the lower baseplate. The main dimensions are shown.}
    \label{fig:upper-BP}
\end{figure}

\begin{figure}[hbt!]
    \centering
    \includegraphics[width=0.7\textwidth]{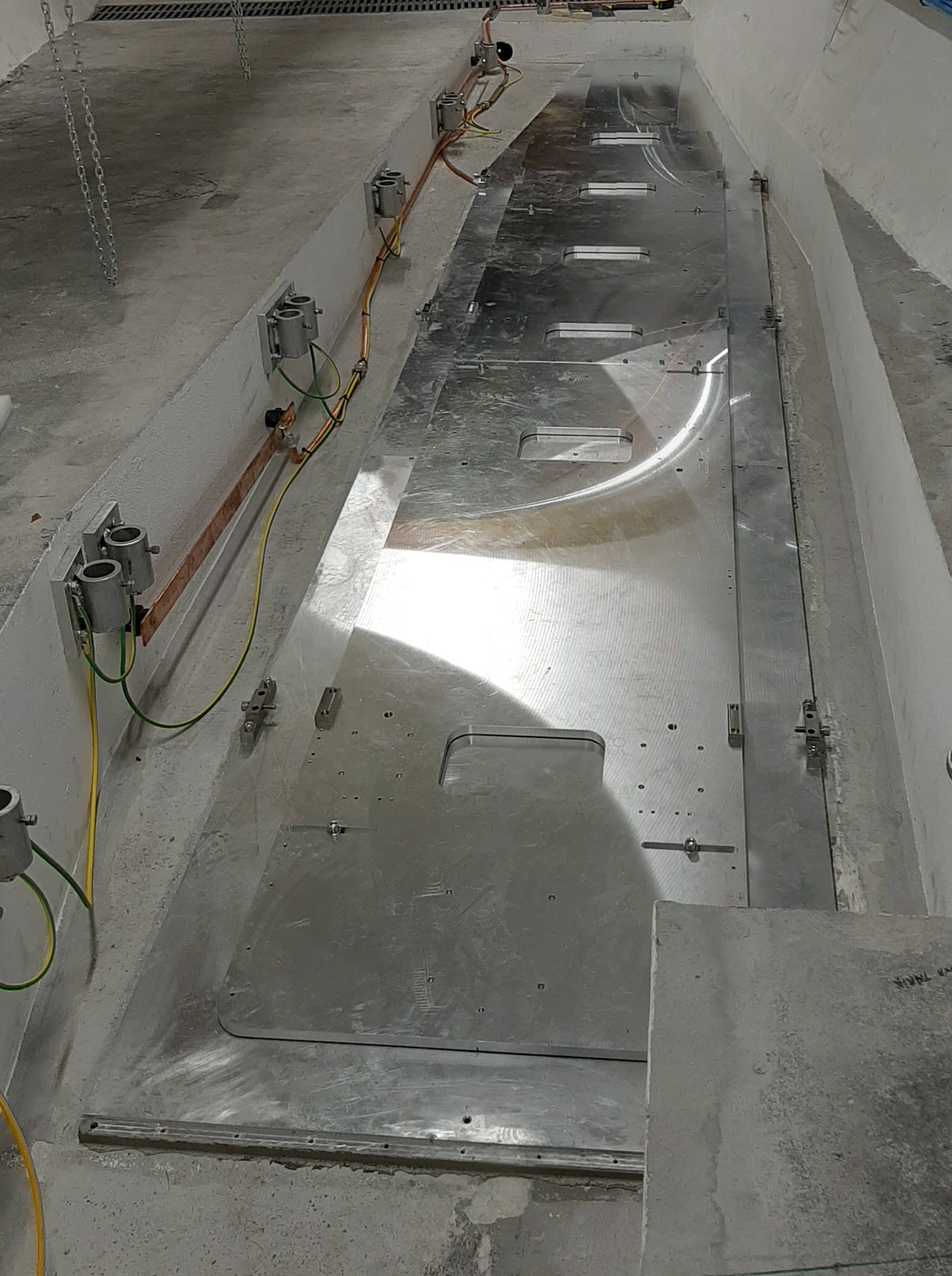}
    \caption{A picture of the upper and lower baseplates installed in the FASER trench, before the detector was installed.}
    \label{fig:BPs}
\end{figure}

\subsubsection{The upper frame}
As discussed in Section~\ref{sec:magnetSupport}, the FASER magnets have their own tunable supports which attach directly to the upper baseplate, and the FASER tracker system is supported by the two short magnets. However the other FASER components (except for FASER$\nu$) are supported by an aluminium profile structure referred to as the upper frame. The upper frame supports three of the scintillator stations (not including those in front of FASER$\nu$) the calorimeter system, as well as on-board electronics, on-detector cables, and cooling manifolds and piping.
The detectors supported by the upper frame do not need a precise alignment, hence their position is only accurate at better than the 10~mm level. The frame is made of aluminum profiles, with 40~mm square cross section, connected together by angle brackets (all materials are non-magnetic). The design of the frame is flexible and allows the calorimeter and scintillator stations to be raised/lowered to be coarsely aligned with the magnets and tracker. Figure~\ref{fig:upper-Frame} gives the main overall dimensions of the upper frame, and Figure~\ref{fig:upperFrame} shows a picture of the upper frame during a test installation in TI12. During this test, the FASER magnets were installed, but most of the other parts of the detector were not yet included. 

\begin{figure}[hbt!]
    \centering
    \includegraphics[width=1\textwidth]{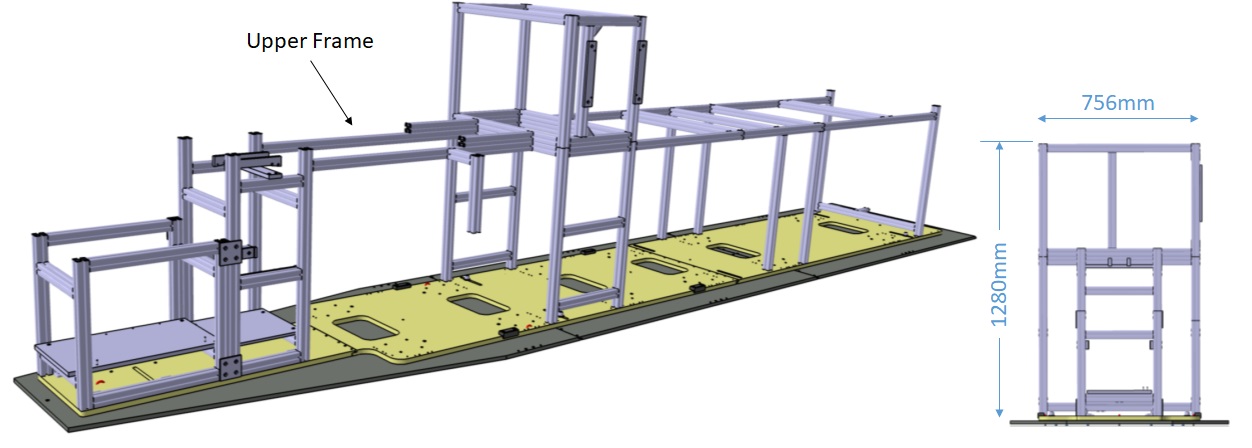}
    \caption{A CAD model of the upper frame with some dimensions.}
    \label{fig:upper-Frame}
\end{figure}

\begin{figure}[hbt!]
    \centering
    \includegraphics[width=0.9\textwidth]{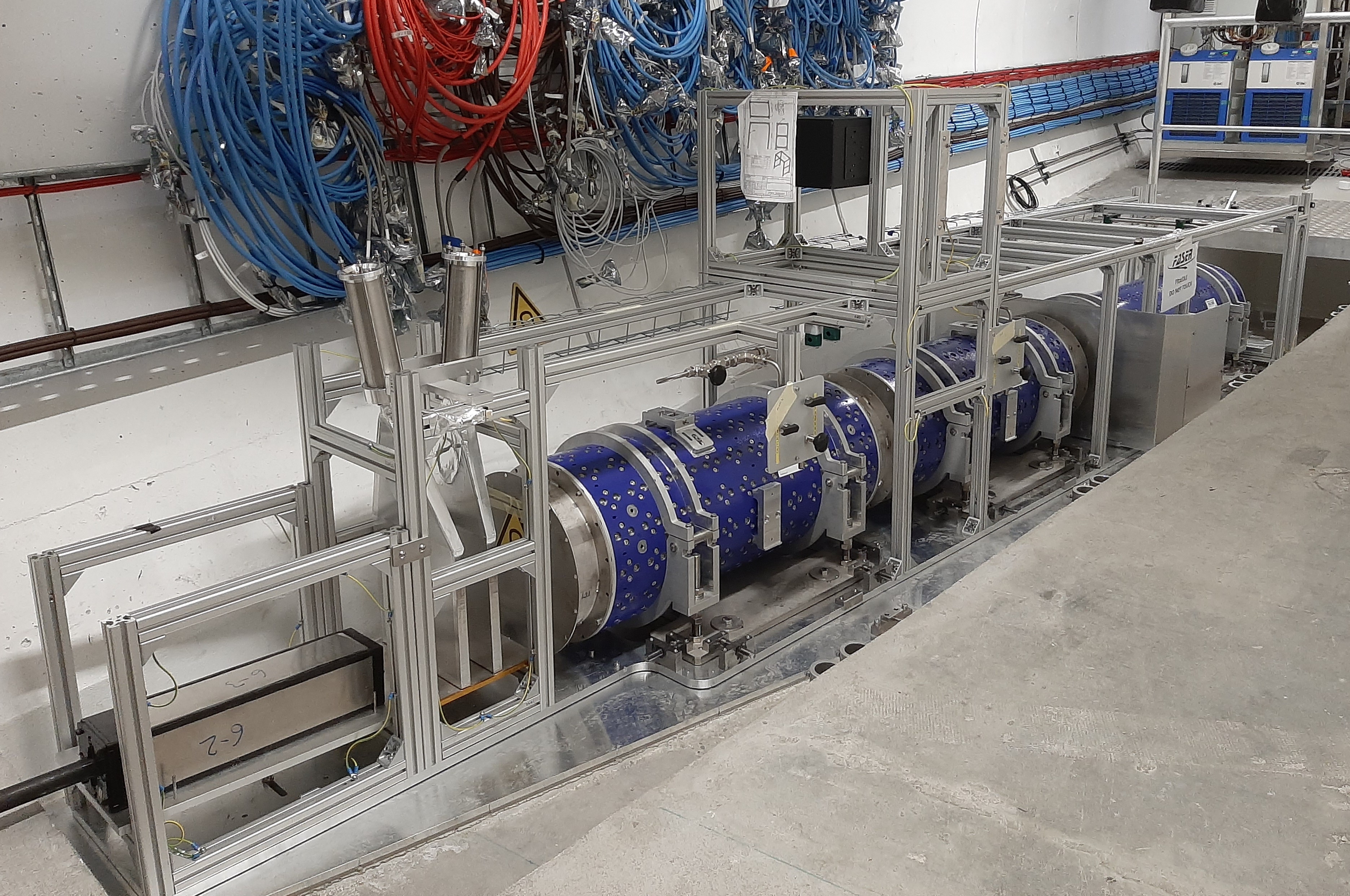}
    \caption{A picture of the upper frame, during a test installation in TI12.}
    \label{fig:upperFrame}
\end{figure}

\subsection{The cooling system}

The cooling system uses two water chillers~\footnote{HRS030 from SMC company} from which two inlet/outlet pipes supply the four tracking stations of FASER. A manifolding system connects each station by means of flexible polyurethane pipes with 8~mm outer diameter and 5.5~mm  inner diameter. A patch panel allows for smooth control of the water flow to each station and balances the flow across the four stations (a flow meter is used once to set the right parameters per circuit). Figure~\ref{fig:cooling-syst} shows the main cooling circuit from cooling unit to tracking stations. The stainless steel pipes are rigidly connected to the upper frame with isolating clamps, to minimize the heat loss along the path and  limit the temperature increase between the set point and the inlet of each stations. The maximum measured water flow per station is about 2.7~$\ell$/min and depends on the pressure drop along the system (which is a function of the pipe length, diameter, manifolds, and bends). The temperature on the chiller is set to a minimum of 15°C, to prevent condensation on the pipes. 
As discussed in Section~\ref{sec:trk_cooling}, the measured temperatures on the tracker stations are well within the tolerance to safely run the electronics.

\begin{figure}[hbt!]
    \centering
    \includegraphics[width=0.9\textwidth]{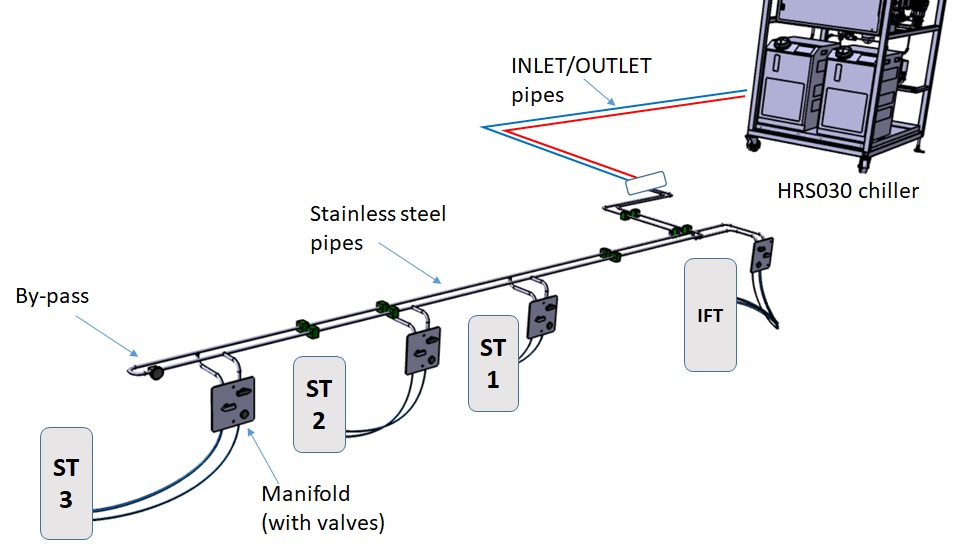}
    \caption{A picture of the cooling system for FASER tracker.}
    \label{fig:cooling-syst}
\end{figure}

\subsection{Detector grounding}
\label{sec:grounding}
The electronic components of the detector are grounded following a meshed common bonded network.
The common ground of each component is connected to the main LHC earth via two copper grounding bars installed on the side of the FASER trench (these can be seen in Figure~\ref{fig:BPs}), with a separate connection for the two electronics racks housing the TDAQ equipment and MPOD power supplies, and for the cooling unit. 
All cable trays are grounded at both ends. The tracker power supplies have individual floating grounds, and are grounded on the tracker patch panels, whereas the PMT power supplies have a common ground. 

For safety reasons all metallic components on the detector are connected to the ground to ensure they are equipotentially bonded.

\subsection{Material inside the active detector volume}
\label{sec:material}
In order to reduce backgrounds to searches for new physics, the amount of material inside the transverse region of the detector active area, and longitudinally from the last veto scintillator until the pre-shower scintillator station should be minimized. The amount of material in this region has been optimized for this purpose and is summarized in Table~\ref{tab:det-mat}.
As expected, the largest fraction of material is in the tracking stations, followed by the material in the timing scintillator station and the covers on the front and back of the decay volume magnet.  

\begin{table}[thb]
    \centering
    \begin{tabular}{|l|c|c|c|c|}
  \hline
Component & Material & \multicolumn{2}{c|}{$X_0$ (\%)}  \\
 & & Central region & Edge region \\
\hline
Scintillator timing station - scintillator & 1~cm polyvinyltoluene  & 2.4\% & 2.4\% \\
Scintillator timing station - foil wrapping & 1~mm Al  & 1.1\% & 1.1\% \\
3 Tracking stations & See Table~\ref{tab:TrackerMaterial} & 6.3\% & 64.5\% \\
Decay volume magnet cover - front & 0.4~mm CFRP & 0.15\% & 0.15\% \\
Decay volume magnet cover - back & 3~mm plastic & 0.75\% & 0.75\% \\
\hline
\bf{Total } & \bf{-} & \bf{10.7\%} & \bf{68.9\%} \\
\hline
    \end{tabular}
    \caption{A summary of the amount of material in the part of the detector in which material is minimized to reduce physics backgrounds. As discussed in Section~\ref{sec:trackerstation}, the material in the tracker stations is minimal in the central region ($|x| < 4$~cm) and larger in the edge region. Both regions are shown in the Table.}
    \label{tab:det-mat}
\end{table}

\subsection{Detector installation}

After the installation of the services in TI12, the main FASER detector (not including the FASER$\nu$ systems) was installed in March 2021. The installation of the detector components proceeded in steps as detailed in the following. 
\begin{itemize}
    \item Installation of the detector cooling unit and the related piping up to the trench.
    \item Installation of the detector baseplates in the trench. 
    \item Pulling of about 150 power and readout cables from the two electronics racks, along the tunnel wall, and up to the FASER trench. The cables were then tested and a careful mapping carried out. The cables were  left coiled up against the wall, until the detector was installed.
    \item Installation of the upper baseplate.
    \item Installation of the two short FASER magnets.
    \item Installation of the timing scintillator station. Since this sits within a few mm of the long FASER magnet, it was necessary for this to be installed before the last magnet.
    \item Installation of the long magnet.
    \item Alignment of all three magnets (see~Section~\ref{align_det_mag}). 
    \item Installation of the FASER tracker. 
    This was installed as a single unit, with the three tracking stations attached to the backbone. As shown in Figure~\ref{fig:trenchwall}, there is not enough clearance for connecting all the cables on the trench wall side of the patch panels. Consequently, some of the cables and the cooling pipes were pre-attached before the installation. The tracker was slowly lowered onto the two short magnets using the crane installed over the FASER trench. Figure~\ref{fig:FASER-trackerInstallation} shows a picture of the tracker being lowered onto the detector. Following the installation, a survey was carried out to precisely measure the position of the tracking stations (see Section~\ref{align_det_mag}).
    \item Installation of the veto and pre-shower scintillator stations. They were then attached to the upper baseplate.
    \item Construction of the upper frame. The frame was constructed around the installed components, and the tracker on-detector electronics mounted on the upper frame. 
    \item Installation of the four cooling manifolds onto the upper frame. They were installed along with the on-detector cooling piping, which was connected to the piping from the cooling unit. The tracker station cooling loops, and dry-air connections, were then  connected and the cooling system tested.
    \item On-detector cabling for the tracker. For the tracker powering, the short cables from the tracker patch panels to the splitter boxes were routed on the upper frame, and the long cables routed from the cable tray along the tunnel wall were connected to the splitter boxes. The TRB cables were routed to the mini-crate, and the DCS cables to the TIM box. At this stage the tracking stations were tested. Calibrations and checks demonstrated similar performance to those seen during testing on the surface, as described in Section~\ref{sec:trk-commissioning}.
    \item Installation of the calorimeter. The scintillator counter PMTs and calorimeter PMTs were also connected and their correct functionality verified.
    \item Installation of the detector grounding. 
\end{itemize}
These installation steps were carried out without problems during a 3-week period. 
A photo of the FASER detector after installation can be seen in Figure~\ref{fig:FASER-afterInstallation}. 
In December 2021, the IFT tracker station was installed, and commissioned. As shown in Figure~\ref{fig:IFT}, the IFT is attached to the upper frame with dedicated mechanics that allows the position to be precisely tuned. 

\begin{figure}[hbt!]
    \centering
    \includegraphics[width=0.9\textwidth]{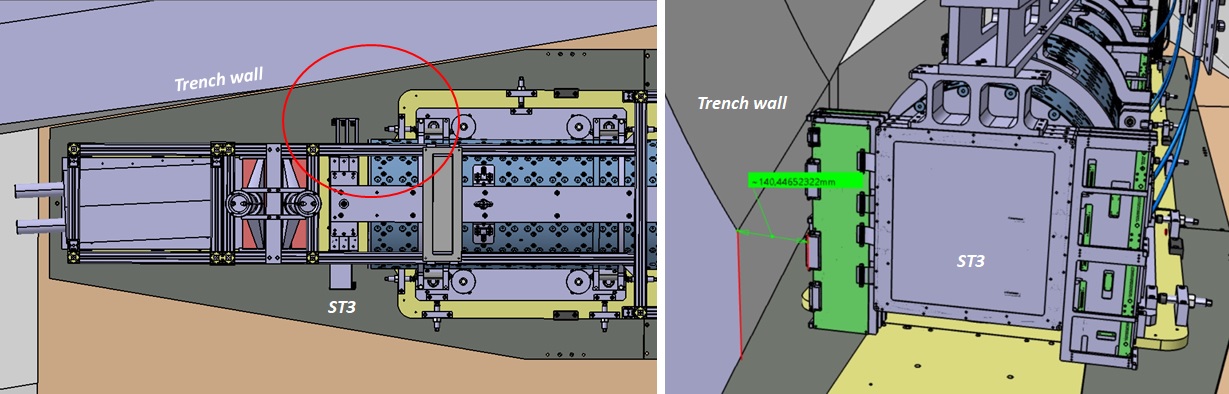}
    \caption{Plan view (left) and view from the back (right) of the CAD model, showing the back region of the FASER detector, and highlighting how close the patch panels on the back tracker station are to the tunnel wall (about 14cm). }
    \label{fig:trenchwall}
\end{figure}

\begin{figure}[hbt!]
    \centering
    \includegraphics[width=0.9\textwidth]{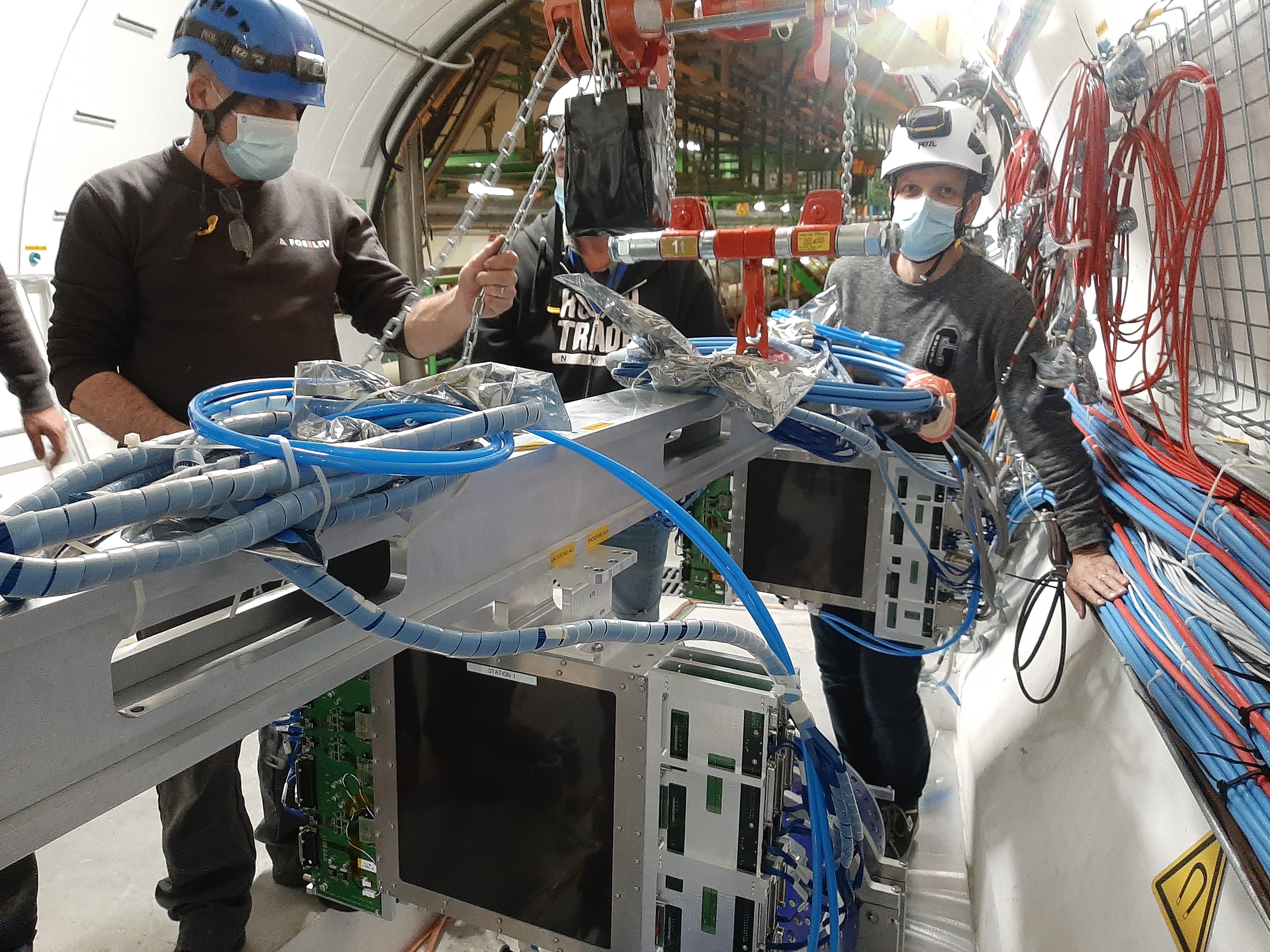}
    \caption{A picture of the FASER tracker being installed onto the detector.}
    \label{fig:FASER-trackerInstallation}
\end{figure}

\begin{figure}[hbt!]
    \centering
    \includegraphics[width=0.9\textwidth]{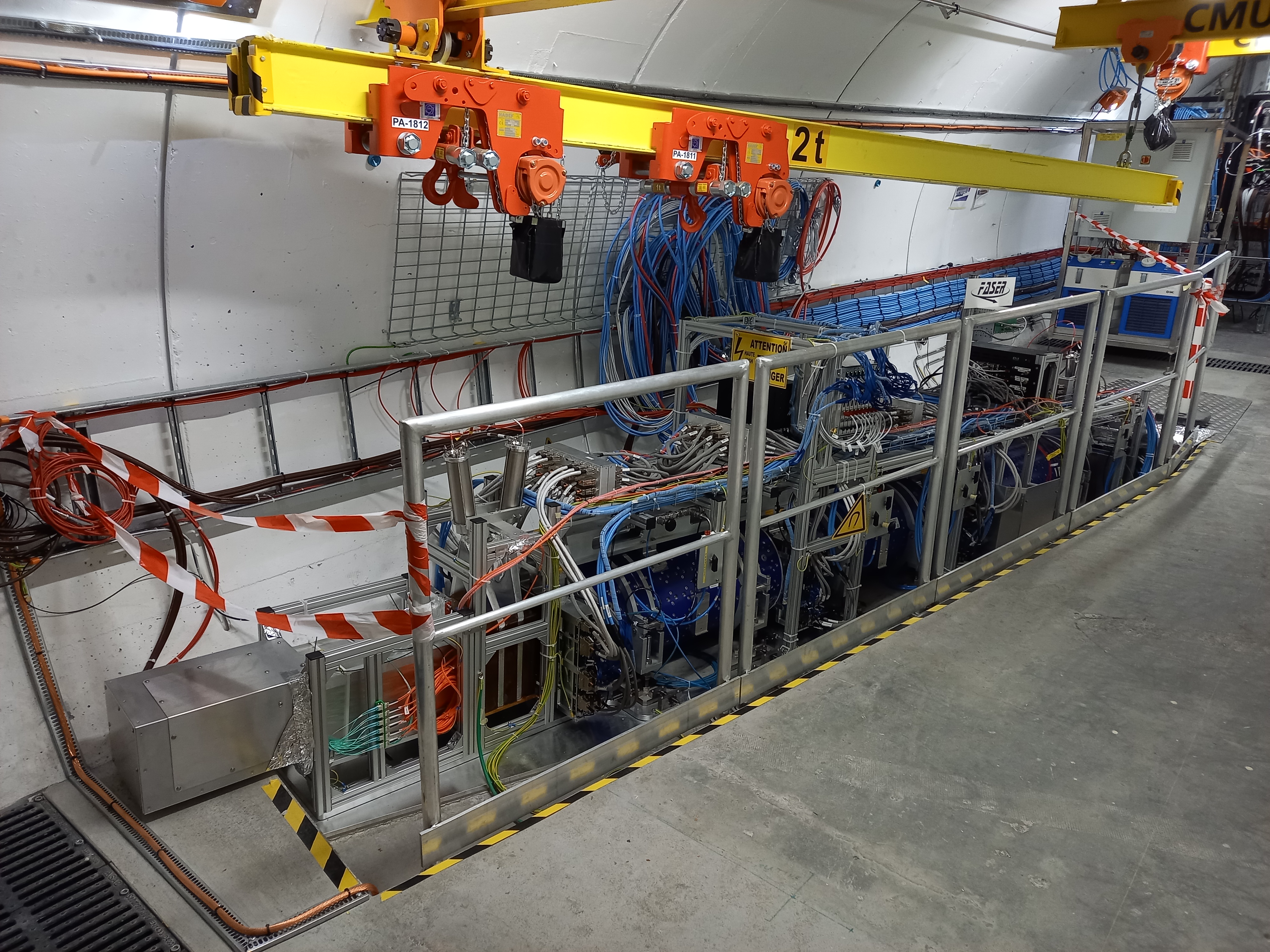}
    \caption{The FASER detector after installation in TI12.}
    \label{fig:FASER-afterInstallation}
\end{figure}

\begin{figure}[hbt!]
    \centering
    \includegraphics[width=0.9\textwidth]{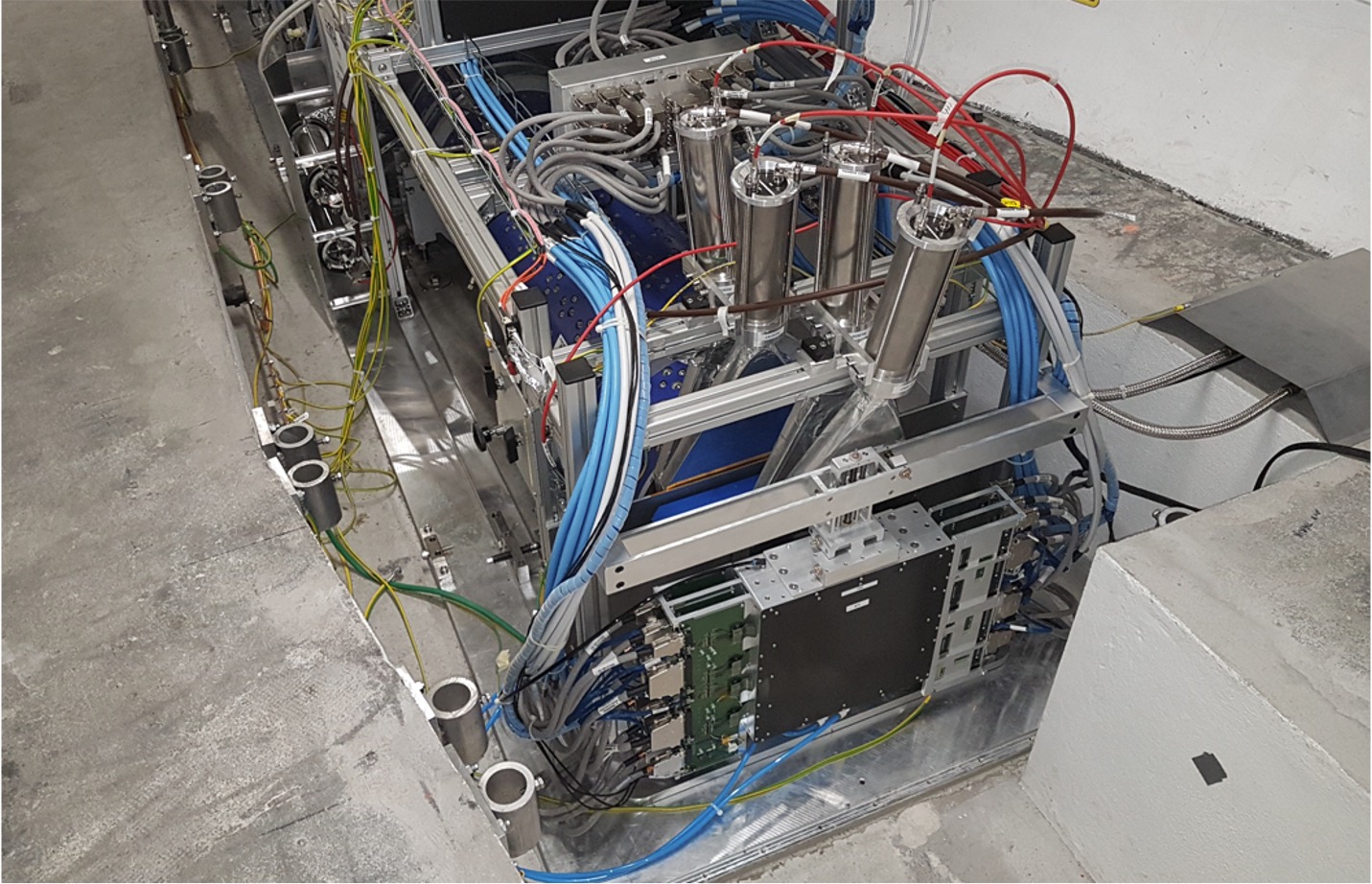}
    \caption{The IFT tracker station, installed at the front of the main FASER detector.}
    \label{fig:IFT}
\end{figure}

\clearpage

\subsubsection{The detector and magnet alignment}
\label{align_det_mag}
There have been several stages in the FASER alignment process and the related survey carried out by the CERN survey group in TI12. As a very first step, a 3D laser scan was performed in  T112 to check the FASER trench and add corrective works where needed. After this, the alignment of the lower baseplates was performed. The core part of the tracker system alignment and related survey was carried out when the three magnets were fixed to the upper baseplate, and again after installation of the three tracker stations. Finally, a dedicated alignment and survey was carried out once the IFT tracker station was integrated. As well as the survey work in TI12, the individual tracker stations were measured during the metrology to ensure the required precision is achieved, as discussed in Section~\ref{sec:trkalignment}. Similarly, the tracker backbone was also measured.   
For each survey campaign, the measurements were recorded as Cartesian coordinate points in the LHC reference system, using a  Leica laser tracker.~\footnote{Leica laser tracker AT 403}  The surveyed points were then compared to the nominal CAD model to compute the deviations in mm. Figures~\ref{fig:Survey_BP} and~\ref{fig:survey_label1}   show the positions and labels of the measured points for the survey of the lower baseplate, and the magnets and tracker backbone, respectively, as well as the coordinate system.~\footnote{Note the coordinate system used here is not the same as the FASER coordinate system shown in Figure~\ref{fig:FASER_labels}.}

The survey measured 16 points on the magnets and 20 points on the tracker backbone, and the results were compared to the nominal CAD model. The magnet measurements are summarized in Table~\ref{table:magnet_survey}. Since the crossing angle for 2022 running pushes the LOS downwards, the magnets were aligned to be as low as possible. During the survey they were measured to be lower than the nominal position (assuming zero crossing angle) by 12.3~mm (+/-0.2~mm). 
The magnets were measured to be within 0.3~mm of the nominal position in the horizontal ($x$) and longitudinal ($z$) directions, and the angle between the magnetic axis and the vertical axis (referred to as roll angle) was measured to be less than 0.2~mrad for all three magnets.

\begin{table}[htbp]
\centering
\begin{tabular}{|c|c|c|c|c|}
\hline
Measurement position & $\Delta x$ &  $\Delta y$  &$\Delta z$  & Roll angle  \\
 & (mm) & (mm) & (mm) & (mrad) \\
 \hline 
Decay Volume magnet: Front         & 0.13   & -12.24 & -0.23 & 0.06 \\
Decay Volume magnet: Back          & 0.19    & -12.48 & -0.23 &  \\
%\hline
1st Spectrometer magnet: Front     & 0.13    & -12.33 & 0.30 & -0.14 \\
1st Spectrometer magnet: Back      & 0.10   & -12.22 & 0.30 &  \\
%\hline
2nd Spectrometer magnet: Front     & 0.03   & -12.13 & -0.12 & 0.17 \\
2nd Spectrometer magnet: Back      & 0.19   & -12.13 & -0.12 & \\
\hline
\end{tabular}
\caption{Positions of the three FASER magnets with respect to the nominal position in mm. The positions are calculated from the magnet geometry and the survey measurements for each magnet. The measurements are in the FASER coordinate system, with the magnets deliberately positioned as low as possible to partially follow the downward movement of the LOS due to the crossing angle, where the final position is 12.3~mm below the nominal location. The measured roll angle of each magnets is also shown (in mrad).
}
\label{table:magnet_survey}
\end{table}

The survey in TI12 of the tracker backbone, including the interface to each tracking station, measured the position of the survey points on the tracker station with $\mathcal{O}$(16~$\mu$m) accuracy. This information, combined with the metrology data taken with the individual tracker planes (described in Section~\ref{sec:trkalignment}) will allow the position of the silicon sensors in the detector frame to be known to $\mathcal{O}$(100~$\mu$m) precision, each measurement is significantly more precise but when taking into account the full system and possible movements after the measurements were taken this is the expected level of precision. The precision of the measurements of the position of the IFT sensors will likely be worse due to the more complex mechanics supporting the IFT. The results will be used as the first alignment in the simulation and reconstruction software, and will be refined by track-based alignment measurements. 
Finally, the survey measurements allow the distance of the FASER detector at $z = 0$ (corresponding to the front of the second tracking station) from the nominal IP1 collision point, which is calculated to be $477.759$~m, to be precisely known.

\begin{figure}[hbt!]
    \centering
    \includegraphics[width=1.\textwidth]{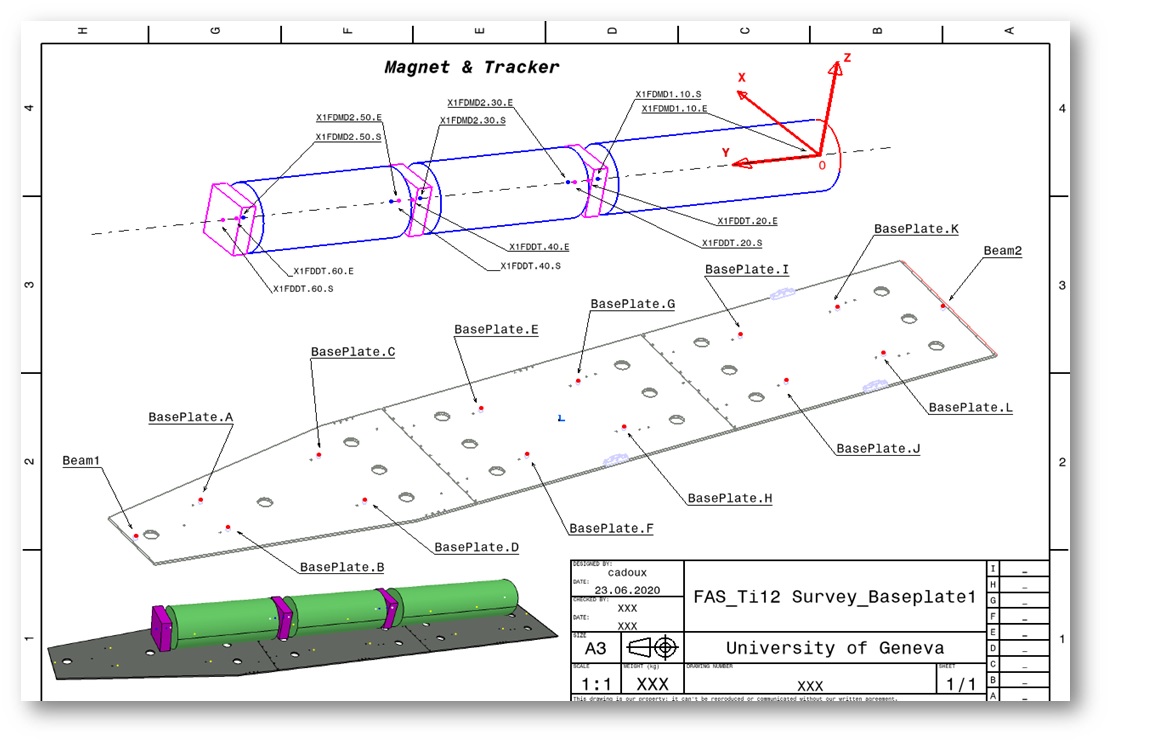}
    \caption{Drawing of the lower baseplate showing the measured survey points and coordinate system used in the survey.}
    \label{fig:Survey_BP}
\end{figure}

\begin{figure}[hbt!]
    \centering
    \includegraphics[width=1.\textwidth]{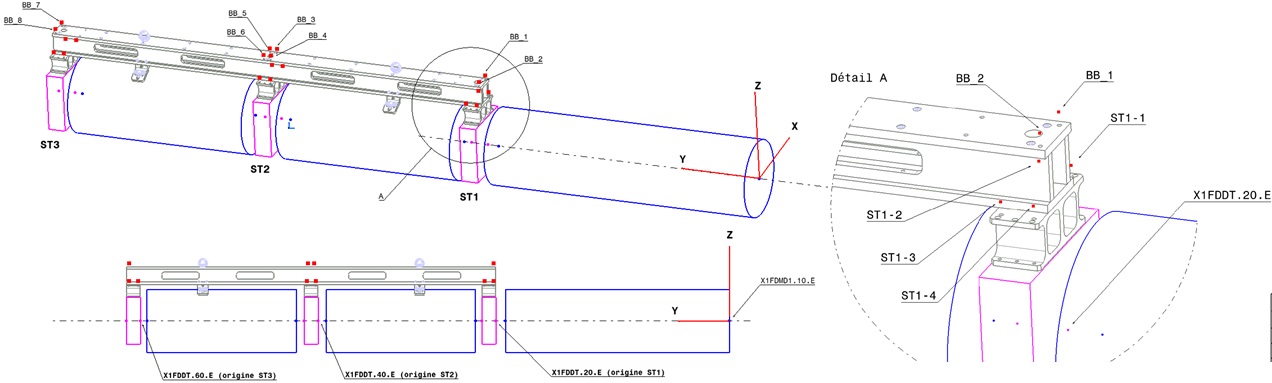}
    \caption{Drawing of the magnets and the trackers stations showing the position and labels of the survey points. The coordinate system used in the survey is also shown.}
    \label{fig:survey_label1}
\end{figure}

\clearpage
\section{Commissioning of the full detector}
\label{sec:commissioning}
All sub-detector components, except the IFT tracking station, the FASER$\nu$ emulsion detector and the FASER$\nu$ veto scintillator systems, were installed in March 2021 in TI12. Following this, the detector was commissioned in situ using cosmic-ray events collected for several months. In October 2021 there was a two week pilot beam test in the LHC, during which beams were circulated and collided, allowing further commissioning of the detector using beam data.  

The commissioning studies done in situ allows to verify the performance/functionality of the different detector components, and to demonstrate that the full detector can operate with the central TDAQ and DCS systems. However, there are intrinsic limitations on the kind of studies that can be performed on these data. 
The rate, angular distribution and random arrival time of cosmic-ray muons does not allow to commission the full combined reconstruction and calibration/alignment of the detector. The pilot beam test proved to be useful for a first timing-in of the detector and trigger, and to assess beam-related backgrounds, but data are not available at a sufficient rate for detailed studies. The full commissioning of the detector and the assessment of the object reconstruction capabilities will only be possible with first high energy collisions. 

\subsection{Commissioning with cosmic rays}

A cosmic-ray dataset of the order of 125 M events was recorded during $\mathcal{O}(10^7)$ s of running between April and September 2021, triggering either on the scintillator counters or on the single calorimeter modules. The overall trigger rate was about 15~Hz, dominated by noise-induced triggers in the veto and timing scintillator stations. The trigger thresholds were generally set to be sensitive to the signal of a minimum ionizing particle. Table~\ref{tab:cosmic-trg-rate} shows the rate of the different triggers during a typical cosmic data taking run.

\begin{table}[thb]
    \centering
    \begin{tabular}{|l|c|}
  \hline
Trigger & Rate (Hz) \\
 \hline
Veto scintillator station & 3.5 \\
Timing scintillator station & 10 \\
Pre-shower scintillator station & 0.25 \\
Calorimeter & 0.25 \\
Random & 1 \\
\hline
Total rate & 15 \\
\hline
    \end{tabular}
    \caption{Summary of the rate of the different triggers during cosmic-ray running of the full detector in TI12. Note that the timing station has a higher noise rate, since the trigger is not given by the AND of signals in separate scintillator counters, but rather by the AND of signals from two PMTs attached to the same scintillator. }
    \label{tab:cosmic-trg-rate}
\end{table}

During cosmic-ray data taking the rate of tracks crossing the front and back tracking stations (which are close to scintillator stations which provide triggers) is $\mathcal{O}(0.01)$ Hz, whereas there is not a suitable trigger for tracks going through the middle tracking station. Because of the angular distribution of the cosmic-ray flux, and particularly at a depth of 80~m under the ground, it is rare to have a cosmic-ray muon travelling at a shallow enough angle to traverse more than one tracking station. An estimate of the expected rate for tracks to traverse one, two and all three tracking stations was made by an analytical integration of the cosmic-rays flux taken from Ref.~\cite{Gaisser:2016uoy}. Muons were assumed to  propagate to TI12 following the model from Ref.~\cite{muonstopping}, taking into account the detector geometry, including the scintillator counters used for triggering. A comparison of the observed and measured rates is shown in Table~\ref{tab:cosmic-trk-rate}: no track passing through all three stations has been observed, in agreement with the very low probability expected from simulation.  

Figure~\ref{fig:event_display_cosmic} shows a simplified event display of a cosmic-ray muon traversing the middle and back tracking stations, and also leaving signals in the pre-shower scintillator station and in one of the calorimeter modules. Over the period April - September 2021 (with the detector turned off for most of August due to magnet training in the LHC close to FASER) more than 100 of such "two-station" events were recorded.

\begin{table}[thb]
    \centering
    \begin{tabular}{|l|c|c|}
  \hline
Event type & Observed Rate & Expected Rate \\
 \hline
Track in 1 tracker station & 0.016~Hz & 0.011~Hz \\
Track in 2 tracker station & 1/(28.6 $\pm$ 2.5)~hrs$^{-1}$ & 1/28~hrs$^{-1}$ \\
Track in 3 tracker station & Not yet observed & 1/82~days$^{-1}$  \\
\hline
    \end{tabular}
    \caption{Summary of the observed and expected rate of tracks traversing different numbers of tracking stations during cosmic-rays running (about 100~days) of the full detector in TI12. No event has been yet seen with a cosmic ray traversing the three tracker stations.}
    \label{tab:cosmic-trk-rate}
\end{table}

The cosmic-ray data taking was used by the different detector systems for various commissioning and performance studies, as briefly detailed below.

\begin{figure}[h]
    \centering
    \includegraphics[angle=0, clip=true, width=\textwidth]{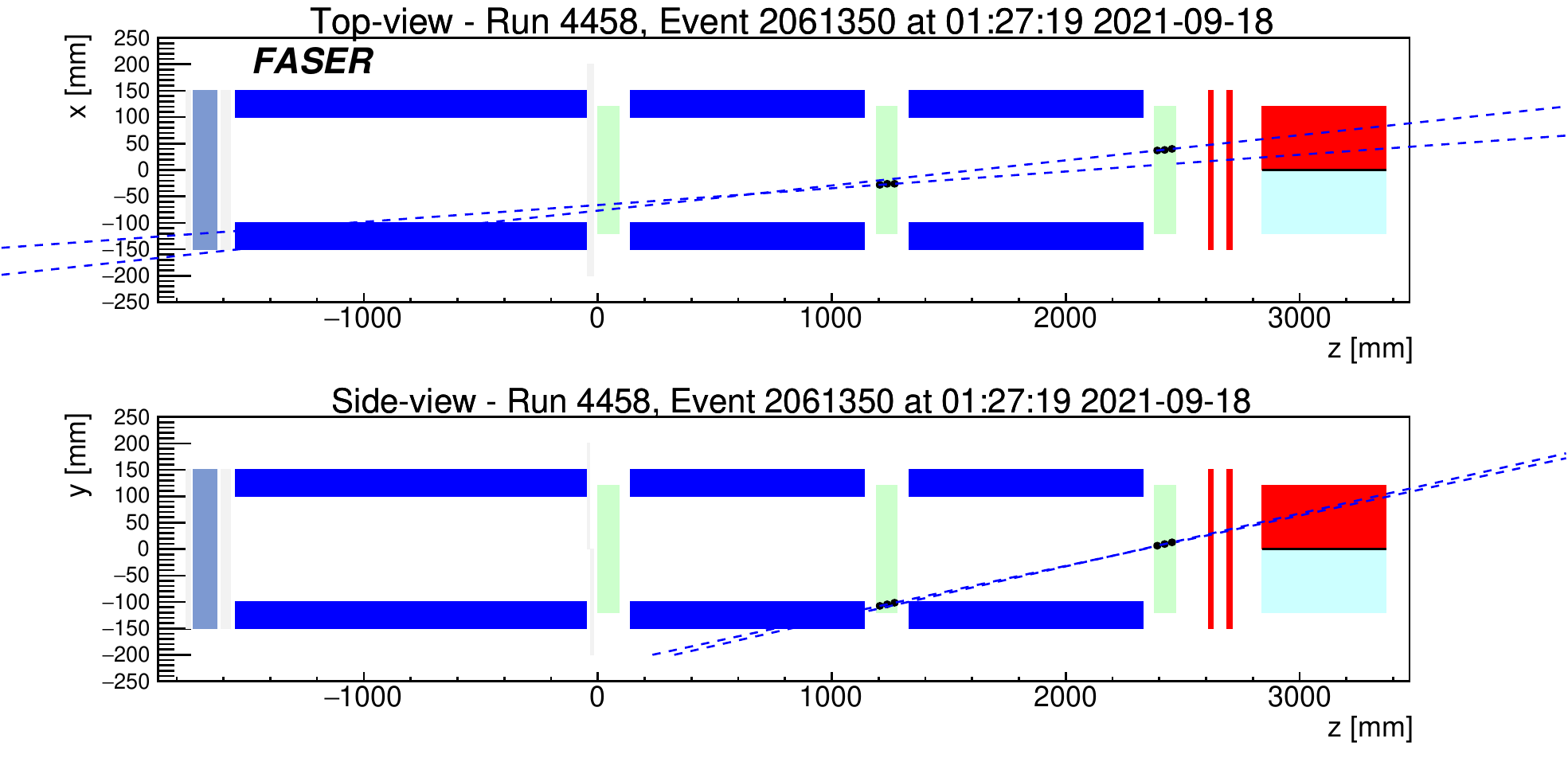}
    \caption{An event display of a cosmic-ray muon traversing part of the  detector. The view is shown both from the top (top row) and the side (bottom row). Non-triggered and triggered scintillator counters (calorimeter modules) are shown in grey (light-blue) and red respectively. Tracker stations are shown in green, while 3D tracker hits are shown as black dots and track segments formed from hits in a single station are indicated with a dashed blue line. The dark blue areas are the magnet material.}
    \label{fig:event_display_cosmic}
\end{figure}

\subsubsection{Tracker}

The noise occupancy of the SCT modules on the tracking stations was measured at the nominal 1\,fC thresholds using randomly triggered events taken during cosmic-ray runs. The strip-by-strip tracker noise occupancy considering all four tracking stations (tracking spectrometer and IFT) is shown in Figure~\ref{fig:tracker_noiseOcc}. Less than 0.4\% of the strips have a noise occupancy above $5 \times 10^{-4}$. The results are in good agreement with the pre-commissioning results on single SCT modules as discussed in Section~\ref{sec:trk-commissioning}. 

The long-term operations of the tracker during the cosmic-ray data taking with the full detector demonstrated that all SCT modules can be kept well below 30 degrees, and that the tracking spectrometer was able to operate efficiently and safely for single runs over many days.

\begin{figure}[tbh]
\begin{center}
\includegraphics[width=0.45\textwidth]{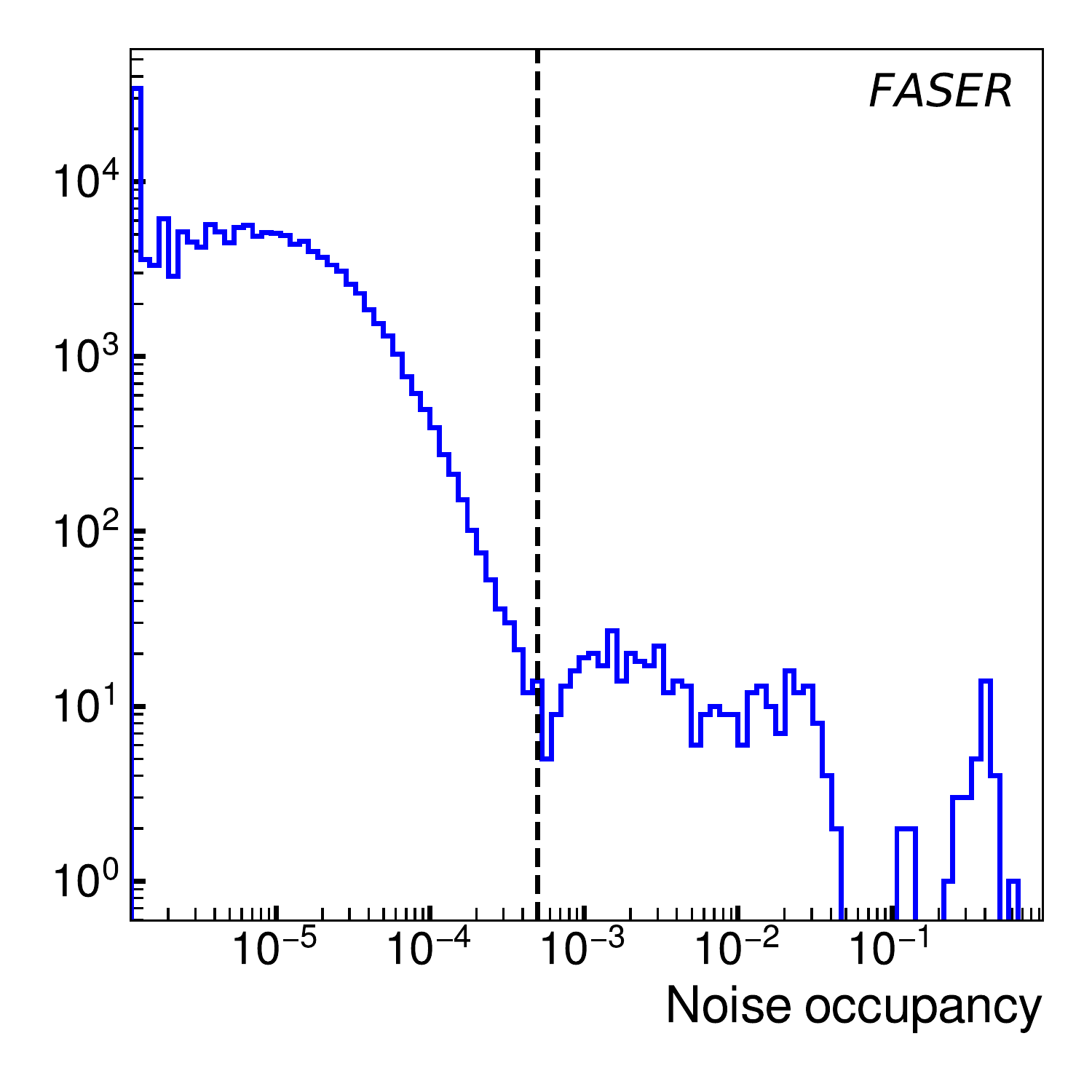}
\caption{Strip-by-strip tracker noise occupancy measured from random triggers during combined runs. The dashed vertical line indicates noise occupancy of $5\times10^{-4}$.}
\label{fig:tracker_noiseOcc}
\end{center}
\end{figure}

\subsubsection{Calorimeter and scintillators systems}

After installation, the first cosmic-ray data were used for an in situ adjustment
of the PMT gains to equalize the MIP signal response of PMTs in the same station 
and to achieve a certain dynamic range, given the 2~V maximum signal voltage of the digitizer. 
The veto stations were adjusted to a dynamic range of a signal equivalent to about 10 MIPs for high single particle detection efficiency. Initially, the same adjustment was done for the pre-shower station, but later the gain was reduced by a factor of ten in order not to saturate on large electromagnetic showers, following evidence from test-beam measurements.  
The timing scintillators initially used a relatively low gain and had a dynamic range of up to about 60 MIPs.
In a later stage, the gain was increased by a factor of six to improve the signal efficiency and the timing resolution. 
These gains will be fine-tuned once a large sample of horizontal muons will be collected from the first collisions.

For the calorimeter system, data were taken at several gain settings without
the optical filters installed in order to maximimize sensitivity in the initial commissioning. 
Cosmic-ray and noise data were taken at a medium gain of about $3\times10^{4}$
for 81.5 days, corresponding roughly to the expected time of stable beam collisions in a full year of the LHC. 
During this time only 15 events were recorded with one calorimeter channel in saturation, corresponding to a deposited energy of more than roughly \SI{60}{\GeV}. This shows that cosmic-rays background will not be significant in physics data analyses: such 
low number of events can easily be suppressed with additional selections on the signal timing and on the pre-shower signals.

As noted in Section~\ref{sec:calodescription}, the calorimeter PMTs were initially affected by noise picked up from a nearby GSM antenna. Before installation of the Faraday cage around the PMTs, the RMS of the PMT pedestal measured in the digitizer was between 0.35 and \SI{0.60}{mV}, depending on the channel. After the installation this was reduced to between 0.33 and \SI{0.40}{mV}, very close to the intrinsic noise level in the digitizer of \SI{0.30}{mV}. The noise level in the scintillator counter PMTs all lie in the same range. At this noise level, it is possible to have a single PMT trigger threshold of \SI{3}{mV} (0.15\% of the full range) for the calorimeter PMTs at medium gain, and maintain a noise trigger rate below \SI{0.5}{Hz}.

The LED calibration system has been used regularly to monitor the calorimeter response versus gain and time. Figure~\ref{fig:calibstability} shows the change in response, measured as the average charge collected in each of the four modules during LED calibration runs, over a period of five weeks. During this time the calorimeter and calibration system settings were kept unchanged. The calibration system and the PMTs were found to be stable to within 1\%. 

\begin{figure}[h]
    \centering
    \includegraphics[angle=0, clip=true, width=0.8\textwidth]{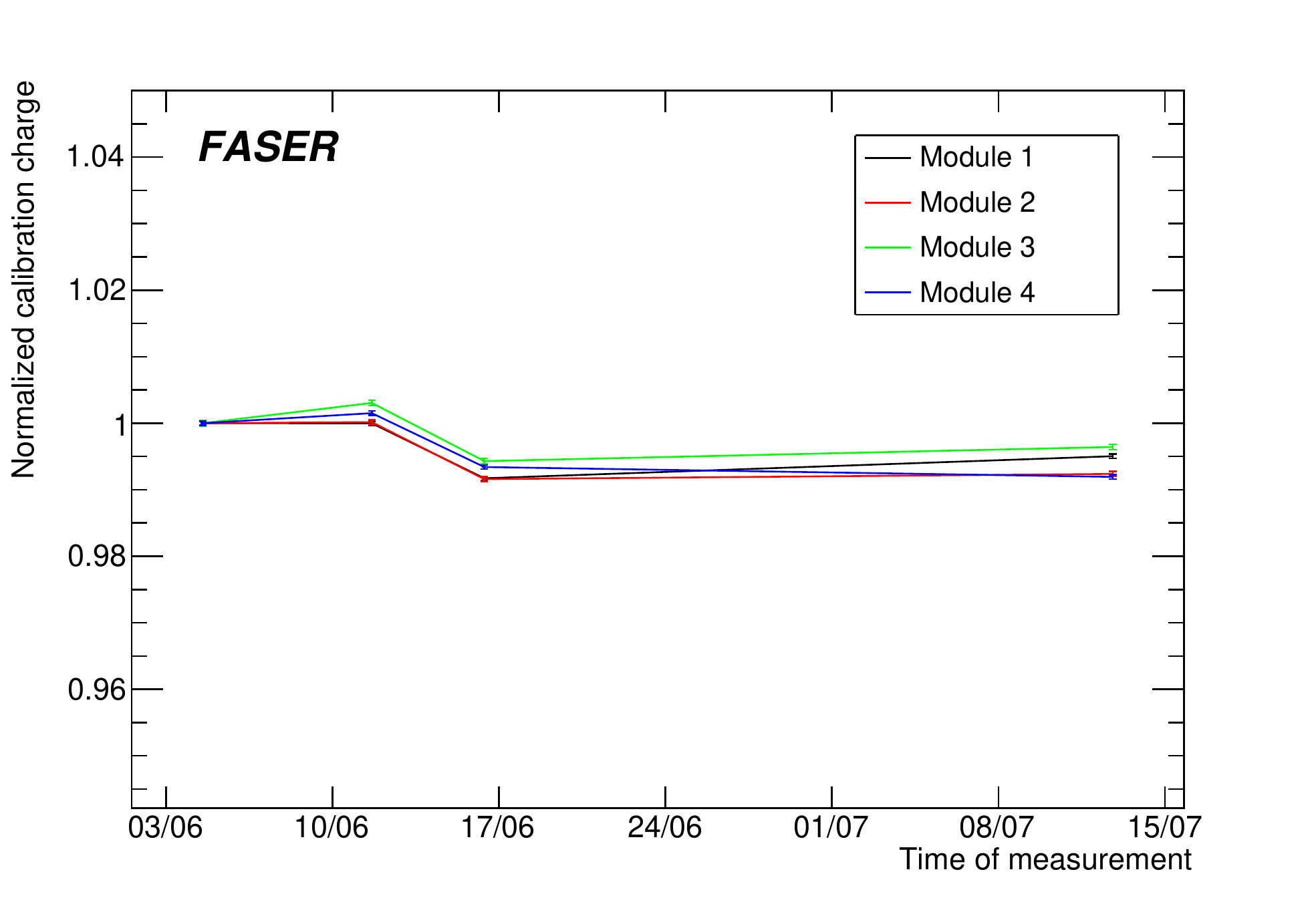}
    \caption{Average charge measured in the four calorimeter modules during
    four separate LED calibration runs with the same settings. The charge has been normalized to 1 for the first run.}
    \label{fig:calibstability}
\end{figure}

\subsubsection{TDAQ system}
The long-term stability of the TDAQ system for the detector in its final location was continuously tested during the combined cosmic-ray data taking. 
The limitations of the TDAQ hardware and software have been probed by running dedicated high-rate tests. Here, noise signals from a signal generator were fed into the digitizer and input trigger rates of 250-5000~Hz were invoked by adjusting the injected signal amplitude. The detector deadtime, buffer occupancy, missing fragment errors and event counts in particular were carefully monitored during high rate tests. Results showed that for rates up to 900 Hz, the deadtime is dominated by the tracker busy signal contributing up to a fraction of 0.7\%. The rate limiter is the dominant source of deadtime from 900~Hz to 4~kHz up to a fraction of 42\%. The first hardware readout limitation is the digitizer which reaches full buffer occupancy at a readout rate of 2.2 kHz, corresponding to 47 MBytes/s. This happens to be the maximum rate of the rate limiter. There are several options to optimise the digitizer configurations that ease this limitation in case rates above 2.2 kHz are met during data taking~\cite{FASERTDAQ:2021}. The FASER DAQ software has been proven to run without signs of limitations during high rate tests. Stress tests have been performed on the DAQ software using emulated trigger and data input at beyond hardware-limited rates.  The main DAQ limitation is the 1 Gbit/s link between the digitizer and the TI12 Ethernet switch due to the large payload size. This link saturates around 5 kHz, but higher rates can be achieved by reducing the readout window for the digitized signals.

\subsection{Commissioning during the LHC pilot beam }
During the last two weeks of October 2021, the LHC carried out a pilot beam test, in which proton beams were circulated in the LHC, and during some periods the beams were brought into collision. The primary purpose of this test was to check the LHC beam pipe aperture, and to test and commission the beam systems after many interventions during the LS2.  However, the pilot beam test provided a valuable opportunity to operate FASER and to take data during beam operations. 

The LHC operated with a maximum of only four circulating bunches per beam, with nominal per-bunch intensity (around $10^{11}$ protons per bunch), and a beam energy of 450 GeV. During periods of collisions, only two of the bunches were colliding at each interaction point, leading to a maximum luminosity of $10^{28}$~cm$^{-2}$s$^{-1}$ (about 4 orders of magnitude lower than the expected luminosity during Run 3 physics operations). 

For the two-week period, the FASER experiment acquired data continuously using both the nominal cosmic triggers, triggering on single signals per scintillator station and calorimeter module, and a coincidence trigger that required a signal in the veto scintillators (positioned at the front of the upstream magnet) and the pre-shower scintillators (positioned after the last tracker station).

Several hundred events were observed containing traversing charged particles that can be traced back to the filled proton bunch interactions in the direction of the IP1 interaction point. 
The LHC circulated two bunches colliding in IP1, and two bunches that were not colliding, and an equal number of events with traversing tracks in FASER were observed for these two sets of bunches, strongly suggesting that the observed signals are arising only from beam backgrounds rather than collision products. This is consistent with the expectation given the low luminosity of these collisions and that the collisions were at low energy ($\sqrt{s}$ = 900~GeV). 
The collected number of events  was sufficient to do an early coarse time adjustment of the trigger and readout signals to have coincident signals from all scintillator counters and calorimeter modules for potential collisions. 

Figure~\ref{fig:event_display_LHCtestbeams} shows an example event containing a charged particle traversing the full detector, triggering each scintillator or calorimeter station and leaving traceable track hits in each tracker station. The scintillator station timings match that of a particle entering the front of the detector, from the direction of IP1.
The arrival signal time as seen in the different PMTs for beam particles can be seen in Figure~\ref{fig:timingspreadbeam} where one should note that the calorimeter PMT signals are expected to be faster than the scintillator PMTs owing to a 23~ns shorter transition time for the calorimeter PMTs.
To adjust for the time-of-flight of particles originating on the LOS upstream of the detector, the veto scintillator and the timing station trigger signals are delayed by $1/2$ a clock cycle, equivalent to $1/2$ a bunch-crossing. 

Events triggering the coincidence trigger occurred in specific BC IDs with a consistent offset to the BC ID of filled proton bunches in the beam approaching FASER from the direction of IP1.

The beam events with a single particle passing through the FASER spectrometer can be used for a first, in situ measurement of the timing resolution of the timing scintillator station. Figure~\ref{fig:timingresolution} shows the time distribution measured in the top and bottom timing scintillator layers. The hit time is calculated as the average time in the two PMTs attached to each layer with respect to the start of the beam bunch crossing. A fixed offset is subtracted. A resolution of just over \SI{400}{ps} is obtained, which is worse than the expected \SI{250}{ps} as discussed in Section~\ref{sec:scintillatormeasurements}, but still well below the requirement for background suppression. 
The overall time resolution of the veto and pre-shower scintillator stations are worse than \SI{400}{ps} due to the intrinsic time-walk. However, selecting events with a reconstructed track pointing to the bottom quarter of the scintillator station, where the time-walk variation is the smallest, a spread of around \SI{250}{ps} is measured. 

\begin{figure}[h]
    \centering
    \includegraphics[trim=0.cm 0.cm 0.cm 0.cm, clip=true, width=\textwidth]{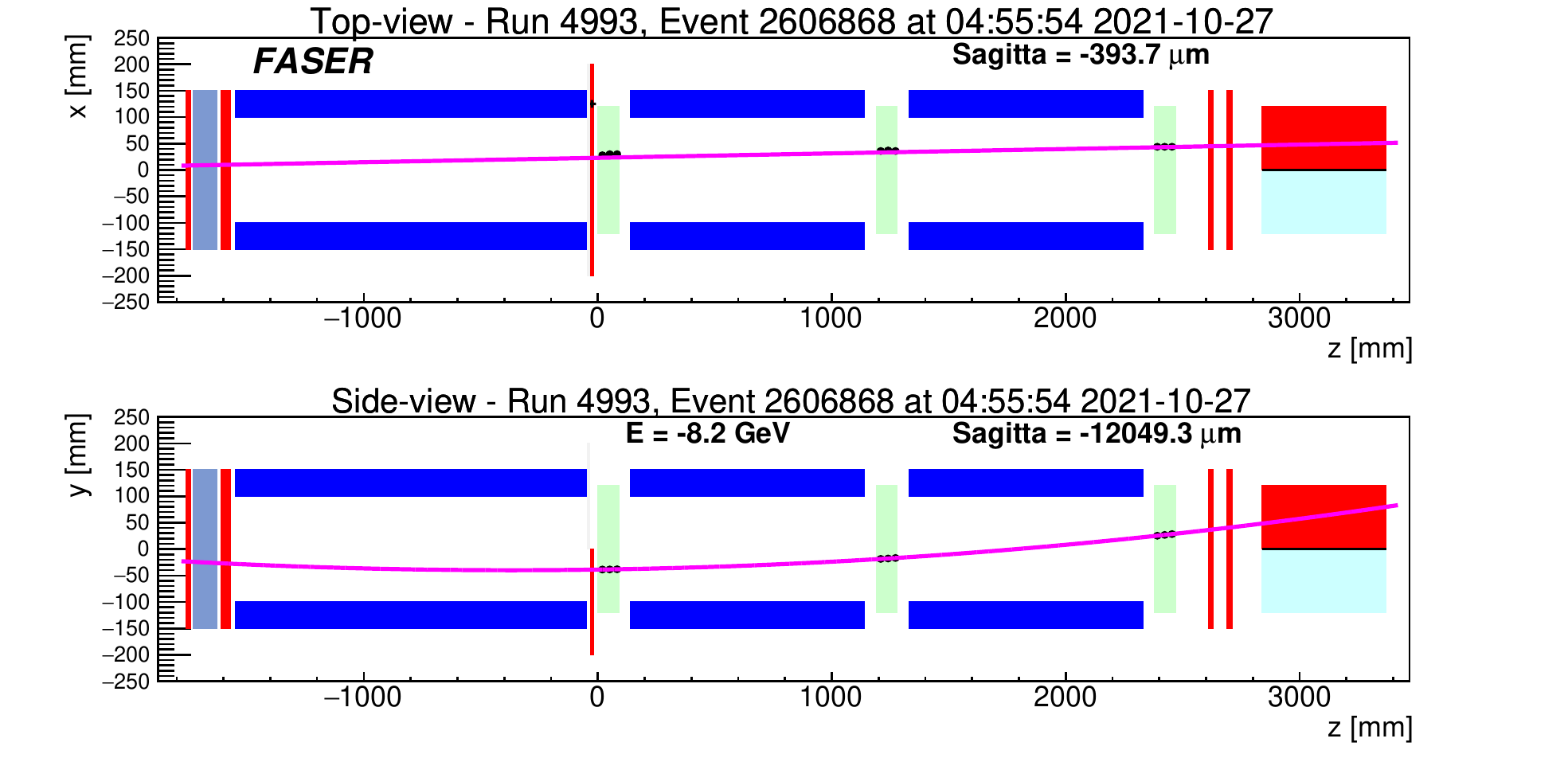}
    \caption{An event display of a charged particle traversing the full detector during an LHC stable beam run with  two 450~GeV colliding beams. The layout of the display and the sub-detector systems are the same as those in Figure~\ref{fig:event_display_cosmic}. The purple line is a combined track fit to the hits in the tracking stations.}
    \label{fig:event_display_LHCtestbeams}
\end{figure}

\begin{figure}[h]
    \centering
    \includegraphics[width=0.65\textwidth]{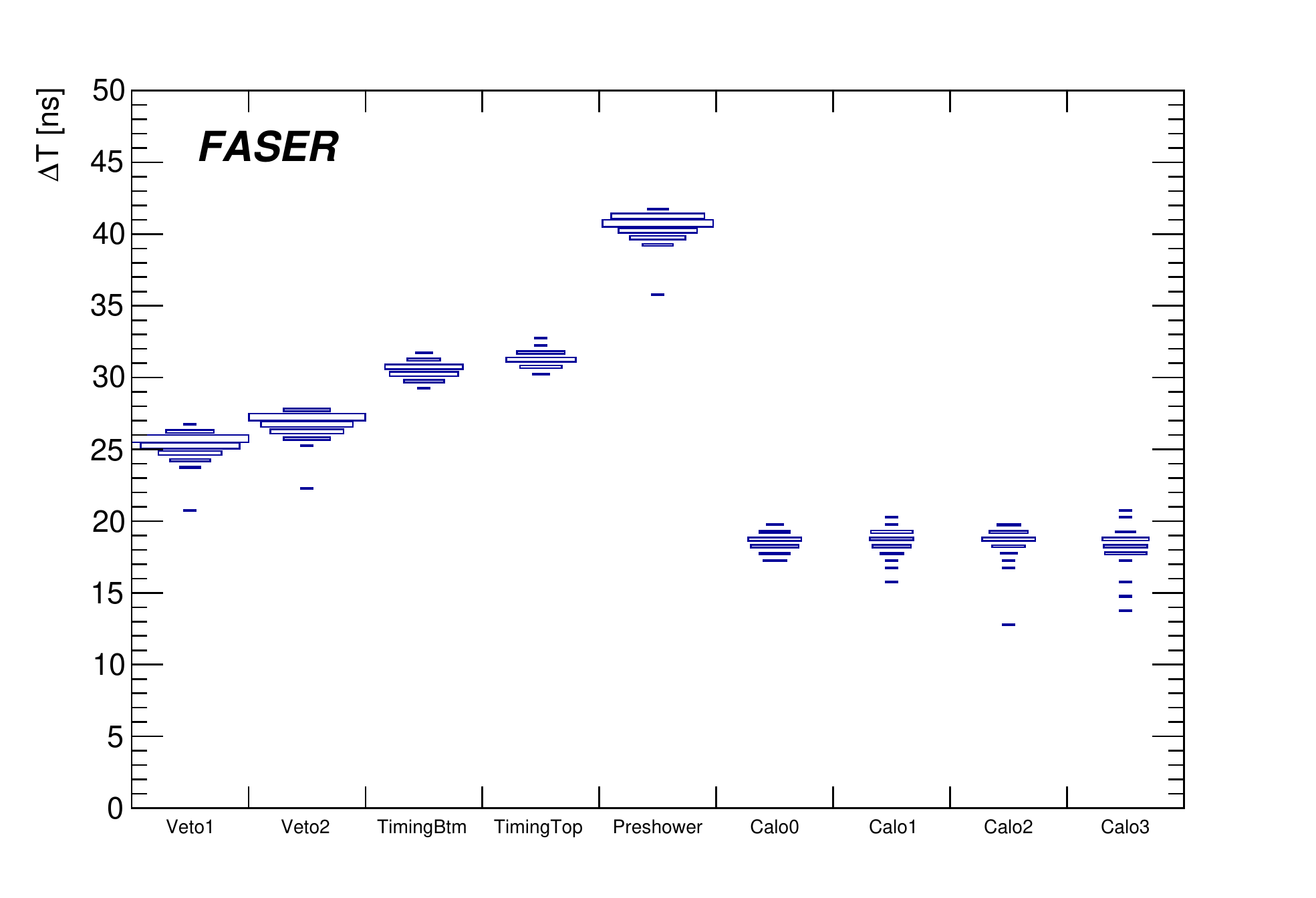}
    \caption{The signal arrival time measured for different scintillator and calorimeter stations/modules with respect to the start of the bunch crossing for single muon tracks recorded during the 2021 pilot beam test.}
    \label{fig:timingspreadbeam}
\end{figure}

\begin{figure}[h]
    \centering
    \includegraphics[width=0.45\textwidth]{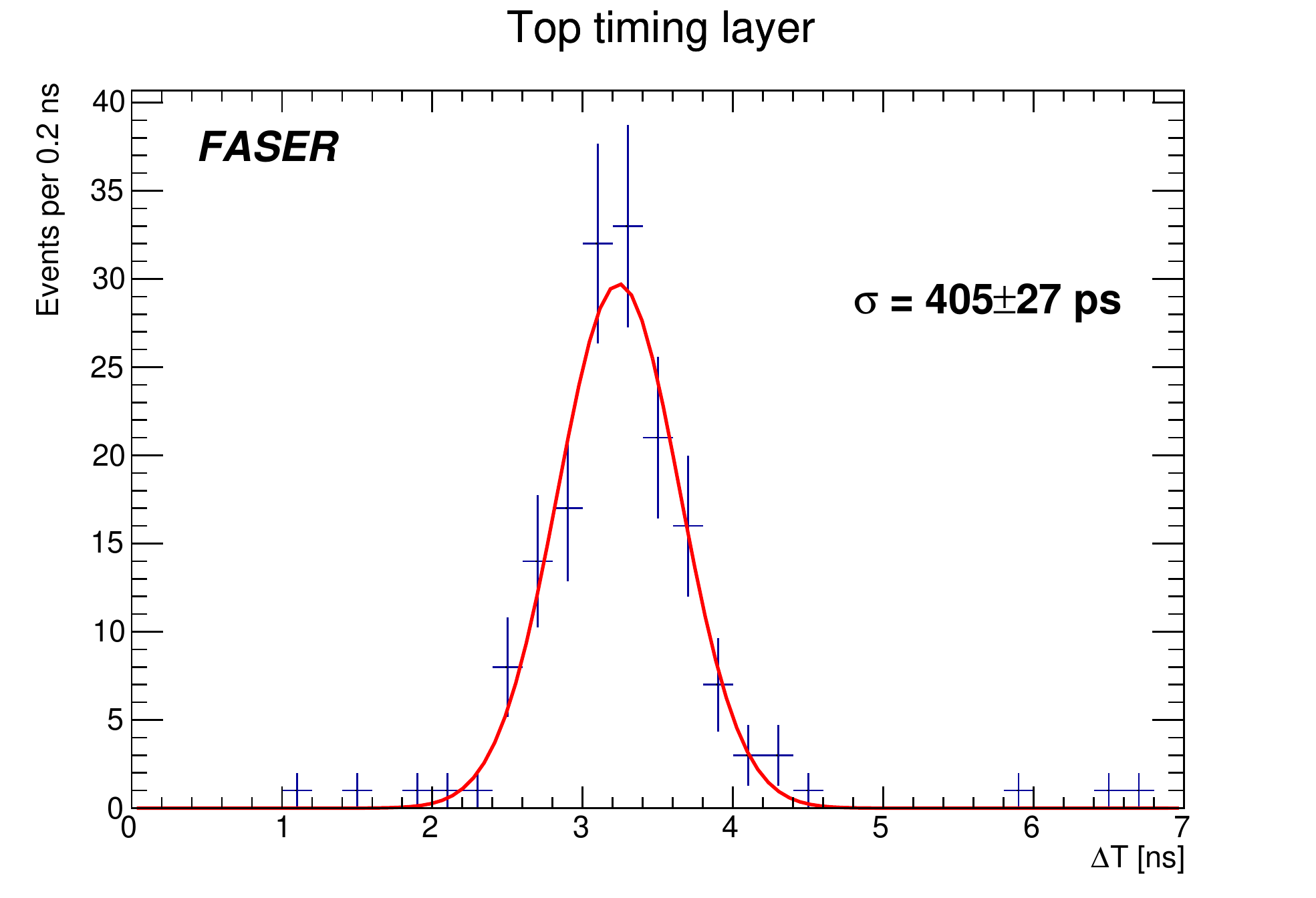}
    \includegraphics[width=0.45\textwidth]{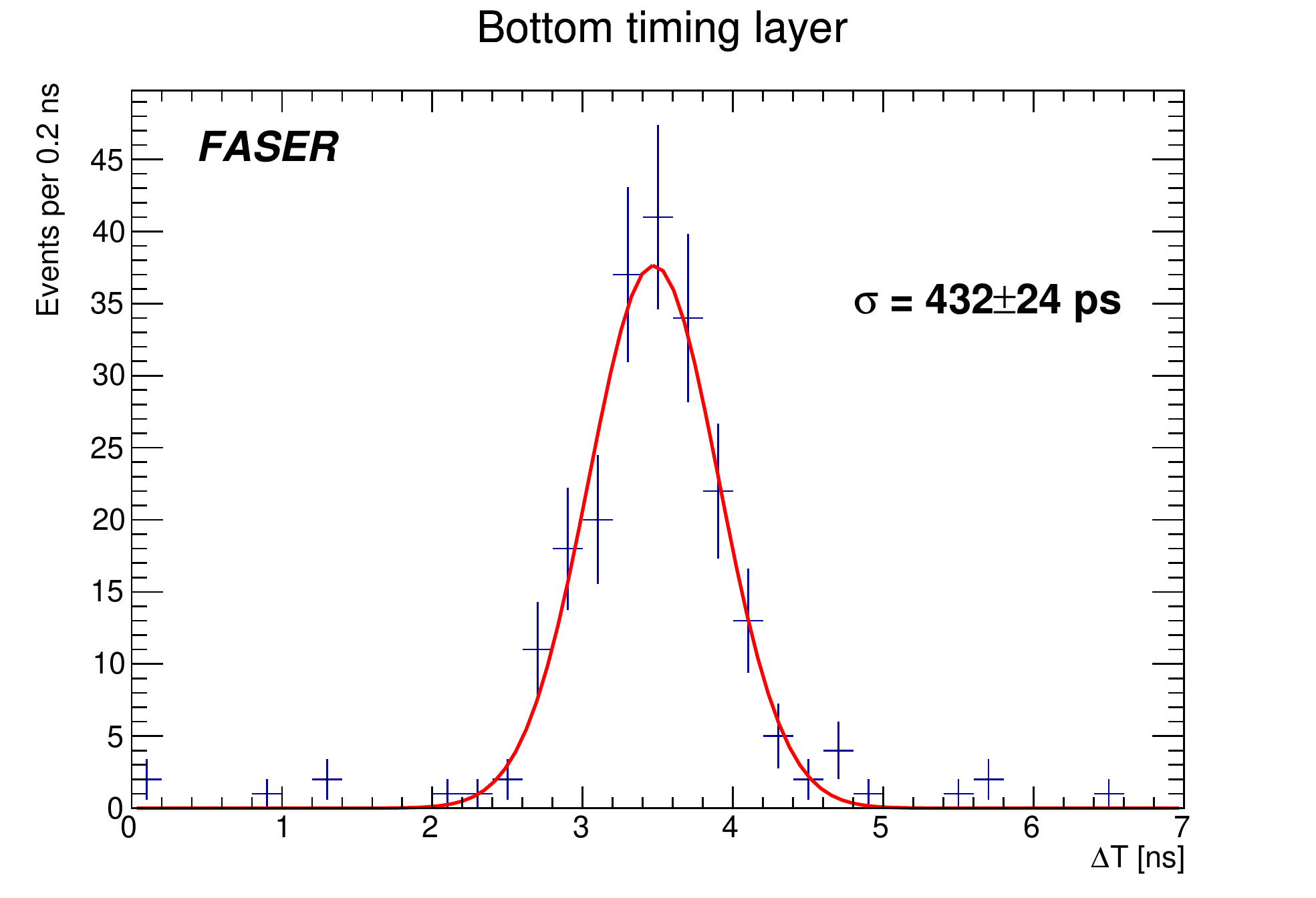}
    \caption{Average signal arrival time for the two PMTs in the top (left) and bottom (right) timing scintillator layer with respect to the start of the bunch crossing for single muon tracks recorded during the 2021 pilot beam test. The data are fitted with a Gaussian function to extract the timing resolution values. }
    \label{fig:timingresolution}
\end{figure}

%%%%%%%%%%%%%%%%%%%%%%%%%%%%%%%
\clearpage
\section{Conclusions}
\label{sec:conclusions}
Following CERN approval of the FASER experiment in March 2019, the collaboration proceeded to design, build and commission its sub-detector components and the needed infrastructure before installation of the full detector in the TI12 tunnel of the LHC complex in March 2021.  This paper presents an overview of the sub-components, including a tracking spectrometer complemented by the Interface Tracker; a decay volume; a set of scintillator systems to veto events, provide timing or act as a pre-shower; an electromagnetic calorimeter system; and the FASER emulsion detector dedicated to the study of neutrinos produced in LHC collisions via hadron decays.  The decay volume and the tracking spectrometer are immersed in a 0.57 T dipole magnetic field, also detailed in the paper.  The FASER detector is triggered by signals in any of the scintillator stations or the calorimeter. The trigger and data acquisition system are also discussed. 
Similarly, details have been provided on the detector control system of the experiment that  ensures safe operation.  Emphasis has been given to the description of the detector integration phase, including the preparation of the TI12 tunnel, the design and manufacturing of the detector support, the installation of the experiment system and its commissioning using cosmic-rays collected between April and September 2021 and during the LHC pilot beam test carried out in October 2021. 

The installed detector is fully operational and shows excellent performance. For the tracker, the number of non-operational channels is less than 0.5\%, and the measured gain and noise from calibration runs are well within the specifications. The individual scintillator counter efficiencies, measured with cosmic ray muons and during testbeam, are higher than 99.9\% and within the experiment's requirements. The electromagnetic calorimeter system has been studied using cosmic ray muons and pilot beam data, and shows good performance, within the experiments specifications. The trigger and data acquisition system has been fully tested, and 
demonstrated to operate with no more than 5\% deadtime up to almost 2~kHz trigger rate, well above the expected rate of 650~Hz, meeting target expectations. The full system has been operated for several months cosmic rays data-taking in TI12, and as part of the LHC pilot beam test in October 2021, accumulating valuable operational experience and data for performance studies.  

The FASER detector is now ready for physics data-taking in proton-proton collisions from the start of LHC Run 3 operations in 2022. It will offer exciting opportunities to complement the LHC's ongoing physics programme on searches for light, long-lived new particles produced in the far-forward region, and to measure neutrino interaction charged-current cross sections for all three neutrino flavours in a previously uncovered energy regime.

\section*{Acknowledgements}
We thank the technical and administrative staff members at all FASER institutions, including CERN, for their contributions to the success of the FASER project. 

Initial studies for FASER were strongly supported by the CERN Physics Beyond Colliders study group, and many groups at CERN, including FLUKA studies (SY-STI), radiation and background particle flux measurements (BE-CEM, SY-BI), civil engineering studies (SCE-DOD), integration studies (EN-ACE, BE-EA), radioprotection studies (HSE-RP), and safety discussions (HSE-OHS, EP-SO). For the FASER site preparation and installation of the services and the detector we acknowledge invaluable support from the following CERN teams: transport and handling (EN-HE), installation of services (EN-EL, EN-CV), cabling (BE-EA, EP-DT), and survey (BE-GM). We acknowledge support from the CERN magnet group (TE-MSC) for the design, construction and measurements of the FASER magnets.

We are extremely grateful to the ATLAS SCT Collaboration for donating spare SCT modules to FASER and to the LHCb Collaboration for the loan of the calorimeter modules. We thank Steve Wotton and Floris Keizer for providing their single module readout system, which enabled us to evaluate the spare SCT modules, and Iouri Guz for help with the initial testing of the calorimeter modules.

The FASER collaboration is grateful to the CERN EP-DT team for their contributions and assistance to the DAQ software developments, to the CERN SY-BI group for their support in setting up the BOBR card, to the CERN IT-CS group for their assistance with the networking, and to the ATLAS TDAQ Collaboration for the loan of old HLT servers which have been used by FASER. 

FASER gratefully acknowledges the use of the publicly available ATLAS offline software as part of the FASER software suite.

This work is supported in part by Heising-Simons Foundation Grant Nos. 2018-1135, 2019-1179,  2020-1840, Simons Foundation Grant No. 623683, U.S. National Science Foundation Grant Nos. PHY-2111427, PHY-2110929, and PHY-2110648, JSPS KAKENHI Grants Nos. JP19H01909, JP20K23373, JP20H01919, JP20K04004, and JP21H00082, and by the Swiss National Science Foundation.

\bibliography{FASER_DetectorPaper}

\end{document}